\documentclass[12pt]{article}
\usepackage{amsmath}
\usepackage{graphicx}
\usepackage{enumerate}
\usepackage[numbers]{natbib}
\usepackage{url} 
\usepackage{amsthm,amsmath,amsfonts,amssymb}

\newcommand{\blind}{1}

\numberwithin{equation}{section}
\theoremstyle{plain}
\newtheorem{thm}{Theorem}[section] 
\newtheorem{lemma}{Lemma}[section] 
\newtheorem{cor}{Corollary}[section]

\newtheorem{definition}{Definition}[section]
\newtheorem{defi}{Definition}[section]
\newcommand{\bed}{\begin{defi}}
\newcommand{\eed}{\end{defi}}

\newcommand{\eps}{\epsilon}

\newcommand{\bitem}{\begin{itemize}}
\newcommand{\eitem}{\end{itemize}}

\newcommand{\goto}{\rightarrow}

\newcommand{\beqn}{\begin{equation}}
\newcommand{\eeqn}{\end{equation}}
\newcommand{\balign}{\begin{align}}
\newcommand{\ealign}{\end{align}}

\newcommand{\tr}{\mathrm{tr}}
\usepackage{amssymb}
\newcommand{\beq}{\begin{equation}}
\newcommand{\eeq}{\end{equation}}

\newcommand{\diag}{\mathrm{diag}}

\usepackage{hyperref}
\hypersetup{
    colorlinks=true,
    linkcolor=blue,
    filecolor=magenta,      
    urlcolor=cyan,
}
\allowdisplaybreaks
\usepackage{xr}
\begin{document}

\def\spacingset#1{\renewcommand{\baselinestretch}%
{#1}\small\normalsize} \spacingset{1}


\if1\blind
{
  \title{\bf Optimal Estimation of the Number of  Communities}
  \author{Jiashun Jin\thanks{
    JJ and SL gratefully acknowledge the support of the NSF grant \textit{DMS-2015469}. ZK gratefully acknowledges the support of the NSF CAREER grant \textit{DMS-1943902}.}\hspace{.2cm}\\
    Department of Statistics, Carnegie Mellon University\\
    and \\
    Zheng Tracy Ke \\
    Department of Statistics, Harvard University\\
    and\\
    Shengming Luo\\
    Department of Statistics, Carnegie Mellon University\\
    and \\
    Minzhe Wang\\
    Department of Statistics, University of Chicago}
  \maketitle
} \fi

\if0\blind
{
  \bigskip
  \bigskip
  \bigskip
  \begin{center}
    {\LARGE\bf Optimal Estimation of the Number of  Communities}
\end{center}
  \medskip
} \fi

\bigskip
\begin{abstract}
In network analysis, how to estimate the number of communities $K$ is a 
fundamental problem.   
We consider a broad  setting where we allow severe degree heterogeneity and a wide range of sparsity levels,  
and propose  Stepwise Goodness-of-Fit (StGoF) as a new approach.   
This is a stepwise algorithm, where for  $m = 1, 2, \ldots$, we alternately use a community detection step  and a goodness-of-fit (GoF) step.  
We adapt SCORE \cite{SCORE} for community detection, and propose a new GoF metric. 
We show that at step $m$, the GoF metric diverges to $\infty$ in probability for all $m < K$ and 
converges to  $N(0,1)$ if $m = K$.  This gives rise to a consistent estimate for $K$.  
Also,  we discover the right way to define the signal-to-noise ratio (SNR) for our problem  and 
show that  consistent estimates  
for $K$ do not exist if $\mathrm{SNR} \goto 0$,  and StGoF is uniformly consistent for $K$ if $\mathrm{SNR} \goto \infty$.  Therefore, StGoF achieves the optimal phase transition.  

Similar stepwise methods (e.g., \cite{wang2017likelihood, ma2018determining}) are known to face analytical challenges.  
We overcome the challenges by using a different stepwise scheme in StGoF and by deriving 
sharp results that are not available before. 
The key to our analysis is to show that SCORE has the  {\it Non-Splitting Property (NSP)}.   
Primarily due to a non-tractable rotation of eigenvectors dictated by the Davis-Kahan $\sin(\theta)$ theorem, 
the NSP is non-trivial to prove and requires new techniques we develop. 
\end{abstract}

\noindent%

\spacingset{1.465} 

\section{Introduction} \label{sec:intro}   
Suppose $A$ is the adjacency matrix for a symmetric and connected network with $n$ nodes: 
\begin{equation} \label{model1a} 
A_{ij} = 
\left\{
\begin{array}{ll} 
1, &\qquad  \mbox{if node $i$ and node $j$ have an edge},  \\
0, &\qquad  \mbox{otherwise},    \\
\end{array}
\right. \qquad 1 \leq i\neq j \leq n. 
\end{equation} 
As a convention,   self-edges are not allowed  so all the diagonal entries of $A$ are $0$.  As usual, we assume the 
network has $K$  (unknown) communities  $\mathcal{N}_1$, $\mathcal{N}_2$, $\ldots, \mathcal{N}_K$.  
 Similar to that of a cluster in multivariate analysis, the precise meaning a community is hard to formalize,  
but frequently and intuitively, communities in a network are groups of nodes that have more edges within than between (e.g., \cite{zhao2012consistency}).

Our primary goal is to estimate $K$.  This is a fundamental problem in network analysis: 
In many recent approaches, 
$K$ is assumed as known a priori (e.g.,  \cite{zhang2020modularity,yuan2018community,huang2020spectral} on community detection,  \cite{Mixed-SCORE,fan2019simple} on mixed-membership estimation, \cite{choi,jiang2020autoregressive,zhang2020mixed} on dynamic networks, and \cite{zhu2019portal,chen2020community,huang2021feature} on network regression analysis).  Unfortunately, $K$ is rarely known in applications,  so the performance of these approaches hinges on how well we can estimate $K$.

Real world networks have several noteworthy features.   
First, a network may have severe degree heterogeneity. Take the Polblog network in Table 1 for example. 
The maximum degree is $351$ and the minimum degree is $1$. Second, the network sparsity (e.g., measured by the average degree) may range significantly from one network to another. Last, frequently, the desired community structure is masked by strong noise, and  
the signal-to-noise ratio (SNR) is usually relatively small. 
Motivated by these features, we adopt the widely-used degree-corrected block model (DCBM) \cite{DCBM}. 
Recall that the network has $K$ communities $\mathcal{N}_1$, $\mathcal{N}_2$, $\ldots, \mathcal{N}_K$. 
For each $1 \leq i \leq n$, we encode the community label of node $i$ by a vector $\pi_i  \in \mathbb{R}^K$ where for $ i \in \mathcal{N}_k$, $\pi_i(k) = 1$ and $\pi_i(m) = 0$ for $m\neq k$.    
Moreover, for a $K \times K$ symmetric nonnegative matrix $P$ which models the community structure and positive parameters $\theta_1, \theta_2, \ldots, \theta_n$ which model  the degree heterogeneity,  
we assume the upper triangular entries of $A$ are independent Bernoulli variables satisfying    
\begin{equation} \label{model1c} 
\mathbb{P}(A_{ij}=1) = \theta_i\theta_j\cdot \pi_i'P\pi_j \equiv \Omega_{ij},   \qquad 1 \leq i < j \leq n,   
\end{equation} 
where $\Omega$ denotes the matrix $\Theta \Pi P \Pi' \Theta$, with $\Theta$ being the $n \times n$ diagonal matrix $\diag(\theta_1, \ldots, \theta_n)$ and $\Pi$ being the $n \times K$ matrix $[\pi_1, \pi_2, \ldots, \pi_n]'$. For identifiability, we assume 
\begin{equation} \label{model1d} 
\mbox{$P$ is non-singular and all diagonal entries of $P$ are $1$}.   
\end{equation} 
Write for short $\diag(\Omega) = \diag(\Omega_{11}, \Omega_{22}, \ldots, \Omega_{nn})$, and let $W$ be the matrix where 
for $1 \leq i, j \leq n$,  $W_{ij} = A_{ij} -  \Omega_{ij}$ if $i \neq j$ and $W_{ij} = 0$ otherwise. In matrix form, we have    
\begin{equation} \label{DCBM-matrixform} 
A = \Omega - \diag(\Omega) + W,   \qquad \mbox{where we recall} \;\;  \Omega = \Theta \Pi P \Pi' \Theta. 
\end{equation}  
When $\theta_1 = \theta_2 = \ldots = \theta_n$, DCBM reduces to the stochastic block model (SBM).  

We let $n$ be the driving asymptotic parameter, and  allow $(\Theta, \Pi, P)$ to depend on $n$, so DCBM is broad enough 
to capture the three features aforementioned.  In detail, let $\theta = (\theta_1,\theta_2, \ldots, \theta_n)'$, $\theta_{max} = \max\{\theta_1, \ldots, \theta_n\}$, 
and $\theta_{min} = \min\{\theta_1,  \ldots, \theta_n\}$.  First, a reasonable metric for the degree heterogeneity is 
$\theta_{max} / \theta_{min}$, so to allow severe degree heterogeneity, we prefer not to put an artificial upper bound on 
$\theta_{max} / \theta_{min}$.
Second, a reasonable metric for network sparsity is $\|\theta\|$ (e.g., see \cite{SP2019, SCORE}).  \spacingset{1}\footnote{An appropriate measure for sparsity is $\|\Omega\|$ (e.g.,\cite{SP2019}). In (1.3), we assume all diagonal entries of $P$ are $1$, 
so if $K$ is finite and some regularity conditions hold, $\|\Omega\| \asymp \|\theta\|^2$. Also, $d_i$ (degree of node $i$) is at the order of $\theta_i \|\theta\|_1$, which is 
$O(n \theta_i^2)$ if all $\theta_i$ are at the same order. Therefore, the range of interest for $\theta_i$ is  
between $1/\sqrt{n}$ and $1$, up to some logarithmic factors (e.g., $\log(n)$).}   \spacingset{1.465}
To cover all sparsity levels of interest,  and especially the very sparse case (e.g., $\theta_i  = O(\sqrt{\log(n)  / n})$ for all $1 \leq i \leq n$) and the very dense case (e.g., $\theta_1 = O(1)$ for all $1 \leq i \leq n$),  we assume ($C > 0$ is a constant)  
\begin{equation} \label{thetarange} 
C \sqrt{\log(n)} \leq \|\theta\| \leq C \sqrt{n}. 
\end{equation}  
Last, let $\lambda_1,\lambda_2,\ldots,\lambda_K$ be the $K$ nonzero eigenvalues of $\Omega$, arranged in the descending order of magnitudes. We will soon see that the signal strength and noise level in our setting are captured by $|\lambda_K|$ and $\|W\|$, respectively, where 
under mild conditions, 
\begin{equation} \label{SNR1} 
 \|W\| =  \mbox{a multi-$\log(n)$ term  $\cdot \sqrt{\lambda}_1$ with high probability, where $\lambda_1 \asymp \|\theta\|^2$}. 
\end{equation} 
Therefore, a reasonable metric for the signal to noise ratio (SNR) is $|\lambda_K| / \sqrt{\lambda_1}$ (see Section \ref{sec:main} for more discussion).   We consider two extreme cases (assuming $n \goto \infty$).
\begin{itemize}
\itemsep0em 
\item {\it Strong signal case}. $|\lambda_1|, |\lambda_2|, \ldots, |\lambda_K|$ are at the same magnitude, and so SNR $\asymp \sqrt{\lambda_1}$.  
\item {\it Weak signal case}. $|\lambda_K| / \sqrt{\lambda_1}$ is much smaller than $\sqrt{\lambda_1}$ and 
grows to $\infty$ slowly.   
\end{itemize}   
For example, in a weak signal case, we may have $\lambda_1 = O(\sqrt{n})$ and $\mbox{SNR} = \log\log(n)$ and $\lambda_1 = \sqrt{n}$.  Section \ref{subsec:LB} suggests that when $\mbox{SNR} = o(1)$,  consistent estimate for $K$ does not exist, so the weak signal case is   very challenging.  
Motivated by the above observations, it is desirable to find a consistent estimate for $K$ that satisfies the following requirements.   
\begin{itemize}  
\itemsep0em
\item (R1) Allow severe degree heterogeneity (i.e., no  artificial bound on $\theta_{max}/\theta_{min}$).     
\item (R2) Optimally adaptive to all sparsity levels of interest (e.g., see (\ref{thetarange})).  
\item (R3) {\it Attain the information lower bound}. Consistent for both the strong signal case  
where SNR is large and the weak signal case where SNR may be as small as $\log\log(n)$.     
\end{itemize}

{\bf Example 1}.  
A frequently considered DCBM  is to assume $P=P_0$ and $\theta_i \asymp \sqrt{\alpha_n}$ for all $1 \leq i \leq n$, where $\alpha_n > 0$ is a scaling parameter and $P_0$ is a fixed matrix.  
It is seen that $\lambda_1, \ldots, \lambda_K$ are at the same order, so the model only considers the strong signal case.

{\bf Example 2}.    Let $e_1,\ldots,e_K$ be  the standard basis vectors   
of $\mathbb{R}^K$.  Fix a positive vector $\theta\in\mathbb{R}^n$ and $b_n\in (0,1)$.   
Consider a DCBM where  each community has $n/K$ nodes, and 
$P = (1 - b_n) I_K + b_n 1_K 1_K'$.  Here, $(1 - b_n)$ measures the ``dis-similarity" of  different communities.  
By basic algebra,  $\lambda_1  \asymp \|\theta\|^2$,  $\lambda_2 = ... = \lambda_K \asymp \|\theta\|^2 (1 - b_n)$,  and $\mbox{SNR}  \asymp \|\theta\| (1 - b_n)$; moreover,   
 $\|\theta\| = O(\sqrt{\log(n)})$ in the very sparse case, and $\|\theta\| = O(\sqrt{n})$ in the dense case. 
When $b_n \leq c_0$ for a constant $c_0 < 1$,  $|\lambda_K| \geq C|\lambda_1|$ and $\mbox{SNR} \asymp \|\theta\|$; we are in the strong signal case if  $\|\theta\| \geq n^a$ for a constant $a > 0$.  When $b_n=1+o(1)$ and $\|\theta\| (1-b_n) = \log\log(n)$ (say), $\mbox{SNR} \asymp \log\log(n)$ and we are in the weak signal case.

{\bf Example 3}.  An SBM can be identifiable even if $P$ is singular (e.g., \cite{Priebe}).  However, 
a DCBM can be non-identifiable if $P$ is singular. For example, 
consider an SBM with parameters $(\widetilde{\Pi}, \widetilde{P})$ where $\widetilde{P} \in \mathbb{R}^{2,2}$,  $\widetilde{P}_{11} =a$, $\widetilde{P}_{22} = c$,  
$\widetilde{P}_{12} = \widetilde{P}_{21} = b$,   and $ac = b^2$  (so the rank of $\widetilde{P}$ is 1). The model is an identifiable SBM with two communities. 
But if we treat it with a DCBM with parameters $(K, \Theta, \Pi, P)$, then we can either take $(K, \Theta, \Pi, P)  = (2, I_n, \widetilde{\Pi}, \widetilde{P})$, or take $(K, \Theta, \Pi, P)  = (1, \Theta, \widetilde{\Pi}, 1)$, so 
it is not identifiable. Here 
$\Theta = \diag(\theta_1, \ldots, \theta_n)$ and $\theta_i = \sqrt{a}$ if $i$ is in community $1$ and $\theta_i = \sqrt{c}$ if $i$ is in community $2$, $1 \leq i \leq n$.

\vspace{-1em}
\subsection{Literature review and our contributions} \label{subsec:review} 
Exiting approaches  for estimating $K$  can 
be roughly divided into the spectral approaches, 
  cross validation approaches,  penalization approaches, and   likelihood ratio approaches.

For spectral approaches,  Le and Levina \cite{le2015estimating} proposed to estimate $K$ using the  eigenvalues of the non-backtracking matrix or Bethe Hessian matrix. The approach uses interesting ideas from 
  graph theory, but unfortunately, it requires relatively strong conditions for consistency. For example,  their Theorem 4.1 only considers  the   very sparse SBM model where $\theta_1 = \theta_2 = \ldots = \theta_n = 1/\sqrt{n}$ and $P = P_0$ for a fixed matrix $P_0$.  Liu {\it et al}. \cite{liu2019community} proposed to estimate $K$ using the scree plot with careful theoretical justification, but the approach is  unsatisfactory for networks with severe degree heterogeneity, for it is hard to derive a sharp bound for the spectral norm of the noise matrix $W$ (e.g., \cite{SCORE}). Therefore,  their approach requires the condition of $\theta_{max} \leq C \theta_{min}$. 
The paper also assumed $\|\theta\| = O(\sqrt{n})$ so it did  not address the settings of sparse networks (e.g., see (\ref{thetarange})).  
For cross-validation approaches, we have \cite{chen2018network, li2020network}, and among the penalization approaches, we have \cite{Salda_a_2017, daudin2008mixture,latouche2012variational}, where  $K$ is estimated by the integer that optimizes some objective functions.  For example, Salda {\it et al}.  \cite{Salda_a_2017}  used a BIC-type objective function and \cite{daudin2008mixture,latouche2012variational} used an objective function of the Bayesian model selection flavor.  However, these methods did not provide  explicit theoretical guarantee on consistency (though a partial result was established in \citep{li2020network}, which stated that under SBM, the proposed estimator $\widehat{K}$ is no greater than $K$ with high probability).

For likelihood ratio approaches,  Wang and Bickel \cite{wang2017likelihood}  
proposed to estimate  $K$ by solving a BIC type optimization problem, where the objective function is the sum of log-likelihood  and model complexity.  The major challenge  
is  that the likelihood is the sum of exponentially many terms and is hard to compute.  In a remarkable paper, Ma {\it et al}.  \cite{ma2018determining}  extended the idea of \cite{wang2017likelihood} by proposing a new  
approach that is computationally more feasible.  

On a high level, we can recast their methods as a  {\it stepwise testing or sequential testing} algorithm. 
Consider a stepwise testing scheme where for $m = 1, 2, \ldots$, they construct a test statistic $\ell_n^{(m)}$ (e.g.  $\log$-likelihood) assuming  $m$ is the correct number of communities. 
They estimate $K$ as the smallest $m$ 
such that the pairwise $\log$-likelihood ratio $(\ell_n^{(m+1)} - \ell_n^{(m)})$ falls below a {\it threshold}.   Call  
$m < K$, $m = K$, and $m > K$ the 
{\it under-fitting}, {\it null}, and {\it over-fitting} cases, respectively.  As mentioned in \cite{wang2017likelihood, ma2018determining}, 
such an approach faces a two-fold challenge. First, one has to analyze $\ell_n^{(m)}$ for both the under-fitting case and the over-fitting case, but 
there are no efficient technical tools to address either case.  
Second, it is hard to derive sharp results on the limiting distribution of $\ell_n^{(m+1)} - \ell_n^{(m)}$ in the null case,  and so 
it is unclear how to pin down the threshold. 
Ma {\it et al}. \cite{ma2018determining} (see also \cite{wang2017likelihood}) made interesting progress  but unfortunately the problems are not resolved satisfactorily.   
For example,  they require hard-to-check conditions on both the under-fitting and over-fitting cases.  
Also,   
it is unclear whether their results are sharp in the over-fitting case    
and  how to standardize $\ell_n^{(m+1)} - \ell_n^{(m)}$ in the under-fitting case as the variance term is unknown (so it  unclear how to  pin down the threshold).   
Most importantly,  both papers focus on the  
setting in Example 1 (see above), where  severe 
degree heterogeneity is not allowed and they only consider the strong signal case.  
 
We propose {\it Stepwise Goodness-of-Fit (StGoF)} as a new approach to estimating $K$.  
Our idea follows a different vein, and is different both in the 
 statistics we developed and in the stepwise scheme we use.  
In detail, for $m = 1, 2, \ldots$, StGoF alternately uses a community detection 
sub-step (where we apply SCORE \cite{SCORE} assuming $m$ is the correct number of communities) and a Goodness-of-Fit (GoF) sub-step. We propose a new GoF approach and let $\psi_n^{(m)}$ be 
the GoF test statistic in step $m$.   
Assuming $\mbox{SNR}  \goto \infty$, we show that 
\begin{equation} \label{intro-null}
\psi_n^{(m)}  
\left\{  
\begin{array}{ll}
\goto  N(0,1),   &\qquad \mbox{when $m = K$ (null case)},  \\
\goto  \infty \;  \mbox{in probability},   &\qquad \mbox{when $1 \leq m < K$ (under-fitting case)}.  
\end{array}  
\right. 
\end{equation} 
For a properly chosen threshold $t$, define the StGoF estimate by $\widehat{K} = \min_{m} \{ \psi_n^{(m)} \leq t\}$. 
By (\ref{intro-null}), $\widehat{K}$ is consistent.  Now,  first, (\ref{intro-null}) shows that $N(0,1)$ is the limiting null.   
Such an explicit limiting null is crucial in pinning down the threshold $t$.  Second, a noteworthy advantage of  StGoF  is that,  
we do not need to analyze the over-fitting case to prove  the consistency of $\widehat{K}$. 
In comparison, if we follow the approaches by \cite{ma2018determining, wang2017likelihood} and 
similarly define $\widehat{K}$ by $\min_{m} \{\ell_n^{(m+1)} - \ell_n^{(m)} \leq t\}$, then we have to derive the limiting distribution 
for $\ell_n^{(m+1)} - \ell_n^{(m)}$ with $m = K$, which is an over-fitting case.  
In this case, how to derive tight bounds is an open problem (even if the limiting distribution of $\ell_n^{(m+1)} - \ell_n^{(m)}$ can be derived theoretically,  it contains unknown parameters, so it is hard to pin down the threshold $t$).
For these reasons, it is unclear how to derive sharp results with these approaches.  

Fortunately, sharp results are possible if we use the StGoF approach. 
In Section \ref{subsec:LB}, we show that when $\mbox{SNR} \goto 0$,  consistent estimates  for $K$ do not exist. Therefore, our consistency result above is sharp in  terms of the rate of $\mbox{SNR}$,  so StGoF achieves the optimal phase transition, in a broad setting (where 
we allow degree heterogeneity, flexible sparsity levels, and weak signals).  The phase transition is a well-known optimality framework. It is related to the minimax framework but can be frequently more informative  \cite{DJ04}.  

Compared with the approaches in \cite{ma2018determining, wang2017likelihood}, 
(a) they focused on more restricted settings, with either strong signals, 
or strong eigen-gap conditions, or the more specific SBM model,    (b) they did not have an explicit limiting null, 
and (c) they have to analyze the over-fitting case but it remains an open problem to derive sharp bounds. 
For these reasons,    
it is unclear whether they are able to achieve the optimal phase transition.

To prove (\ref{intro-null}), the key is to show that when $m \leq K$,  SCORE has the so-called {\it Non-Splitting Property (NSP)},  meaning that with high probability all nodes in each (true) community are always clustered together.    The proof of NSP is non-trivial. It depends on the row-wise distances of the matrix $\Xi$ consisting of the first $m$ columns of $[\xi_1, \ldots, \xi_K] \Gamma$, where $\xi_k$ is the $k$-th eigenvector of $\Omega$ and $\Gamma$ is an orthogonal matrix dictated by the Davis-Kahan $\sin(\theta)$ theorem \cite{sin-theta}. $\Gamma$ is data dependent and hard to track, and when it  ranges, the row-wise distances of $\Xi$ are the same if $m = K$ but may vary significantly 
if $m < K$. This is why SCORE is much harder to study in the under-fitting case than in the null case. 
To overcome the challenge, we need new and non-trivial proof ideas; see Section~\ref{sec:NSP}.  

While our paper uses SCORE, it is very different from \cite{SCORE}.  The goal of \cite{SCORE} is community detection where $K$ is  known, focusing on the null case ($m = K$).  Here, the goal  is to estimate $K$: SCORE is only used as part of our stepwise algorithm, and the focus is on the under-fitting case $(m < K)$,  where the property of SCORE  is largely unknown,  and our results on the NSP of SCORE are new.   
Our contributions are two fold. First, we propose StGoF as a new approach to estimating $K$.  We show that StGoF has  $N(0,1)$ as the limiting null, achieves the optimal phase transition, and is uniformly consistent in broad settings (so it satisfies all requirements (R1)-(R3) as desired).   
Second, we overcome the technical challenges for stepwise algorithms of this kind by (a) developing a new stepwise scheme as in StGoF, (b) 
deriving sharp results as in (\ref{intro-null}), and (c) developing new techniques to prove the NSP of SCORE.

\vspace{-1em}
 
\subsection{Content} 
Section~\ref{subsec:rSgnQ} introduces the StGoF algorithm, and 
Section~\ref{sec:main}  shows that StGoF is consistent for $K$ uniformly in a broad setting, and achieves the  
optimal phase transition. Section \ref{sec:NSP} shows that SCORE has the Non-Splitting Property (NSP) for $1\leq m\leq K$, 
which is one of the keys to our study in Section~\ref{subsec:rSgnQ}. 
Section \ref{sec:Simul} presents simulation results, and Section~\ref{sec:realdata} contains real data analysis. The supplementary material contains the proofs of all theorems and lemmas.

\section{The stepwise Goodness-of-Fit (StGoF) algorithm}   \label{subsec:rSgnQ} 
StGoF is a stepwise algorithm where for $m = 1, 2, \ldots$, 
we alternately use a community detection step and a 
Goodness-of-Fit (GoF) step. We may view StGoF 
as a general framework, where for either step, we can 
use a different algorithm. However, for most existing community 
detection algorithms (e.g., \cite{DCBM,rohe2011}),  it is unclear whether they 
have the desired theoretical properties (especially the NSP), so we may face analytical challenges. For this reason, we choose to use 
SCORE \cite{SCORE}, which we prove to have the NSP. 
For GoF, existing algorithms (e.g., \cite{ZhuGoF, LeiGoF}) 
do not apply to the current setting,  so we propose  Refitted Quadrilateral (RQ) as a new 
GoF metric  (a quadrilateral in a graph is a length-$4$ cycle \citep{Bollobas}; see details below).

\begin{figure}[htb!]
\begin{center}
\includegraphics[width = 4.5 in]{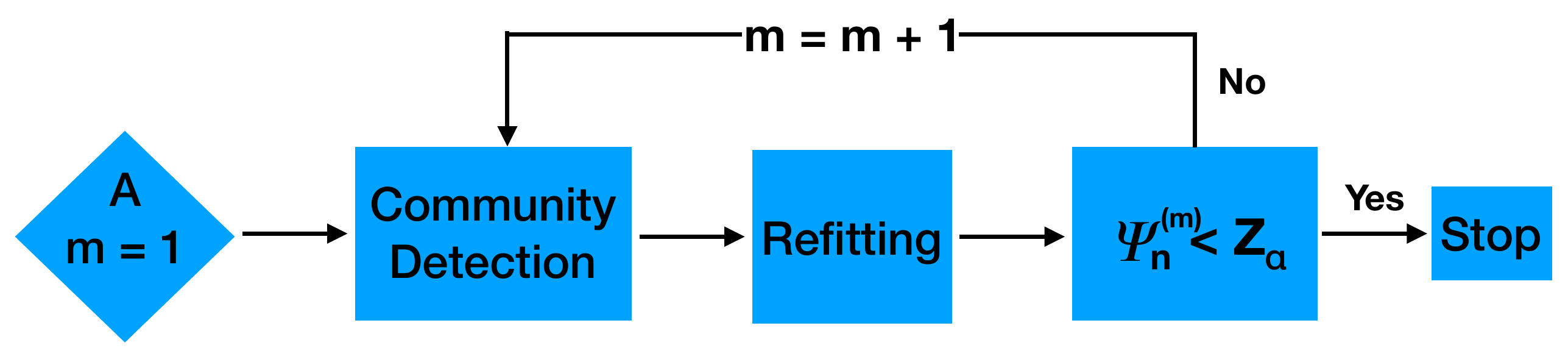} 
\end{center} 
\vspace{-1.5 em} 
\caption{The flow chart of StGoF.} 
\label{fig:flow} 
\end{figure}   
  
In detail, fix $0 < \alpha < 1$ (e.g., $\alpha = 1\%$ or $5\%$).  Let $z_{\alpha}$ 
be the $\alpha$ upper-quantile of $N(0,1)$,  StGoF runs as follows.    
Input: adjacency matrix $A$ (initialize with $m = 1$; see Figure \ref{fig:flow}).  
\begin{itemize} 
\itemsep0em
\item {\it (a). Community detection}. If $m=1$, let $\widehat{\Pi}^{(m)}$ be the $n$-dimensional vector of $1$'s. If $m>1$, apply SCORE to $A$ assuming $m$ is the correct number of communities and obtain 
an $n \times m$ matrix $\widehat{\Pi}^{(m)}$ for the estimated community labels.  
\item {\it (b). Goodness-of-Fit}.  Pretending $\widehat{\Pi}^{(m)}$ is the matrix of true community labels, we obtain an estimate $\widehat{\Omega}^{(m)}$ for $\Omega$  by refitting the DCBM,  following \eqref{refitting1a}-\eqref{refitting1b} below.  Obtain the Refitted Quadrilateral test score $\psi_n^{(m)}$ as in \eqref{refitting2a}-\eqref{refitting2d}. 
\item {\it (c). Termination}. If $\psi_n^{(m)} \geq z_{\alpha}$, repeat (a)-(b) with $m = m + 1$. Otherwise, output $m$ as the estimate for $K$.  Denote the final estimate by $\hat{K}^*_\alpha$. 
\end{itemize}

We now fill in the details for steps (a)-(b).  Consider (a) first.  The case of $m=1$ is trivial so we only consider the case of $m > 1$. 
Let $\hat{\lambda}_k$ be the $k$-th largest (in magnitude) eigenvalue of $A$, and let $\hat{\xi}_k$ be the corresponding eigenvector.   For each $m > 1$, we apply SCORE as follows.  Input:  $A$ and $m$.  Output:   estimated community label matrix $\widehat{\Pi}^{(m)} \in \mathbb{R}^{n,m}$.    
\begin{itemize} 
\itemsep0em
\item Obtain the first $m$ eigenvectors $\hat{\xi}_1, \hat{\xi}_2, \ldots, \hat{\xi}_m$ of $A$. Define the $n \times (m-1)$ matrix of entry-wise ratios $\widehat{R}^{(m)}$ by 
$\widehat{R}^{(m)}(i, k) = \hat{\xi}_{k+1}(i) / \hat{\xi}_1(i)$, $1 \leq i \leq n, 1 \leq k \leq m-1$.   
\spacingset{1}\footnote{As the network is connected, $\hat{\xi}_1$ is uniquely defined with all positive entries, by Perron's theorem \cite{SCORE}.}  \spacingset{1.465}
\item Cluster the rows of $\widehat{R}^{(m)}$ by the  $k$-means assuming we have $m$ clusters.  Output $\widehat{\Pi}^{(m)} = [\hat{\pi}_1^{(m)}, \ldots, \hat{\pi}_n^{(m)}]'$ 
($\hat{\pi}_i^{(m)}(k) = 1$ if node $i$ is clustered to cluster $k$ and $0$ otherwise).  
\end{itemize}

Consider (b).  The idea is to {\it pretend} that the SCORE estimate  
$\widehat{\Pi}^{(m)}$ is accurate. We then use it to 
estimate $\Omega$ by re-fitting, and check how well 
the estimated $\Omega$ fits with the adjacency matrix $A$.  
In detail, let $d_i$ be the degree of node $i$, $1 \leq i \leq n$, and let $\widehat{\cal N}_k^{(m)}$ be the set of nodes that SCORE assigns to group $k$, $1 \leq k \leq m$.  We decompose ${\bf 1}_n$ as follows 
\begin{equation} \label{refitting1aADD}
{\bf 1}_n = \sum_{k = 1}^m \hat{\bf 1}_k^{(m)},  \qquad \mbox{where $\hat{\bf 1}_k^{(m)}(j) = 1$ if $j \in \widehat{\cal N}_k^{(m)}$ and 0 otherwise}.  
\end{equation} 
For most quantities that have superscript $(m)$, we may only include the superscript when introducing these quantities for the first time, and omit it later for notational simplicity when there is no confusion. Introduce a 
vector $\hat{\theta}^{(m)} = (\hat{\theta}_1^{(m)}, \hat{\theta}_2^{(m)}, \ldots, \hat{\theta}_n^{(m)})' \in \mathbb{R}^n$ and a matrix $\widehat{P}^{(m)}  \in \mathbb{R}^{m, m}$ where for all $1 \leq i \leq n$ and $1 \leq k, \ell \leq m$, 
\begin{equation} \label{refitting1a}
\hat{\theta}_i^{(m)} = [d_i /(\hat{\bf 1}_k' A {\bf 1}_n)] \cdot\sqrt{\hat{\bf 1}_k' A \hat{\bf 1}_k},  \qquad
\widehat P_{k\ell}^{(m)}  = (\hat{\bf 1}_k' A \hat{\bf 1}_{\ell}) / 
{\sqrt{(\hat{\bf 1}_k' A \hat{\bf 1}_k) (\hat{\bf 1}_{\ell}' A \hat{\bf 1}_{\ell})}}. 
\end{equation} 
Let $\widehat{\Theta}^{(m)}  = \diag(\hat{\theta})$. We refit $\Omega$ by 
\begin{equation} \label{refitting1b}
\widehat{\Omega}^{(m)}   =   \widehat{\Theta}^{(m)}    \widehat{\Pi}^{(m)}    \widehat{P}^{(m)}  (\widehat{\Pi}^{(m)})'  \widehat{\Theta}^{(m)}.  
\end{equation}  
Recall that $\Omega = \Theta \Pi P \Pi' \Theta$ and $P$ has unit diagonal entries.  
In the ideal case where $m = K$, $\widehat{\Pi}^{(m)} = \Pi$, and $A = \Omega$,  we have $(\widehat{\Theta}^{(m)}, \widehat{P}^{(m)}, \widehat{\Omega}^{(m)})  = (\Theta, P, \Omega)$. This suggests that the refitting in (\ref{refitting1b}) is reasonable.  
The Refitted Quadrilateral (RQ) test statistic is then 
\begin{equation}  \label{refitting2b} 
Q_n^{(m)} = \sum_{i_1, i_2, i_3, i_4 (dist)} (A_{i_1 i_2} - \widehat{\Omega}_{i_1 i_2}^{(m)}) (A_{i_2 i_3} - \widehat{\Omega}_{i_2 i_3}^{(m)}) (A_{i_3 i_4} - \widehat{\Omega}_{i_3 i_4 }^{(m)}) (A_{i_4 i_1} - \widehat{\Omega}_{i_4 i_1}^{(m)}),  
\end{equation} 
(``dist" means the indices are distinct).  Without the refitted matrix $\widehat{\Omega}^{(m)}$, $Q_n^{(m)}$ 
reduces to 
\begin{equation} \label{refitting2a}  
C_n = \sum_{i_1, i_2, i_3, i_4 (dist)} A_{i_1 i_2} A_{i_2i_3} A_{i_3i_4} A_{i_4 i_1} = \mbox{total number of quadrilaterals}. 
\end{equation} 
 
In the null case of $m = K$, first, $\mathrm{Var}(Q_n^{(m)})$ can be well-approximated by $8C_n$.  Second,  while  the mean of $Q_n^{(K)}$ is $0$ in the ideal case of  $\widehat{\Omega}^{(K)} = \Omega$, 
in the real case, it is comparable to $[\mathrm{Var}(Q_n^{(K)})]^{1/2}$ and is not negligible, so we need bias correction. 
Motivated by these, for any $m \geq 1$,  we introduce two vectors $\hat{g}^{(m)}, \hat{h}^{(m)} \in \mathbb{R}^m$ where 
\begin{equation}\label{refitting2gh} 
\hat{g}_k^{(m)} = (\hat{\bf 1}_k' \hat{\theta}) / \|\hat{\theta}\|_1, \qquad  \hat{h}_k^{(m)} = (\hat{\bf 1}_k' \widehat{\Theta}^2 \hat{\bf 1}_k)^{1/2} / \|\hat{\theta}\|, \qquad 1 \leq k \leq m. 
\end{equation}
Write for short $\widehat{V}^{(m)} =  \diag(\widehat{P} \hat{g})$ and $\widehat{H} ^{(m)} = \diag(\hat h)$.  
We estimate the mean of $Q_n^{(m)}$ by 
\begin{equation} \label{refitting2c} 
{B}^{(m)}_n = 2 \|\hat{\theta}\|^4 \cdot [ \hat{g}' \widehat{V}^{-1} (\widehat{P}\widehat{H}^2\widehat{P}\circ \widehat{P}\widehat{H}^2\widehat{P}) \widehat{V}^{-1} \hat{g}],
\end{equation} 
where for matrixes $A$ and $B$, $A\circ B$ is their Hadamard product \cite{HornJohnson}.   
Here, in the null case, $B_n^{(m)}$ is a good estimate for $\mathbb{E}[Q_n^{(m)}]$, and in the under-fitting case,  it is much smaller than the leading term of $Q_n^{(m)}$ and so is negligible.   Finally, the StGoF statistic is defined by 
\begin{equation} \label{refitting2d} 
\psi_n^{(m)} = [Q_n^{(m)} - {B}^{(m)}_n]/\sqrt{8{C}_n}.   
\end{equation}

For each $m$, StGoF has a SCORE step (consisting of a PCA step and a $k$-means step) \cite{SCORE} 
and a GoF step. The complexity of PCA step is $O(n^2 m)$ if we use the power method, and 
the complexity of the GoF step is $O(n^2 \bar{d})$, where $\bar{d}$ is the average node degree. 
In Section~\ref{sec:main}, we show that under mild conditions, StGoF terminates in $K$ steps with high probability. So aside from running $K$ times of $k$-means, the complexity of StGoF is $O(n^2 K^2 + n^2 K \bar{d})$. Note that many real networks are sparse, where the factor $\bar{d}$ is relatively small. 
Similarly, \citep{ma2018determining}  iterates for $m = 1, 2, \ldots, k_{max}$ ($k_{max}$ is a prescribed upper bound for $K$),   where for each $m$, it runs PCA once, $k$-means for $(m+1)$ times, and then computes a quantity with a cost of $O(n^2 m)$. Therefore, aside from running $k$-means for $O(k_{max}^2)$ times, 
the cost is $O(n^2 k_{max}^2)$. The approach by \cite{wang2017likelihood}  also iterates for $m = 1, 2, \ldots, k_{max}$, where for each $m$, they need an exhaustive search step which is NP hard. 
To overcome the challenge, they use a spectral clustering approach to approximate the solution, where 
the cost  (aside from running $k$-means for $k_{max}$ times) is $O(n^2 k_{max}^2)$.  
In theory, as the complexity of $k$-means is relatively high,  the main costs of three algorithms come from the $k$-means part, and StGoF is less expensive (the times it runs for $k$-means 
is fewer than those of the others).   In practice, we usually implement the $k$-means with the (relatively fast) Lloyd's algorithm \cite{HTF},  so all three algorithms are reasonably fast.  For example,  for a typical setting in Experiment 5a of Section 6 with $(n, K) = (600, 6)$, the computing time of three methods for $100$ repetitions  are $1, 10, 8$ minutes, respectively, 
and for a typical setting in Experiment 4b of Section 6 with $(n, K) = (1200, 3)$, 
the computing time of three methods for $100$ repetitions are $2, 30, 40$ minutes, respectively.

\begin{lemma} \label{lemma:complexity}
Suppose $K=O(\bar{d})$, where $\bar{d}$ is the average degree of the network. 
For each $m = 1, 2, \ldots, K$,  the complexity for computing $\psi_n^{(m)}$  by \eqref{refitting1a}-\eqref{refitting2a}  is $O(n^2{\bar d})$. 
\end{lemma}

Our StGoF procedure is new.  Existing stepwise algorithms (e.g., those in \cite{wang2017likelihood, ma2018determining}) iterate by comparing $\ell_n^{(m+1)} - \ell_n^{(m)}$ with 
a benchmark (which unfortunately has unknown parameters) for 
$m = 1, 2, \ldots, K$, and can not avoid the over-fitting case.  
StGoF iterates by comparing $\psi_n^{(m)}$ with $N(0,1)$ for $m = 1, 2, \ldots, K$, 
and successfully avoids the over-fitting case. Such a difference is crucial for obtaining 
sharp theoretical result; see Section~\ref{sec:main}.   

Comparing with \cite{SCORE}, though we use  SCORE in the clustering step,  
 but is for a different purpose: The orthodox SCORE is for community detection in the null case of $m=K$. We use SCORE to  construct a low-rank matrix $\widehat{\Omega}^{(m)}$ in the under-fitting case of $m<K$, where the analysis is quite different and requires new technical tools; see Section \ref{sec:NSP}.

  The RQ test $\psi_n^{(m)}$ is connected to the SgnQ test  \cite{SP2019} (a recent idea for global testing, which can be viewed as an improved version of the GC test by \cite{OGC} and the EZ test by 
\cite{gao2017testing}),  but there are major differences. First,  the SgnQ test is for global testing where we test  $K=1$ v.s. $K>1$, and it is unclear how to use it for goodness-of-fit in each step of StGoF.  
Second, SgnQ is not a stepwise algorithm and does not depend on any intermediate clustering results. The RQ critically depends on the intermediate clustering results by SCORE, where the NSP of SCORE plays a key role. Third, SgnQ does not need re-fitting, but RQ requires a re-fitting step. The re-fitting errors cause a non-negligible bias in $Q_n^{(m)}$.  To obtain a  tractable limiting null (where $m = K$),  we need to figure out the right bias correction as in \eqref{refitting2c}, with long and careful calculations.   
At the same time, by similar proofs as in our main theorems, we can show that $\psi_n^{(1)} \goto N(0,1)$ if $K = 1$ and $\psi_n^{(1)} \goto \infty$ in probability if $K > 1$ and $|\lambda_2| / \sqrt{\lambda_1} \goto \infty$, where $\lambda_k$ is the $k$-th largest (in magnitude) eigenvalue of $\Omega$. 
Comparing with the lower bound in \cite{SP2019}, $\psi_n^{(1)}$ is optimal for global testing.

{\bf Remark 1}.  Existing GoF algorithms include \cite{ZhuGoF, LeiGoF}, but 
they only address narrower settings (e.g., dense networks  that follow SBM and have strong signals). 
As mentioned in \cite{ZhuGoF}, it remains unclear how to generalize these approaches to the DCBM setting here. In principle,  a GoF approach only focuses on the null case, and can not be used for estimating $K$ without sharp results in the under-fitting case,  or the over-fitting case, or both.

{\bf Remark 2}.  
 For SBM settings where $P$ is singular (see Example 3), $r < K$ ($r = \mathrm{rank}(\Omega))$.  In this case, StGoF can consistently estimate $r$. To estimate $K$,   we may revise StGoF by replacing the 
SgnQ test in the GoF step by a degree-based $\chi^2$-test (the success of which 
was shown for global testing with SBM; e.g. \cite{CK2021, JKLiang2021}).   By the NSP of SCORE,  we can show that the new estimator is consistent under similar 
regularity conditions. Though the $\chi^2$-tests may be powerful in some SBM settings, they usually lose 
power in more general DCBM settings, as suggested by the following result on {\it degree matching}.   Consider a DCBM setting where we test $K = 1$ vs. $K > 1$ (i.e., global testing).  
It was shown in \cite{SP2019, JKLiang2021}  that for any alternative (i.e., $K > 1$),  we can pair it with a null such that for  each node, the expected degrees under the two models in the pair match with each other.  Therefore, a naive degree-based test may lose power in separating the two models in the pair.

\section{The consistency and optimality of StGoF} \label{sec:main}
\setcounter{equation}{0}  
In this section, we discuss  the consistency and optimality of StGoF. The NSP of SCORE (one of the key components in our proofs and a second part of our main results) is deferred to Section \ref{sec:NSP}.  
Consider a DCBM with $K$ communities as in (\ref{DCBM-matrixform}).   We assume 
\begin{equation} \label{condition1a}
\|P\| \leq C ,   \qquad \|\theta\| \goto \infty, \qquad \mbox{and} \qquad \theta_{\max} \sqrt{\log(n)} \goto 0. 
\end{equation} 
The first one is a mild regularity condition on the $K \times K$ community structure matrix $P$. 
The other two are mild conditions on sparsity. See (\ref{thetarange}) for the interesting 
range of $\|\theta\|$. We exclude the case where $\theta_i = O(1)$ for all $1 \leq i \leq n$ for convenience, 
but our results continue to hold in this case provided that we make some small changes in our proofs.   
Moreover, for $1 \leq k \leq K$,   let ${\cal N}_k$ be the set of nodes belonging to community $k$,  let $n_k$ be the cardinality of ${\cal N}_k$, and let 
$\theta^{(k)}$ be the $n$-dimensional vector where $\theta_i^{(k)} = \theta_i$ if $i \in {\cal N}_k$ and 
$\theta_i^{(k)} = 0$ otherwise. We assume the $K$ communities are balanced in the sense that 
\begin{equation} \label{condition1b} 
\min_{\{1 \leq k \leq K\}} \{n_k/n,  \; \|\theta^{(k)}\|_1 / \|\theta\|_1,  \;  \|\theta^{(k)}\|^2/\|\theta\|^2\}  \geq C. 
\end{equation} 
In the presence of severe degree heterogeneity, the valid SNR for SCORE is  
\[
s_n = a_0(\theta) (|\lambda_K| / \sqrt{\lambda_1}), \qquad \mbox{where} \;\;  a_0(\theta) = (\theta_{min}/\theta_{max}) \cdot (\|\theta\| /  \sqrt{\theta_{max} \|\theta\|_1}) \leq 1. 
\] 
In the special case of $\theta_{max}\leq C \theta_{min}$, it is true that $a_0(\theta) \asymp 1$ and  $s_n \asymp  |\lambda_K | / \sqrt{\lambda_1}$. In this case,  $s_n$ is the  SNR introduced in (\ref{SNR1}).  
We assume  
\begin{equation} \label{condition1c} 
s_n\geq C_0\sqrt{\log(n)}, \qquad\mbox{for a sufficiently large constant $C_0>0$}. 
\end{equation} 
In the special case $\theta_{max}\leq C\theta_{min}$, (\ref{condition1c}) is 
equivalent to $|\lambda_K| / \sqrt{\lambda_1} \geq C\sqrt{\log(n)}$, which is mild.   
Define a diagonal matrix $H\in\mathbb{R}^{K,K}$ by 
$H_{kk} = \|\theta^{(k)}\|/\|\theta\|$, $1 \leq k \leq K$.  
For the matrix $HPH$ and $1 \leq k \leq K$,  let  
$\mu_k$ be the $k$-th largest eigenvalue (in magnitude) and $\eta_k$ be the corresponding eigenvector.  
By Perron's theorem \cite{HornJohnson}, if $P$ is irreducible, then the multiplicity of $\mu_1$ is $1$, 
and all entries of $\eta_1$ are  strictly positive. Note also the size of the matrix $P$ is small.   It is therefore only a mild condition to assume that for a constant $0 < c_0 < 1$,   
\begin{equation}  \label{condition1d} 
\min_{2\leq k\leq K}|\mu_1-\mu_k|\geq c_0|\mu_1|, \qquad \mbox{and}  \qquad  \frac{\max_{1\leq k\leq K}\{\eta_1(k)\}}{\min_{1\leq k\leq K}\{\eta_1(k)\}}  \leq C. 
\end{equation} 
In fact,  (\ref{condition1d}) holds if all entries of $P$ are lower bounded by a positive constant or $P\to P_0$ for a fixed irreducible matrix $P_0$. We also note that the most challenging case for network analysis is when $P$ is close to the matrix of $1$'s (where it is hard to distinguish one community from 
another), and \eqref{condition1d} always holds in such a case.  In this paper, we implicitly assume $K$ is fixed.  
 Our method can be extended to the case where $K$ diverges with $n$ at a speed not too fast, but 
the right hand side of (\ref{condition1b}) needs to be replaced by $C / K$. See Section \ref{sec:Discu} 
for discussions.

\subsection{The null case and a confidence lower bound for $K$}  \label{subsec:CLB} 
In the null case, $m = K$, so if we apply SCORE 
to the rows of $\widehat{R}^{(m)}$ assuming $m$ clusters, then we have perfect community recovery with overwhelming probability, and StGoF provides a confidence lower bound for $K$.  
The next theorem is proved in the supplement. 
\begin{thm} \label{thm:CLB} 
Fix $0 < \alpha < 1$. Suppose we apply StGoF to a  DCBM model where (\ref{condition1a})-(\ref{condition1d}) hold. As $n \goto \infty$, up to a permutation of the columns of $\widehat{\Pi}^{(K)}$,  
$\mathbb{P}(\widehat{\Pi}^{(K)} \neq \Pi) \leq C n^{-3}$, 
$\psi_n^{(K)}  \goto N(0,1)$ in law, and 
$\mathbb{P}(\widehat{K}_{\alpha}^* \leq K) \geq (1 - \alpha) + o(1)$.  
\end{thm}  
\noindent 
Theorem \ref{thm:CLB} allows for severe degree heterogeneity. If the degree heterogeneity is moderate, $s_n \asymp |\lambda_K| / \sqrt{\lambda_1}$, and we have the following corollary.   
\begin{cor} \label{cor:CLB} 
Fix $0 < \alpha < 1$. Suppose we apply StGoF to a  DCBM model where (\ref{condition1a})-(\ref{condition1b}) and (\ref{condition1d}) hold. 
Suppose $\theta_{max} \leq C \theta_{min}$  and  $|\lambda_K|/\sqrt{\lambda_1}\geq C_0\sqrt{\log(n)}$ for a sufficiently large constant $C_0>0$.   
 As $n \goto \infty$, up to a permutation of the columns of $\widehat{\Pi}^{(K)}$, 
$\mathbb{P}(\widehat{\Pi}^{(K)} \neq \Pi) \leq C n^{-3}$,  $\psi_n^{(K)}  \goto N(0,1)$ in law,  
and $\mathbb{P}(\hat{K}_{\alpha}^* \leq K) \geq (1 - \alpha) + o(1)$. 
\end{cor} 
\noindent 
It follows that $\widehat{K}_{\alpha}^*$ is a level-$(1-\alpha)$ confidence lower bound for $K$.  If $\alpha$ depends on $n$ and tends to $0$ slowly enough, these results continue to hold. In this case, $\mathbb{P}(\hat{K}_{\alpha}^* \leq K) = 1- o(1)$.  
When perfect community recovery is impossible but the faction  
of misclassified nodes is small with high probability (e.g., for a slightly smaller SNR),  the asymptotic normality continues to hold. 
Similar comments apply to Theorem \ref{thm:CUB} and Corollary \ref{cor:CUB}. 
As far as we know, this is the first time in the literature that we have derived a (completely) explicit limiting null.   
The result can be used to derive $p$-values in settings such as Goodness-of-Fit  \cite{ZhuGoF, LeiGoF}. If the assumed model in GoF is DCBM with $K$ communities, 
then by Theorem \ref{thm:CLB}, we can apply StGoF with $m = K$ and derive an approximate $p$-value 
as  $\mathbb{P}(N(0,1) \geq \psi_n^{(K)})$.  The proof of Theorem \ref{thm:CLB} is 
non-trivial and tedious. The main reason is that $\Omega$ is unknown and we must estimate it with refitting (see  
 $\widehat{\Omega}^{(K)}$ in  (\ref{refitting1b})).    The 
refitting errors are non-negligible  even when $\Pi$ is given: we must choose a 
bias correction term as in (\ref{refitting2b}) and analyze $Q_n^{(K)}$ carefully.

\subsection{The under-fitting case of $m < K$ and consistency of StGoF}  \label{subsec:SCORE} 
Fixing an $m$ such that $1 < m < K$ (the case of $m = 1$ is trivial),      
suppose we apply SCORE to the rows of $\widehat{R}^{(m)}$ assuming $m$ is the correct number of communities. Let $\widehat{\Pi}^{(m)}$ be the matrix of estimated community labels. In this case,  we underestimate the number of clusters, so perfect community recovery is impossible. Fortunately,  SCORE satisfies the  {\it Non-Splitting Property (NSP)}. 
Recall that $\Pi$ is the matrix of true community labels.  
\bed
Fix $K > 1$ and $m \leq K$. We say that a realization of the $n \times m$ matrix of estimated labels $\widehat{\Pi}^{(m)}$ satisfies the NSP if  
for any pair of nodes in the same (true) community, the estimated community labels are the same (i.e.,   each community in $\Pi$ is  contained in a community in the realization of $\widehat{\Pi}^{(m)}$).  
When this happens, we write  $\Pi \preceq \widehat{\Pi}^{(m)}$. 
\eed 
\begin{thm} \label{thm:SCORE2} 
Consider a DCBM where (\ref{condition1a})-(\ref{condition1d}) hold. 
With probability at least $1 - 
O(n^{-3})$, for each $1 < m \leq K$,  $\Pi \preceq \widehat{\Pi}^{(m)}$ up to a permutation in the columns.    
\end{thm} 
\noindent 
By Theorem \ref{thm:SCORE2}, SCORE has the NSP (with high probability).  Theorem \ref{thm:SCORE2} is the key to our upper bound study below. 
In Section \ref{sec:NSP}, we explain the main technical challenges for proving Theorem \ref{thm:SCORE2}, and present the key theorems and lemmas required for the proof. 
\begin{thm} \label{thm:CUB} 
Fix $0 < \alpha < 1$. Suppose we apply StGoF to a DCBM model where (\ref{condition1a})-(\ref{condition1d}) hold. As $n \goto \infty$,   
$\min_{1 \leq m < K} \{\psi_n^{(m)}\}   \goto \infty$ in probability and $\mathbb{P}(\widehat{K}_{\alpha}^* \neq K)  \leq \alpha + o(1)$. 
\end{thm}  
\noindent 
Theorem \ref{thm:CUB} allows for severe degree heterogeneity. 
When the degree heterogeneity is moderate, $\mbox{SNR} \asymp |\lambda_K| / \sqrt{\lambda_1}$ and we have the following corollary.

\begin{cor} \label{cor:CUB} 
Fix $0 < \alpha < 1$. Suppose we apply StGoF to a  DCBM model where (\ref{condition1a})-(\ref{condition1b}) and (\ref{condition1d}) hold, $\theta_{max} \leq C \theta_{min}$,  and $|\lambda_K|/\sqrt{\lambda_1}\geq C_0\sqrt{\log(n)}$ for a sufficiently large constant $C_0>0$. As $n \goto \infty$,   $\min_{1 \leq m < K}\{\psi_n^{(m)}\} \goto \infty$  in probability and $\mathbb{P}(\hat{K}_{\alpha}^* \neq K) \leq \alpha + o(1)$.   
\end{cor} 
\noindent 
Now in Theorem \ref{thm:CUB} and Corollary \ref{cor:CUB}, if we let $\alpha$ depend on $n$ and tend to $0$ slowly enough, then we have $\mathbb{P}(\hat{K}_{\alpha}^* = K)\to 1$.    Theorem \ref{thm:CUB} is proved in the supplement. The proof is non-trivial and long, so 
for instruction,  we explain (a) what are the technical challenges and especially 
why the NSP is critical, and (b) why StGoF provides a consistent estimate.

Consider (a) first.  
The main technical challenge is how to analyze $\psi_n^{(m)}$ where we 
not only need sharp row-wise large deviation bounds for the matrix $\widehat{R}^{(m)}$, but also need 
to establish the NSP of SCORE, where we note $m \leq K$.  To see why NSP is important,  note that 
$Q^{(m)}_n$ depends on $\widehat{\Omega}^{(m)}$ (see (\ref{refitting2b})),  where $\widehat{\Omega}^{(m)}$ is 
obtained by refitting using the SCORE estimate $\widehat{\Pi}^{(m)}$, 
and depends on $A$ in a complicate way. The dependence poses challenges for analyzing $Q_n^{(m)}$,   to overcome which, 
a conventional approach is to use concentrations.  However, $\widehat{\Pi}^{(m)}$ 
has $\mathrm{exp}(O(n))$ possible realizations, and how  to characterize the concentration of $\widehat{\Pi}^{(m)}$ is a challenging problem (e.g., \cite{wang2017likelihood, ma2018determining}). \spacingset{1}\footnote{To shed light on why $\widehat{\Pi}^{(m)}$ has so many possible realizations,  suppose we wish to group $n$ $iid$ samples from $N(0,1)$ into two  clusters with the same size.  We have $\mathrm{exp}(O(n))$ possible clustering results.}\spacingset{1.465}  
Fortunately,  if  SCORE has the NSP,  then $\widehat{\Pi}^{(m)}$ only has $\binom{K}{m}$ possible realizations. In fact,  $\widehat{\Pi}^{(m)}$ may have even fewer possible realizations if we impose some mild conditions. 
Therefore,  for each $1 \leq m \leq K$, $\widehat{\Omega}^{(m)}$ only concentrates on a few non-stochastic matrices. 
Using this and union bound, we can therefore remove the technical hurdle for analyzing $\psi_n^{(m)}$ in the under-fitting case.  

The proof of NSP is non-trivial, partially due to the intractable rotation of eigenvectors dictated by the 
Davis-Kahan $\sin(\theta)$ theorem. See Section~\ref{sec:NSP} for detailed explanations.

Consider (b).   Fix $1\leq m\leq K$. By the NSP of SCORE, except for a small probability, 
the estimated membership matrix $\widehat{\Pi}^{(m)} \in \mathbb{R}^{n, m}$  only has finitely many realizations. 
Fixing a realization $\widehat{\Pi}^{(m)} = \Pi_0$, let $\mathcal{N}_1^{(m,0)},  \cdots, \mathcal{N}_m^{(m,0)}$ be 
the clusters defined by $ \Pi_0$. Let $\theta^{(m,0)}$, $\Theta^{(m,0)}$ and $P^{(m,0)}$ be constructed 
similarly as in (\ref{refitting1aADD})-(\ref{refitting1a}), except that $(A, \widehat{\Pi}^{(m)})$ and the vector $d=( d_1,d_2,\ldots,d_n)'$ are replaced by $(\Omega, \Pi_0)$ and $\Omega {\bf 1}_n$, respectively. 
Let $\Omega^{(m,0)} = \Theta^{(m,0)} \Pi_0 P^{(m,0)}\Pi_0' \Theta^{(m,0)}$.  
Then, on the event $\widehat{\Pi}^{(m)}=\Pi_0$,   $\Omega^{(m,0)}$ is a non-stochastic proxy of the refitted matrix $\widehat{\Omega}^{(m)}$. Recall that $\Omega$ is a non-stochastic proxy of the adjacency matrix $A$. We thus expect the RQ statistic in \eqref{refitting2b} to satisfy that
\begin{align} \label{insight-add1}
Q_n^{(m)} &\approx \sum_{i_1, i_2, i_3, i_4 (dist)} (\Omega_{i_1i_2} - {\Omega}^{(m,0)}_{i_1i_2}) (\Omega_{i_2i_3} - {\Omega}^{(m,0)}_{i_2i_3}) (\Omega_{i_3i_4} - {\Omega}_{i_3i_4}^{(m,0)}) (\Omega_{i_4i_1} - {\Omega}^{(m,0)}_{i_4i_1})\cr
&\approx \tr((\Omega-\Omega^{(m,0)})^4), \qquad\quad \mbox{on the event of $\widehat{\Pi}^{(m)}=\Pi_0$.} 
\end{align}
Now, when $m=K$, it can be shown that $\widehat{\Pi}^{(m)}=\Pi$  except for a small  probability. Note also that when $\Pi_0=\Pi$, our re-fitting procedure guarantees that $\theta^{(m,0)}=\theta$, $P^{(m,0)}=P$, and so $\Omega^{(m,0)}=\Omega$. It follows that $ \tr((\Omega-\Omega^{(m,0)})^4)=0$. When $m<K$, $\Omega^{(m,0)}$ has a rank $m<K$ and $\Omega$ has a rank $K$. Recall that $\lambda_1,\ldots,\lambda_K$ are the nonzero eigenvalues of $\Omega$ (arranged in the descending order of magnitudes). 
By Weyl's theorem, the $k$th largest absolute eigenvalue of $\Omega-\Omega^{(m,0)}$ is always lower bounded by $|\lambda_{k+m}|$, for all $1\leq k\leq K-m$. It follows that  
$\tr((\Omega-\Omega^{(m,0)})^4) = \sum_{k=1}^{K-m}|\lambda_k(\Omega-\Omega^{(m,0)})|^4 \geq \sum_{k=1}^{K-m}\lambda_{m+k}^4$.
In summary, 
\beq  \label{insight-add2}
\mbox{$\tr((\Omega-\Omega^{(m,0)})^4)= 0$ if $m = K$,  and $\tr((\Omega-\Omega^{(m,0)})^4) 
\geq \sum_{k=m+1}^K \lambda_k^4$ if $m < K$}.    \;\;  \spacingset{1}\footnote{This explains why in StGoF we do not use the  refitted triangle (RT)  
$T_n^{(m)} = \sum_{i_1, i_2, i_3 (dist)} (A_{i_1 i_2} - \widehat{\Omega}_{i_1i_2}^{(m)}) (A_{i_2 i_3} - \widehat{\Omega}_{i_2i_3}^{(m)}) (A_{i_3 i_1} - \widehat{\Omega}_{i_3 i_1}^{(m)})$, which is comparably easier to analyze.  While we may similarly derive $T_n^{(m)} \gtrsim \sum_{k = m+1}^K \lambda_k^3$ with large probability,  
$\lambda_{m+1}, \ldots, \lambda_K$ may have different signs and so may cancel with each other. 
We can not use $B_n^{(m)} = \sum_{i_1, i_2 (dist)} (A_{i_1 i_2} - \widehat{\Omega}_{i_1i_2}^{(m)}) (A_{i_2 i_1} - \widehat{\Omega}_{i_2i_1}^{(m)})$ either. The variance of $B_n^{(m)}$ is 
unappealingly large so the resultant procedure can not achieve the optimal phase transition; see Section \ref{subsec:LB}. Also, see \cite{SP2019} for discussion on statistics similar to $T_n^{(m)}$ and $B_n^{(m)}$.} \spacingset{1.465}
\eeq 
Recall that $\psi_n^{(m)}$ is the standardized version of $Q_n^{(m)}$, and that except for a small probability, 
$\widehat{\Pi}$ has only one possible realization for $\widehat{\Pi}$ in the null case and has only finite realizations in the alternative case. Using the above and  union bounds, we can show that 
\begin{equation} \label{whywork} 
\left\{ 
\begin{array}{ll} 
\psi_n^{(m)} \goto N(0,1),  &\;\;\;   \mbox{if $m = K$},  \\
\mathbb{E}[\psi_n^{(m)}]  \asymp (\sum_{k = m+1}^K \lambda_k^4) / \lambda_1^2 \;   \mbox{and so $\psi_n^{(m)} \goto \infty$ in prob.}, & \;\;\;   \mbox{if $1 \leq m < K$},  
\end{array} 
\right.  
\end{equation} 
where $(\sum_{k = m+1}^K \lambda_k^4) / \lambda_1^2 \geq (\lambda_K / \sqrt{\lambda_1})^4$ when $m < K$.   
Therefore, with a proper threshold on $\psi_n^{(m)}$, StGoF stops at $m=K$ with an overwhelming probability and outputs a consistent estimate for $K$.  The proofs for the NSP and (\ref{insight-add2})-(\ref{whywork}) are technically  demanding. See Section~\ref{sec:NSP} and Section~\ref{sec:proof} of the supplement for detailed explanations and proofs.

\subsection{Information lower bound and phase transition}  \label{subsec:LB} 
In Theorem~\ref{thm:CUB} and Corollary~\ref{cor:CUB}, 
we require the SNR, $|\lambda_K|/\sqrt{\lambda_1}$, to tend to $\infty$ 
at a speed of at least $\sqrt{\log(n)}$.  
We show that such a condition cannot be significantly relaxed.
 There are relatively few studies on the lower bound for estimating $K$, and our results are new.

We say two DCBM models are asymptotically 
indistinguishable if for any test that tries to decide which model is true, the sum of Type I and Type II errors is no smaller than $1 + o(1)$, as $n \goto \infty$. Given a DCBM with $K$ communities, 
our idea is to construct a DCBM with $(K+m)$ communities for any $m \geq 1$, and show 
that two DCBM are asymptotically indistinguishable, provided that the SNR of the latter is $o(1)$. 
 
Fixing $K_0 \geq 1$, we consider a DCBM with $K_0$ communities that satisfies (\ref{model1a})-(\ref{model1d}). Let $(\Theta,\widetilde{\Pi}, \widetilde{P})$ be the parameters of this DCBM, and let $\widetilde{\Omega} = \Theta \widetilde{\Pi} \widetilde{P} \widetilde{\Pi}' \Theta$. 
When $K_0 > 1$, let $(\beta', 1)'$ be the last column of $\widetilde{P}$, and let $S \in \mathbb{R}^{K_0-1,  K_0-1}$ be the sub-matrix of $\widetilde{P}$ excluding the last row and the last column. 
Given $m \geq 1$ and $b_n \in (0,1)$, we construct a  DCBM model with $(K_0+m)$ communities as follows. We define a $(K_0+m)\times (K_0+m)$ matrix $P$: 
\beq \label{LB-construct1}
P=
\begin{bmatrix}
S & \beta {\bf 1}_{m+1}'\\
{\bf 1}_{m+1}\beta' & \frac{m+1}{1+mb_n} M
\end{bmatrix}, \qquad 
\mbox{where}\quad M =(1-b_n)I_{m+1} + b_n {\bf 1}_{m+1} {\bf 1}_{m+1}'.
\eeq
When $K_0=1$, we simply let $P=\frac{m+1}{1+mb_n}M$. 
Let $\tilde{\ell}_i\in\{1, \ldots,K_0\}$ be the community label of node $i$ defined by $\widetilde{\Pi}$. We generate labels $\ell_i\in \{1, \ldots,K_0+m\}$ by
\beq \label{LB-construct2}
\ell_i = \begin{cases}
\tilde{\ell}_i, &\mbox{if }\tilde{\ell}_i\in\{1, \ldots,  K_0-1\},\\
\mbox{uniformly drawn from $\{K_0, K_0+1,\ldots K_0+m\}$}, &\mbox{if }\tilde{\ell}_i=K_0. 
\end{cases}
\eeq
Let $\Pi$ be the corresponding community label matrix.  This gives rise to a DCBM model with $(K_0+m)$ communities, where $\Omega = \Theta \Pi P \Pi' \Theta$. Though $P$ does not have unit diagonals,  we can re-parametrize 
so that it has unit diagonals: Let $D$ be the $(K_0+m) \times (K_0+m)$ diagonal matrix 
with $D_{kk} = \sqrt{P_{kk}}$, $1 \leq k \leq K_0 + m$. 
Now, if we let $P^* = D^{-1} P D^{-1}$, $\theta_i^* = \theta_i  \|D \pi_i\|_1$,   and  $\Theta^* = \diag(\theta_1^*, \ldots, \theta_n^*)$,  then $P^*$ has unit-diagonals and $\Omega = \Theta^* \Pi P^* \Pi' \Theta^*$.  

Here some rows of $\Pi$ are random (so we may call the corresponding model the random-label DCBM), but this is conventional in the study of 
lower bounds.  Let $\lambda_k$ be the $k$th largest eigenvalue (in magnitude) of $\Omega$. Since $\Omega$ is random,  $\lambda_k$'s are also random (but we can 
bound $|\lambda_K|/\sqrt{\lambda_1}$ conveniently).    
The following theorem is 
proved in the supplement. 

\begin{thm} \label{thm:LB} 
Fix $K_0\geq 1$ and consider a DCBM model with $n$ nodes and $K_0$ communities, whose parameters $(\theta,\widetilde{\Pi}, \widetilde{P})$ satisfy (\ref{condition1a})-(\ref{condition1b}). Let $(\beta', 1)'$ be the last column of $\widetilde{P}$, and let $S$ be the sub-matrix of $\widetilde{P}$ excluding the last row and last column. 
We assume $|\beta'S^{-1}\beta-1|\geq C$. 
\begin{itemize}
\itemsep0em
\item Fix $m\geq 1$. Given any $b_n \in (0,1)$, we can construct a random-label DCBM model with $K=K_0+m$ communities as in (\ref{LB-construct1})-(\ref{LB-construct2}). Then, as $n \goto \infty$,  $|\lambda_{K}|/\sqrt{\lambda_1}\leq C\|\theta\|(1-b_n)$ with probability $1-o(n^{-1})$. Moreover, if $(1-b_n)/|\lambda_{\min}(S)|=o(1)$, where $\lambda_{\min}(S)$ is the minimum eigenvalue (in magnitude) of $S$, then $|\lambda_{K}|/\sqrt{\lambda_1}\geq C^{-1}\|\theta\|(1-b_n)$ with probability $1-o(n^{-1})$. Here $C>1$ is a constant that does not depend on $b_n$. 
\item Fix $m_1,m_2\geq 1$ with $m_1\neq m_2$. As $n \goto \infty$, if $\|\theta\|(1-b_n)\to 0$, then the two random-label DCBM models associated with $m_1$ and $m_2$ are asymptotically indistinguishable. 
\end{itemize}
\end{thm}
Here, the condition $|\beta' S^{-1} \beta - 1| \geq C$ is used to bound the last diagonal entry of $\widetilde{P}^{-1}$, which is $1/(\beta' S^{-1} \beta -1)$.  
By Theorem~\ref{thm:LB}, starting from a (fixed-label) DCBM with $K_0$ communities, we can construct a collection of random-label DCBM, with $K_0+1, K_0+2,\ldots,K_0+m$ communities, respectively, where 
(a)  for the model with $(K_0 + m)$ communities, $|\lambda_{K_0 +m}| / \sqrt{\lambda_1} \asymp \|\theta\| (1 - b_n)$, with an overwhelming probability,  and (b) each pair of models are asymptotically indistinguishable if $\|\theta\|(1-b_n)=o(1)$. 
Therefore, for a broad class of DCBM with unknown $K$ where $\mbox{SNR} = o(1)$ for some models,   a consistent estimate for $K$ does not exist.  

Fixing $m_0 > 1$ and a sequence of numbers  $a_n > 0$,  let ${\cal M}_n(m_0, a_n)$ be the collection of DCBM for an $n$-node network with $K$ communities,  where $1 \leq K \leq m_0$, $|\lambda_K|/\sqrt{\lambda}_1 \geq a_n$, and (\ref{condition1a})-(\ref{condition1b}) hold.  In Section \ref{subsec:SCORE}, we show that if  $a_n \geq C_0 \sqrt{\log(n)}$ for a sufficiently large constant $C_0$, then for each DCBM in ${\cal M}_n(m_0, a_n)$, StGoF provides a consistent estimate for $K$.  
The following theorem says that, if we allow $a_n \goto 0$, then ${\cal M}_n(m_0, a_n)$ is  too broad, and a consistent estimate for $K$ does not exist. 
\begin{thm} \label{thm:LB2}
Fix $m_0 > 1$ and let ${\cal M}_n(m_0, a_n)$ be the class of DCBM as above.  As $n\to\infty$, if $a_n \goto 0$, then $\inf_{\hat{K}} \bigl\{ \sup_{{\cal M}_n(m_0, a_n)} \mathbb{P}(\hat{K}\neq K) \bigr\} \geq (1/6 + o(1))$,  where the probability is evaluated at any given model in ${\cal M}_n(m_0, a_n)$ and the supremum is  over all such models. 
\end{thm}

Combining Theorems \ref{thm:CLB}, \ref{thm:LB2},   and Corollary \ref{cor:CUB}, we have a phase transition result (phase transition is a recent theoretical framework (e.g., \cite{DJ04, JKW}). It is closely related to the classical minimax framework but can be more informative in many cases).  
\begin{itemize} 
\itemsep0em
\item {\it Impossibility}. If $a_n \goto 0$, then ${\cal M}_n(m_0, a_n)$ defines 
 a class of  DCBM that is too broad where  some pairs of models in the class are asymptotically 
indistinguishable. Therefore,   no estimator can consistently estimate the number of communities  for each model in the class (and we say ``a consistent estimate for $K$ does not exist" for short).   
\item {\it Possibility}. If  $a_n\geq C_0 \sqrt{\log(n)}$ for a sufficiently large $C_0$,  then for every DCBM in ${\cal M}_n(m_0, a_n)$,  StGoF 
provides a consistent estimate for the number of communities if the model only has moderate degree heterogeneity (i.e., $\theta_{max} \leq C\theta_{min})$. StGoF continues to be consistent in the presence of severe degree heterogeneity if the adjusted SNR satisfies that $s_n  \geq C_0 \sqrt{\log(n)}$ with a sufficiently large $C_0$.     
\end{itemize} 
The case of $C \leq a_n < C_0 \sqrt{\log(n)}$ is more delicate.  
Sharp results are possible if we consider more specific models (e.g., for a scaling parameter $\alpha_n > 0$,  $(\theta_i / \alpha_n)$ are $iid$ from a fixed distribution $F$, and the off-diagonals of $P$ are the same). 
We leave  this to the future.

Comparing with existing works,  we have the following comments:  (a) StGoF is the first method that is proved to achieve  the optimal transition, (b) StGoF is the first method that is proved to have  a (completely) explicit limiting null,  (c) we prove 
the NSP of SCORE, and use it to derive sharp results that are not available before,   (d) our settings are much broader and our regularity conditions are much weaker, and (e) 
we overcome the challenges of stepwise algorithms of this kind by using the 
sharp results we derive and by using a different stepwise scheme (so to avoid the analysis of the over-fitting case where the 
NSP does not hold).  
We now compare   with \cite{wang2017likelihood, ma2018determining} with more details.  

First,  their approaches require a signal strength much stronger than ours, and so do not achieve the phase transition. 
When $\theta_{\max}\leq C\theta_{\min}$, our result requires $|\lambda_K|/\sqrt{\lambda_1}\geq C_0\sqrt{\log(n)}$, which matches the lower bound  in Section~\ref{subsec:LB}. However,  \cite{wang2017likelihood} needs  $|\lambda_K|/\sqrt{\lambda_1}\gg n^{1/4} \sqrt{\log(n)}$ (see their Section 2.5), which is non-optimal. Also, \cite{ma2018determining} proves consistency under the condition of $\lambda_1\geq C \log(n)$. Recall that they assume $P=\rho_n P_0$. 
In their setting, $|\lambda_1|, \ldots, |\lambda_K|$ are at the same order, and $\lambda_1\geq C \log(n)$ indeed
translates to $|\lambda_K|/\sqrt{\lambda_1}\geq C_0\sqrt{\log(n)}$. However, for general settings where $|\lambda_1|, \cdots, |\lambda_K|$ are at different orders,  
it is unclear whether their method is optimal (because the SNR is captured by $|\lambda_K|/\sqrt{\lambda_1}$, not $\sqrt{\lambda_1}$).
In comparison, our result matches with the lower bound for all settings. 
Second,  \cite{wang2017likelihood} only studies the SBM where $\theta_i$'s are all equal, and \cite{ma2018determining} assumes that $\theta_{\max}\leq C\theta_{\min}$ and $P=\rho_n P_0$, for a fixed matrix $P_0$; in this setting, the degree heterogeneity is only moderate, and $|\lambda_1|, \cdots, |\lambda_K|$ are at the same order. This excludes many practical cases of interest. 
Last, besides the very mild condition of \eqref{condition1b}, we do not need any hard-to-check conditions on $\Pi$. In contrast, \cite{wang2017likelihood,ma2018determining} impose stringent conditions.  
For example, \cite{ma2018determining} defines a quantity $Q_K(k)$ by applying spectral clustering to $\Omega$ and then evaluating the change of residual sum of squares by further splitting one cluster. They impose conditions on $Q_K(k)$ for every $1\leq k\leq K-1$ (see their Assumption 3). These conditions are hard to check in practice. Moreover, when $P=\rho_n P_0$ does not hold for a fixed $P_0$, the conditions on $\{Q_K(k)\}_{1\leq k\leq K-1}$ are easy to violate (e.g., in our Example 1).

The advantage of our theory partially comes from the way our algorithm is designed. StGoF only assesses one candidate of $K$ in each step, instead of comparing two adjacent values of $K$. It helps avoid the analysis of the over-fitting case, and it also avoids imposing stringent conditions on $\Pi$. Another advantage comes from our new proof ideas. We do not need $\widehat{\Pi}^{(m)}$ to converge to a non-stochastic matrix, because our proof is {\it not} based on Taylor expansion. For example, a key component of our analysis is the NSP of SCORE. We develop the NSP under very weak conditions where $\widehat{\Pi}^{(m)}$ can be non-tractable, non-unique, and depending on a data-driven rotation matrix (see Section~\ref{sec:NSP}).

\section{The non-splitting property (NSP) of SCORE}  \label{sec:NSP}
To prove the NSP of SCORE, we face technical challenges. 
In the SCORE step of StGoF,  for each $2 \leq m \leq K$,    
we cluster $n$ rows of the matrix $\widehat{R}^{(m)}$ into $m$ clusters.  
We find that for any two rows of $\widehat{R}^{(m)}$, the  
distance critically depends on a non-tractable data-dependent rotation matrix $\widehat{\Gamma}$ dictated by the 
David-Kahn sin($\theta$) theorem \cite{sin-theta}, 
and it may vary significantly as $\widehat{\Gamma}$ changes from one realization to another.  
This poses an unconventional setting for clustering.  
To overcome the challenge, we first discover a new distance-based quantity that is {\it semi-invariant} 
with respect to $\widehat{\Gamma}$:   the quantity remains at the same order of $O(1)$ 
as $\widehat{\Gamma}$ varies from one realization to another. We then develop a new $k$-means theorem (Theorem \ref{thm:kmeans}) and use it to prove the NSP. The proof of Theorem \ref{thm:kmeans} is non-trivial: our setting is an unconventional clustering setting and we do not want to impose unrealistic and  strong conditions. 
Note that the literature on SCORE has been focused on the null case of $m = K$,  but 
our primary interest is in the under-fitting case of $m < K$.

\subsection{Row-wise large-deviation bounds and Ideal polytope}  \label{subsec:NSP-geometry} 
Recall that for $2 \leq m \leq K$,    $\widehat{R}^{(m)}$ is an $n \times (m-1)$  matrix  constructed 
from the eigenvectors $\hat{\xi}_1, \hat{\xi}_2, \ldots,\hat{\xi}_m$ by taking entry-wise ratios between 
$\hat{\xi}_2, \ldots, \hat{\xi}_m$ and $\hat{\xi}_1$;  
see Section~\ref{subsec:rSgnQ}.  Let $\lambda_k$ be the $k$-th largest (in magnitude) eigenvalue of $\Omega$ and let $\xi_k$ 
be the corresponding eigenvector. Under our assumptions (e.g., see condition \eqref{condition1d}),  
there exists a $(K-1)\times (K-1)$ orthogonal matrix $\widehat{\Gamma}$ such that 
$[\hat{\xi}_1, \hat{\xi}_2, \ldots, \hat{\xi}_K] \approx [\xi_1, \xi_2, \ldots, \xi_K]\cdot\diag(1,\widehat{\Gamma})$.   
The rotation matrix $\widehat{\Gamma}$ is dictated by the Davis-Kahan $\sin(\theta)$ theorem in spectral analysis.  It is well-known that the matrix is data dependent and 
hard to track.  Even if $\lambda_1, \ldots, \lambda_K$ are distinct (so each vector in $\{\xi_1, \ldots,\xi_K\}$ is unique up to 
a $\pm 1$ factor), $\widehat{\Gamma}$ can still take arbitrary values on the Stiefel manifold and does not concentrate on any non-stochastic rotation matrix on the manifold. \spacingset{1}\footnote{$\widehat{\Gamma}$ is tractable only if we impose a strong eigen-gap condition. However, this excludes many practical settings of interest, especially when the signals are weak and $|\lambda_1|, \ldots,|\lambda_K|$ are at different orders.}\spacingset{1.465}

Therefore, we must consider all possible realizations  $\widehat{\Gamma} = \Gamma$.  This not only poses analytical challenges (see Section \ref{subsec:NSP-core}) but also makes notations more complicate. 
Fix a (non-stochastic) orthogonal matrix $\Gamma$. 
For $2\leq k\leq K$, let $\xi_k(\Gamma)$ be the $k$th column of $[\xi_1,\xi_2,\ldots,\xi_K]\cdot\mathrm{diag}(1, \Gamma)$,\spacingset{1}\footnote{By Perron's theorem, $\xi_1$ is uniquely defined and a strictly positive vector. Vectors $\xi_2, \ldots, \xi_K$ are not necessarily unique, but we can select an arbitrary candidate of $\xi_2,\ldots,\xi_K$ to define $\xi_2(\Gamma), \ldots,\xi_K(\Gamma)$.}\spacingset{1.465}
and let $\xi_k(j,\Gamma)$ be the $j$th entry of $\xi_k(\Gamma)$, $1\leq j\leq n$. Define $R^{(m)}(\Gamma)\in \mathbb{R}^{n, m-1}$ by 
\begin{equation} \label{DefineR}
R^{(m)}(i, \ell; \Gamma) =  \xi_{\ell+1}(i; \Gamma) / \xi_1(i), \qquad 1 \leq  i  \leq n, \; 1 \leq \ell \leq m-1.  
\end{equation} 
Comparing \eqref{DefineR} with the definition of $\widehat{R}^{(m)}$ in Section~\ref{subsec:rSgnQ}, it is seen that $R^{(m)}(\Gamma)$ is the population counterpart of $\widehat{R}^{(m)}$ on the event of $\widehat{\Gamma}=\Gamma$.  
Lemma \ref{lem:entrywise-bound} provides a sharp row-wise large-deviation bound for $\widehat{R}^{(m)}- R^{(m)}(\Gamma)$ and is proved in the supplemental material.
\begin{lemma}[Row-wise bounds] \label{lem:entrywise-bound}
Consider a DCBM model where (\ref{condition1a})-(\ref{condition1d}) hold. Let $s_n = a_0(\theta) (|\lambda_K| / \sqrt{\lambda_1})$, where $a_0(\theta)$ is as in Section~\ref{sec:main}. 
For each $1 < i \leq n$, let $(r_i^{(m)}(\Gamma))'$ and $(\hat{r}_i^{(m)})'$ denote the $i$-th row of $R^{(m)}(\Gamma)$ and $\widehat{R}^{(m)}$, respectively.
As $n\to\infty$, with probability $1-O(n^{-3})$,  for all $1\leq m\leq K$ and $1\leq i\leq n$ 
and all $(K-1) \times (K-1)$ orthogonal matrix $\Gamma$,  $\| \hat{r}_i^{(m)} - r_i^{(m)}(\Gamma)\|  \leq \| \hat{r}_i^{(K)} - r_i^{(K)}(\Gamma)\| \leq   Cs_n^{-1}\sqrt{\log(n)}$ over the event $\widehat{\Gamma}=\Gamma$. 
\end{lemma}
\noindent
Under our assumptions, $s_n^{-1}\sqrt{\log(n)}$ is upper bounded by a sufficiently small constant. It implies that each $\hat{r}_i^{(m)}$ is sufficiently close to $r_i^{(m)}(\Gamma)$ on the event $\widehat{\Gamma}=\Gamma$.

It remains to study the geometry underlying $\{ r_i^{(m)}(\Gamma)\}_{1\leq i\leq n}$ for {\it an arbitrary} rotation matrix 
$\Gamma \in \mathbb{R}^{K-1, K-1}$.  
Recall that $H\in\mathbb{R}^{K,K}$ is the diagonal matrix with $H_{kk}=\|\theta^{(k)}\|/\|\theta\|$, $1\leq k\leq K$.
For each $1 \leq k \leq K$, let $\mu_k$ be the 
$k$-th largest (in magnitude) eigenvalue of $HPH$  and let $\eta_k \in \mathbb{R}^K$ be the associated (unit-norm) eigenvectors, respectively.     By  Lemma~\ref{lemma:eigenvector-relation} of the supplement,   $\eta_1$ is unique and all entries are strictly positive.  Also, while $(\eta_2,\ldots,\eta_K)$ may be  non-unique,  there is a one-to-one correspondence between the choice of $(\eta_2,\ldots,\eta_K)$ and the choice of $(\xi_2,\ldots,\xi_K)$; see the paragraph above (\ref{DefineR}). 
Fix $\Gamma$. 
For each $2\leq k\leq K$, let $\eta_k(\Gamma)$ be the $(k-1)$-th column of $[\eta_2, \eta_3, \ldots, \eta_K] \Gamma$, and let $\eta_k(i, \Gamma)$ denote the $i$-th entry of $\eta_k(\Gamma)$, $1 \leq i \leq K$. Define a $K\times (m-1)$ matrix $V^{(m)}(\Gamma)$ by   
\begin{equation} \label{DefineV} 
V^{(m)}(k, \ell; \Gamma) = \eta_{\ell+1}(k; \Gamma) / \eta_1(k), \qquad 1 \leq  k  \leq K, \; 1 \leq \ell \leq m-1.   
\end{equation} 
Let $(v_k^{(m)}(\Gamma))'$ be the $k$-th row of $V^{(m)}(\Gamma)$. 
Lemma \ref{lem:matR}  is proved in the supplement.  

\spacingset{1}
\begin{figure}[tb!]
\begin{center}
\includegraphics[width = 5 in, height = 1.25 in]{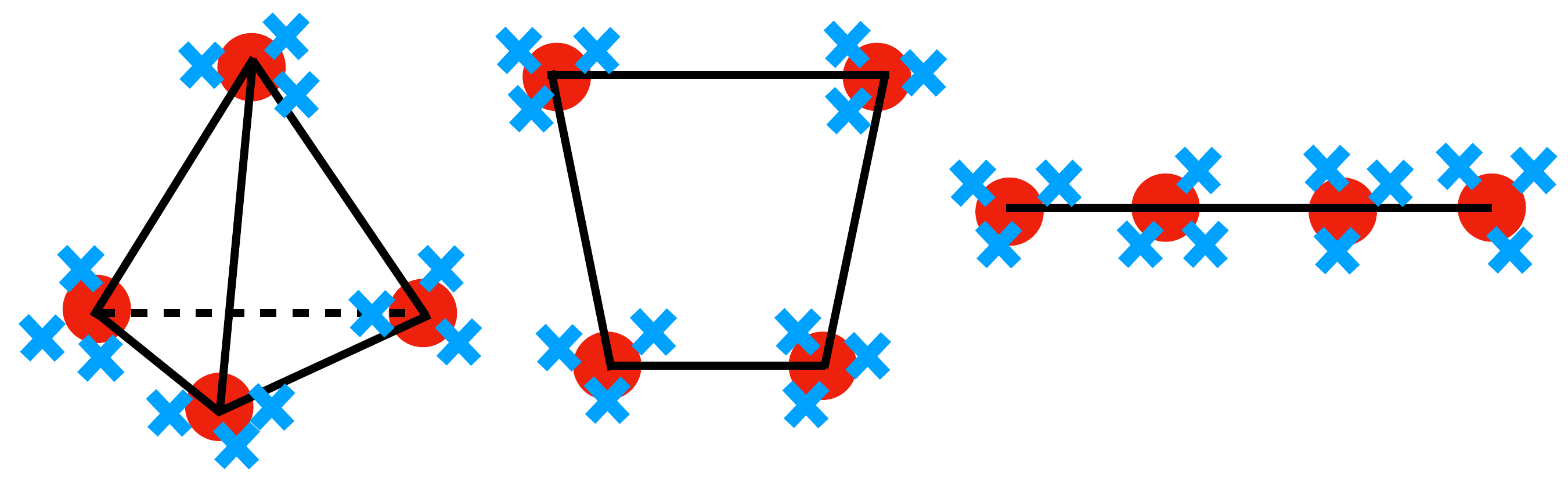} 
\end{center} 
\vspace{-1.25 em} 
\caption{An example ($K = 4$). From left to right: $m = 4, 3, 2$. Red dots: the $4$ distinct rows of $R^{(m)}$,  which are $v_1^{(m)}, v_2^{(m)}, v_3^{(m)}, v_4^{(m)}$. Blue crosses: the rows of $\widehat{R}^{(m)}$.    The red dots are the vertices of a tetrahedron when $m = 4$, vertices of a quadrilateral when $m = 3$, and scalars when $m = 2$.     
For each $m$, the $n$ rows of $\widehat{R}^{(m)}$ form $K$ clusters, each corresponding to a true community. The figure is only for illustration, and we should not have the wrong 
impression that the $K$ clusters are always well-separated.} 
\label{fig:simplex2} 
\end{figure}   
\spacingset{1.465}

\begin{lemma}[The ideal polytope] \label{lem:matR} 
Consider a DCBM model where (\ref{condition1d}) holds. Fix $1<m\leq K$. We have that $r_i^{(m)}(\Gamma) = v_k^{(m)}(\Gamma)$  for all $i\in {\cal N}_k$ and $1\leq k\leq K$. 
\end{lemma} 

Combining Lemmas~\ref{lem:entrywise-bound}-\ref{lem:matR} gives the 
following claim. Viewing $\{\hat{r}_i^{(m)}\}_{1\leq i\leq n}$ as a point cloud in $\mathbb{R}^{m-1}$, we have that with overwhelming probability, for any realization of $\widehat{\Gamma}=\Gamma$ and each $1<m\leq K$, there are $K$ clusters in the point cloud,  corresponding to $K$ true communities,  where $v_1^{(m)}(\Gamma), \ldots, v_K^{(m)}(\Gamma)$ are the cluster centers (see Figure~\ref{fig:simplex2}).

\subsection{Challenges in proving NSP and our approach} \label{subsec:NSP-core}
Given the results in previous section, one may think that NSP  is easy to prove.  
Unfortunately, this is not the case:  even with the results in the previous section,  how to prove NSP remains a non-trivial problem, especially when $m < K$.  
We now provide a detailed explanation.

Recall that $\widehat{\Gamma}$ is data dependent and hard to track, so 
we have to consider all realizations of $\widehat{\Gamma} = \Gamma$. 
Therefore, to prove the claim, we need to show that the NSP holds 
{\it uniformly for all $\Gamma$ in the Stiefel manifold ${\cal O}_{K-1}$}.  
Given a realization of $\widehat{\Gamma} = \Gamma$, let $\widehat{B}^{(m)}$ and $B^{(m)}$ 
be the sub-matrices of $\widehat{\Gamma}$ and $\Gamma$, consisting of the first $(m-1)$ columns.  
Introduce a matrix $V_0\in\mathbb{R}^{K,K-1}$ by $V_0(i,k) = \eta_{k+1}(i) / \eta_1(i)$, $1\leq i\leq K,1\leq k\leq K-1$, where $\eta_k$'s are 
as in the previous section. 
For each $1 < m \leq K$,  by our notations, the $K$ cluster centers in Lemma \ref{lem:matR} are the $K$ rows of the 
matrix $V^{(m)}(\Gamma)\in\mathbb{R}^{K,m-1}$, and $V^{(m)}(\Gamma)$ is related to $V_0$ by $
V^{(m)}(\Gamma) = V_0  B^{(m)}$.   
For any $K \times (m-1)$ matrix $M$, let $d_K(M)$ be the minimum pairwise Euclidean distance 
of the $K$ rows of $M$.  In \cite{SCORE}, it was shown that 
$d_K(V_0)  \geq \sqrt{2}$.   
We now discuss the null case and the under-fitting case separately. 

In the null case, $m = K$, and $B^{(m)} = \Gamma$ is a rotation matrix.    
Since the Euclidean distances remain unchanged for rotation, the relative position of the $K$ cluster centers is {\it invariant} with respect 
to $\Gamma$, and especially,  $d_K(V^{(m)}(\Gamma)) = d_K(V_0) \geq \sqrt{2}$. 
Combining this with Lemmas \ref{lem:entrywise-bound}-\ref{lem:matR}, we have: (a)  With high probability,
the $n$ rows of $\widehat{R}^{(m)}$ split into $K$ clusters; for each row, the distance to the 
closest cluster center is $\leq O(s_n \sqrt{\log(n)}) = o(1)$. (b)  The $K$ cluster centers are well-separated by a distance of $\sqrt{2}$. (c) $m = K$, so the number of clusters assumed in $k$-means matches the number of true clusters. 
In this case,  the cluster labels estimated by $k$-means match with the true cluster labels (up to a permutation) so the NSP follows.

The under-fitting case is unfortunately much harder to prove. For $m<K$, $B^{(m)}$ is not a square matrix (it is not a rotation matrix even in the simplest case where $\Gamma$ is the identity matrix).     
Compared to the null case, we have some major differences. 
First, even in the case where $\Gamma$ is the identity matrix, we may have $d_K(V^{(m)}) = 0$ so the $K$ cluster centers are not well-separated.   
Second, for any two rows of $V^{(m)}(\Gamma)$ (each is one of the $K$ cluster centers), the Euclidean distance critically depends on 
$\Gamma$. As $\Gamma$ varies continuously in the Stiefel manifold, the distance may vary from $O(1)$ to $0$.   
Therefore, the relative positions of the $K$ cluster centers critically depend on $\Gamma$, and 
may vary significantly from one case to another (we may have $d_K(V^{(m)}(\Gamma) )\geq C$ for one $\Gamma$ and $d_K(V^{(m)}(\Gamma))  = 0$ for another $\Gamma$).   
Note also that since $m < K$, the number of clusters fed into the $k$-means algorithm is smaller than the number of true clusters.  
Seemingly, this is an unconventional clustering setting,  especially as our goal is to show that when we apply $k$-means, the 
NPS holds uniformly for all $\Gamma$.   
To overcome the challenges, (1) 
 we propose a new metric for the relative positions of the $K$ cluster centers, and (2)  
we develop a new $k$-means theorem specifically for the setting we have. 
Here, (1) is motivated by the observation that,  the main reason NSP is 
easier to prove in the null case  is that $d_K(V^{(m)}(\Gamma))$ (minimum pairwise distance of the $K$ cluster centers)
  is invariant to $\Gamma$ and so 
the $K$ clusters are always well-separated, uniformly for all $\Gamma$. 
In the under-fitting case, $d_K(V^{(m)}(\Gamma))$ 
is not invariant to $\Gamma$, but there may exist a different measure 
that is invariant to  $\Gamma$. This motivates 
us to define $d_m(V^{(m)}(\Gamma))$ as a new measure for the relative positions 
of the $K$ cluster centers, which is {\it semi-invariant} to $\Gamma$ (i.e., 
  there are constants $c_2 > c_1 > 0$ such that $c_1 \leq d_m(V^{(m)}(\Gamma)) \leq c_2$ 
for all $\Gamma$ in the Stiefel manifold).   
In detail, for any $1 < m \leq K$ and any given $K$ points in $\mathbb{R}^{m-1}$,  
we have the following definition, which is  an extension of the {\it minimum pairwise distance}. 
\begin{definition}[Distance-based metrics defined by bottom up pruning] \label{def:bottom-up-pruning} 
Fixing $K > 1$ and $1 <  m \leq K$, consider a $K \times (m-1)$ matrix $U = [u_1, u_2, \ldots, u_K]'$. First, let $d_K(U)$ be the minimum pairwise distance of all $K$ rows.  Second, let $u_k$ and $u_{\ell}$ ($k < \ell$) be the pair that satisfies $\|u_k - u_{\ell}\| = d_K(U)$ (if this holds for multiple pairs, pick the first pair in the lexicographical order). Remove row $\ell$ from the matrix  $U$ and let $d_{K-1}(U)$ be the minimum pairwise distance for the remaining $(K-1)$ rows.  Repeat this step and define 
$d_{K-2}(U), d_{K-3}(U),  \ldots, d_2(U)$ recursively.  Note that $d_{K}(U) \leq d_{K-1}(U) \leq \ldots \leq d_2(U)$. 
\end{definition} 
For each fixed $\Gamma$,  $d_K(V^{(m)}(\Gamma))$ is the minimum pairwise distance between the $K$ cluster centers $v_1^{(m)}(\Gamma), \ldots, v_K^{(m)}(\Gamma)$, 
and $d_m(V^{(m)}(\Gamma))$ is the minimum pairwise distance of the $m$ remaining cluster centers after 
we prune $(m-K)$ cluster centers in the bottom-up fashion as above. When $\Gamma$ ranges continuously in ${\cal O}_{K-1}$, 
$d_K(V^{(m)}(\Gamma))$ may range continuously from $O(1)$ to $0$, but fortunately $d_m(V^{(m)}(\Gamma))$ remains 
at the same order of $O(1)$, and so is {\it semi-invariant}. This is the following lemma, which is proved in the supplement. 
\begin{lemma} \label{lem:d_m}
Consider a DCBM model where \eqref{condition1b} and \eqref{condition1d} hold. Fix $1 \leq m \leq K$. There is a constant $C>0$ (which may depend on $m$), such that $\min_{\Gamma\in{\cal O}_{K-1}}\bigl\{ d_m(V^{(m)}(\Gamma)) \bigr\} \geq C$.
\end{lemma}
 
We now discuss (2). 
To prove that NSP holds uniformly for all $\Gamma$, it remains to develop a new 
$k$-means theorem.  We can have two versions of the $k$-means theorem:   
a ``weaker" version where we assume $d_K(V^{(m)}(\Gamma)) \geq C$, for a constant $C > 0$, and a ``stronger" version where we only require $c_1 \leq d_m(V^{(m)}(\Gamma) \leq c_2$,  and $d_K(V^{(m)}(\Gamma))$ may be as large as $O(1)$ or as small as $0$.   
As $d_K(V^{(m)}(\Gamma)) = 0$ for many $\Gamma$, the ``weaker" version is inadequate for our setting. 
Theorem \ref{thm:kmeans} is a ``stronger" version of the $k$-means theorem, and is proved in the supplement. 
The ``weaker" version is implied by Theorem \ref{thm:kmeans} and so the proof is skipped. 
\begin{thm}[The ``stronger" version of the $k$-means theorem] \label{thm:kmeans} 
Fix $1 <  m \leq K$ and let $n$ be sufficiently large. 
Consider the non-stochastic vectors $x_1,\ldots,x_n$ that take only $K$ values in $u_1,\ldots,u_K$.  Write $U = [u_1, \ldots, u_K]'$. Let $F_k=\{1\leq i\leq n: x_i=u_k\}$, $1\leq k\leq K$. Suppose for some constants $0 < \alpha_0 < 1$ and $C_0>0$, $\min_{1\leq k\leq K}|F_k|\geq \alpha_0 n$ and $\max_{1\leq k\leq K}\|u_k\|\leq C_0\cdot d_m(U)$.  
We apply the $k$-means clustering to a set of $n$ points $\hat{x}_1, \hat{x}_2, \ldots, \hat{x}_n$ assuming $\leq m$ clusters, and denote by $\hat{S}_1, \hat{S}_2, \ldots, \hat{S}_m$ the obtained clusters (if the solution is not unique, pick any of them). There exists a constant $c>0$, which only depends on $(\alpha_0, C_0, m)$, such that, if 
$\max_{1\leq i\leq n}\|\hat{x}_i-x_i\|\leq c\cdot d_m(U)$, 
then $\#\bigl\{1\leq j\leq m: \hat{S}_j\cap F_k\neq\emptyset \bigr\} =1$,  for each $1\leq k\leq K$.
\end{thm}
To prove the NSP  of SCORE,   we apply Theorem \ref{thm:kmeans} with $U =V^{(m)}(\Gamma)$, $x_i=r_i^{(m)}(\Gamma)$, and $\hat{x}_i=\hat{r}_i^{(m)}$, and the main condition we need is $c_1 \leq d_m(V^{(m)}(\Gamma)) \leq c_2$ uniformly for all $\Gamma$. 
But by Lemma \ref{lem:d_m}, this is implied, so we do not need extra conditions to show the NSP. 
If however we use a ``weaker" version of the $k$-means theorem, then we need 
conditions such as $d_K(V^{(m)}(\Gamma)) \geq C$ for all $\Gamma$ (as explained above, the condition can be violated easily).    
The formal proof of the NSP (i.e., Theorem \ref{thm:SCORE2}) is given in Section~\ref{subsec:thmscore2pf} of the supplement, where 
we combine Lemmas \ref{lem:entrywise-bound}-\ref{lem:d_m}, Theorem \ref{thm:kmeans},  and some elementary probability.  

Theorem \ref{thm:kmeans}  is quite general and may be useful for many other unsupervised learning settings (e.g., \cite{han2021eigen}).  
The proof of the theorem is non-trivial and we now briefly explain the reason. As the objective function of the 
$k$-means is nonlinear and we do not have an explicit formula for the $k$-means solution, we prove 
by contradiction. Let $\hat{\ell}$ be the estimated cluster label vector by $k$-means and $RSS(\hat{\ell})$ be the associated objective function, we aim to show that, when NSP does not hold for $\hat{\ell}$, we can always find a cluster label vector $\ell$ such that $RSS(\ell)<RSS(\hat{\ell})$ (a contradiction). The key is finding such an $\ell$ and evaluating $RSS(\ell)$. 
However, except for a lower bound on $d_m(U)$, we have little information about the $K$ true cluster centers. Since $d_K(U)$ can take any value in $[0, d_m(U)]$, a pair of true cluster centers may be well-separated, moderately close, sufficiently close, or exactly overlapping (correspondingly, their distance is much larger than, comparable with, or much smaller than $\max_{1\leq i\leq n}\|\hat{x}_i-x_i\|$, or exactly zero). 
With the {\it infinitely many} configurations of true cluster centers, the main challenge in the proof is pinning down a
strategy of constructing $\ell$ that guarantees a decrease of RSS for {\it every} possible configuration. 
One might think that the oracle k-means solution $\ell^*$ (k-means applied to $x_1,x_2,\ldots,x_n$) can help guide the construction of $\ell$, but unfortunately this does not work: first, we do not have an explicit form of $\ell^*$; second, in some of our settings, $\hat{\ell}$ can be significantly different from $\ell^*$. 
The way we construct $\ell$ and evaluate $RSS(\ell)$ subtly utilizes the definition of $d_m(U)$ and properties of k-means objective, which is highly non-trivial (see the supplemental material).  Note that while \cite{wang2017likelihood,ma2018determining} proved special cases of the ``weaker" version of the k-means theorem,  they used assumptions (i) true cluster centers are mutually well separated, (ii) the oracle solution $\ell^*$ is mathematically tractable, and (iii) $\hat{\ell}$ is exactly the same as $\ell^*$. As none of (i)-(iii) holds in our setting, it is unclear how to generalize their proofs. We deal with a much harder setting (the ``stronger" version), and our proof  is  different.

We conjecture that Theorem~\ref{thm:kmeans} (and so the NSP of SCORE)  continues to hold if we replace the $k$-means step in SCORE by (say) the $\epsilon$-approximation $k$-means (e.g., 
\cite{Kumar2004}).  Let $\hat{\ell}$ be the $\epsilon$-approximate $k$-means solution. We have $RSS(\hat{\ell})\leq (1+\epsilon)\min_{\ell}RSS(\ell)$. For an appropriately small $\eps$, if the NSP does not hold, then by a similar proof as that of Theorem~\ref{thm:kmeans},  we can first construct an $\tilde{\ell}$ such that $RSS(\tilde{\ell})<RSS(\hat{\ell})-O(d_m(U))$, and then use it to deduce a contradiction.   For reasons of space,
we leave this to future.


\section{Simulations}  \label{sec:Simul} 
 In Experiments 1-3,  we compare StGoF with the BIC approach \cite{wang2017likelihood},\spacingset{1}\footnote{\cite{wang2017likelihood} primarily focused on the SBM model. Their algorithm has an ad-hoc extension to DCBM, which has no theoretical guarantee. We use this extension, instead of the original BIC approach.} \spacingset{1.465}
the ECV approach \cite{li2020network}, and the NCV approach  \cite{chen2018network}. We use the R package ``randnet'' to implement these other methods. 
 In Experiment 4, we compare StGoF with the RPLR approach \cite{ma2018determining}. In Experiment 5, we consider settings with comparably larger values of $K$. In all simulations, we fix $\alpha=0.05$ in StGoF. 
Given $(n,K)$, a scalar $\beta_n > 0$ that controls the sparsity, a symmetric non-negative matrix $P\in\mathbb{R}^{K\times K}$, a distribution $f(\theta)$ on $(0, \infty)$, and a distribution $g(\pi)$ on the standard simplex of $\mathbb{R}^K$, we generate the adjacency matrix $A\in \mathbb{R}^{n, n}$
as follows: First, generate $\tilde\theta_1, ...,\tilde\theta_n$ $iid$ from $f(\theta)$. Let $\theta_i = \beta_n\tilde\theta_i/\|\tilde\theta\|$ and $\Theta = \diag(\theta_1, ..., \theta_n)$.  
Next, generate $\pi_1,  ..., \pi_n$ iid from $g(\pi)$, and let $\Pi = [\pi_1, ..., \pi_n]'$.  
Last, let $\Omega = \Theta\Pi P\Pi'\Theta$ and generate $A$ from $\Omega$,  for $100$ times independently.  
For each algorithm, we measure the performance by the fraction of times it correctly estimates $K$ (i.e., accuracy). 
Note that $\|\theta\| = \beta_n$,  and $\text{SNR} \asymp \|\theta\|(1-b_n)$.  
For the experiments, we let $\beta_n$ range so to cover many different sparsity levels, but keep $\|\theta\|(1-b_n)$ fixed (so the problem  of 
estimating $K$ is not too difficult or too easy; see details below). 

\spacingset{1}
\begin{figure}[tb] 
\centering
\vspace{-0mm}
\includegraphics[width=.32\textwidth, height = .26\textwidth, trim = 0mm 10mm 0mm 0mm, clip=true]{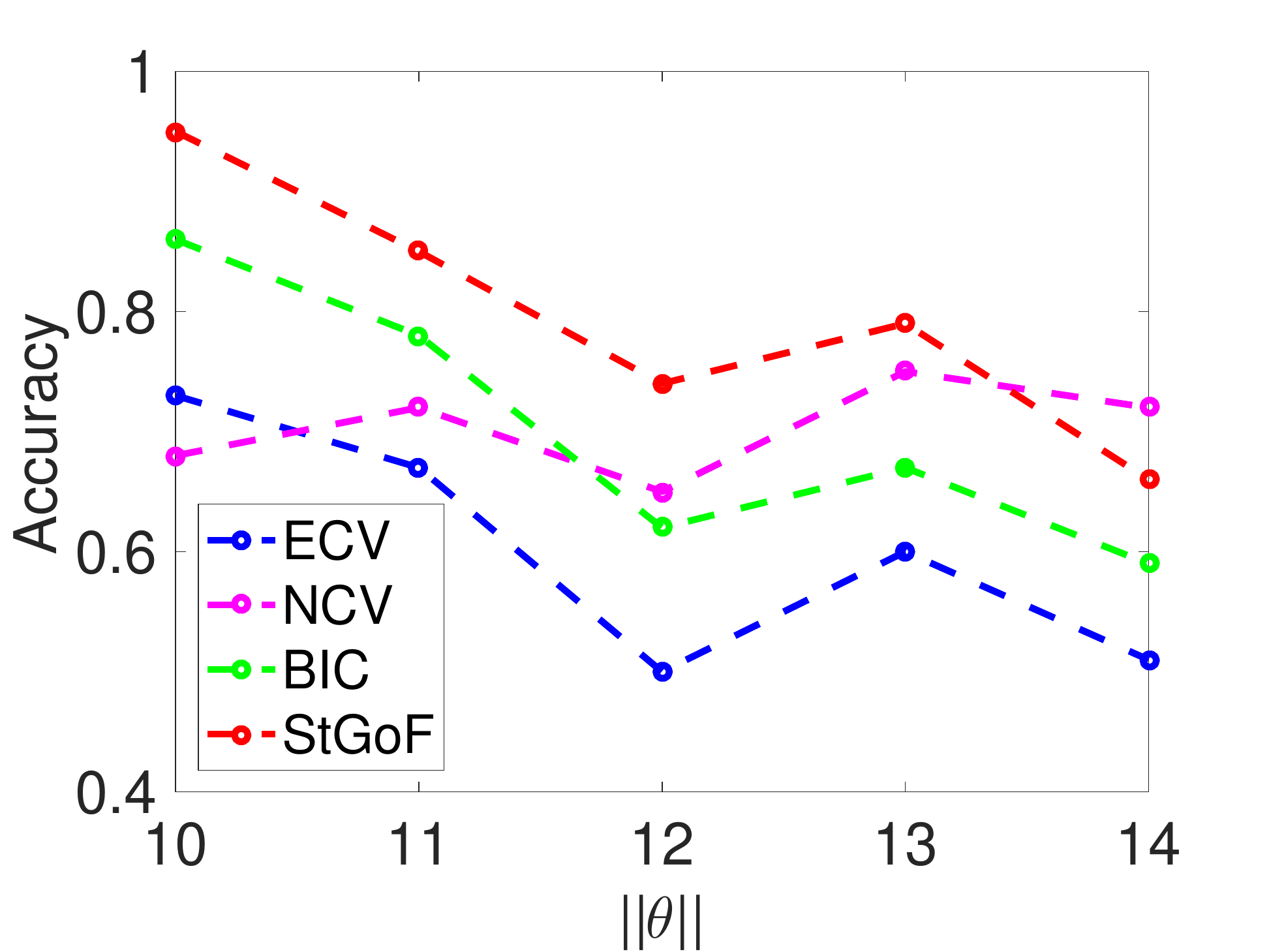}
\includegraphics[width=.32\textwidth, height = .26\textwidth, trim = 0mm 10mm 0mm 0mm, clip=true]{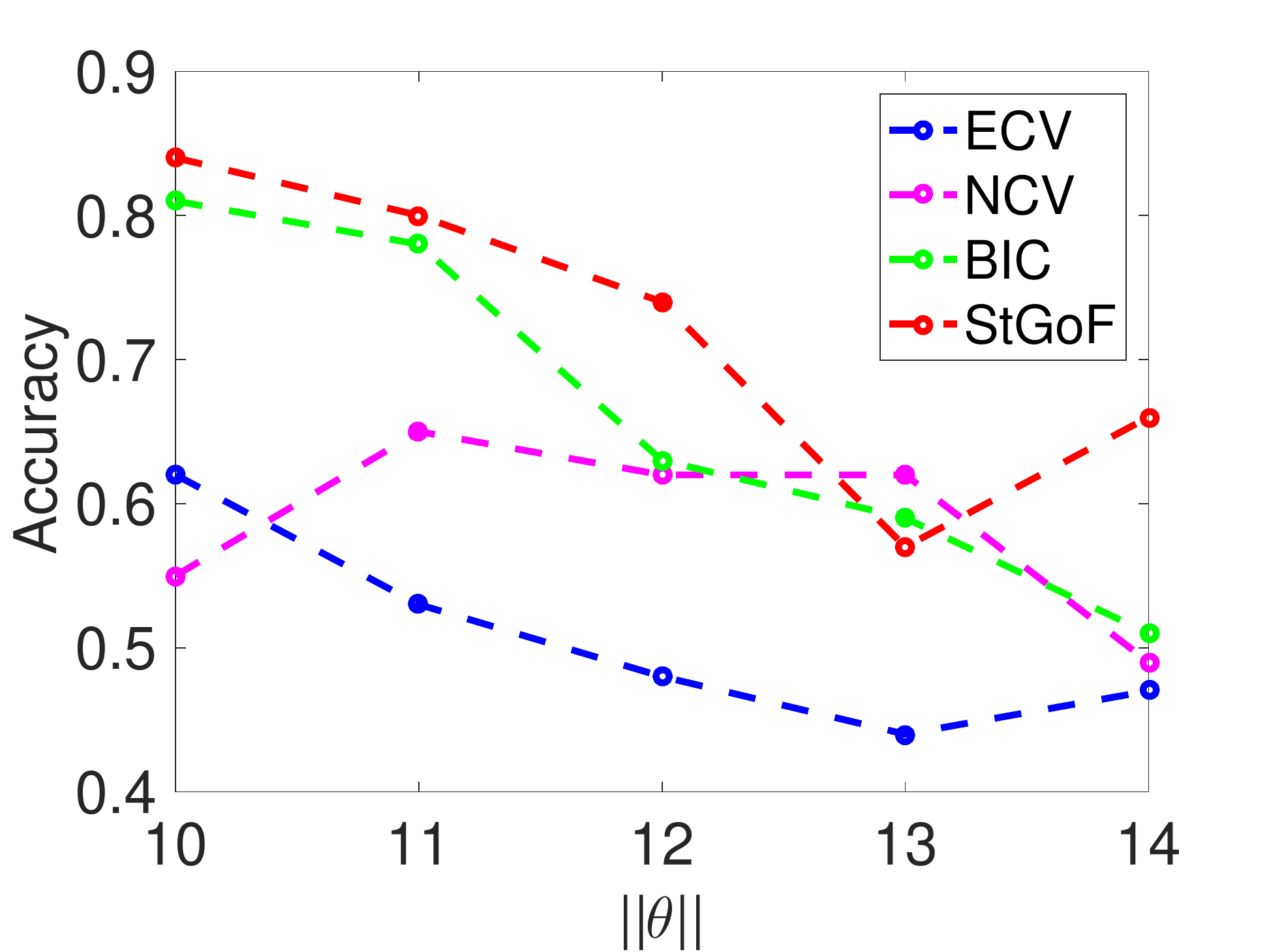}
\includegraphics[width=.32\textwidth, height = .26\textwidth, trim = 0mm 10mm 0mm 0mm, clip=true]{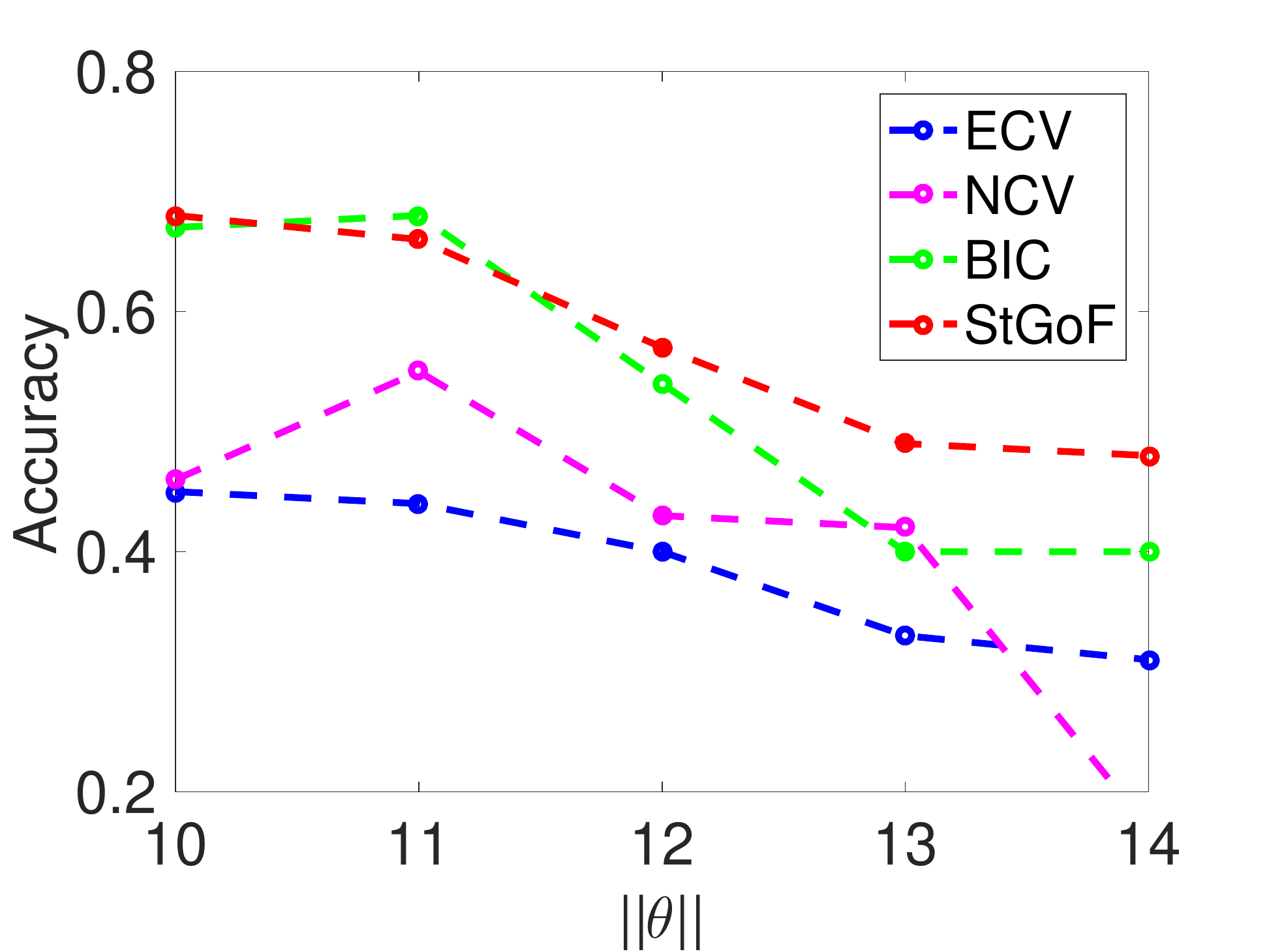}\\
\caption{Experiment 1a (left), 1b (middel) and 1c (right), where  from 1a to 1c, the degree heterogeneity is increasingly more severe. For all three panels, x-axis is $\|\theta\|$ (sparsity level), and $y$-axis is the estimation accuracy over 100 repetitions ($(n,K)=(600,4)$).}  \label{fig:Experiment1}
\end{figure}
\spacingset{1.465}

{\bf Experiment 1}.  We study how degree heterogeneity affect the results and comparisons. Fixing $(n, K) = (600, 4)$, we let $P\in\mathbb{R}^{4,4}$ be a Toeplitz matrix with $P(k,\ell) = 1 - [(1 - b_n)(|k - \ell| + 1)]/K$ in the off-diagonal and $1$ in the diagonal.  
Let $g(\pi)$ be the uniform distribution over $e_1, e_2, e_3, e_4$ (standard basis vectors).   
We consider three sub-experiments, Exp 1a-1c. In these sub-experiments, 
we keep $(1 -b_n)\|\theta\|$ fixed at $9.5$ so the SNR's are roughly at the same level. We let $\beta_n$ range from $10$ to $14$ so to cover both the more sparse and  the more dense cases.   
Moreover, for the three sub-experiments, we take $f(\theta)$ to be  $\mathrm{Unif}(2,3)$,  Pareto($8, .375$) ($8$ is the shape parameter and $.375$ is the scale parameter), and two point mixture $0.95\delta_1 + 0.05\delta_2$ ($\delta_a$ is a point mass at $a$), respectively (from Exp 1a to Exp 1c, the degree heterogeneity gets increasingly more severe).    
See Figure~\ref{fig:Experiment1}. 
StGoF consistently outperforms other approaches. 
%
%

\spacingset{1}
\begin{figure}[tb] 
\centering
\vspace{-0mm}
\includegraphics[width=.32\textwidth, height = .26\textwidth, trim = 0mm 10mm 0mm 0mm, clip=true]{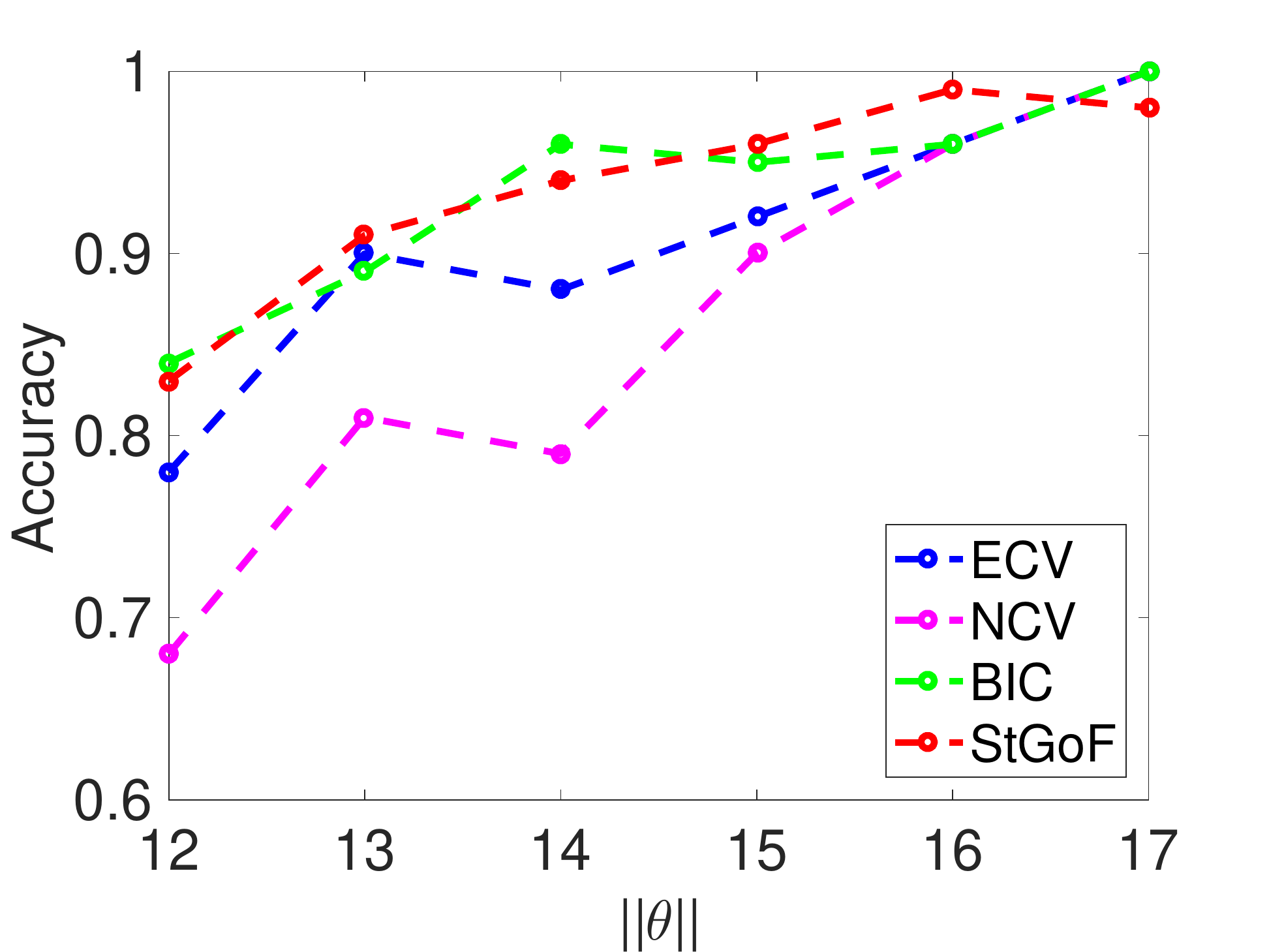}
\includegraphics[width=.32\textwidth, height = .26\textwidth, trim = 0mm 10mm 0mm 0mm, clip=true]{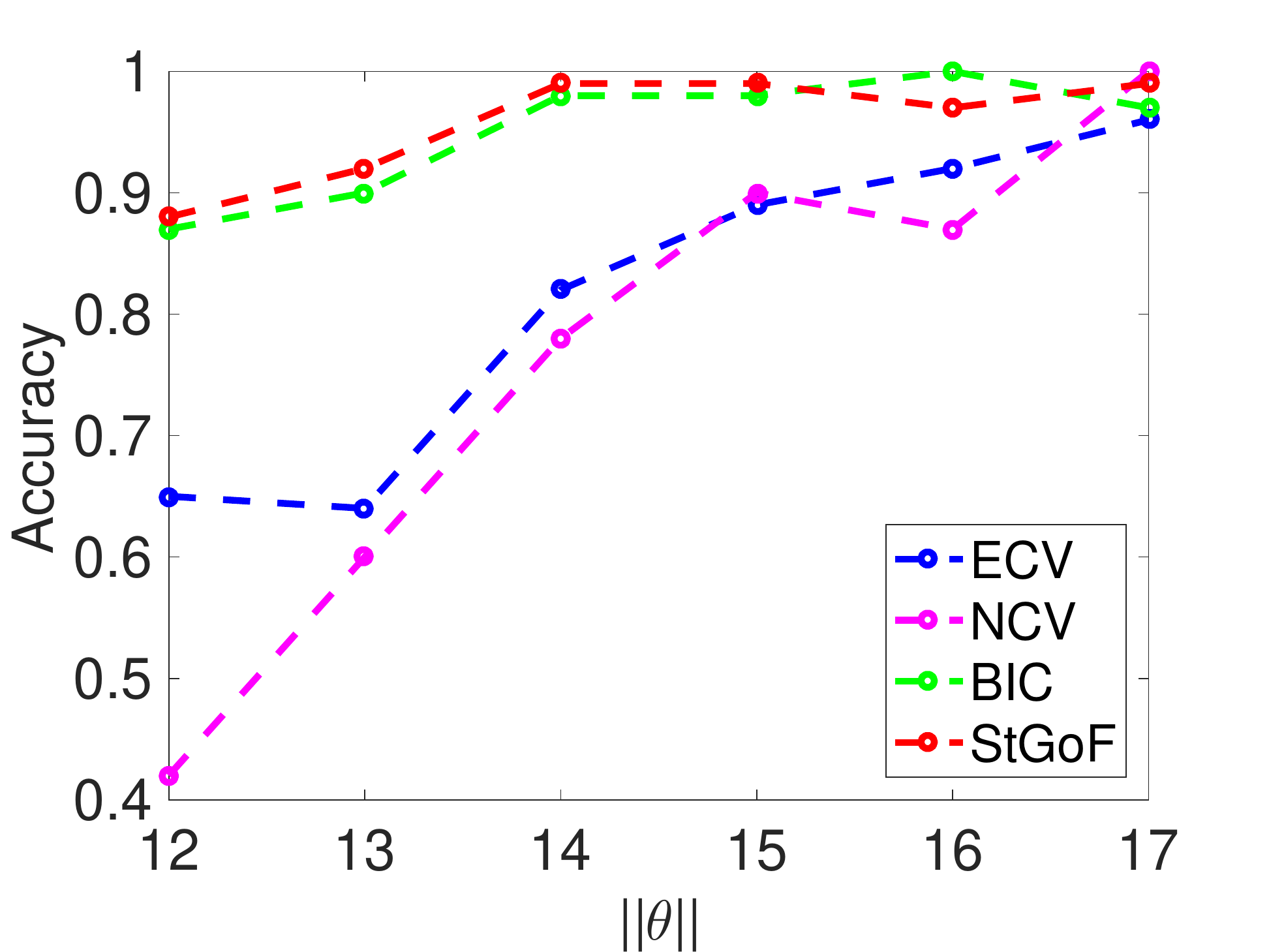}
\includegraphics[width=.32\textwidth, height = .26\textwidth, trim = 0mm 10mm 0mm 0mm, clip=true]{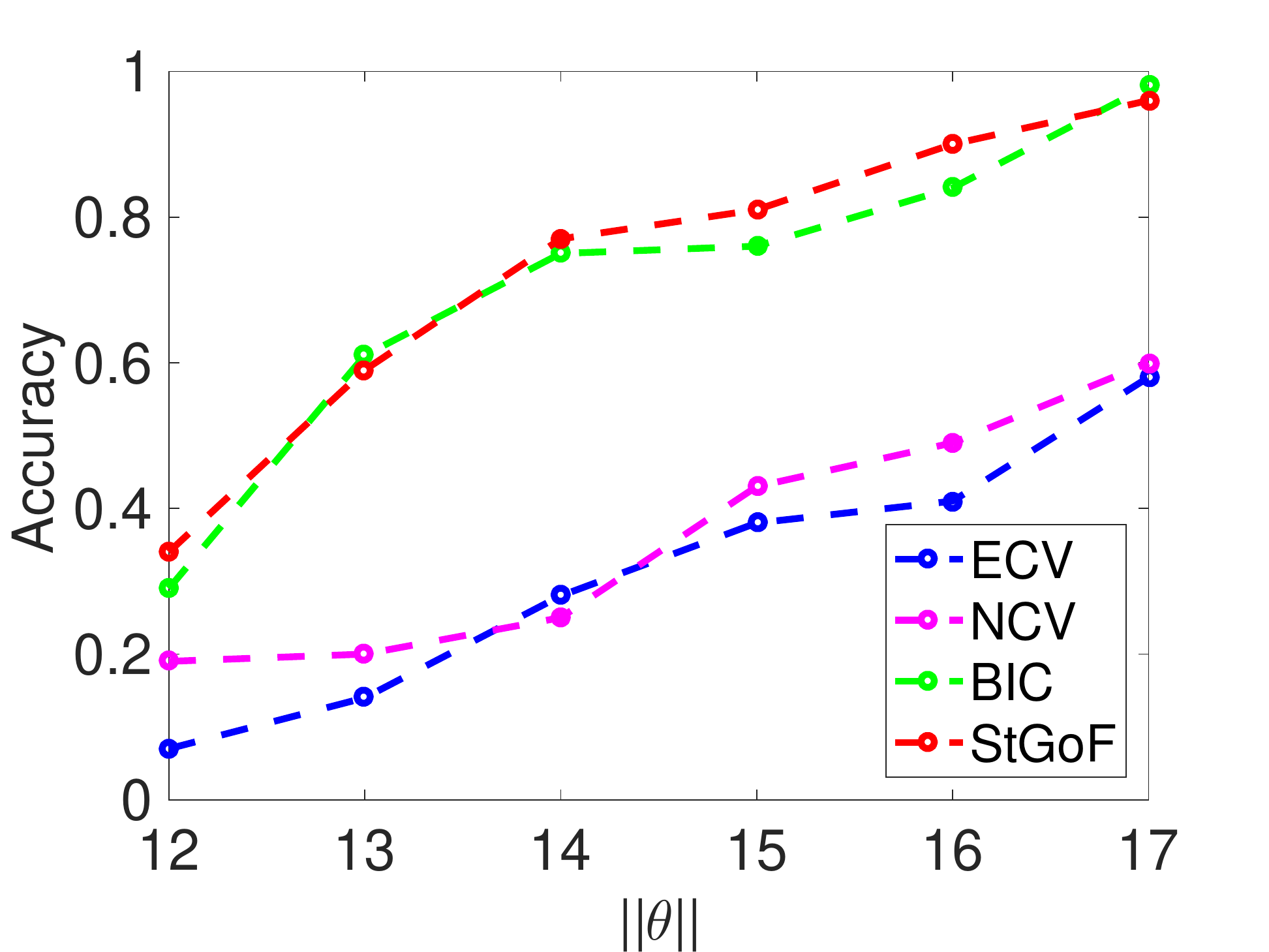}
\caption{Experiment 2a (left),  2b (middle), and 2c (right), where from 2a to 2c,  the communities sizes are more and more unbalanced. For all three panels,   $x$-axis is $\|\theta\|$ (sparsity level), and $y$-axis is the estimation accuracy over 100 repetitions ($(n,K)=(1200, 3))$.}  \label{fig:Experiment2}
\end{figure}
\spacingset{1.465}

{\bf Experiment 2}.   We study how the relative sizes of different communities affect the results and comparisons.  Given $b_n > 0$,   we set $(n,K) = (1200, 3)$,  $f(\theta)$ as $\mathrm{Pareto}(10,0.375)$, and  let $P$ be such that $P(k, \ell) = 1 -\frac{|k-\ell|(1-b_n)}{2}$, for $1\leq k,\ell\leq 3$. 
We 
let $\beta_n$ range in $\{12, 13, ..., 17\}$ and keep $(1 -b_n)\|\theta\|$ fixed at  $10$ so the SNR's are roughly at the same level.  
We take $g(\pi)$ as the distribution with weights $a$, $b$, and $(1 - a - b)$ 
on vectors $e_1, e_2, e_3$,  respectively.  Consider three sub-experiments, Exp 2a-2c, 
where we take 
$(a, b) = (.30, .35),  (.25, .375)$, and $(.20, .40)$, respectively, 
so three  communities  are slightly unbalanced,  moderately unbalanced, 
and very unbalanced, respectively.  See Figure~\ref{fig:Experiment2}.  First, StGoF consistently outperforms NCV, ECV and BIC. 
Second,  when three communities get increasingly unbalanced, 
all methods become less accurate,  suggesting that estimating $K$ gets increasingly harder.  Last, the performances of ECV and NCV are close to that of StGoF when communities are 
relatively balanced (e.g., Exp 2a), but is more unsatisfactorily when communities are more unbalanced (e.g., Exp 2b-2c).

\spacingset{1}
\begin{figure}[tb] 
\centering
\vspace{-0mm}
\includegraphics[width=.32\textwidth, height = .26\textwidth, trim = 0mm 10mm 0mm 0mm, clip=true]{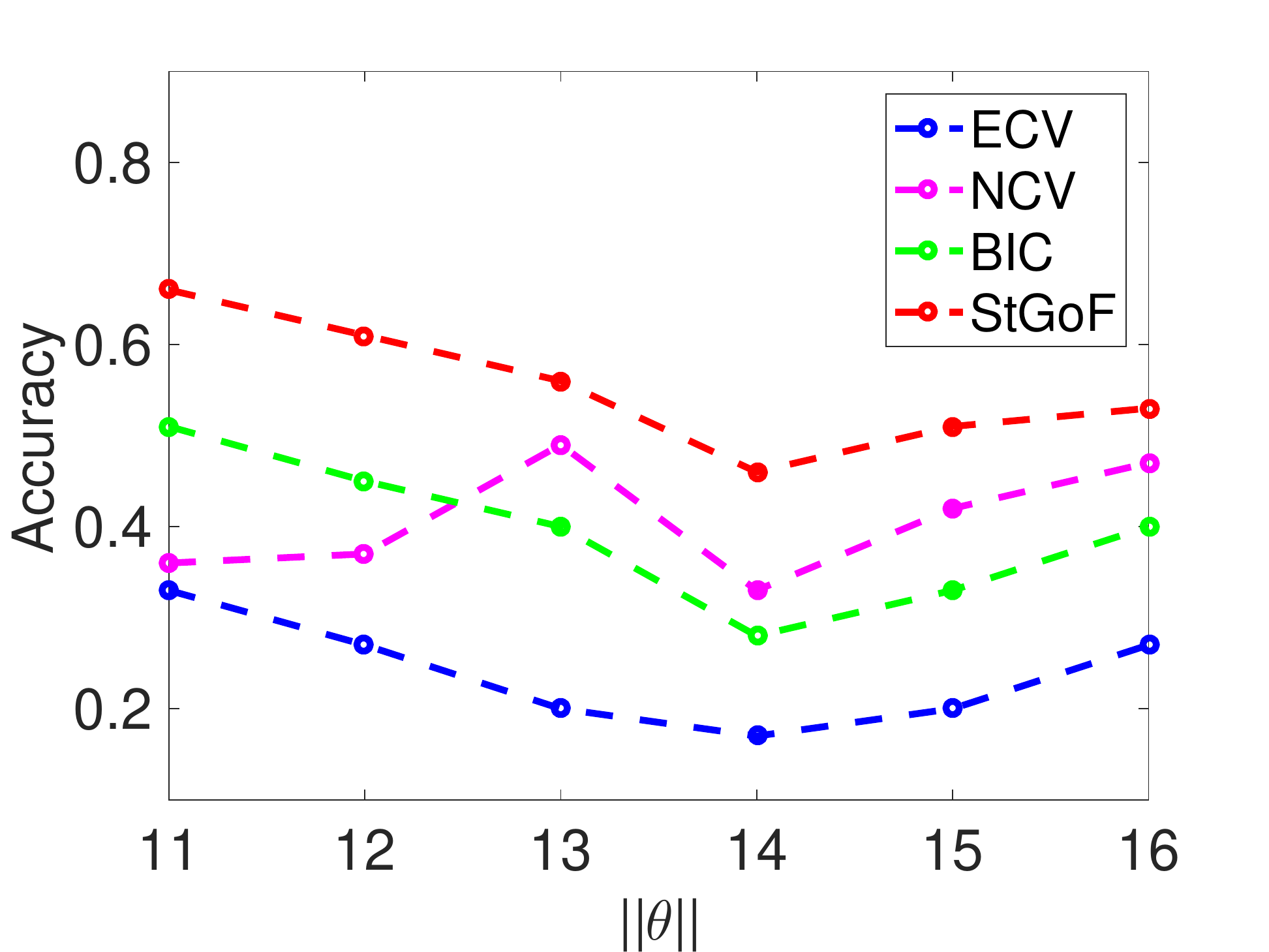}
\includegraphics[width=.32\textwidth, height = .26\textwidth, trim = 0mm 10mm 0mm 0mm, clip=true]{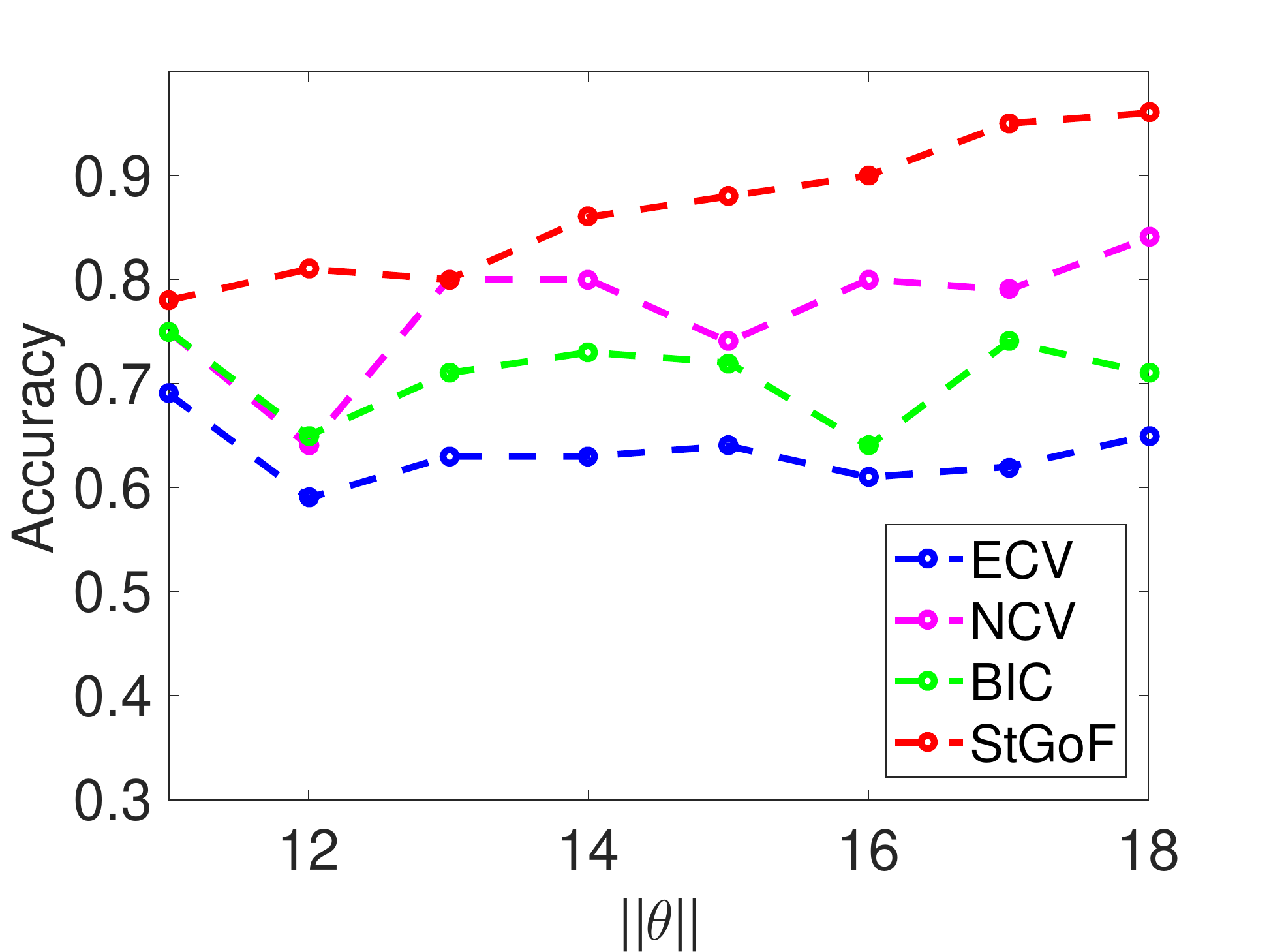}
\caption{Experiment 3a (left) and 3b (right), where 3a allows for mixed memberships and 3b allows for outlier nodes. For both panels,  $x$-axis is $\|\theta\|$ (sparsity level), and $y$-axis is the  estimation accuracy over 100 repetitions ($(n,K)=(600,4))$.}  \label{fig:Experiment3}
\end{figure}
\spacingset{1.465}


{\bf Experiment 3}.   We study robustness of the algorithms under model misspecification.  Fix $(n, K) = (600, 4)$. Let $P$ have unit diagonals and $P(k,\ell) = 1 -  \frac{(1 - b_n)(|k - \ell| + 1)}{K}$ as off-diagonals. Let $f(\theta)$ be $\mathrm{Unif}(2, 3)$.  We consider two sub-experiments, Exp 3a-3b. For sparsity, we let $\beta_n$ range from $11$ to $16$ in Exp 3a and from $11$ to $18$ in Exp 3b, while fixing $(1 -b_n)\|\theta\|=10.5$. In Exp 3a, 
we allow mixed-memberships.  Let
$g(\pi)$ to be the mixing distribution with probability $.2$ on each of $e_1, e_2, e_3, e_4$ and probability $.2$ on $\mathrm{Dirichlet}({\bf 1}_4)$.  
Once we have $\theta_i$, $\pi_i$, and $P$, let $\Omega_{ij} = \theta_i\theta_j \pi_i'  P\pi_j$, similar to that in DCBM.   In Exp 3b, we allow outliers. Let $g(\pi)$ be the mixing distribution with a point mass $.25$ on each of $e_1, e_2, e_3, e_4$, and obtain $\Omega$ as in DCBM. Let $\rho_n = \frac{1}{n} \sum_{1 \leq i, j \leq n}\Omega_{ij}$. We then  randomly select $10\%$ of nodes and re-set
$\Omega_{ij}=\rho_n$ if either of $(i,j)$ is selected.  
ECV and NCV are not model based so should be less sensitive to model misspecification; we use their results  as benchmarks to evaluate StGoF and BIC.  Figure~\ref{fig:Experiment3} shows that StGoF is not sensitive to model misspecification, and that  BIC  behaves less satisfactory here than in Experiment 1-2, and so 
is more sensitive to model misspecification.

\spacingset{1}
\begin{figure}[tb] 
\centering
\vspace{-0mm}
\includegraphics[width=.32\textwidth, height = .28\textwidth, trim = 0mm 10mm 0mm 0mm, clip=true]{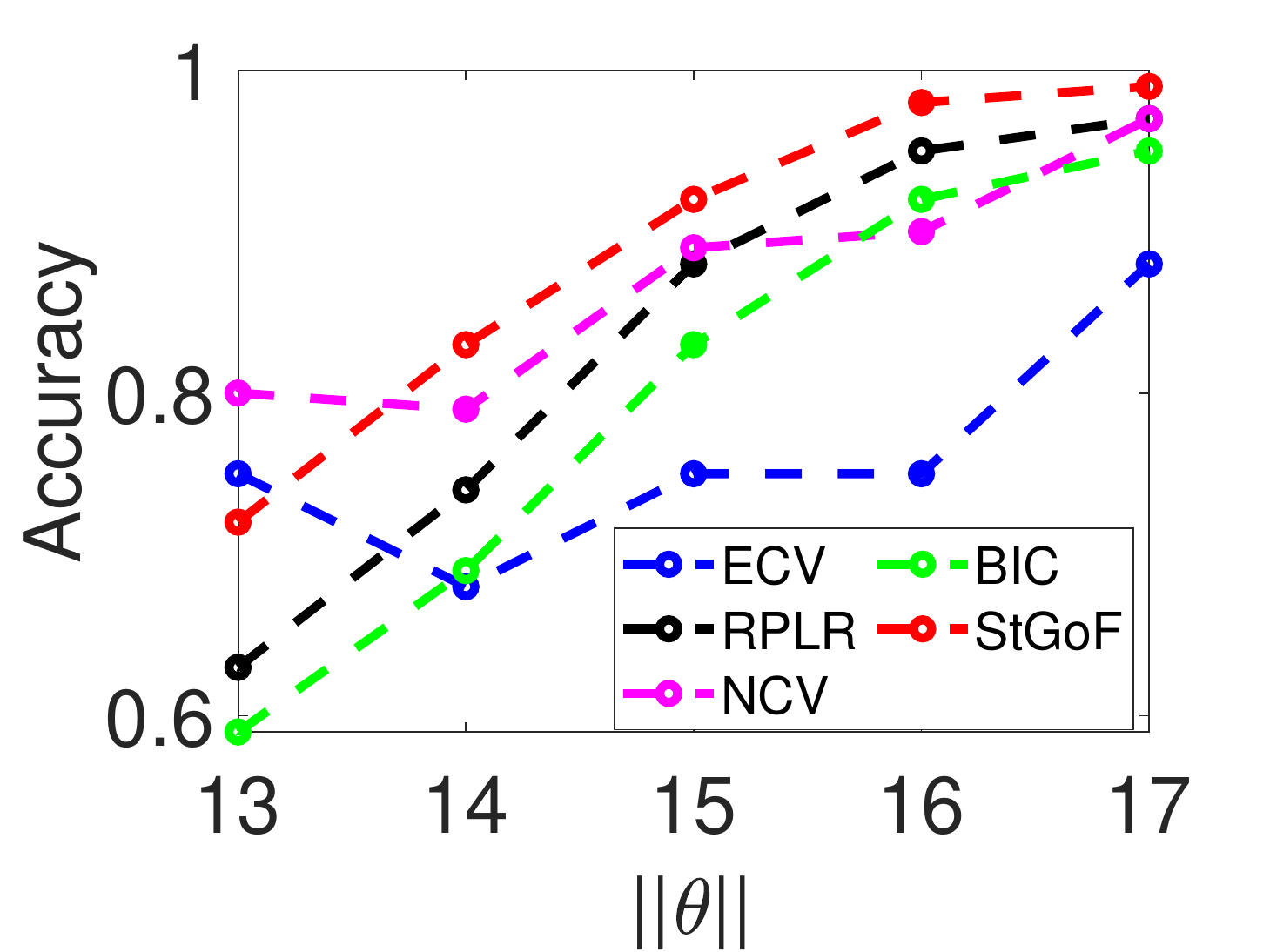}
\includegraphics[width=.32\textwidth, height = .28\textwidth, trim = 0mm 10mm 0mm 0mm, clip=true]{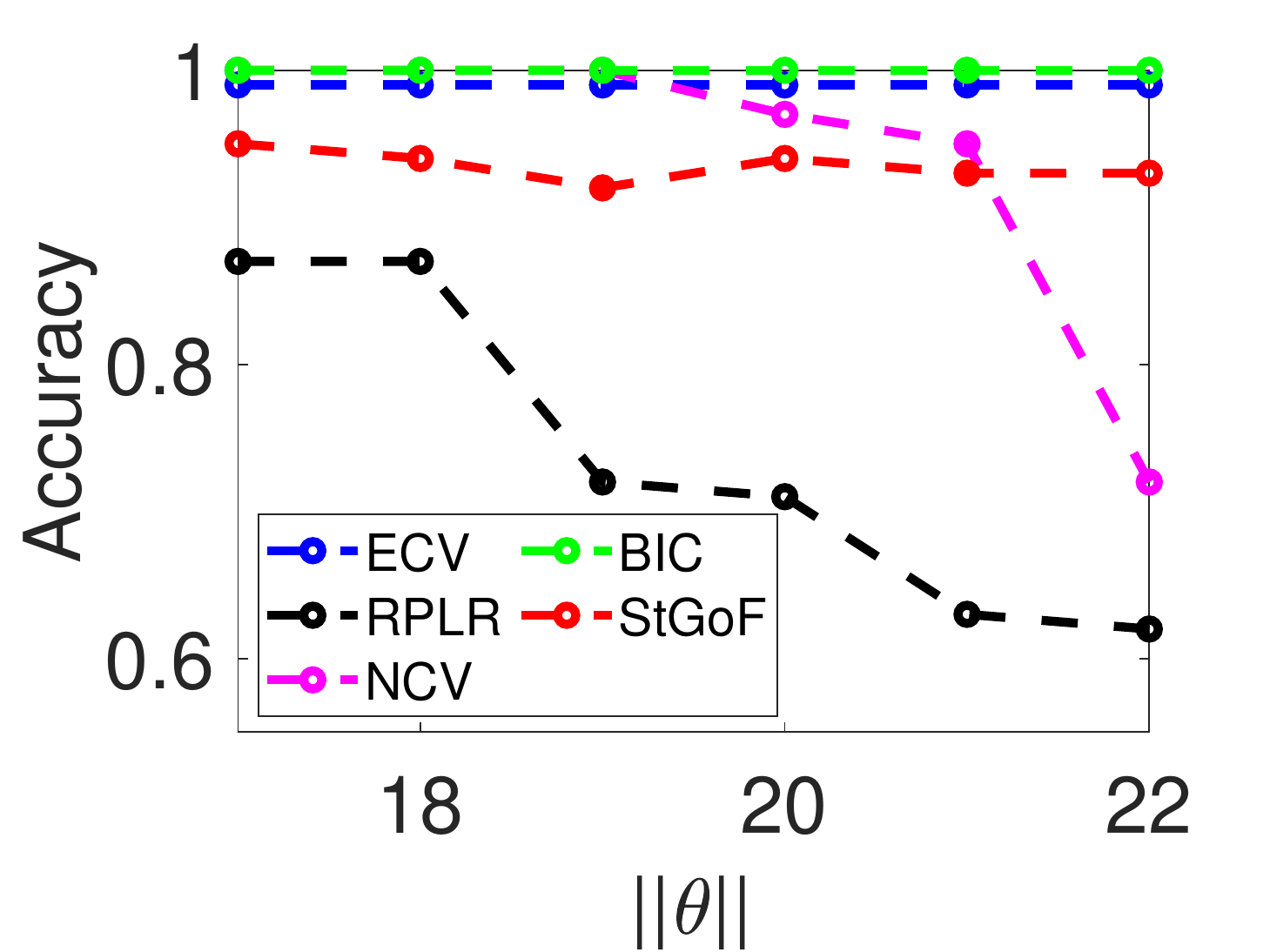}
\includegraphics[width=.32\textwidth, height = .28\textwidth, trim = 0mm 10mm 0mm 0mm, clip=true]{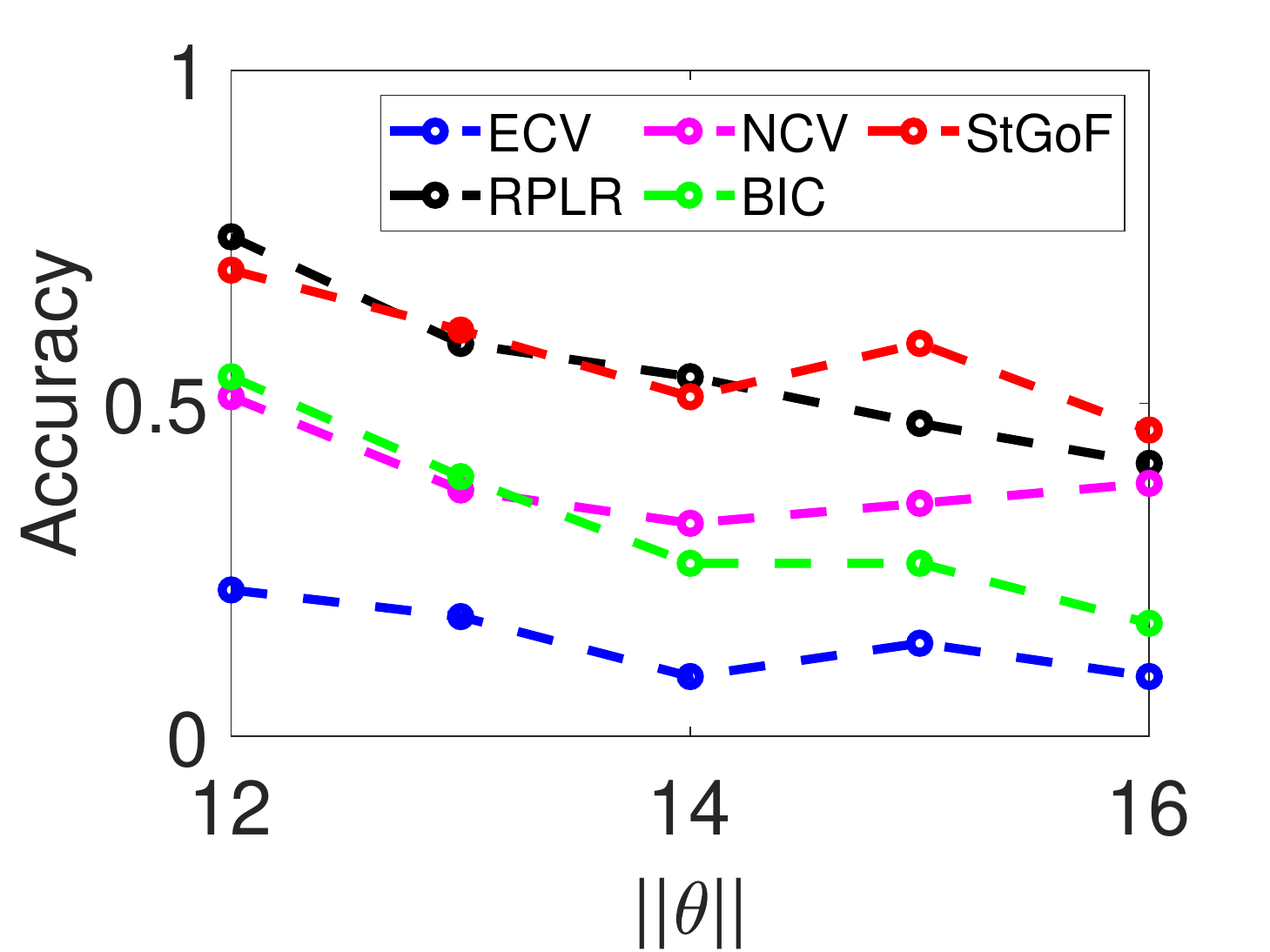}
\caption{Experiment 4a (left),  4b (middle), and 4c (right),  where $(n,K)=(600, 3), (1200,3), (1200,4)$  respectively.   For all three panels, $x$-axis is $\|\theta\|$ (sparsity level), and $y$-axis is the estimation accuracy over 100 repetitions. }  \label{fig:Experiment4}
\end{figure}
\spacingset{1.465}

{\bf Experiment 4}. We compare StGoF with RPLR  \cite{ma2018determining} (and also BIC, ECV, and NCV). RPLR has tuning parameters $(K_{\max}, c_{\eta}, h_n)$. Following \cite{ma2018determining}, we  set $(K_{\max}, c_{\eta}, h_n)=(10, 1, \bar{d}^{-1/2})$, where $\bar{d}$ is the average node degree. Consider three sub-experiments, Exp 4a-4b, covering different combinations of $(n, K, \Theta, P)$.   In Exp 4a,   $(n, K)=(600, 3)$,  $P(k,\ell) = 1 - \frac{(1 - b_n)(|k - \ell| + 1)}{K}$  if $k \neq \ell$ and 
$1$ otherwise.  We let $\|\theta\|$ vary and  
select $b_n$ such that $(1-b_n)\|\theta\|=9$, and let $f(\theta)$  be $\mathrm{Unif}(2, 3)$. In Exp 4b, $(n, K)=(1200, 3)$,   $P(k, \ell) = 1$ if $k = \ell$ and $b_n$ otherwise.    We let $\|\theta\|$ vary while keeping $(1-b_n)\|\theta\|=4.75$, and let $f(\theta)$ be $\mathrm{Unif}(3,4)$. In Exp 4c, $(n, K)=(1200, 4)$, and $P$ is the same as in Exp 4a.  We let $\|\theta\|$ vary while keeping $(1-b_n)\|\theta\|=10.5$, and let $f(\theta)$ be Pareto($10, .375$). We take 
$g(\pi)$ to be the mixing distribution which puts probability $.2$ on each of $e_1, e_2, e_3, e_4$ and $.2$ on $\mathrm{Dirichlet}({\bf 1}_4)$ (the model does not satisfy 
DCBM so we have a model misspecification).   See Figure~\ref{fig:Experiment4}. 
RPLR underperforms StGoF, especially in Exp 4b (where the first two eigenvalues of 
$\Omega$ have a relatively large gap).  This is because RPLR tends to 
estimate $K$ as the index that has the largest eigen-gap. If the largest 
eigen-gap happens at an index smaller than $K$, RPLR tends to underestimate (see 
Section \ref{sec:realdata} for more discussion).

\spacingset{1}
\begin{figure}[tb] 
\centering
\vspace{-0mm}
\includegraphics[width=.32\textwidth, height = .26\textwidth, trim = 0mm 10mm 0mm 0mm, clip=true]{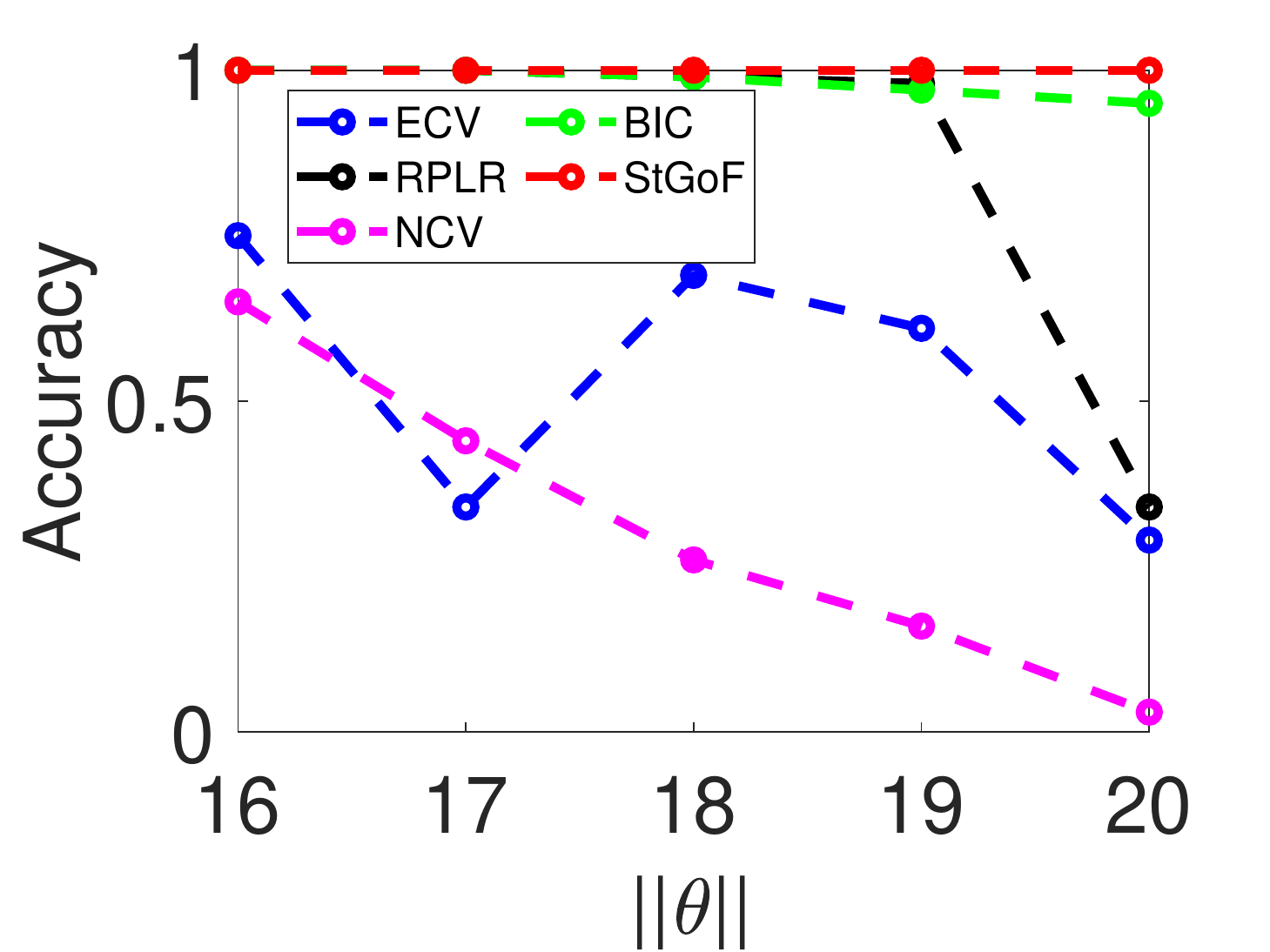}
\includegraphics[width=.32\textwidth, height = .26\textwidth, trim = 0mm 10mm 0mm 0mm, clip=true]{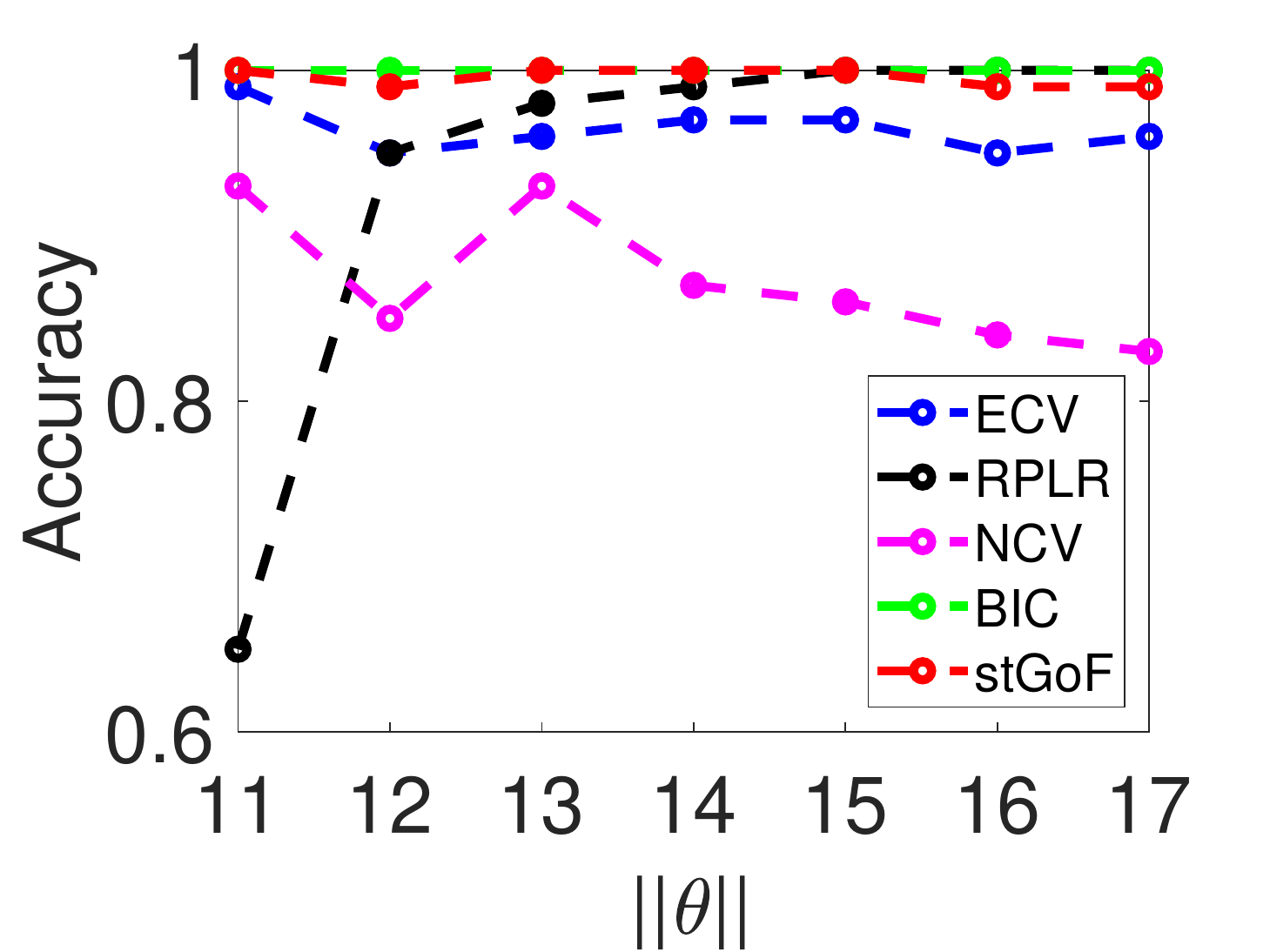}
\caption{Experiment 5a (left) and 5b (right), where $(n, K)=(600, 6)$ for 5a and $(n,K)=(600, 8)$ for 5b. For both panels,  $x$-axis is $\|\theta\|$ (sparsity level), and $y$-axis is the  estimation accuracy over 100 repetitions.}  \label{fig:Experiment5}
\end{figure}
\spacingset{1.465}

{\bf Experiment 5}. We study two sub-experiments, Exp 5a-5b, for settings with a larger $K$. In Exp 5a, $(n,K)=(600, 6)$. We let $P$ have $1$ in the diagonal and $P(k,\ell) = 1 - \frac{(1 - b_n)(|k - \ell| + K - 1}{2K}$ in the off-diagonal, and take $f(\theta)$ as the two-point mixture $0.95\delta_1 + 0.05\delta_{2}$. We vary $\|\theta\|$ and select $b_n$ such that $(1-b_n)\|\theta\|=15.5$. In Exp 5b, $(n, K)=(600, 8)$, $P$ has unit diagonals and $b_n$ in the off-diagonal, and $f(\theta)$ is the same as in Exp 5a.  As $\|\theta\|$ vary, we select $b_n$ such that  $(1-b_n)\|\theta\|=10.5$. See Figure~\ref{fig:Experiment5}. Note that ECV and NCV are cross-validation approaches, which may be less satisfactory for larger $K$. 
%

{\bf Remark 3}.  StGoF may estimate $K$ incorrectly if some regularity conditions are violated. 
If this happens, we may either underestimate or overestimate $K$, depending 
on the data set.  E.g., 
if the network has very weak signals (i.e., $|\lambda_K| / \sqrt{\lambda_1}$ is small), StGoF may underestimate 
$K$, and if the model is misspecified (say, due to many outliers), StGoF may overestimate $K$.

\section{Real data analysis}  \label{sec:realdata} 

In theory,  a good approximation for the null distribution 
of $\psi_n^{(m)}$ is $N(0,1)$ (see Theorem \ref{thm:CLB}), but     
such a result requires some model assumptions, which may be violated in 
real applications (e.g., outliers, artifacts). We thus propose a modification of StGoF using the idea of empirical null \cite{Efron}.  
Under model misspecification, a good approximation 
for the null distribution of $\psi_n^{(m)}$ is no longer $N(0,1)$ (i.e., theoretical null), but $N(u, \sigma^2)$ (i.e., empirical null) 
for some $(u, \sigma) \neq (0,1)$.
Efron \cite{Efron} 
argued that due to artifacts or model misspecification,  the {\it empirical null} frequently 
works better for real data than the {\it theoretical null}.  
The problem is then how to estimate the parameters $(u, \sigma^2)$ of the empirical null.  

We propose a bootstrap approach to estimating $(u, \sigma^2)$.  Recall that 
$\hat{\lambda}_k$ is the $k$-th largest eigenvalue of $A$ and $\hat{\xi}_k$ is the 
corresponding eigenvector. Fixing $N > 1$ and $m > 1$,  letting $\widehat{M}^{(m)} =  \sum_{k =1}^m \hat{\lambda}_k \hat{\xi}_k \hat{\xi}_k'$ and 
let $\widehat{S}^{(m)} = A - \widehat{M}^{(m)}$.  
For $b = 1, 2, \ldots, N$,  we simultaneously permute the rows and columns of $\widehat{S}^{(m)}$ and denote the resultant matrix by $\widehat{S}^{(m, b)}$.   Truncating all entries of $(\widehat{M}^{(m)} + \widehat{S}^{(m, b)})$ at $1$ at the top and $0$ at the bottom, and denote the resultant matrix by $\widehat{\Omega}^{(b)}$.  Generate an adjacency matrix $A^{(b)}$ such that for all $1 \leq i < j \leq n$,  $A_{ij}^{(b)}$ are independent Bernoulli samples with parameters $\widehat{\Omega}_{ij}^{(b)}$  
(we may need to repeat this step until the network is connected).  
 Apply StGoF to $A^{(b)}$ and denote the  resultant statistic by $Q_n^{(b)}$.  We estimate $u$ 
and $\sigma$ by the empirical mean and standard deviation of $\{Q_n^{(b)}\}_{b = 1}^{N}$, respectively.   
Denote the estimates by $\hat{u}^{(m)}$ and $\hat{\sigma}^{(m)}$, respectively.  
 The bootstrap  StGoF statistic  is then $\psi_n^{(m,*)}  = [Q_n^{(m)} - \hat{u}^{(m)}] / \hat{\sigma}^{(m)}$, $m = 1, 2, \ldots$,   
where $Q_n^{(m)}$ is the same as in (\ref{refitting2d}).   
Similarly, we estimate $K$ as the smallest integer $m$ such that $\psi_n^{(m, *)} \leq z_{\alpha}$, for the same $z_{\alpha}$ in StGoF.   We recommend $N = 25$, as it usually gives stable estimates for $\hat{u}^{(m)}$ and $\hat{\sigma}^{(m)}$.   
We call this method the bootstrap StGoF (StGoF*).

We consider $6$ data sets as in Table \ref{tab:realdata1}, which can be downloaded from \url{http://www-personal.umich.edu/~mejn/netdata/}.   We now discuss the true $K$.   
For the dolphin network, it was argued in \cite{dolphin5} that both $K=2$ or $K=4$ are reasonable.  For UKfaculty network, we symmetrize the network by ignoring the directions of the edges.  There are $4$ school affiliations for the faculty members so we take $K = 4$. 
For the Football network, we take $K = 11$. The network was manually labelled as $12$ groups, but the $12^{th}$ group only consists of the $5$ ``independent''   teams that do not belong to any conference and do not form a conference themselves.  For the Polbooks network,   Le and Levina  \cite{le2015estimating} suggest that $K = 3$, but  it was argued by \cite{Mixed-SCORE} that  a more appropriate model for the network is a degree corrected mixed-membership (DCMM) model with two communities, so $K = 2$ is also appropriate.

We compare StGoF and StGoF* with the BIC \cite{wang2017likelihood}, BH \cite{le2015estimating}, ECV \cite{li2020network}, NCV \cite{chen2018network}  and RPLR \cite{ma2018determining}.  The first $4$ methods are implemented via  the R package ``randnet''.  Among them,  ECV and NCV are cross validation (CV) approaches and the results vary from one repetition to the other. Therefore,  we run each method for $25$ times and report the mean and SD. 
The StGoF* uses bootstrapping mean and standard deviation and  is also random, but the SDs are $0$ for five data sets.   Most methods require a feasible range of $K$ a priori (say, $\{1, 2, \ldots, k_{max}\}$, where $k_{max}$ is a prescribed upper bound for $K$.  For the $6$ data sets considered here, the largest (true) $K$ is $11$,  so we take $k_{max} = 15$. 

In Section \ref{sec:Simul}, we mention that RPLR tends to underestimate $K$ 
if the largest eigen-gap of $\Omega$ happens at an index smaller than $K$. 
This seems to be the case for Football, where RPLR significantly underestimates. 
RPLR also has a (seemingly fixable) coding issue:  the code (generously shared by the authors)  
may report an error message and does not output an estimate for $K$ (e.g., if we apply it to the  
Dolphins,  Karate, and UKfaculty with $k_{max} = 15$ for $500$ times, then   
in $63\%$,  $100\%$, and $96\%$ of the times respectively, the code reports an error and does not 
output an estimate for $K$).   If we take $k_{max} = 10$ for Dolphins and UKfaculty and take 
$k_{max} = 5$ for Karate, then the error  
messages disappear and the estimated $K$ are $2, 4, 2$ for Dophis, UKfaculty, and Karate, respectively.

\spacingset{1}
\begin{table}[tb!]
\centering
\scalebox{.85}{
\setlength\tabcolsep{4pt}
\hspace*{-1cm}
\begin{tabular}{| lcc |ccccccc|}
\hline
Name & $n$  & $K$ & BIC & BH & ECV      & NCV  & RPLR & StGoF & StGoF* \\
\hline
Dolphins &62  & 2, 4 & 2  & 2  &      3.08(.91) [2,5]     &    2.20(2.71)  [1,15]   & 2       & 2 & 3 \\
Football & 115  & 11 & 10 & 10 &       11.28(.61) [11,13]  &    12.36(1.15) [11,15]    & 2  & 10 &  10  \\
Karate & 34  &2 &2  & 2  &     2.60(1.0) [1,6] &    2.56(.58) [2,4]    & *      & 2  &  2  \\
UKfaculty & 81 & 4 & 4 & 3 &      5.56(1.61) [3,11]  &    2.40(.28)  [2,3]    & 4    & 4  & 4  \\
Polblogs  & 1222 & 2 &6 & 8 &     4.88(1.13) [4, 8]&      2(0)  [2, 2]    & 2  & 2* &  2 \\ 
Polbooks  & 105 &2, 3 &  3 & 4 &     7.56(2.66) [2,15] &  2.08(.71) [2,5]   &  3     & 5 & 2.4(.25) [2,3] \\
 \hline
\end{tabular}}\label{tab:realdata1}
\caption{Comparison of estimated $K$. Take ECV for Dolphins for example: for $25$ independent repetitions,  the mean and SD of estimated $K$ are $3.08$ and  $0.91$,  ranging from $2$ to $5$ (the SDs of StGoF* are $0$ for the first $5$ data sets).  For Karate, RPLR (with $k_{max} = 15$) reports an error message without  an estimated $K$; the error messages disappear if we take $k_{max} = 5$, where the estimated $K$ is $2$.  See the text for more discussion.}
\end{table}
\spacingset{1.465}

The Polblogs network is suspected to have outliers,  so most methods do not work well. 
For this particular network, the mean of StGoF is much larger than expected,  
so we choose to estimate $K$ by the $m$ that minimizes $\psi_n^{(m)}$ for $1 \leq m \leq 15$ (for this reason, we put a $*$ next to $2$ in the table).  Note that StGoF* correctly estimates $K$ as $2$.  
The Polbooks network is suspected to have a signifiant faction of mixed nodes \cite{Mixed-SCORE},  which explains 
why StGoF overestimates $K$. Fortunately,  for both data sets, StGoF* estimates $K$ correctly, 
suggesting that the bootstrapping means and standard deviations help standardize $Q_n^{(m)}$.

\vspace{-0.5 em} 

\section{Discussions}  \label{sec:Discu} 
How to estimate $K$ is a fundamental problem in network analysis. 
We propose StGoF as a new stepwise algorithm for estimating $K$, which (a) 
has $N(0,1)$ as its limiting null, (b) is uniformly consistent in a setting 
much broader than those considered in the literature,  and (c) achieves the optimal phase transition. 
The  results, especially (a) and (c), do not exist before.  Analysis of stepwise algorithms of this kind 
is known to face challenges. We overcome them by using a different stepwise scheme and by deriving 
sharp results, where the key is to prove the NSP of SCORE;  we prove the NSP with new ideas and techniques.  

We discuss some open questions.  First, in this paper, we are primarily interested in DCBM, but the idea can be extended to the broader DCMM, where mixed-memberships exist. 
To this end, we need to replace SCORE by 
Mixed-SCORE \cite{Mixed-SCORE} (an adapted version of 
SCORE for networks with mixed memberships),  and modify the 
refitting step accordingly.  In this case, whether NSP continues to hold is unclear,  but 
we may have a revised version of NSP that holds for all pure nodes (i.e., nodes 
without mixed-memberships) and we can then use it to study the mixed nodes. 
The analysis of the resultant procedure is much 
more challenging so we leave it to the future.  
Second, in this paper, we assume $K$ is fixed. For diverging $K$, 
the main idea of our paper continues to be valid, but we need to 
revise several things (e.g., definition of consistency and SNR, 
some regularity conditions, phase transition) to reflect the role of $K$. 
The proof for the case of diverging $K$ can be much more tedious, but 
aside from that, we do not see a major technical hurdle. 
Especially, the NSP of SCORE continues to hold for a diverging $K$. Then,  with some mild 
conditions, we can show that $\widehat{\Pi}^{(m)}$ has very few realizations, 
so the analysis of StGoF is readily extendable. That we assume $K$ as fixed is not only for simplicity but 
also for practical relevance.  For example, real networks may have hierarchical tree structure, and in each layer,  the number of leaves (i.e., clusters) is small (e.g.,  \cite{SCC-JiJin}).  
Therefore, we have small $K$ in each layer when we perform hierarchical network analysis.  Also, the goal of real applications is to have interpretable results. For example, for community detection, results with a large $K$ is hard to interpret, so we may prefer a DCBM with a small $K$ to an 
SBM with a large $K$. In this sense, a small $K$ is practically more relevant. 
Last,  while the NSP of SCORE largely facilitates the analysis, it does not mean that  StGoF ceases to work well once NSP does not hold; 
it is just harder to analyze in such cases. Our study suggests 
that StGoF continues to behave well even when NSP does not hold exactly. 
How to analyze StGoF in such cases is an interesting problem for 
  the future.

\newpage
  \begin{center}
    {\LARGE\bf Supplement of ``Optimal Estimation of the Number of  Communities"}
\end{center}

%
%
%
\spacingset{1.48} 

This supplementary material contains the proofs of theorems and lemmas in the main article \cite{EstK}. 
Section~\ref{sec:proof} proves properties of the RQ test statistic, including Theorems \ref{thm:CLB} and \ref{thm:CUB}; the proof uses many supplementary lemmas, whose proofs are also included in this section. Section~\ref{sec:proof-B} proves the non-splitting property of SCORE, including Theorem~\ref{thm:SCORE2}, Lemmas~\ref{lem:entrywise-bound}-\ref{lem:d_m}, and Theorem~\ref{thm:kmeans}. 
Section~\ref{sec:proof-A} proves the lower bound arguments, including Theorems~\ref{thm:LB} and \ref{thm:LB2}.
Section~\ref{sec:proof-compu} 
proves Lemma~\ref{lemma:complexity}. 
 

\tableofcontents

\appendix

\section{Proofs of properties of the RQ test statistic}  \label{sec:proof}
In this section, we prove Theorems \ref{thm:CLB} and \ref{thm:CUB}.  
Corollaries~\ref{cor:CLB}-\ref{cor:CUB} follow directly from Theorems \ref{thm:CLB} and \ref{thm:CUB}, 
respectively, so the proofs are omitted.


\subsection{Proof of Theorem \ref{thm:CLB} (the null case of $m = K$)}  \label{subsec:SgnQproof1} 
First, it is seen that the first item is a direct result of Theorem \ref{thm:SCORE2}. Second, by definitions, 
\[
\mathbb{P}(\widehat{K}_{\alpha}^* \leq K) \geq \mathbb{P}(\psi_n^{(K)} \leq z_{\alpha}), 
\] 
and so the last item follows once the second item is proved. Therefore, we only need to show the second item. 
Recall that when $m=K$,  
\[ 
\psi_n^{(K)} = [Q_n^{(K)} - B_n^{(K)}]/\sqrt{8 C_n},  
\] 
where $Q_n^{(K)}$, $B_n^{(K)}$, and $C_n$ are defined  in \eqref{refitting2a}, \eqref{refitting2b} and \eqref{refitting2c}, respectively,  which we reiterate below:
\[
Q_n^{(K)} = \sum_{i_1, i_2, i_3, i_4 (dist)} (A_{i_1i_2} - \widehat\Omega_{i_1i_2}^{(K)}) (A_{i_2i_3} - \widehat\Omega_{i_2i_3}^{(K)} ) (A_{i_3i_4} - \widehat\Omega_{i_3i_4}^{(K)}) (A_{i_4i_1} - \widehat\Omega_{i_4i_1}^{(K)}),  
\]
\[
C_n = \sum_{i_1, i_2, i_3, i_4 (dist)} A_{i_1i_2}  A_{i_2i_3} A_{i_3i_4} A_{i_4i_1}, 
\quad B_n^{(K)} = 2 \|\hat{\theta}\|^4 \cdot [ \hat{g}' \widehat{V}^{-1} (\widehat{P}\widehat{H}^2\widehat{P}\circ \widehat{P}\widehat{H}^2\widehat{P}) \widehat{V}^{-1} \hat{g}]. 
\]
In the first equation here, $\widehat{\Omega}^{(K)}$ depends on the estimated community label matrix  $\widehat{\Pi}^{(K)}$.  To facilitate the analysis, it's desirable to replace $\widehat{\Pi}^{(K)}$ by  the true  membership matrix $\Pi$. By the first claim of the current theorem,  this replacement only has a  negligible effect. 

Formally,  we introduce  $\widehat{\Omega}^{(K,0)}$ to be the proxy of $\widehat{\Omega}^{(K)}$ with $\widehat{\Pi}^{(K)}$ in its definition replaced by $\Pi$. Moreover, define  $Q_n^{(K,0)}$ to be the proxy of $Q_n^{(K)}$ with $\widehat{\Omega}^{(K)}$ replaced by $\widehat{\Omega}^{(K,0)}$ in its definition, and 
define the corresponding counterpart of $\psi_{n}^{(K)}$ as
\[
  \psi_n^{(K,0)} = [Q_n^{(K,0)} - B_n^{(K)}]/\sqrt{8 C_n}.
\]
Then, for any fixed number $t\in\mathbb{R}$ we have
\[
  \left| \mathbb{P}(\psi_n^{(K)}\leq t) - \mathbb{P}(\psi_n^{(K,0)}\leq t) \right|\leq \mathbb{P}(\widehat{\Pi}^{(K)}\neq \Pi)\to0,\qquad
  \text{as }n\to\infty,
\]
where the last step follows from the first claim in the current theorem. 
Hence by elementary probability,  to prove $\psi_n^{(K)}\to N(0,1)$ in law, it suffices to  show $\psi_n^{(K,0)}\to N(0,1)$ in law.

Recall that if we neglect the difference in the main diagonal entries, then $A - \Omega = W$.  By definition, we expect that $\widehat{\Omega}^{(K, 0)}   \approx \Omega$, and so  $(A - \widehat{\Omega}^{(K, 0)}) \approx W$.  This motivates us to define 
\begin{equation} \label{DefinewidetildeQ}
\widetilde{Q}_n = \sum_{i_1, i_2, i_3, i_4 (dist)} W_{i_1i_2} W_{i_2i_3} W_{i_3i_4} W_{i_4i_1}. 
\end{equation}
At the same time, for short, let  $b_n$ and $c_n$ be the oracle counterparts of $B_n^{(K)}$ and $C_n$
\beq\label{def:bncn}
c_n = \sum_{i_1, i_2, i_3, i_4 (dist)}\Omega_{i_1i_2}\Omega_{i_2i_3}\Omega_{i_3i_4}\Omega_{i_4i_1}, \qquad b_n = 2 \|{\theta}\|^4 \cdot [ {g}' {V}^{-1} ({P}{H}^2{P}\circ {P}{H}^2{P}){V}^{-1} {g}].
\eeq
Here, two vectors $g, h\in\mathbb{R}^K$ are defined as $g_k = ({\bf{1}}_k' \theta)/ \|\theta\|_1$ and $h_k = ({\bf{1}}_k' \Theta {\bf{1}}_k)^{1/2}/ \|\theta\|$, where ${\bf 1}_k$ is for short of ${\bf 1}^{(K)}_k$, which is defined as
\[
  {\bf 1}_k^{(K)}(i) = 1\mbox{ if $i \in \mathcal{N}_k$ and 0 otherwise}.  
\] 
Moreover, $V = \diag(Pg)$, and $H = \diag(h)$.
We have the following lemmas, whose proofs can be found in Sections~\ref{subsec:SgnQ1a-proof}-\ref{subsec:SgnQ1d-proof}, respectively. 
\begin{lemma} \label{lemma:SgnQ1a}  
Under the conditions of Theorem \ref{thm:CLB}, we have $\mathbb{E}[C_n] = c_n\asymp\|\theta\|^8$ and  $\mathrm{Var}(C_n) \leq C \|\theta\|^8\cdot[1+\|\theta\|_3^6]$, and so
$C_n / c_n \goto 1$ in probability for $c_n$ defined in \eqref{def:bncn}. 
\end{lemma} 
\begin{lemma} \label{lemma:SgnQ1b}  
Under the conditions of Theorem \ref{thm:CLB},  
$\widetilde{Q}_n/\sqrt{8c_n} \goto N(0,1)$ in law.  
\end{lemma} 
\begin{lemma} \label{lemma:SgnQ1c}  
Under the conditions of Theorem \ref{thm:CLB},  
$\mathbb{E}(Q_n^{(K,0)}-\widetilde{Q}_n -b_n)^2 = o(\|\theta\|^8)$.  
\end{lemma} 
\begin{lemma} \label{lemma:SgnQ1d}  
Under the conditions of Theorem \ref{thm:CLB}, we have $b_n\asymp\|\theta\|^4$ and $B_n^{(K)}/b_n\goto 1$  in probability for $b_n$ defined in \eqref{def:bncn}. 
\end{lemma} 

We now prove Theorem \ref{thm:CLB}. Rewrite $\psi_n^{(K,0)}$ as 
\begin{equation} \label{SgnQ1pf2} 
\sqrt{\frac{c_n}{C_n}} 
\biggl[\frac{\widetilde{Q}_n }{\sqrt{8c_n}} + 
\frac{(Q_n^{(K,0)}  - \widetilde{Q}_n - b_n)}{\sqrt{8c_n}} + 
\frac{(b_n - B^{(K)}_n)}{\sqrt{8c_n}} \biggr]  = \sqrt{\frac{c_n}{C_n}} \cdot [(I) + (II) + (III)],  
\end{equation} 
where $(I) = \widetilde{Q}_n / \sqrt{8c_n}$, $(II) = (Q_n^{(K,0)}  - \widetilde{Q}_n-b_n)/\sqrt{8c_n}$, and 
$(III) = (b_n - B_n^{(K)})/\sqrt{8c_n}$. 
Now, first by Lemmas~\ref{lemma:SgnQ1a}-\ref{lemma:SgnQ1b}, 
\begin{equation} \label{SgnQ1pf2a} 
c_n / C_n \goto 1  \;\; \mbox{in probability}, \qquad \mbox{and} \qquad 
 (I) \goto N(0,1) \;\; \mbox{in law}.  
\end{equation} 
Second,  by Lemma \ref{lemma:SgnQ1b}, 
\begin{equation} \label{SgnQ1pf2b} 
\mathbb{E}[(II)^2] \leq (8c_n)^{-1} \cdot\mathbb{E}[(Q^{(K,0)}_n - \widetilde{Q}^{(K,0)}_n - b_n)^2] \leq c_n^{-1}\cdot o(\|\theta\|^8),   
\end{equation} 
where the right hand side is $o(1)$ for $c_n\asymp\|\theta\|^8$ by Lemma~\ref{lemma:SgnQ1a}.  
Last,  by Lemma \ref{lemma:SgnQ1a}- \ref{lemma:SgnQ1d}, we have $b_n \asymp \sqrt{c_n}\asymp\|\theta\|^4$ and $B_n^{(K)}/b_n\stackrel{p}{\goto}1$, 
and so 
\begin{equation} \label{SgnQ1pf2c} 
(III) = \bigg(\frac{b_n}{\sqrt{8c_n}}\bigg) \cdot \bigg(\frac{B^{(K)}_n}{b_n} - 1\bigg)\stackrel{p}{\goto} 0.
\end{equation} 
Inserting (\ref{SgnQ1pf2a})-(\ref{SgnQ1pf2c}) into (\ref{SgnQ1pf2}) gives the claim and concludes the proof 
of Theorem \ref{thm:CLB}.    \qed

\subsection{Proof of Theorem \ref{thm:CUB} (the under-fitting case of $m < K$)}  \label{subsec:SgnQproof2} 
In the proof of Theorem~\ref{thm:CLB}, we start from replacing $\widehat{\Pi}^{(K)}$ with the true community label matrix $\Pi$. However, when $m<K$, $\widehat{\Pi}^{(m)}$ does not concentrate on one particular label matrix. Below, we introduce a collection of label matrices, ${\cal G}_m$, consisting of all possible realizations of $\widehat{\Pi}^{(m)}$ when NSP holds. We then study the GoF statistic on the event that $\widehat{\Pi}^{(m)}=\Pi_0$, for a fixed $\Pi_0\in {\cal G}_m$. 

Recall that $\Pi$ is the true community label matrix.  Fix $1 \leq m < K$.   
Let ${\cal G}_m$ be the class of  $n \times m$ matrices  $\Pi_0$, where each $\Pi_0$ is formed as follows: 
let $\{1, 2, \ldots, K\} = S_1 \cup S_2 \ldots \cup S_m$ be a partition, 
column $\ell$ of $\Pi_0$ is the sum of all columns of $\Pi$ in $S_{\ell}$, $1 \leq \ell \leq m$.   
Let $L_0$ be the $K \times m$ matrix of $0$ and $1$ where   
\begin{equation} \label{def:Gm} 
L_0(k, \ell) = 1 \;  \mbox{if and only if} \;   k \in S_{\ell}, \qquad 1 \leq k \leq K, \; 1 \leq \ell \leq m.  
\end{equation} 
Therefore, for each $\Pi_0 \in {\cal G}_m$, we can find an $L_0$ such that $\Pi_0  = \Pi L_0$. 
Note that each $\Pi_0$ is the community label matrix where each community implied by it (i.e., ``pseudo community'')  is formed by merging one or more (true) communities of the original network.  

Fix a $\Pi_0$ and let $\mathcal{N}_1^{(m,0)}, \mathcal{N}_2^{(m,0)}, \cdots, \mathcal{N}_m^{(m,0)}$ be the $m$ ``pseudo  communities" associated with $ \Pi_0$. 
Recall that $\hat{\theta}^{(m)}$, $\widehat{\Theta}^{(m)}$ and $\widehat{P}^{(m)}$ are refitted quantities obtained by  using the adjacency matrix $A$ and  $\widehat{\Pi}^{(m)}$; see (\ref{refitting1aADD})-(\ref{refitting1a}).  To misuse the notations a little bit, 
  let $\hat{\theta}^{(m,0)}$, $\widehat{\Theta}^{(m,0)}$ and $\widehat{P}^{(m,0)}$ be the  proxy of 
$\hat{\theta}^{(m)}$, $\widehat{\Theta}^{(m)}$ and $\widehat{P}^{(m)}$ respectively, constructed 
similarly by (\ref{refitting1aADD})-(\ref{refitting1a}), but with $\widehat{\Pi}^{(m)}$ replaced by $\Pi_0$.
Introduce 
\begin{equation} \label{DefinehatOmegaL} 
\widehat{\Omega}^{(m,0)}   = \widehat{\Theta}^{(m,0)}\Pi_0 \widehat{P}^{(m,0)}\Pi_0'\widehat{\Theta}^{(m,0)},  
\end{equation} 
\[ 
Q_n^{(m,0)}  = \sum_{i_1, i_2, i_3, i_4 (dist)} (A_{i_1i_2} - \widehat{\Omega}^{(m,0)}_{i_1i_2}) (A_{i_2i_3} - \widehat{\Omega}^{(m,0)}_{i_2i_3}) (A_{i_3i_4} - \widehat{\Omega}_{i_3i_4}^{(m,0)}) (A_{i_4i_1} - \widehat{\Omega}^{(m,0)}_{i_4i_1}), 
\] 
and 
\beq\label{def:psim0}
  \psi_n^{(m,0)} = [Q_n^{(m,0)} - B_n^{(m)}]/\sqrt{8 C_n}.
\eeq
These are the proxies of $\Omega^{(m)}$, $Q_n^{(m)}$, and $\psi_n^{(m)}$, respectively, 
where $\widehat{\Pi}^{(m)}$ is now frozen at a non-stochastic matrix $\Pi_0$.

In the under-fitting case,   $m<K$, and we do not expect $\widehat{\Omega}^{(m,0)}$ to be close to $\Omega$. We define a non-stochastic counterpart of $\widehat{\Omega}^{(m,0)}$ as follows. Let $\theta^{(m,0)}$, $\Theta^{(m,0)}$ and $P^{(m,0)}$ be constructed 
similarly by (\ref{refitting1aADD})-(\ref{refitting1a}), except that $(A, \widehat{\Pi}^{(m)})$ and the vector $d=( d_1,d_2,\ldots,d_n)'$ are replaced with $(\Omega, \Pi_0)$ and $\Omega {\bf 1}_n$, respectively. Let
\beq \label{def:tOmega}
\Omega^{(m,0)} = \Theta^{(m,0)} \Pi_0 P^{(m,0)}\Pi_0' \Theta^{(m,0)}. 
\eeq
The following lemma gives an equivalent expression of $\Omega^{(m,0)}$ and is proved in Section~\ref{subsec:SgnQ20-proof}. 
\begin{lemma} \label{lemma:SgnQ2-0} 
Fix $K > 1$ and $1 \leq m \leq K$.  Let $\Pi_0=\Pi L_0 \in {\cal G}_m$ and  ${\Omega}^{(m,0)}$ be as above. Write $D = \Pi'\Theta\Pi\in\mathbb{R}^{K,K}$ and $D_0 = \Pi_0'\Theta \Pi\in\mathbb{R}^{m,K}$.  
Let $P_0$ be the $K\times K$ matrix given by 
\[
P_0 = \diag(P D {{\bf 1}_K})\cdot L_0 \cdot \diag(D_0 P D {{\bf 1}_K})^{-1}({D}_0 PD_0') \diag(D_0 P D {{\bf 1}_K})^{-1} \cdot  L_0'\cdot\diag(PD {{\bf 1}_K}), 
\]
where the rank of $P_0$ is $m$. Then, $\Omega^{(m,0)}=\Theta\Pi P_0\Pi'\Theta$. 
\end{lemma} 

\noindent
This lemmas says that ${\Omega}^{(m,0)}$ has a similar expression as $\Omega$, with $P$ replaced by a rank-$m$ matrix $P_0$. When $m=K$, ${\cal G}_m$ has only one element $\Pi$; then $(P_0, \Omega^{(m,0)})$ reduces to $(P, \Omega)$.

We expect $\widehat{\Omega}^{(m,0)}$ to concentrate at $\Omega^{(m,0)}$. This motivates the following proxy of $Q_n^{(m,0)}$.  
\beq \label{def:widetildeQ}
\widetilde{Q}_n^{(m,0)}  =  \hspace{-1.3 em} \sum_{i_1, i_2, i_3, i_4 (dist)} (A_{i_1i_2} - {\Omega}^{(m,0)}_{i_1i_2}) (A_{i_2i_3} - {\Omega}^{(m,0)}_{i_2i_3}) (A_{i_3i_4} - {\Omega}_{i_3i_4}^{(m,0)}) (A_{i_4i_1} - {\Omega}^{(m,0)}_{i_4i_1}).
\eeq
Introduce 
\begin{equation} \label{DefineOmegaL}  
\widetilde{\Omega}^{(m,0)} = \Omega - \Omega^{(m,0)}. 
\end{equation} 
Recall that $A = (\Omega - \diag(\Omega))  + W$,  we rewrite $\widetilde{Q}_n^{(m,0)}$ as
\beq\label{def:tildeOmega}
\widetilde{Q}_n^{(m,0)} = \hspace{-1.4 em}  \sum_{i_1, i_2, i_3, i_4 (dist)}  \hspace{-1 em} (W_{i_1i_2} + \widetilde{\Omega}^{(m,0)}_{i_1i_2}) (W_{i_2i_3} + \widetilde{\Omega}^{(m,0)}_{i_2i_3}) (W_{i_3i_4} + \widetilde{\Omega}_{i_3i_4}^{(m,0)}) (W_{i_4i_1} + \widetilde{\Omega}^{(m,0)}_{i_4i_1}).
\eeq 
Note that when $m = K$ and $\Pi_0 = \Pi$, the statistic $\widetilde{Q}_n^{(m,0)} $ reduces to $\widetilde{Q}_n$ defined in \eqref{DefinewidetildeQ}. 

The matrix $\widetilde{\Omega}^{(m,0)}$ captures the signal strength in $\widetilde{Q}_n^{(m,0)}$. From now on, for notation simplicity, we write $\widetilde{\Omega}^{(m,0)}=\widetilde{\Omega}$ in the rest of the proof. Let $\tilde{\lambda}_k$  be the $k$-th largest (in magnitude) eigenvalue of $\widetilde{\Omega}$ 
and recall that $\lambda_k$ is the $k$-th largest (in magnitude) eigenvalue of $\Omega$.   
In light of \eqref{DefineOmegaL}, we write $\Omega= \Omega^{(m,0)}+\widetilde{\Omega}$ and apply Weyl's theorem for singular values (see equation (7.3.13) of \cite{HornJohnson}). Note that $\Omega^{(m,0)}$ has a rank $m$ and $\Omega$ has a rank $K$. By Weyl's theorem, for all $1\leq k\leq K-m$, $|\lambda_{m+k}|\leq |\lambda_{m+1}(\Omega^{(m,0)})|+ |\tilde{\lambda}_k|=|\tilde{\lambda}_k|$. It follows that
\[
\tr(\widetilde{\Omega}^4)\geq \sum_{k=1}^{K-m}|\tilde{\lambda}_k|^4 \geq \sum_{k=m+1}^{K}|\lambda_k|^4. 
\] 
As we will see in Lemma~\ref{lemma:SgnQ2b} below, $\tr(\widetilde{\Omega}^4)$ is the dominating term of $\mathbb{E}[\widetilde{Q}_n^{(m,0)}]$.   
Define 
\beq\label{def:tau}  
\tau^{(m,0)} = |\tilde{\lambda}_1|/{\lambda}_1. 
\eeq
For notation simplicity, we write $\tau^{(m,0)}=\tau$, but keep in mind both $\widetilde{\Omega}$ and $\tau$ actually depend on $m$ and $\Pi_0\in\mathcal{G}_m$. The following lemmas are proved in Sections~\ref{subsec:SgnQ2a-proof}-\ref{subsec:SgnQ2d-proof}, respectively. 
\begin{lemma} \label{lemma:SgnQ2a} 
Under the conditions of Theorem \ref{thm:CUB}, for each $1\leq m\leq K$, let $\widetilde{\Omega}$ and $\tau$ be defined as in \eqref{def:tOmega} and \eqref{def:tau}. The following statements are true:
\begin{itemize}
\item There exists a constant $C>0$ such that $|\widetilde{\Omega}_{ij} | \leq C\tau\theta_i\theta_j$, for all $1\leq i,j\leq n$. 
\item $c_n\asymp \|\theta\|^8$, $\lambda_1\asymp\|\theta\|^2$, and $\tau=O(1)$. 
\item $\tr(\widetilde{\Omega}^4)\geq C\tau^4\|\theta\|^8$, and $\tau\|\theta\|\to\infty$. 
\end{itemize}
\end{lemma}

\begin{lemma} \label{lemma:SgnQ2b} 
Under the condition of Theorem~\ref{thm:CUB},  for $1\leq m<K$, 
\[
\mathbb{E}[\widetilde{Q}_n^{(m,0)}] = \tr(\widetilde{\Omega}^4) + o(\tau^4\|\theta\|^8),\quad \mathrm{Var}(\widetilde Q_n^{(m,0)})\leq C(\|\theta\|^8 + \tau^6\|\theta\|^8\|\theta\|_3^6).
\] 
\end{lemma}

\begin{lemma} \label{lemma:SgnQ2c} 
Under the condition of Theorem~\ref{thm:CUB}, for $1\leq m<K$, 
\[
\mathbb{E}[Q_n^{(m,0)} - \widetilde{Q}_n^{(m,0)}] = o(\tau^4\|\theta\|^8),\quad \mathrm{Var}(Q_n^{(m,0)} - \widetilde{Q}_n^{(m,0)})\leq o(\|\theta\|^8) +C \tau^6\|\theta\|^8\|\theta\|_3^6.
\]
\end{lemma} 
\begin{lemma} \label{lemma:SgnQ2d} 
Under the conditions of Theorem \ref{thm:CUB}, for $1\leq m<K$, there exists a constant $C>0$, such that $\mathbb{P}(B_n^{(m)} \leq C\|\theta\|^4) \geq 1+o(1).$
\end{lemma}

We now prove  Theorem \ref{thm:CUB}.  Note that by Theorem \ref{thm:CLB}, the second item of Theorem \ref{thm:CUB} follows once the first item is proved. Therefore we only consider the first item, where it is sufficient to show that for all $1 < m <  K$, 
\[
\psi_n^{(m)} \goto \infty, \qquad \mbox{in probability}. 
\]  
By the NSP of the solutions produced by SCORE, which is shown in Theorem \ref{thm:SCORE2}, there exists an event $A_n$ with $\mathbb{P}(A_n^c)\leq Cn^{-3}$ as $n\to\infty$, such that on event $A_n$ we have $\widehat{\Pi}^{(m)}\in{\cal G}_m$. This further indicates that on event $A_n$ we have
\beq\label{psi-m}
\psi_n^{(m)}  \geq  \min_{\Pi_0\in{\cal G}_m} \psi_n^{(m,0)} ,
\eeq
where $\psi_n^{(m,0)}$ is defined in \eqref{def:psim0}.
The LHS is hard to analyze, but the RHS is relatively easy to analyze. 
Then further notice that the cardinality of ${\cal G}_m$ is $|{\cal G}_m|=m^K$, which is of constant order as long as $K$ is constant. Therefore to prove $\psi_n^{(m)}  \goto \infty$ in probability, it suffices to show that for any fixed $\Pi_0\in\mathcal{G}_m$,  
\begin{equation} \label{SgnQ2pf0}  
\psi_n^{(m,0)} \goto \infty, \qquad \mbox{in probability}. 
\end{equation} 
We now show (\ref{SgnQ2pf0}). Rewrite $\psi_n^{(m,0)}$ as 
\begin{equation} \label{SgnQ2pf1} 
\sqrt{\frac{c_n}{C_n}} \cdot \biggl[\frac{{Q}_n^{(m,0)} }{\sqrt{8c_n}}   -  \frac{B_n^{(m)}}{\sqrt{8c_n}} \biggr]  = \sqrt{\frac{c_n}{C_n}} \cdot [(I) -  (II) ],  
\end{equation} 
where $(I) = {Q}_n^{(m,0)} / \sqrt{8c_n}$, and 
$(II) =  B_n^{(m)}/\sqrt{8c_n}$.  
First, by Lemma~\ref{lemma:SgnQ1a} (since $C_n$ and $c_n$ do not depend on $m$, this lemma applies to both the null case and the under-fitting case), 
\begin{equation} \label{SgnQ2pf2a} 
c_n/C_n\goto1 \qquad \mbox{in probability}.
\end{equation} 
Second, by Lemma~\ref{lemma:SgnQ2a}, $c_n\asymp \|\theta\|^8$. Combining it with Lemma~\ref{lemma:SgnQ2d} gives that there is a constant $C > 0$ such that 
\begin{equation} \label{SgnQ2pf2b}  
\mathbb{P}( (II) \leq C) \geq 1 +o(1).
\end{equation} 
Last,  by Lemma~\ref{lemma:SgnQ2a}-\ref{lemma:SgnQ2c},  
\[
\mathbb{E}[(I)] \geq C\tau^4\|\theta\|^4\cdot [1+o(1)]\goto\infty, \quad \mathrm{Var}((I))\leq C(1 + \tau^6\|\theta\|_3^6). 
\]
Therefore, by Chebyshev's inequality, for any constant $M>0$, 
\begin{equation} \label{SgnQ2pf2c}
\mathbb{P}((I) <M )\leq (\mathbb{E}[(I)] - M)^{-2} \mathrm{Var}((I)) \leq C\bigg[\frac{1+\tau^6\|\theta\|_3^6}{(\tau^4\|\theta\|^4[1+ o(1)] - M)^2}\bigg],
\end{equation} 
where on the denominator, $\tau \|\theta\| \goto \infty$ by Lemma~\ref{lemma:SgnQ2a}.  
Note that under our conditions, $\|\theta\|_3^3  = o(\|\theta\|^2)$ and $\|\theta\| \goto \infty$.    Combining these, the RHS of (\ref{SgnQ2pf2c}) tends to $0$ as $n \goto \infty$. Inserting (\ref{SgnQ2pf2a})-(\ref{SgnQ2pf2c}) into (\ref{SgnQ2pf1}) proves the claim, and concludes the proof of Theorem \ref{thm:CUB}. \qed

%
%



\subsection{Proof of Lemma~\ref{lemma:SgnQ1a}} \label{subsec:SgnQ1a-proof}
Consider the first two claims. 
It is easy to see that $\mathbb{E}[C_n]=c_n$. In the proof of Theorem 3.1 of \cite{OGC}, it has been shown that
\[
c_n=\tr(\Omega^4) +O(\|\theta\|_4^4\|\theta\|^4) = \tr(\Omega^4)+o(\|\theta\|^8). 
\]
Moreover, $\lambda_1^4\leq \tr(\Omega^4)\leq K\lambda_1^4$. In the proof of Theorem~\ref{thm:LB}, we have seen that $\lambda_1=\|\theta\|^2\cdot \lambda_1(HPH')$. Using the condition \eqref{condition1b} and the fact that $P$ has unit diagonals, we have $\lambda_1(HPH')\geq C\lambda_1(P)\geq C$. Similarly, since we have assumed $\|P\|\leq C$ in \eqref{condition1a}, $\lambda_1(HPH')\leq C\lambda_1(P)\leq C$. Here, $C$ is a generic constant. We have proved that
\[
\mathbb{E}[C_n] = c_n\asymp\|\theta\|^8. 
\]
To compute the variance of $C_n$, write
\[
C_n = \widetilde{Q}_n + \Delta, \qquad\mbox{where} \quad \widetilde{Q}_n=\sum_{i_1, i_2, i_3, i_4 (dist)} W_{i_1i_2} W_{i_2i_3} W_{i_3i_4} W_{i_4i_1}.
\]
The variance of $\Delta$ is computed in the proof of Lemma B.2 of \cite{OGC}. Using the upper bound of the variance of $\bigl(\sum_{CC(I_n)}\Delta_{i_1i_2i_3i_4}^{(k)}\bigr)$ for $k=1,2,3$ there, we have
\[
\mathrm{Var}(\Delta)\leq C\|\theta\|_3^6\|\theta\|^8. 
\]
Furthermore, we show in the proof of Lemma~\ref{lemma:SgnQ1b} that $\mathrm{Var}(\widetilde{Q}_n)=8 c_n\cdot [1+o(1)]$. It follows that $\mathrm{Var}(\widetilde{Q}_n)\asymp c_n\asymp \|\theta\|^8$. Combining these results gives
\[
\mathrm{Var}(C_n)\leq C\|\theta\|^8\cdot [1+\|\theta\|_3^6]. 
\]

Consider the last claim. For any $\eps>0$, using Chebyshev's inequality, we have 
\[
\mathbb{P}(|C_n/c_n - 1|\geq \eps)\leq (c_n\eps)^{-2}\mathrm{Var}(C_n)\leq \frac{C(1+\|\theta\|_3^6)}{\eps^2\, \|\theta\|^8}.
\]
Here we have used the first two claims. Since $\|\theta\|_3^3\leq\theta_{\max}\|\theta\|^2=o(\|\theta\|^8)$, the rightmost term is $o(1)$ as $n\to\infty$. This proves that $C_n/c_n\to 1$ in probability.\qed

\subsection{Proof of Lemma~\ref{lemma:SgnQ1b}}  \label{subsec:SgnQ1b-proof}

Recall that $\widetilde{Q}_n = \sum_{i_1, i_2, i_3, i_4 (dist)} W_{i_1i_2} W_{i_2i_3} W_{i_3i_4} W_{i_4i_1}$. This variable was analyzed in \cite{OGC} (see their proof of Theorem 3.2, where $\widetilde{Q}_n$ is denoted as $S_{n,n}$ there). It was shown that 
\[
\widetilde{Q}_n/\sqrt{\mathrm{Var}(\widetilde{Q}_n)}\goto N(0,1), \qquad\mbox{in law}.
\]
It remains to prove $\mathrm{Var}(\widetilde{Q}_n)=8c_n\cdot [1+o(1)]$. 

Note that for each ordered quadruple $(i,j,k,\ell)$ with four distinct indices, there are 8 summands in the definition of $\widetilde{Q}_n$ whose values are exactly the same; these summands correspond to $(i_1, i_2, i_3, i_4)\in \{(i,j,k,\ell)$, $(j,k,\ell, i)$, $(k,\ell,i,j)$, $(\ell, i,j,k)$, $(k,j,i,\ell)$, $(j,i,\ell,k)$, $(i,\ell,k,j)$, $(\ell,k,j,i)\}$, respectively. We treat these 8 summands as in an equivalent class. Denote by $CC_4$ the collection of all such equivalent classes. Then, for any doubly indexed sequence $\{x_{ij}\}_{1\leq i\neq j\leq n}$ such that $x_{ij}=x_{ji}$, it is true that $\sum_{i_1, i_2, i_3, i_4 (dist)}x_{i_1i_2}x_{i_2i_3}x_{i_3i_4}x_{i_4i_1}=8\sum_{CC_4}x_{i_1i_2}x_{i_2i_3}x_{i_3i_4}x_{i_4i_1}$. In particular, 
\[
\widetilde{Q}_n=8\sum_{CC_4}W_{i_1i_2}W_{i_2i_3}W_{i_3i_4}W_{i_4i_1}.
\]
The summands are independent of each other, and the variance of $W_{i_1i_2}W_{i_2i_3}W_{i_3i_4}W_{i_4i_1}$ is equal to $\Omega^*_{i_1i_2}\Omega^*_{i_2i_3}\Omega^*_{i_3i_4}\Omega^*_{i_4i_1} $, where $\Omega^*_{ij}=\Omega_{ij}(1-\Omega_{ij})$. As a result, 
\begin{align*}
\mathrm{Var}(\widetilde{Q}_n)= 64 \sum_{CC_4}\Omega^*_{i_1i_2}\Omega^*_{i_2i_3}\Omega^*_{i_3i_4}\Omega^*_{i_4i_1} =8\sum_{i_1,i_2,i_3,i_4(dist)}\Omega^*_{i_1i_2}\Omega^*_{i_2i_3}\Omega^*_{i_3i_4}\Omega^*_{i_4i_1}. 
\end{align*} 
Recall that $c_n=\sum_{i_1, i_2, i_3, i_4 (dist)}\Omega_{i_1i_2}\Omega_{i_2i_3}\Omega_{i_3i_4}\Omega_{i_4i_1}$. Then,
\begin{align*}
|\mathrm{Var}(\widetilde{Q}_n)-8 c_n|&\leq 8\sum_{i_1, i_2, i_3, i_4 (dist)}|\Omega_{i_1i_2}\Omega_{i_2i_3}\Omega_{i_3i_4}\Omega_{i_4i_1}-\Omega^*_{i_1i_2}\Omega^*_{i_2i_3}\Omega^*_{i_3i_4}\Omega^*_{i_4i_1}|\cr
&\leq 8\sum_{i_1,i_2,i_3,i_4 (dist)}\Omega_{i_1i_2}\Omega_{i_2i_3}\Omega_{i_3i_4}\Omega_{i_4i_1}\cdot C\|\Omega\|_{\max}\cr
& = 8c_n\cdot O(\theta_{\max}^2).  
\end{align*}
Since $\theta_{\max}=o(1)$ by the condition \eqref{condition1a}, we immediately have $\mathrm{Var}(\widetilde{Q}_n)=8c_n\cdot [1+o(1)]$. \qed


\subsection{Proof of Lemma~\ref{lemma:SgnQ1c}}   \label{subsec:SgnQ1c-proof}

The proof is combined with the proof of Lemma~\ref{lemma:SgnQ2c}; see below. 


\subsection{Proof of Lemma~\ref{lemma:SgnQ1d}} \label{subsec:SgnQ1d-proof}

Consider the first claim. Since $ b_n  = 2\|\theta\|^4\cdot [ g' {V}^{-1} ( P H^2 P\circ P H^2  P)  {V}^{-1} g ]$ (see \eqref{def:bncn}), it suffices to show that
\[
g' {V}^{-1} ( P H^2 P\circ P H^2  P)  {V}^{-1} g\asymp 1. 
\]
The vectors $g, h\in\mathbb{R}^K$ are defined by $g_k = ({\bf{1}}_k' \theta)/ \|\theta\|_1$ and $h_k = ({\bf{1}}_k' \Theta^2 {\bf{1}}_k)^{1/2}/ \|\theta\|$, where ${\bf 1}_k$ is for short of ${\bf 1}^{(K)}_k$. 
By condition \eqref{condition1b}, $c_1\leq g_k\leq 1$ and $c_1\leq h_k^2\leq 1$ for $1\leq k\leq K$, and $\|P\|\leq c_2$, for some constants $c_1, c_2\in (0,1)$. 
 
For the upper bound, by $h_k^2\leq 1$ and $\|P\|\leq c_2$, we have $\| (PH^2P)\circ (PH^2P)\| \leq C.$ Since $P$ has unit diagonals and $g_k\geq c_1$, the diagonal elements of  $V = \diag(Pg)$ is no less than $c_1$. Hence \beq\label{lemma:SgnQ1d-claim-1-lb}
g' {V}^{-1} ( P H^2 P\circ P H^2  P)  {V}^{-1} g \leq \|g' {V}^{-1}\|^2 \cdot\|P H^2 P\circ P H^2  P\| \leq C. 
\eeq
For the lower bound, since $P$ has unit diagonals and $h_k^2\geq c_1$, we can lower bound diagonal elements of $P H^2 P\circ P H^2  P$ by $c_1^2$. Since $g\in\mathbb{R}^K$ is a non-negative vector with entries summing to $1$, the diagonal elements of $V = \diag(Pg)$ is no more than $\max_{k,l}P_{k,\ell}\leq \|P\|\leq c_2.$ Therefore each entry of vector $gV^{-1}$ is at least $c_1/c_2.$
Since $P H^2 P\circ P H^2  P\in\mathbb{R}^{(K,K)}$ is non-negative matrix and $g'V^{-1}\in\mathbb{R}^{(K}$ is non-negative vector, we can lower bound 
\beq\label{lemma:SgnQ1d-claim-1-ub}
g' {V}^{-1} ( P H^2 P\circ P H^2  P)  {V}^{-1} g\geq c_1^2\|g' {V}^{-1} \|^2 \geq C,
\eeq
Combining \eqref{lemma:SgnQ1d-claim-1-lb}-\eqref{lemma:SgnQ1d-claim-1-ub}, we completes the proof of the first claim.

Consider the second claim. Introduce the following event
\beq\label{lemma:SgnQ1d-def-event-D}  
A_n = \big\{ \widehat{\Pi}^{(K)} = \Pi, \text{ up to a permutation in the columns of $\widehat{\Pi}^{(K)}$}  \big\}. 
\eeq
By Theorem~\ref{thm:CLB}, when $m=K$, SCORE exactly recovers $\Pi$ with probability $1-o(n^{-3})$, i.e.,   
\[
\mathbb{P}(A_n^c) \leq C n^{-3} = o(1).
\]
This means if we replace every $\widehat{\Pi}^{(K)}$ in the definition of $B_n^{(K)}$ with $\Pi$, and denote the resulting quantity as $B_n^{(K,0)}$, the above inequality immediately implies that $B_n^{(K)}/B_n^{(K,0)}\stackrel{p}{\to}1$. So we only need to prove $B_n^{(K,0)}/b_n\stackrel{p}{\to}1$. Since we will never use the original definition of $B_n^{(K)}$ in the rest of the proof, without causing any confusion we will suspend the original definitions of $B_n^{(K)}$ and the quantities used to define $B_n^{(K)}$, including $(\hat{\theta}, \hat{g},\widehat{V}, \widehat{P}, \widehat{H})$, and use them to actually denote the correspondents with every $\widehat{\Pi}^{(K)}$ replaced by $\Pi$.

Recall the formulas for $B_n^{(K)}$ and $b_n$ in \eqref{refitting2c} and \eqref{def:bncn}, we have
\beq  \label{lemma:SgnQ1d-add}
\frac{B_n^{(K)}}{b_n} = \frac{\|\hat\theta\|^4}{\|\theta\|^4}\cdot\frac{\hat g' \widehat{V}^{-1} (\widehat P\widehat H^2 \widehat P\circ \widehat P\widehat H^2 \widehat P) \widehat{V}^{-1}\hat g}{g' {V}^{-1} ( P H^2 P\circ P H^2  P)  {V}^{-1} g}.
\eeq
To show that $B_n^{(K)}/b_n\to 1$, we need the follow lemma, which is proved in Section~\ref{subsec:prop-converge-proof}.

\begin{lemma}\label{prop:converge}
Suppose the conditions of Theorem \ref{thm:CLB} hold. Let ${\bf 1}_n\in\mathbb{R}^n$ be the vector of $1$'s, and let ${\bf 1}_k\in\mathbb{R}^n$ be the vector such that ${\bf 1}_k(i)=1\{i\in {\cal N}_k\}$, for $1\leq i\leq n$ and $1\leq k\leq K$.  As $n\to\infty$, for all $1\leq k\leq K$, 
 \[
 \frac{{\bf{1}}_n' A {\bf{1}}_n }{{\bf{1}}_n' \Omega {\bf{1}}_n }\stackrel{p}{\goto}1,\qquad
  \frac{{\bf{1}}_k' A {\bf{1}}_n }{{\bf{1}}_k' \Omega {\bf{1}}_n }\stackrel{p}{\goto}1, \qquad \frac{{\bf{1}}_k' A {\bf{1}}_k }{{\bf{1}}_k' \Omega {\bf{1}}_k }\stackrel{p}{\goto}1.
\]
Moreover, let $d_i$ be the degree of node $i$ and let $d_i^*=(\Omega{\bf 1}_n)_i$, for $1\leq i\leq n$. Write $D = \diag(d)\in\mathbb{R}^{n,n}$ and  $D^* = \diag(d^*)\in\mathbb{R}^{n,n}$. As $n\to\infty$, for all $1\leq k\leq K$, 
\[
\frac{\|\hat\theta\|_1}{\|\theta\|_1} \stackrel{p}{\goto}1, \qquad \frac{\|\hat\theta\|}{\|\theta\|} \stackrel{p}{\goto}1, \qquad \frac{{\bf{1}}_k' D^2{\bf{1}}_k}{{\bf{1}}_k' (D^*)^2{\bf{1}}_k} \stackrel{p}{\to} 1. 
\]
\end{lemma}

First, by Lemma~\ref{prop:converge}, $\|\hat\theta\|/\|\theta\|\overset{p}{\to}1$. It follows from the continuous mapping theorem that
\beq\label{lemma:SgnQ1d-1-claim-toshow-1.5}  
\|\hat\theta\|^4/\|\theta\|^4 \stackrel{p}{\goto}1. 
\eeq
Second, recall that $g_k = ({\bf{1}}_k' \theta)/\|\theta\|_1$ and $\hat{g}_k=({\bf{1}}_k' \hat{\theta})/\|\hat{\theta}\|_1$, where by \eqref{refitting1a}, we have the equality ${\bf 1}_k'\hat{\theta} = ({\bf 1}_k'd) \cdot\sqrt{{\bf 1}_k' A {\bf 1}_k}/ ({\bf 1}_k' A {\bf 1}_n)$. Here, keep in mind that we have replaced $\widehat{\Pi}^{(K)}$ with $\Pi$, which implies that $\hat{\bf 1}_k={\bf 1}_k$. The vector $d$ is such that $d=A{\bf 1}_n$. It follows that ${\bf 1}_k'\hat{\theta} = \sqrt{{\bf 1}_k' A {\bf 1}_k}$. Furthermore, ${\bf 1}_k'\Omega {\bf 1}_k=({\bf 1}_k'\theta)^2$, because $P$ has unit diagonals. Combining the above gives 
\beq \label{lemma:SgnQ1d-1-claim-toshow-2} 
\frac{\hat g_k}{g_k} = \frac{ {\bf{1}}_k'  \hat\theta  }{ {\bf{1}}_k'  \theta}\cdot\frac{\|\theta\|_1}{\|\hat\theta\|_1} = \frac{\sqrt{{\bf{1}}_k' A {\bf{1}}_{k}}}{\sqrt{{\bf{1}}_k' \Omega {\bf{1}}_{k}}}  \cdot\frac{\|\theta\|_1}{\|\hat\theta\|_1} \stackrel{p}{\goto}1, \qquad 1\leq k\leq K. 
\eeq
Third, note that by definition and basic algebra, both $P$ and $\widehat P$ have unit diagonals. 
We compare their off-diagonals. By \eqref{refitting1a}, $\widehat{P}_{k\ell}={\bf 1}_k'A{\bf 1}_\ell/\sqrt{({\bf 1}_k'A{\bf 1}_k)({\bf 1}_\ell'A{\bf 1}_\ell)}$. At the same time, it can be easily verified that $P_{k\ell}={\bf 1}_k'\Omega {\bf 1}_\ell/\sqrt{({\bf 1}_k'\Omega{\bf 1}_k)({\bf 1}_\ell'\Omega{\bf 1}_\ell)}$. 
Introduce
\[
X=\frac{\sqrt{({\bf 1}_k'\Omega{\bf 1}_k)({\bf 1}_\ell'\Omega{\bf 1}_\ell)}}{\sqrt{({\bf 1}_k'A{\bf 1}_k)({\bf 1}_\ell'A{\bf 1}_\ell)}}. 
\]
By Lemma~\ref{prop:converge}, $X\overset{p}{\to }1$. 
We re-write
\begin{align*} 
\widehat{P}_{k\ell}-P_{k\ell}& = \frac{{\bf 1}_k'A{\bf 1}_\ell-{\bf 1}_k'\Omega{\bf 1}_\ell}{\sqrt{({\bf 1}_k'A{\bf 1}_k)({\bf 1}_\ell'A{\bf 1}_\ell)}} +P_{k\ell} (X-1) = \frac{{\bf 1}_k'W{\bf 1}_\ell}{({\bf 1}_k'\theta)({\bf 1}_\ell'\theta)} X+ P_{k\ell}(X-1), 
\end{align*}
where in the last inequality we have used the fact that ${\bf 1}_k'\Omega{\bf 1}_k=({\bf 1}_k'\theta)^2$ for all $1\leq k\leq K$. 
Note that $\mathbb{E}[{\bf 1}_k'W{\bf 1}_\ell]=0$. Moreover, $\mathrm{Var}(W_{ij})\leq \|P\|_{\max}\theta_i\theta_j\leq C\theta_i\theta_j$. It follows that $\mathrm{Var}({\bf 1}_k'W{\bf 1_\ell})\leq C({\bf 1}_k'\theta)({\bf 1}_\ell'\theta)$. Therefore, 
\[
    \mathbb{E}\bigg[\frac{{\bf 1}_k'W{\bf 1}_\ell}{({\bf 1}_k'\theta)({\bf 1}_\ell'\theta)}\bigg]^2 \leq \frac{C}{({\bf 1}_k'\theta)({\bf 1}_\ell'\theta)}=O(\|\theta\|_1^{-2}) =o(1). 
 \]
Hence, $\frac{{\bf 1}_k'W{\bf 1}_\ell}{({\bf 1}_k'\theta)({\bf 1}_\ell'\theta)}\overset{p}{\to }0$. 
Combining the above results, we have
\beq  \label{lemma:SgnQ1d-1-claim-toshow-3}
\widehat{P}_{k\ell}-P_{k\ell}\stackrel{p}{\goto} 0, \qquad 1\leq k,\ell\leq K. 
\eeq
Fourth, since $V=\diag(Pg)$ and $\widehat{V}=\diag(\widehat{P}\hat{g})$, it follows from \eqref{lemma:SgnQ1d-1-claim-toshow-2} and \eqref{lemma:SgnQ1d-1-claim-toshow-3} that
\beq \label{lemma:SgnQ1d-1-claim-toshow-3.5}
\widehat{V}_{kk}/V_{kk} \stackrel{p}{\goto} 1, \qquad 1\leq k \leq K. 
\eeq
Last, note that $H^2, \widehat{H}^2\in\mathbb{R}^{K,K}$ are diagonal matrices, with $k$-th diagonal elements being $h_k^2$ and $\hat h_k^2$, respectively. By \eqref{refitting2gh}, $\hat{h}_k^2=({\bf 1}_k'\widehat{\Theta}^2{\bf 1}_k)/\|\hat{\theta}\|^2$. In addition, by \eqref{refitting1a}, for any $i\in {\cal N}_k$, we have $\hat{\theta}_i^2=d_i^2({\bf 1}_k'A{\bf 1}_k)/({\bf 1}_k'A{\bf 1}_n)^2$. We thus re-write
\[
\widehat{H}_{kk}\equiv \hat{h}_k^2 = \frac{({\bf 1}_k'D^2{\bf 1}_k )\cdot ({\bf 1}_k'A{\bf 1}_k)}{ ({\bf 1}_k'A{\bf 1}_n)^2\cdot \|\hat{\theta}\|^2}. 
\]
Additionally, $h_k=({\bf 1}_k'\Theta^2{\bf 1}_k)/\|\theta\|^2$, as defined in the paragraph below \eqref{def:bncn}. By direct calculations, $({\bf 1}_k'\Omega {\bf 1}_n)/\sqrt{{\bf 1}_k'\Omega {\bf 1}_k}=\bigl[({\bf 1}_k'\theta)\sum_{\ell}P_{k\ell}({\bf 1}_\ell'\theta)\bigr]/({\bf 1}_k'\theta)=\sum_{\ell}P_{k\ell}({\bf 1}_\ell'\theta)$. Also, for any $i\in {\cal N}_k$, we have $d_i^*=(\Omega {\bf 1}_n)_i=\theta_i[\sum_{\ell}P_{k\ell}({\bf 1}_\ell'\theta)]$. It implies that ${\bf 1}_k'(D^*)^2{\bf 1}_k=({\bf 1}_k'\Theta^2 {\bf 1}_k)[\sum_{\ell}P_{k\ell}({\bf 1}_\ell'\theta)]^2$. We can use these expressions to verify that
\[
H_{kk}\equiv h_k^2 = \frac{[{\bf 1}_k'(D^*)^2{\bf 1}_k ]\cdot ({\bf 1}_k'\Omega{\bf 1}_k)}{ ({\bf 1}_k'\Omega{\bf 1}_n)^2\cdot \|\theta\|^2}. 
\]
We apply Lemma~\ref{prop:converge} to obtain that
\beq \label{lemma:SgnQ1d-1-claim-toshow-4}
\widehat{H}_{kk}/H_{kk} \stackrel{p}{\goto}1, \qquad 1\leq k\leq K. 
\eeq
We plug \eqref{lemma:SgnQ1d-1-claim-toshow-1.5}, \eqref{lemma:SgnQ1d-1-claim-toshow-2}, \eqref{lemma:SgnQ1d-1-claim-toshow-3}, \eqref{lemma:SgnQ1d-1-claim-toshow-3.5} and \eqref{lemma:SgnQ1d-1-claim-toshow-4} into \eqref{lemma:SgnQ1d-add}. It follows from elementary probability that $B_n^{(K)}/b_n\to 1$. This gives the second claim.\qed

\subsection{Proof of Lemma~\ref{lemma:SgnQ2-0} } \label{subsec:SgnQ20-proof}
Recall  $\mathcal{N}_1^{(m,0)}, \mathcal{N}_2^{(m,0)}, ..., \mathcal{N}_m^{(m,0)}$ are ``fake'' communities associated with $\Pi_0$, and we decompose the vector ${\bf{1}}_n\in\mathbb{R}^n$ as follows
\beq\label{def:1m0}
{\bf{1}}_n = \sum_{k=1}^m {\bf{1}}^{(m,0)}_k, \qquad\mbox{ where ${\bf{1}}^{(m,0)}_k(j) = 1$ if $j\in  \mathcal{N}_k^{(m,0)}$ and $0$ otherwise.  }
\eeq
Notice for $\Pi_0\in\mathcal{G}_m$ defined in \eqref{def:Gm}, there exists an $K\times m$ matrix $L_0$ such that $\Pi_0 = \Pi L_0$.

By definitions, $ \Omega^{(m,0)} = \Theta^{(m,0)}\Pi_0 P^{(m,0)}\Pi_0' \Theta^{(m,0)}$. Here $\Theta^{(m,0)}$ and $P^{(m,0)}$ are obtained by replacing $(d_i, \hat{\bf 1}_k, A)$ by $(d_i^*, {\bf 1}_k^{(m,0)}, \Omega)$ in the definition \eqref{refitting1a}. It yields that, for $1\leq k,\ell\leq m$ and $i\in\mathcal{N}_k^{(m,0)}$, 
\[
\theta_i^{(m, 0)} = \frac{d^*_i}{ ( {\bf 1}_k^{(m,0)})' \Omega {\bf 1}_n} \cdot\sqrt{ ({\bf 1}_k^{(m,0)})' \Omega  {\bf 1}_k^{(m,0)}},  \qquad
P_{k\ell}^{(m, 0)}  =   \frac{({\bf 1}_k^{(m,0)})' \Omega   ({\bf 1}_{\ell}^{(m,0)}) }{\sqrt{ ({\bf 1}_k^{(m,0)})' \Omega {\bf 1}^{(m,0)}_k}\sqrt{  ({\bf 1}_{\ell}^{(m,0)})' \Omega {\bf 1}_{\ell}^{(m,0)}}}. 
\]
As a result, for $i\in {\cal N}_k^{(m,0)}$ and $j\in {\cal N}_\ell^{(m,0)}$, 
\beq\label{def:Omega-ij-m0}
\Omega^{(m,0)}_{ij} = \theta_i^{(m, 0)}\theta_j^{(m, 0)}P_{k\ell}^{(m, 0)} = d_i^* d_j^* \cdot\frac{({\bf 1}_k^{(m,0)})' \Omega   {\bf 1}_{\ell}^{(m,0)} }{[ ({\bf 1}_k^{(m,0)})' \Omega {\bf 1}_n] \cdot [( {\bf 1}_\ell^{(m,0)})' \Omega {\bf 1}_n]}. 
\eeq
Note that $({\bf 1}_k^{(m,0)})'\Omega{\bf 1}_\ell^{(m,0)}=(\Pi_0'\Omega\Pi_0)_{k\ell}$. Since $\Omega=\Theta\Pi P\Pi'\Theta$ and $D_0=\Pi_0'\Theta\Pi$, we immediately have $\Pi_0'\Omega\Pi_0=\Pi_0'\Theta\Pi P\Pi'\Theta'\Pi_0=D_0PD'_0$. It follows that 
\[
({\bf 1}_k^{(m,0)})' \Omega   {\bf 1}_{\ell}^{(m,0)} = (D_0PD'_0)_{k\ell}, \qquad 1\leq k,\ell\leq m. 
\] 
Similarly, $({\bf 1}_k^{(m,0)})'\Omega {\bf 1}_n=(e_k'\Pi_0')\Omega (\Pi{\bf 1}_K)=e_k'\Pi_0'\Theta\Pi P\Pi'\Theta \Pi {\bf 1}_K=e_k'D_0PD{\bf 1}_K$. This gives
\[
({\bf 1}_k^{(m,0)})'\Omega {\bf 1}_n = \diag(D_0PD{\bf 1}_K)_{kk}, \qquad 1\leq k,\ell\leq m.
\]
We plug the above equalities into \eqref{def:Omega-ij-m0}. It follows that, for $i\in {\cal N}_k^{(m,0)}$ and $j\in {\cal N}_{\ell}^{(m,0)}$, 
\beq \label{Omega-ij-m0-expression2}
\Omega_{ij}^{(m,0)}= d_i^* d_j^*\cdot  \bigl[(\diag(D_0PD{\bf 1}_K))^{-1}D_0PD'_0(\diag(D_0PD{\bf 1}_K))^{-1}\bigr]_{k\ell}. 
\eeq
Write for short
\beq \label{Omega-ij-m0-notationM}
M = [\diag(D_0PD{\bf 1}_K)]^{-1}(D_0PD'_0)[\diag(D_0PD{\bf 1}_K)]^{-1}. 
\eeq
Then, \eqref{Omega-ij-m0-expression2} can be written equivalently as 
\[
\Omega_{ij}^{(m,0)}=d_i^*d_j^*\cdot\sum_{k,\ell=1}^m M_{k\ell}\cdot 1\{i\in {\cal N}_k^{(m,0)}\}\cdot 1\{j\in {\cal N}_\ell^{(m,0)}\}.
\] 
By definition, $L_0(u,k)=1\{{\cal N}_{u}\subset{\cal N}_k^{(m,0)}\}$, for $1\leq u\leq K$ and $1\leq k\leq m$. Therefore, we have the equalities: $1\{i\in {\cal N}_k^{(m,0)}\}=\sum_{u=1}^K L_0(u,k)\cdot 1\{ i\in {\cal N}_u\}$ and $1\{j\in {\cal N}_\ell^{(m,0)}\}=\sum_{v=1}^K L_0(v,\ell)\cdot 1\{ j\in {\cal N}_v\}$. Combining them with the above equation gives
\begin{align}  \label{Omega-ij-m0-expression3}
\Omega_{ij}^{(m,0)}& =d_i^*d_j^*\cdot\sum_{u,v=1}^K 1\{ i\in {\cal N}_u\}\cdot 1\{ j\in {\cal N}_v\}\sum_{k,\ell=1}^m L_0(u,k)L_0(v,\ell)M_{k\ell}\cr
&=d_i^*d_j^*\cdot \sum_{u,v=1}^K 1\{ i\in {\cal N}_u\}\cdot 1\{ j\in {\cal N}_v\}\cdot (L_0ML_0')_{uv}.  
\end{align}
By definition, $d^* = \Omega {{\bf 1}_n}=\Omega(\Pi {\bf 1}_K)$. Since $\Omega=\Theta\Pi P\Pi'\Theta$,  we immediately have 
\[
d_i^* = \theta_i\cdot \pi_i' P \Pi' \Theta \Pi {\bf 1}_K = \theta_i\cdot \pi_i'PD{\bf 1}_K=\theta_i\cdot \sum_{u=1}^K \diag(PD{\bf 1}_K)_{uu}\cdot 1\{ i\in {\cal N}_{u}\}. 
\]
Similarly, we have $d_j^*=\theta_i\cdot \sum_{v=1}^K \diag(PD{\bf 1}_K)_{vv}\cdot 1\{ j\in {\cal N}_{v}\}$. Plugging the expressions of $(d_i^*, d_j^*)$ into \eqref{Omega-ij-m0-expression3} gives 
\begin{align}  \label{Omega-ij-m0-expression4}
\Omega_{ij}^{(m,0)}&=\theta_i\theta_j\sum_{u,v=1}^K 1\{i\in {\cal N}_u\}\,1\{j\in {\cal N}_v\}\, \diag(PD{\bf 1}_K)_{uu} (L_0ML_0')_{uv} \diag(PD{\bf 1}_K)_{vv}\cr
&= \theta_i\theta_j\cdot \pi_i'\bigl[ \diag(PD{\bf 1}_K)L_0ML_0'\diag(PD{\bf 1}_K)\bigr]\pi_j. 
\end{align}
Combining it with the expression of $M$ in \eqref{Omega-ij-m0-notationM} gives the claim. \qed


\subsection{Proof of Lemma~\ref{lemma:SgnQ2a}}  \label{subsec:SgnQ2a-proof}
The claim of $c_n\asymp \|\theta\|^8$ is proved in Lemma~\ref{lemma:SgnQ1a}. To prove the claim of $\lambda_1\asymp\|\theta\|^2$, we note that by Lemma~\ref{lemma:eigenvector-relation}, $\lambda_k=\|\theta\|^2\cdot\lambda_k(HPH)$, where $H$ is the diagonal matrix such that $H_{kk}=\|\theta^{(k)}\|^2/\|\theta\|^2$. By the condition \eqref{condition1b}, all the diagonal entries of $H$ are between $[c, 1]$, for a constant $c\in (0,1)$. It follows that $\lambda_1(HPH)\asymp \lambda_1(P)$. Since $\lambda_1\geq P_{11}=1$ and $\lambda_1\leq \|P\|\leq C$, we have $\lambda_1(P)\asymp 1$. Combining the above gives 
\[
\lambda_1\asymp \|\theta\|^2 \lambda_1(P)\asymp \|\theta\|^2.
\] 

We then prove the claims related to the matrix $\widetilde{\Omega}$. First, we show the upper bound of $|\widetilde{\Omega}_{ij}|$ and the lower bound of $\tr(\widetilde{\Omega}^4)$. Recall that $\widetilde{\Omega}=\Omega-\Omega^{(m,0)}$. By Lemma~\ref{lemma:SgnQ2-0}, $\Omega^{(m,0)}=\Theta\Pi P_0\Pi'\Theta$ for a rank-$m$ matrix $P_0$. It follows that 
\beq \label{lemma:SgnQ2a-1}
\widetilde{\Omega} = \Theta\Pi (P-P_0)\Pi'\Theta. 
\eeq
Let $H$ be the same diagonal matrix as above. It can be easily verified that $\|\theta\|^2\cdot H^2=\Pi'\Theta^2\Pi$. This means that the matrix $U=\|\theta\|^{-1}\Theta\Pi H^{-1}$ satisfies the equality $U'U=I_K$. As a result, we can write $\widetilde{\Omega} = U\cdot (\|\theta\|^2\cdot H(P-P_0)H)\cdot U'$. 
Since $U$ contains orthonormal columns, the nonzero eigenvalues of $\widetilde{\Omega}$ are the same as the nonzero eigenvalues of $\|\theta\|^2\cdot H(P-P_0)H$, i.e.,
\[
\tilde{\lambda}_k=\|\theta\|^2\cdot \lambda_k(H(P-P_0)H), \qquad 1\leq k\leq m. 
\] 
In particular, $|\tilde{\lambda}_1|=\|\theta\|^2\cdot \|H(P-P_0)H\|\asymp \|\theta\|^2\cdot \|P-P_0\|\asymp \lambda_1\|P-P_0\|$, where we have used $\|H\|\asymp \|H^{-1}\|\asymp 1$, and $\lambda_1\asymp\|\theta\|^2$. Combining it with the definition of $\tau$ gives
\beq \label{lemma:SgnQ2a-2}
\tau \asymp \|P-P_0\|. 
\eeq
Consider $|\widetilde{\Omega}_{ij}|$. 
By \eqref{lemma:SgnQ2a-1}, $|\widetilde{\Omega}_{ij}|=\theta_i\theta_j\cdot |\pi_i'(P-P_0)\pi_j|\leq\theta_i\theta_j\cdot C\|P-P_0\|$. We plug in \eqref{lemma:SgnQ2a-2} to get $|\widetilde{\Omega}_{ij}|\leq C\tau\theta_i\theta_j$, for $1\leq i,j\leq n$. Consider $\tr(\widetilde{\Omega}^4)$. We have seen that $|\tilde{\lambda}_1|\asymp \|\theta\|^2\cdot \|P-P_0\|\asymp \tau\|\theta\|^2$. As a result, $\tr(\widetilde{\Omega}^4)\geq \tilde{\lambda}_1^4\geq C\tau^4\|\theta\|^8$.

Next, we study the order of $\tau$. Note that $\Omega= \Omega^{(m,0)}+\widetilde{\Omega}$. We aim to apply Weyl's inequality.  In our notation, $\lambda_k(\cdot)$ refers to the $k$th largest eigenvalue (in magnitude) of a symmetric matrix. As a result, $|\lambda_k(\cdot)|$ is the $k$th singular value. 
By Weyl's inequality for singular values (equation (7.3.13) of \cite{HornJohnson}), we have 
\[
|\lambda_{r+s-1}(\Omega)| \leq |\lambda_r(\Omega^{(m,0)})|+|\lambda_s(\widetilde\Omega)|,  \quad \text{ for $1\leq r, s \leq n-1$}. 
\]
Since $\Omega^{(m,0)}$ only has $m$ nonzero eigenvalues, by taking $r=m+1$ and $s=k$ in the above, we immediately have
\beq \label{lemma:SgnQ2a-3}
|\lambda_{m+k}(\Omega)|\leq |\lambda_k(\widetilde{\Omega})|=|\tilde{\lambda}_k|, \qquad 1\leq k\leq K-m. 
\eeq
In particular, $|\tilde{\lambda}_1|\geq |\lambda_{m+1}|\geq |\lambda_K|$. At the same time, $\lambda_1\asymp\|\theta\|^2$ and by definition, $\tau=|\tilde{\lambda}_1|/\lambda_1$. It follows that
\[
\tau \|\theta\|\geq (|\lambda_K|/\lambda_1)\cdot \|\theta\|\geq C\bigl(|\lambda_K|/\sqrt{\lambda_1}\bigr)\to\infty. 
\]
This gives $\tau\|\theta\|\to\infty$. We then prove $\tau\leq C$. In  light of  \eqref{lemma:SgnQ2a-2}, it suffices to show $\|P_0\|\leq C$. Consider the expression of $P_0$ in Lemma~\ref{lemma:SgnQ2-0}. It is easy to see that $\|L_0\|\leq C$, $\|D_0PD'_0\|\leq C\|\theta\|_1^2$, and $\|\mathrm{diag}(PD{\bf 1}_K)\|\leq C\|\theta\|_1$. As a result,
\beq \label{lemma:SgnQ2a-4}
\|P_0\|\leq C\|\theta\|_1^4 \cdot \|\diag(D_0PD{\bf 1}_K)^{-1}\|^2. 
\eeq
Since $D_0=\Pi'_0\Theta\Pi$ and $D=\Pi'\Theta\Pi$, it is true that $D_0PD{\bf 1}_K=\Pi_0'\Theta\Pi P\Pi'\Theta\Pi {\bf 1}_K=\Pi_0'\Theta\Pi P\Pi'\Theta{\bf 1}_n=\Pi_0'\Omega {\bf 1}_n$. 
Then, for each $1\leq k\leq m$, 
\[
[\diag(D_0PD{\bf 1}_K)]_{kk}=(\Pi_0'\Omega {\bf 1}_n)_k= \sum_{i\in {\cal N}_k^{(m,0)}}d^*_i, \qquad\mbox{where}\quad d^*=\Omega {\bf 1}_n. 
\]
Here $\mathcal{N}_1^{(m,0)}, \mathcal{N}_2^{(m,0)}, ..., \mathcal{N}_m^{(m,0)}$ are the pseudo-communities defined by $\Pi_0$. 
Suppose $i\in {\cal N}_\ell$ for some true community ${\cal N}_\ell$. Then, $d_i^*\geq \sum_{j\in {\cal N}_\ell}\theta_i\theta_jP_{\ell\ell}=\theta_i\|\theta^{(\ell)}\|_1\geq C\theta_i\|\theta\|_1$. Moreover, for any $\Pi_0\in {\cal G}_m$, each pseudo-community ${\cal N}_k^{(m,0)}$ is the union of one or more true community. It yields that $\sum_{i\in {\cal N}_k^{(m,0)}}\theta_i\geq \min_{1\leq \ell\leq K}\{\|\theta^{(\ell)}\|_1\}\geq C\|\theta\|_1$. Combining these results  gives $\sum_{i\in {\cal N}_k^{(m,0)}}d_i^*\geq C\|\theta\|_1^2$. This shows that each diagonal entry of $\diag(D_0PD{\bf 1}_K)$ is lower bounded by $C\|\theta\|_1^2$. We immediately have
\beq  \label{lemma:SgnQ2a-5}
\|\diag(D_0PD{\bf 1}_K)^{-1}\|\leq C\|\theta\|_1^{-2}. 
\eeq
Combining \eqref{lemma:SgnQ2a-4} and \eqref{lemma:SgnQ2a-5} gives $\|P_0\|\leq C$. The claim $\tau\leq C$ then follows from \eqref{lemma:SgnQ2a-2}. \qed


\subsection{Proof of Lemma~\ref{lemma:SgnQ2b}} \label{subsec:SgnQ2b-proof}
Recall that $W=A-\Omega$. Given an $n\times n$ symmetric matrix $T$, we define a random variable:
\beq \label{lem-SgnQ2b-generalform}
{\cal Q}_W(T) = \sum_{i_1, i_2, i_3, i_4 (dist)}(W_{i_1i_2}+T_{i_1i_2})(W_{i_2i_3}+T_{i_2i_3})(W_{i_3i_4}+T_{i_3i_4})(W_{i_4i_1}+T_{i_4i_1}). 
\eeq
Then, $\widetilde{Q}_n^{(m,0)}$ is a special case with $T=\widetilde{\Omega}^{(m,0)}$, where $\widetilde{\Omega}^{(m,0)}$ is defined in \eqref{DefineOmegaL}.

We aim to study the general form of ${\cal Q}_W(T)$ and prove the following lemma:
\begin{lemma} \label{lemma:idealSQ}
Consider a DCBM model where (\ref{condition1a})-(\ref{condition1b}) and (\ref{condition1d}) hold. Let $W=A-\Omega$ and let ${\cal Q}_W(T)$ be the random variable defined in \eqref{lem-SgnQ2b-generalform}. As $n\to\infty$, suppose there is a constant $C>0$ and a scalar $\alpha_n>0$ such that $\alpha_n\leq C$, $\alpha_n\|\theta\|\to\infty$, and $|T_{ij}|\leq C\alpha_n\theta_i\theta_j$ for all $1\leq i,j\leq n$. Then, $\mathbb{E}[{\cal Q}_W(T)]=\tr(T^4)+o(\alpha^4\|\theta\|^8)$ and $\mathrm{Var}({\cal Q}_W(T))\leq C(\|\theta\|^8+\alpha_n^6\|\theta\|^8\|\theta\|_3^6)$. 
\end{lemma}

We now set $T=\widetilde{\Omega}^{(m,0)}$ and verify the conditions of Lemma~\ref{lemma:idealSQ}. Recall that $\tau=\tilde{\lambda}_1/\lambda_1$, where $\tilde{\lambda}_1$ and $\lambda_1$ are the respective largest (in magnitude) eigenvalue of $\widetilde{\Omega}^{(m,0)}$ and $\Omega$. By Lemma~\ref{lemma:SgnQ2a},
\[
\tau\leq C, \qquad \tau\|\theta\|\to\infty, \qquad |\widetilde{\Omega}_{ij}^{(m,0)}|\leq C\tau\theta_i\theta_j, \qquad\mbox{for all }1\leq i,j\leq n. 
\]
Therefore, we can apply Lemma~\ref{lemma:idealSQ} with $\alpha_n=\tau$. The claim follows immediately.

It remains to show Lemma~\ref{lemma:idealSQ}. By an expansion of \eqref{lem-SgnQ2b-generalform}, we write ${\cal Q}_W(T)$ as the sum of $2^4=16$ post-expansion sums. Each post-expansion sum takes a form
\beq\label{thm:IdealSgnQ-def:termX1}
X = \sum_{i_1, i_2, i_3, i_4 (dist)} a_{i_1i_2}b_{i_2i_3}c_{i_3i_4}d_{i_4 i_1}, 
\eeq
where each of $a_{ij}, b_{ij}, c_{ij}, d_{ij}$ may take value in $\{W_{ij}, T_{ij}\}$. 
Then, $\mathbb{E}[X]$ is equal to the sum of means of these post-expansion sums, and $\mathrm{Var}(X)$ is bounded by a constant times the sum of variances of these post-expansion sums. It suffices to study the means and variances of these post-expansion sums.

\begin{table}[tb!]   
\centering
\scalebox{0.85}{
\begin{tabular}{lclcr}
Type  & $\;\#\;\;$  &     Examples  &  Mean & Variance \\
\hline 
I   & 1  & $X_1 = \sum_{i_1, i_2, i_3, i_4 (dist)} W_{i_1i_2} W_{i_2i_3} W_{i_3i_4} W_{i_4i_1}$ & 0 & $ \asymp \|\theta\|^8$    \\  
II  & 4  & $X_2 = \sum_{i_1, i_2, i_3, i_4 (dist)} T_{i_1i_2} W_{i_2i_3} W_{i_3i_4} W_{i_4i_1}$ & 0 & $\leq C\alpha_n^2\|\theta\|^4\|\theta\|_3^6 = o(\|\theta\|^8)$    \\ 
IIIa & 4   & $X_3 = \sum_{i_1, i_2, i_3, i_4 (dist)} T_{i_1i_2} T_{i_2i_3} W_{i_3i_4} W_{i_4i_1}$ & 0 & $\leq C\alpha_n^4 \|\theta\|^6\|\theta\|_3^6 = o( \alpha_n^6\|\theta\|^8\|\theta\|_3^6 )$  \\   
IIIb             & 2   &  $X_4 = \sum_{i_1, i_2, i_3, i_4 (dist)} T_{i_1i_2} W_{i_2i_3} T_{i_3i_4} W_{i_4i_1}$ & 0 &  
$\leq C  \alpha_n^4\|\theta\|_3^{12} = o(\|\theta\|^8)$ \\  
IV & 4   & $X_5 = \sum_{i_1, i_2, i_3, i_4 (dist)} T_{i_1i_2} T_{i_2i_3} T_{i_3i_4}W_{i_4i_1}$    & 0 & $\leq \alpha_n^6 \|\theta\|^8\|\theta\|_3^6$   \\  
V & 1  & $X_6 = \sum_{i_1, i_2, i_3, i_4 (dist)} T_{i_1i_2} T_{i_2i_3}T_{i_3i_4} T_{i_4i_1}$ & $ \tr(T^4)$+$o(\alpha_n^4\|\theta\|^8)$  & 0 \\ 
\hline 
\end{tabular} 
} 
\caption{The post-expansion sums of ${\cal Q}_W(T)$ have 6 different types. We present the mean and variance of each type. Note that  $\|\theta\|^{-1} \ll \alpha_n \leq C$  and $ \|\theta\|_3^3\ll \|\theta\|^2 \ll \|\theta\|_1$.}  
 \label{tab:IdealSgnQsums}
\end{table}

We divide 16 post-expansion sums into 6 common types (see Table~\ref{tab:IdealSgnQsums}) and compute the mean and variance of each type. 
For example, for $
X_6=\sum_{i_1,i_2,i_3,i_4 (dist)}T_{i_1i_2}T_{i_2i_3}T_{i_3i_4}T_{i_4i_1}$,  
it is easy to see that 
\[
\mathrm{Var}(X_6)=0, \qquad\mbox{and}\qquad \mathbb{E}[X_6] =\tr(T^4)-\Delta,  
\]
where $\Delta$ is the sum of $T_{i_1i_2}T_{i_2i_3}T_{i_3i_4}T_{i_4i_1}$ over those $(i_1,i_2,i_3,i_4)$ such that some of the four indices are equal. Under the assumption of $|T_{ij}|\leq C\alpha_n\theta_i\theta_j$, we have $|T_{i_1i_2}T_{i_2i_3}T_{i_3i_4}T_{i_4i_1}|\leq C\alpha_n^4\theta_{i_1}^2\theta_{i_2}^2\theta_{i_3}^2\theta_{i_4}^2$. 
It follows that 
\[
\Delta\leq \sum_{i_2,i_3,i_4} C\alpha_n^4 \theta_{i_2}^4\theta_{i_3}^2\theta_{i_4}^2 =  O\bigl(\alpha_n^4\|\theta\|_4^4\|\theta\|^4\bigr)=o(\alpha_n^4\|\theta\|^8), 
\]
where the last equation uses $\|\theta\|_4^4\leq \theta_{\max}^2\|\theta\|^2=o(\|\theta\|^4)$. This gives the last row of Table~\ref{tab:IdealSgnQsums}. 
As another example, for $X_1=
\sum_{i_1,i_2,i_3,i_4 (dist)}W_{i_1i_2}W_{i_2i_3}W_{i_3i_4}W_{i_4i_1}$, it is seen that
\[
\mathbb{E}[X_1]=0.
\] 
Since the summands are mutually uncorrelated, we have  
\[
\mathrm{Var}(X_1)\leq \sum_{i_1,i_2,i_3,i_4 (dist)}\mathrm{Var}(W_{i_1i_2}W_{i_2i_3}W_{i_3i_4}W_{i_4i_1})\leq \sum_{i_1,i_2,i_3,i_4}C\theta_{i_1}^2\theta_{i_2}^2\theta_{i_3}^2\theta_{i_4}^2=O(\|\theta\|^8).
\]
We can also show that $\mathrm{Var}(X_1)\geq C\|\theta\|^8$ (this is very similar to the analysis in the proof of Lemma~\ref{lemma:SgnQ1b}). Therefore, we have $\mathrm{Var}(X_1)\asymp \|\theta\|^8$. This gives the first row of Table~\ref{tab:IdealSgnQsums}. 
The means and variances of other terms in Table~\ref{tab:IdealSgnQsums} are calculated in a similar way, and the details are omitted. \footnote{These calculations use some similar tricks as those in the proof of \cite{SP2019}. Interesting readers may refer to \cite{SP2019} for a more detailed guideline of conducting such calculations.}

Once we have the arguments in Table~\ref{tab:IdealSgnQsums},  Lemma~\ref{lemma:idealSQ} follows immediately. \qed

\subsection{Proof of Lemma~\ref{lemma:SgnQ2c}} \label{subsec:SgnQ2c-proof}
Before proceed, recall \eqref{def:tildeOmega} that
\[
\widetilde{Q}_n^{(m,0)} = \sum_{i_1, i_2, i_3, i_4 (dist)} (W_{i_1i_2} + \widetilde{\Omega}^{(m,0)}_{i_1i_2}) (W_{i_2i_3} + \widetilde{\Omega}^{(m,0)}_{i_2i_3}) (W_{i_3i_4} + \widetilde{\Omega}_{i_3i_4}^{(m,0)}) (W_{i_4i_1} + \widetilde{\Omega}^{(m,0)}_{i_4i_1}).
\]
Here $\widetilde{\Omega}^{(m,0)} = \Omega - \Omega^{(m,0)}$ and $\Omega^{(m,0)}$ is as in \eqref{def:tOmega}. By Lemma~\ref{lemma:SgnQ2-0}, $\Omega^{(m,0)}=\Theta\Pi P_0\Pi'\Theta$, for a rank-$m$ matrix $P_0$. If $m=K$ and $\Pi_0=\Pi$, it can be verified that $P_0=P$. Therefore, $\Omega^{(m,0)}=\Omega$, and $\widetilde{\Omega}^{(m,0)} $ reduces to a zero matrix. In this case, $\widetilde{Q}_n^{(m,0)}$ reduces to $\widetilde{Q}_n $ in \eqref{DefinewidetildeQ}. 
It means that we can treat Lemma~\ref{lemma:SgnQ1c} as a ``special case" of Lemma~\ref{lemma:SgnQ2c}, with $\widetilde{\Omega}^{(m,0)}$ being a zero matrix. We thus combine the proofs of two lemmas.  

We now show the claim. First, we introduce two proxies of $Q_n^{(m,0)}$. 
By definition, 
\[
Q_n^{(m,0)}  = \sum_{i_1, i_2, i_3, i_4 (dist)} (A_{i_1i_2} - \widehat{\Omega}^{(m,0)}_{i_1i_2}) (A_{i_2i_3} - \widehat{\Omega}^{(m,0)}_{i_2i_3}) (A_{i_3i_4} - \widehat{\Omega}_{i_3i_4}^{(m,0)}) (A_{i_4i_1} - \widehat{\Omega}^{(m,0)}_{i_4i_1}). 
\]
By \eqref{DefinehatOmegaL}, $\widehat{\Omega}^{(m,0)}$ is defined by $\hat{\theta}$, $\Pi_0$, and $\widehat{P}$. For $1\leq k\leq m$, let ${\cal N}_k^{(m,0)}$ and ${\bf 1}_k^{(m,0)}$ be the same as in \eqref{def:1m0}. Then, $(\hat{\theta}, \widehat{P})$ are obtained by replacing $\hat{\bf 1}_k$ with ${\bf 1}_k^{(m,0)}$ in \eqref{refitting1a}. For the rest of the proof, we write ${\bf 1}_k={\bf 1}_k^{(m,0)}$ for short. It follows that, for $1\leq k,\ell\leq K$ and $i\in {\cal N}_k^{(m,0)}$, 
\[
\hat{\theta}_i^{(m,0)}= d_i\frac{\sqrt{{\bf 1}_k'A{\bf 1}_k}}{{\bf 1}_k'A{\bf 1}_n}, \quad\widehat{P}_{k\ell}^{(m,0)} = \frac{{\bf 1}_k'A{\bf 1}_\ell}{\sqrt{({\bf 1}_k'A{\bf 1}_k)({\bf 1}_\ell'A{\bf 1}_\ell)}}, \quad \mbox{with }{\bf 1}_k={\bf 1}_k^{(m,0)} \mbox{ (for short)}. 
\] 
We plug it into \eqref{DefinehatOmegaL} and note that $d=A{\bf 1}_n$. It yields that, for $i\in {\cal N}_k^{(m,0)}$ and $j\in {\cal N}_{\ell}^{(m,0)}$, 
\beq\label{def:hatU}
\widehat\Omega_{ij}^{(m,0)}   = d_id_j\cdot \widehat{U}_{k\ell}^{(m,0)}, \qquad \mbox{where}\quad \widehat{U}_{k\ell}^{(m,0)} = \frac{{\bf 1}_k'A {\bf 1}_\ell }{ ({\bf 1}_k'd)({\bf 1}_\ell'd)}.
\eeq 
At the same time, in \eqref{def:Omega-ij-m0}, we have seen that (recall: $d^*=\Omega {\bf 1}_n$)
\beq\label{def:Ustar}
\Omega^{(m,0)}_{ij} = d_i^* d_j^* \cdot U_{k\ell}^{*(m,0)}, \qquad\mbox{where}\quad 
U_{k\ell}^{*(m,0)} =  \frac{ {\bf 1}_k'\Omega {\bf 1}_\ell }{ ({\bf 1}_k'd^*)({\bf 1}_\ell'd^*)}. 
\eeq
Note that $(\Omega, d^*)$ are approximately $(\mathbb{E}[A], \mathbb{E}[d])$ but there is subtle difference. We thus introduce an intermediate quantity:
\beq\label{def:U}
U_{k\ell}^{(m,0)} =  \frac{{\bf 1}_k'\mathbb{E}[A] {\bf 1}_\ell }{ ({\bf 1}_k'\mathbb{E}[d])({\bf 1}_\ell'\mathbb{E}[d])}.
\eeq
We now use \eqref{def:hatU}-\eqref{def:U} to decompose $(A_{ij}-\widehat{\Omega}^{(m,0)}_{ij})$. 
Recall that $\widetilde{\Omega}_{ij}^{(m,0)}=\Omega_{ij}-\Omega_{ij}^{(m,0)}$. We immediately have  
\beq\label{decompose-1}
A_{ij}-\widehat{\Omega}^{(m,0)}_{ij}=W_{ij} + \widetilde{\Omega}^{(m,0)}_{ij} + (\Omega^{(m,0)}_{ij}-\widehat{\Omega}^{(m,0)}_{ij}). 
\eeq
From now on, we omit the superscript ``$(m,0)$" in $\widehat{U}^{(m,0)}_{k\ell}$, $U^{*(m,0)}_{k\ell}$ and $U_{k\ell}^{(m,0)}$, and rewrite them as $\widehat{U}_{k\ell}$, $U^*_{k\ell}$, and $U_{k\ell}$, respectively.  
By \eqref{def:hatU}-\eqref{def:U}, $\Omega_{ij}^{(m,0)}-\widehat{\Omega}^{(m,0)}_{ij}=d_i^*d_j^*U^*_{k\ell}-d_id_j\widehat{U}_{k\ell}=[d_i^*d_j^*U^*_{k\ell}-(\mathbb{E}d_i)(\mathbb{E}d_j)U_{k\ell}]+U_{k\ell}[ (\mathbb{E}d_i)(\mathbb{E}d_j)-d_id_j]+ (U_{k\ell}-\widehat{U}_{k\ell}) d_id_j$. It turns our that the term $U_{k\ell}[ (\mathbb{E}d_i)(\mathbb{E}d_j)-d_id_j]$ is the ``dominating" term. This term does not have an exactly zero mean, and so we introduce a proxy to this term as 
\beq \label{def:delta}
\delta^{(m,0)}_{ij} = U_{k\ell}\big[ (\mathbb{E}d_i)(\mathbb{E}d_j - d_j) + (\mathbb{E}d_j)(\mathbb{E}d_i - d_i)\big].
\eeq
Note that $U_{k\ell}[ (\mathbb{E}d_i)(\mathbb{E}d_j)-d_id_j]=\delta^{(m,0)}_{ij}-U_{k\ell}(d_i-\mathbb{E}d_i)(d_j-\mathbb{E}d_j)$. 
We then have
\begin{align*}
\Omega_{ij}^{(m,0)}-\widehat{\Omega}^{(m,0)}_{ij}&=[d_i^*d_j^*U^*_{k\ell}-(\mathbb{E}d_i)(\mathbb{E}d_j)U_{k\ell}]+ [\delta^{(m,0)}_{ij}-U_{k\ell}(d_i-\mathbb{E}d_i)(d_j-\mathbb{E}d_j)] + (U_{k\ell}-\widehat{U}_{k\ell}) d_id_j\cr
&=\delta^{(m,0)}_{ij}+ [d_i^*d_j^*U^*_{k\ell}-(\mathbb{E}d_i)(\mathbb{E}d_j)U_{k\ell}]-U_{k\ell}(d_i-\mathbb{E}d_i)(d_j-\mathbb{E}d_j)\cr
&\qquad +(U_{k\ell}-\widehat{U}_{k\ell}) (\mathbb{E}d_i)(\mathbb{E}d_j)+(U_{k\ell}-\widehat{U}_{k\ell})[(\mathbb{E}d_i) (d_j-\mathbb{E}d_j)+(\mathbb{E}d_j)(d_i-\mathbb{E}d_i)]\cr
&\qquad + (U_{k\ell}-\widehat{U}_{k\ell})(d_i-\mathbb{E}d_i)(d_j-\mathbb{E}d_j)\cr
&= \delta^{(m,0)}_{ij}+\tilde{r}^{(m,0)}_{ij} +\epsilon^{(m,0)}_{ij}, 
\end{align*}
where
\beq \label{def:tilder}
\tilde{r}^{(m,0)}_{ij} = -\widehat{U}_{k\ell}(d_i-\mathbb{E}d_i) (d_j-\mathbb{E}d_j)
\eeq
and 
\begin{align} \label{def:eps}
\epsilon^{(m,0)}_{ij} &= [d_i^*d_j^*U^*_{k\ell}-(\mathbb{E}d_i)(\mathbb{E}d_j)U_{k\ell}] +(U_{k\ell}-\widehat{U}_{k\ell}) (\mathbb{E}d_i)(\mathbb{E}d_j)\cr
&\qquad + (U_{k\ell}-\widehat{U}_{k\ell})[(\mathbb{E}d_i) (d_j-\mathbb{E}d_j)+(\mathbb{E}d_j)(d_i-\mathbb{E}d_i)]. 
\end{align}
We plug the above results into \eqref{decompose-1} to get
\beq \label{decompose-2}
A_{ij}-\widehat{\Omega}_{ij}^{(m,0)}=\widetilde{\Omega}^{(m,0)}_{ij} + W_{ij} + \delta^{(m,0)}_{ij}+\tilde{r}^{(m,0)}_{ij}+\epsilon^{(m,0)}_{ij}. 
\eeq
We then use \eqref{decompose-2} to define two proxies of $Q_n^{(m,0)}$. For any $1\leq i\neq j\leq n$, let 
\begin{align}  \label{def:X}
X_{ij}&= \widetilde{\Omega}_{ij}^{(m,0)} + W_{ij} + \delta_{ij}^{(m,0)} + \tilde r_{ij}^{(m,0)} + \eps_{ij}^{(m,0)},\cr
\widetilde X^*_{ij} &= \widetilde{\Omega}_{ij}^{(m,0)} + W_{ij} + \delta_{ij}^{(m,0)} + \tilde r_{ij}^{(m,0)},\cr
X^*_{ij} &= \widetilde{\Omega}_{ij}^{(m,0)} + W_{ij} + \delta_{ij}^{(m,0)},\cr
\widetilde{X}_{ij} &= \widetilde{\Omega}_{ij}^{(m,0)} + W_{ij}. 
\end{align}
Correspondingly, we introduce 
\begin{align} \label{def:proxies}
Q_n^{(m,0)} &= \sum_{i_1, i_2, i_3, i_4 (dist)} X_{i_1i_2}X_{i_2i_3}X_{i_3i_4}X_{i_4i_1}\cr
\widetilde{Q}_n^{*(m,0)} &= \sum_{i_1, i_2, i_3, i_4 (dist)}  \widetilde{X}^*_{i_1 i_2} \widetilde{X}^*_{i_2 i_3} \widetilde{X}^*_{i_3 i_4}  \widetilde{X}^*_{i_4 i_1},\cr
Q_n^{*(m,0)}&= \sum_{i_1, i_2, i_3, i_4 (dist)} X_{i_1 i_2}^* X_{i_2 i_3}^* X_{i_3 i_4}^* X_{i_4 i_1}^*,\cr
\widetilde{Q}_n^{(m,0)} &= \sum_{i_1, i_2, i_3, i_4 (dist)}  \widetilde{X}_{i_1 i_2} \widetilde{X}_{i_2 i_3} \widetilde{X}_{i_3 i_4}  \widetilde{X}_{i_4 i_1}. 
\end{align}
By comparing it with \eqref{def:tildeOmega}, 
we can see that the above expression of $\widetilde{Q}_n^{(m,0)}$ is the same as before. Additionally, by \eqref{decompose-2}, the above expression of $Q_n^{(m,0)}$ is also equivalent to the definition. The other two quantities, $Q_n^{*(m,0)}$ and $\widetilde{Q}_n^{*(m,0)}$, are the two proxies we introduce here.    

Next, we decompose 
\[ 
  Q_n^{(m,0)} -  \widetilde{Q}_n^{(m,0)} = (Q_n^{*(m,0)} - \widetilde{Q}_n^{(m,0)})  +  (\widetilde Q_n^{*(m,0)} -  Q_n^{*(m,0)}) + (Q_n^{(m,0)} - \widetilde Q_n^{(*,m,0)}). 
\]
For any random variables $X,Y,Z$, we know that $\mathbb{E}[X+Y+Z]=\mathbb{E}X+\mathbb{E}Y+\mathbb{E}Z$ and $\mathrm{Var}(X+Y+Z)\leq 3\mathrm{Var}(X)+3\mathrm{Var}+3\mathrm{Var}(Z)$. 
Therefore, to show the claim, we only need to study the mean and variance of each term in the above equation. The next three lemmas are proved in Sections~\ref{subsec:proof-remainder0}-\ref{subsec:proof-remainder2}, respectively.

\begin{lemma}\label{prop:remainder0}
Let $b_n = 2 \|{\theta}\|^4 \cdot [ {g}' {V}^{-1} ({P}{H}^2{P}\circ {P}{H}^2{P}){V}^{-1} {g}]$ be the same as in \eqref{def:bncn}. 
Under conditions of Lemma~\ref{lemma:SgnQ1c}, it is true that \[
\mathbb{E}[Q_n^{*(m,0)} - \widetilde{Q}_n^{(m,0)} ] = b_n + o(\|\theta\|^4), \quad \mbox{ and  }\quad\mathrm{Var}\bigl(Q_n^{*(m,0)} - \widetilde{Q}_n^{(m,0)}\bigr) = o(\|\theta\|^8),
\]
Let $\tau=\tilde{\lambda}_1/\lambda_1$ be the same as in \eqref{def:tau}.
Under conditions of Lemma~\ref{lemma:SgnQ2c}, it is true that   
\[
\mathbb{E}[Q_n^{*(m,0)} - \widetilde{Q}_n^{(m,0)} ] = o(\tau^4\|\theta\|^8), \quad \mbox{ and  }\quad\mathrm{Var}\bigl(Q_n^{*(m,0)} - \widetilde{Q}_n^{(m,0)}\bigr) \leq C\tau^6\|\theta\|^8\|\theta\|_3^6 + o(\|\theta\|^8).
\]
\end{lemma}

\begin{lemma}\label{prop:remainder1}
Under conditions of Lemma~\ref{lemma:SgnQ1c}, it is true that
\[
\mathbb{E}[\widetilde{Q}_n^{*(m,0)}-{Q}_n^{*(m,0)}]=o(\|\theta\|^4),\quad \mbox{ and  }\quad \mathrm{Var}\bigl(\widetilde{Q}_n^{*(m,0)}-{Q}_n^{(*,m,0)}\bigr)=o(\|\theta\|^8).
\]
Under conditions of Lemma~\ref{lemma:SgnQ2c}, it is true that 
\[
\mathbb{E}[\widetilde{Q}_n^{*(m,0)}-{Q}_n^{*(m,0)}]=o\bigl(\|\theta\|^4 + \tau^4\|\theta\|^8\bigr),\quad \mbox{ and  }\quad \mathrm{Var}\bigl(\widetilde{Q}_n^{*(m,0)}-{Q}_n^{*(m,0)}\bigr)=o\bigl(\|\theta\|^8 + \tau^6\|\theta\|^8\|\theta\|_3^6\bigr).
\] 
\end{lemma}

\begin{lemma}\label{prop:remainder2}
Under conditions of Lemma~\ref{lemma:SgnQ1c}, it is true that 
\[
\mathbb{E}[Q_n^{(m,0)} - \widetilde Q_n^{*(m,0)}]=o(\|\theta\|^4),\quad \mbox{ and  }\quad \mathrm{Var}\bigl(Q_n^{(m,0)} - \widetilde Q_n^{*(m,0)}\bigr)=o(\|\theta\|^8).
\]
Under conditions of Lemma~\ref{lemma:SgnQ2c}, it is true that 
\[
\mathbb{E}[Q_n^{(m,0)} - \widetilde Q_n^{*(m,0)}]=o\bigl(\|\theta\|^4 + \tau^4\|\theta\|^8\bigr),\quad \mbox{ and  }\quad \mathrm{Var}\bigl(Q_n^{(m,0)} - \widetilde Q_n^{*(m,0)}\bigr)=o\bigl(\|\theta\|^8 + \tau^6\|\theta\|^8\|\theta\|_3^6\bigr).
\] 
\end{lemma}

We are now ready to prove Lemma~\ref{lemma:SgnQ1c} and Lemma~\ref{lemma:SgnQ2c}. By Lemmas~\ref{prop:remainder0}-\ref{prop:remainder2}, under the conditions of Lemma~\ref{lemma:SgnQ1c}, 
\[
\mathbb{E}[  Q_n^{(m,0)} -  \widetilde{Q}_n^{(m,0)}] = b_n + o(\|\theta\|^4), \quad\mbox{ and }\quad \mathrm{Var}(Q_n^{(m,0)} -  \widetilde{Q}_n^{(m,0)}) = o(\|\theta\|^8),
\]
which implies $\mathbb{E}(Q_n^{(m,0)} -  \widetilde{Q}_n^{(m,0)} - b_n)^2 = o(\|\theta\|^8)$ and completes the proof of Lemma~\ref{lemma:SgnQ1c}. Under the conditions of Lemma~\ref{lemma:SgnQ2c}, it follows from Lemmas~\ref{prop:remainder0}-\ref{prop:remainder2} that
\[
\mathbb{E}[  Q_n^{(m,0)} -  \widetilde{Q}_n^{(m,0)}]  = o(\tau^4\|\theta\|^8)\quad \mbox{ and }\quad \mathrm{Var}(Q_n^{(m,0)} -  \widetilde{Q}_n^{(m,0)}) \leq C\tau^6\|\theta\|^8\|\theta\|_3^6 +  o(\|\theta\|^8),
\]
which completes the proof of Lemma~\ref{lemma:SgnQ2c}. \qed

\subsection{Proof of Lemma~\ref{lemma:SgnQ2d}} \label{subsec:SgnQ2d-proof}
Let ${\cal G}_m$ be the class of $n\times m$ membership matrices that satisfy NSP (the definition of ${\cal G}_m$ is in Section~\ref{subsec:SgnQproof2}). 
By Theorem~\ref{thm:SCORE2}, $\widehat{\Pi}^{(m)}\in {\cal G}_m$ with probability $1-O(n^{-3})$. 
Given any $\Pi_0\in {\cal G}_m$, Let $B_n^{(m)}(\Pi_0)$ be defined in the same way as in \eqref{refitting2c}, except that $(\hat{\theta}, \hat{g},\widehat{V}, \widehat{P}, \widehat{H})$ are defined based on $\Pi_0$ instead of  $\widehat\Pi^{(m)}$. Then, with probability $1-O(n^{-3})$, 
 \[
B_n^{(m)} \leq \max_{\Pi_0\in\mathcal{G}_m} B_n(\Pi_0). 
\]
It follows from the probability union bound that 
\[
\mathbb{P}\bigl(B_n^{(m)}>C\|\theta\|^4\bigr)\leq \sum_{\Pi_0\in {\cal G}_m}\mathbb{P}\bigl(B_n(\Pi_0)>C\|\theta\|^4\bigr)+O(n^{-3}). 
\]
Since $m<K$ and $K$ is finite, ${\cal G}_m$ has only a bounded number of elements. Therefore, it suffices to show that
\beq \label{lem-SgnQ2d-goal}
\mathbb{P}\bigl(B_n(\Pi_0)>C\|\theta\|^4\bigr)=o(1), \qquad \mbox{for each }\Pi_0\in {\cal G}_m. 
\eeq

We now show \eqref{lem-SgnQ2d-goal}. From now on, we fix $\Pi_0\in {\cal G}_m$ and write $B_n(\Pi_0)=B_n$ for short. By \eqref{refitting2c} and direct calculations, 
\[
B_n = 2\|\hat{\theta}\|^4\cdot \hat{g}' \widehat{V}^{-1} (\widehat P\widehat H^2 \widehat P\circ \widehat P\widehat H^2 \widehat P) \widehat{V}^{-1}\hat g =2\|\hat{\theta}\|^4\cdot \sum_{1\leq k, \ell \leq m} \frac{\hat{g}_k \hat{g}_\ell [(\widehat{P}\widehat{H}^2\widehat{P})_{k ,\ell}]^2}{ (\widehat{P}_k' \hat{g}) \cdot (\widehat{P}_{\ell}' \hat{g})}, 
\]
where $\widehat{P}_k$ denotes the $k$th column of $\widehat{P}$. We have mis-used the notations $(\hat{\theta}, \hat{g},\widehat{V}, \widehat{P}, \widehat{H})$, using them to refer to the counterparts of original definitions with $\widehat{\Pi}^{(m)}$ replaced by $\Pi_0$. 
Denote by ${\cal N}_1^{(m,0)}, {\cal N}_2^{(m,0)},\ldots,{\cal N}_m^{(m,0)}$ the pseudo-communities defined by $\Pi_0$. Let ${\bf 1}_k^{(m,0)}\in\mathbb{R}^n$ be such that ${\bf 1}_k^{(m,0)}(i)=1\{i\in {\cal N}_k^{(m,0)}\}$. We write ${\bf 1}_k={\bf 1}_k^{(m,0)}$ when there is no confusion. By \eqref{refitting2gh}, 
\[
\hat{g} = ({\bf 1}_k'\hat{\theta})/\|\hat{\theta}\|_1, \qquad \hat{h}_k^2 = ({\bf 1}_k'\widehat{\Theta}^2{\bf 1}_k)/\|\hat{\theta}\|^2, \qquad 1\leq k\leq m. 
\]
Note that $\hat g$, $\hat h$ and $\widehat P$ all have non-negative entries, with all entries of $\hat g$ and $\hat h$ are further bounded by $1$. Moreover, the diagonals of $\widehat{P}$ are all equal to 1. 
It follows that, for all $1\leq k, \ell\leq m$,
\[
0\leq \hat g_k \leq \widehat{P}_{k}' \hat{g}, \quad\mbox{ and }\quad 0\leq (\widehat{P}\widehat{H}^2\widehat{P})_{k\ell} \leq (\widehat{P}^2)_{k\ell}.  
\] 
As a result,
\beq  \label{lem-SgnQ2d-bound1}
B_n\;\; \leq\;\; 2\|\hat{\theta}\|^4\sum_{k,\ell=1}^m [(\widehat{P}^2)_{k\ell}]^2 \;\;\leq\;\; 2\|\hat{\theta}\|^4\cdot m^4 \|\widehat{P}\|_{\max}^4, 
\eeq
where $\|\cdot\|_{\max}$ is the element-wise maximum norm. Below, we study $\|\widehat{P}\|_{\max}$ and $\|\hat{\theta}\|$ separately. 

First, we bound $\|\widehat{P}\|_{\max}$. By \eqref{refitting1a}, 
\[
\widehat{P}_{k\ell}=({\bf 1}_k'A{\bf 1}_\ell)/\sqrt{({\bf 1}_k'A{\bf 1}_k)({\bf 1}_\ell'A{\bf 1}_\ell)}.
\]
Write
${\bf 1}_k'A{\bf 1}_\ell=\sum_{i\in {\cal N}_k^{(m,0)}, j\in {\cal N}_\ell^{(m,0)}}A_{ij}$, where $\mathbb{E}[A_{ij}]=\Omega_{ij}$, and $\sum_{i\in {\cal N}_k^{(m,0)}, j\in {\cal N}_\ell^{(m,0)}}\mathrm{Var}(A_{ij})\leq\sum_{i\in {\cal N}_k^{(m,0)}, j\in {\cal N}_\ell^{(m,0)}} C\theta_i\theta_j\leq C({\bf 1}_k'\theta)({\bf 1}_\ell'\theta)$. We apply the Bernstein's inequality \cite{Wellner86} to get
\[
\mathbb{P}\bigl( |{\bf 1}_k'A{\bf 1}_\ell-{\bf 1}_k'\Omega {\bf 1}_\ell|>t \bigr)\leq 2\exp\Bigl( -\frac{t^2/2}{C({\bf 1}_k'\theta)({\bf 1}_\ell'\theta)+t/3} \Bigr), \quad\mbox{for all }t>0. 
\]
By NSP, each pseudo-community ${\cal N}_k^{(m,0)}$ contains at least one true community, say, ${\cal N}_{k^*}$. Combining it with the condition \eqref{condition1b} gives ${\bf 1}_k'\theta\geq \sum_{i\in {\cal N}_{k^*}}\theta_i \geq C\|\theta\|_1$. At the same time, ${\bf 1}_k'\theta\leq \|\theta\|_1$. We thus have ${\bf 1}_k'\theta\asymp \|\theta\|_1\gg\sqrt{\log(n)}$. Similarly, we can show that ${\bf 1}_k\Omega {\bf 1}_\ell \asymp \|\theta\|_1^2$. In the above equation, if we choose $t=C_1\|\theta\|_1\sqrt{\log(n)}$ for a properly large constant $C_1>0$, then the right hand side is $O(n^{-3})$. In other words, with probability $1-O(n^{-3})$, 
\[
|{\bf 1}_k'A{\bf 1}_\ell-{\bf 1}_k'\Omega {\bf 1}_\ell|\leq C\|\theta\|_1\sqrt{\log(n)}. 
\] 
Since ${\bf 1}_k'\Omega {\bf 1}_\ell\asymp\|\theta\|_1^2\gg \|\theta\|_1\sqrt{\log(n)}$, the above implies ${\bf 1}_k'A{\bf 1}_\ell\asymp \|\theta\|_1^2$. We combine this result with the probability union bound. It follows that there exists a constant $C_2>1$ such that with probability $1-O(n^{-3})$,
\beq \label{lem-SgnQ2d-bound2(0)}
C_2^{-1}\|\theta\|_1^2\leq \min_{1\leq k,\ell\leq m}\{{\bf 1}_k'A{\bf 1}_\ell\}\leq \max_{1\leq k,\ell\leq m}\{{\bf 1}_k'A{\bf 1}_\ell\}\leq C_2\|\theta\|_1^2
\eeq
We plug it into the expression of $\widehat{P}_{k\ell}$ above and can easily see that
\beq  \label{lem-SgnQ2d-bound2}
\|\widehat{P}\|_{\max}\leq C, \qquad \mbox{with probability }1-O(n^{-3}). 
\eeq

Second, we bound $\|\hat{\theta}\|$. 
By \eqref{refitting1a}, $\hat{\theta}_i=d_i\sqrt{{\bf 1}_k'A{\bf 1}_k}/({\bf 1}_k'A{\bf 1}_n)$ for $i\in {\cal N}_k^{(m,0)}$. It follows that
\[
\|\hat{\theta}\|^2 = \sum_{k=1}^m \frac{({\bf 1}_k'D^2{\bf 1}_k)({\bf 1}_k'A{\bf 1}_k)}{({\bf 1}_k'A{\bf 1}_n)^2}, \qquad\mbox{where}\quad D=\diag(d_1,d_2,\ldots,d_n). 
\]
Note that ${\bf 1}_k'A{\bf 1}_n=\sum_{\ell=1}^m{\bf 1}_k'A{\bf 1}_\ell$. It follows from \eqref{lem-SgnQ2d-bound2(0)} that ${\bf 1}_k'A{\bf 1}_k\asymp \|\theta\|_1^2$ and ${\bf 1}_k'A{\bf 1}_n\asymp \|\theta\|_1^2$. As a result, $\|\hat{\theta}\|^2\leq C\|\theta\|_1^{-2}\sum_{k=1}^m({\bf 1}_k'D^2{\bf 1}_k)$. Since $\sum_{k=1}^m({\bf 1}_k'D^2{\bf 1}_k)=\|d\|^2$, we immediately have 
\beq \label{lem-SgnQ2d-bound3}
\|\hat{\theta}\|^2  \leq C\|\theta\|_1^{-2}\|d\|^2, \qquad\mbox{with probability }1-O(n^{-3}). 
\eeq
Recall that $d_i=\sum_{j:j\neq i}A_{ij}=\sum_{j:j\neq i}(\Omega_{ij}+W_{ij})$. Then, 
\begin{align*}  
\|d\|^2& =\sum_{i=1}^n\sum_{j,s: j\neq i, s\neq i}(\Omega_{ij}+W_{ij})(\Omega_{is}+W_{is})\cr
&=\sum_{i,j,s:j\neq i, s\neq i}\Omega_{ij}\Omega_{is} + \underbrace{2\sum_{i\neq j} \Bigl(\sum_{s\notin\{i,j \}}\Omega_{is}\Bigr) W_{ij}}_{\equiv X_1} +\underbrace{\sum_{i\neq j}W^2_{ij}}_{\equiv X_2} + \underbrace{\sum_{i,j,s(dist)}W_{ij}W_{is}}_{\equiv X_3}. 
\end{align*}
Since $\sum_{s\notin\{i,j \}}\Omega_{is}\leq C\theta_i\|\theta\|_1$, we have $\mathbb{E}[X_1^2] \leq\sum_{i\neq j}C\theta_i^2\|\theta\|_1^2 \cdot\mathbb{E}[W^2_{ij}]\leq C\|\theta\|_3^3 \|\theta\|_1^3$. Moreover, $X_2\geq 0$ and $\mathbb{E}[X_2]=\sum_{i\neq j}\mathbb{E}[W^2_{ij}]\leq C\|\theta\|_1^2$. Last, $\mathbb{E}[X_3^2]= 2\sum_{i,j,s(dist)}\mathrm{Var}(W_{ij}W_{is})\leq C\sum_{i,j,s}\theta_i^2\theta_j\theta_s\leq C\|\theta\|^2\|\theta\|_1^2$. By Markov's inequality, for any sequence $\epsilon_n\to 0$,
\[
|X_1|\leq C\sqrt{\epsilon_n^{-1}\|\theta\|_3^3 \|\theta\|_1^3}, \qquad |X_2|\leq C\epsilon_n^{-1}\|\theta\|_1^2, \qquad  |X_3|\leq C\sqrt{\epsilon_n^{-1}\|\theta\|^2\|\theta\|^2_1}. 
\]
It is not hard to see that we can choose a property $\epsilon_n\to 0$ so that all the right hand sides are $o(\|\theta\|_1^2\|\theta\|^2)$. Then, with probability $1-\epsilon_n$, 
\[
\|d\|^2= \sum_{i,j,s:j\neq i, s\neq i}\Omega_{ij}\Omega_{is}+o(\|\theta\|_1^2\|\theta\|^2)\leq C\|\theta\|^2\|\theta\|_1^2. 
\]
We plug it into \eqref{lem-SgnQ2d-bound3} to get
\beq \label{lem-SgnQ2d-bound4}
\|\hat{\theta}\|^2\leq C\|\theta\|^2, \qquad\mbox{with probability }1-o(1). 
\eeq
Then, \eqref{lem-SgnQ2d-goal} follows from plugging \eqref{lem-SgnQ2d-bound2} and \eqref{lem-SgnQ2d-bound4} into \eqref{lem-SgnQ2d-bound1}. This proves the claim.\qed


\subsection{Proof of Lemma~\ref{prop:converge}} \label{subsec:prop-converge-proof}
Recall that ${\bf 1}_k\in\mathbb{R}^n$ is such that ${\bf 1}_k(i)=\{i\in {\cal N}_k\}$,  $D=\diag(d_1,d_2,\ldots,d_n)$, and $d^*=\Omega {\bf 1}_n$. We re-state the claims as 
 \beq \label{lem-converge-claim1}
 \frac{{\bf{1}}_n' A {\bf{1}}_n }{{\bf{1}}_n' \Omega {\bf{1}}_n }\stackrel{p}{\goto}1,\qquad
  \frac{{\bf{1}}_k' A {\bf{1}}_n }{{\bf{1}}_k' \Omega {\bf{1}}_n }\stackrel{p}{\goto}1, \qquad \frac{{\bf{1}}_k' A {\bf{1}}_k }{{\bf{1}}_k' \Omega {\bf{1}}_k }\stackrel{p}{\goto}1.
\eeq
and
\beq \label{lem-converge-claim2}
\frac{\|\hat\theta\|_1}{\|\theta\|_1} \stackrel{p}{\goto}1, \qquad \frac{\|\hat\theta\|}{\|\theta\|} \stackrel{p}{\goto}1, \qquad \frac{{\bf{1}}_k' D^2{\bf{1}}_k}{{\bf{1}}_k' (D^*)^2{\bf{1}}_k} \stackrel{p}{\to} 1. 
\eeq
We note that convergence in $\ell^2$-norm implies convergence in probability. Hence, to show $X\stackrel{p}{\to}1$ for a random variable $X$, it is sufficient to show $\mathbb{E}[(X-1)^2]\to 0$. Using the equality $\mathbb{E}[(X-1)^2]= (\mathbb{E}X-1)^2+\mathrm{Var}(X)$, we only need to prove that $\mathbb{E}[X]\to 1$ and $\mathrm{Var}(X)\to 0$, for each variable $X$ on the left hand sides of \eqref{lem-converge-claim1}-\eqref{lem-converge-claim2}. 

First, we prove the three claims in \eqref{lem-converge-claim1}. Since the proofs are similar, we only show the proof of the first claim. Note that ${\bf 1}_n'\Omega {\bf 1}_n=\sum_{k,\ell}({\bf 1}_k'\theta)({\bf 1}_\ell'\theta)P_{k\ell}$. Under the conditions \eqref{condition1a}-\eqref{condition1b}, ${\bf 1}_n'\Omega {\bf 1}_n\asymp \|\theta\|_1^2$. Additionally, ${\bf 1}_n'\diag(\Omega){\bf 1}_n=\|\theta\|^2$. It follows that
\[
\Bigl|\frac{\mathbb{E}[{\bf{1}}_n' A {\bf{1}}_n] }{{\bf{1}}_n' \Omega {\bf{1}}_n}-1\Bigr|=\frac{{\bf{1}}_n'\diag(\Omega){\bf{1}}_n }{{\bf{1}}_n' \Omega {\bf{1}}_n}\asymp \frac{\|\theta\|^2}{\|\theta\|_1^2}
  = o(1),
\] 
where the last inequality is because $\|\theta\|^2\leq \theta_{\max}\|\theta\|_1\leq C\|\theta\|_1$ and $\|\theta\|_1\to\infty$. Also, since the upper triangular entries of $A$ are independent, $\mathrm{Var}({\bf 1}_n'A{\bf 1}_n)=4\mathrm{Var}(\sum_{i<j}A_{ij})\leq 4\sum_{i<j}\Omega_{ij}\leq C\|\theta\|_1^2$. It follows that 
\[
\frac{\mathrm{Var}({\bf 1}_n'A{\bf 1}_n)}{({\bf 1}_n'\Omega {\bf 1}_n)^2}\leq \frac{C\|\theta\|_1^2}{\|\theta\|_1^4}=o(1). 
\]
Combining the above gives $({\bf 1}_n'A{\bf 1}_n)/({\bf 1}_n'\Omega {\bf 1}_n)\stackrel{p}{\goto}1$. 

Second, we show the first claim in \eqref{lem-converge-claim2}. By Theorem~\ref{thm:SCORE2}, $\widehat{\Pi}^{(K)}=\Pi$, with a probability of $1-O(n^{-3})$. It is sufficient to consider the re-defined $\hat{\theta}$ where $\widehat{\Pi}^{(K)}$ is replaced with $\Pi$.  
Combining it with the definition in \eqref{refitting1a}, we have $\hat{\theta}_i=d_i \sqrt{{\bf 1}_k'A{\bf 1}_k}/({\bf 1}_k'A{\bf 1}_n)$. It follows that
\[
\|\hat{\theta}\|_1=\sum_{k=1}^K \frac{({\bf 1}_k'd) \sqrt{{\bf 1}_k'A{\bf 1}_k}}{{\bf 1}_k'A{\bf 1}_n} = \sum_{k=1}^K \sqrt{{\bf 1}_k'A{\bf 1}_k},
\]
where the last equality is because of $d=A{\bf 1}_n$. At the same time, it is easy to see that ${\bf 1}_k'\Omega {\bf 1}_k=({\bf 1}_k'\theta)P_{kk}({\bf 1}_k'\theta)=({\bf 1}_k'\theta)^2$, which implies $
\|\theta\|_1 = \sum_{k=1}^K \sqrt{{\bf 1}_k'\Omega{\bf 1}_k}$. We thus have
\[
  \frac{\|\hat{\theta}\|_1}{\|\theta\|_1} 
  =\sum_{k=1}^K \delta_k X_k, \qquad \mbox{where}\quad \delta_k=\frac{\sqrt{{\bf 1}_k'\Omega{\bf 1}_k}}{\sum_{\ell=1}^K \sqrt{{\bf 1}_\ell'\Omega{\bf 1}_\ell}}, \;\; X_k =\sqrt{\frac{{\bf 1}_k'A{\bf 1}_k}{{\bf 1}_k'\Omega{\bf 1}_k}}. 
  \]
By the last claim in \eqref{lem-converge-claim1} and the continuous mapping theorem,   $X_k\stackrel{p}{\goto}1$ for each $1\leq k\leq K$. Also, $\sum_{k=1}^K \delta_k=1$. It follows immediately that $\sum_{k=1}^K\delta_kX_k\stackrel{p}{\goto}1$. This proves $\|\hat{\theta}\|_1/\|\theta\|_1\stackrel{p}{\goto}1$. 

Next, we show the last claim in \eqref{lem-converge-claim2}. Recall that $d^*=\Omega {\bf 1}_n$ and $D^*=\diag(d^*)$. Then, for $i\in {\cal N}_k$, $\sum_{i\in {\cal N}_k}(d_i^*)^2\leq C\sum_{i\in {\cal N}_k}(\theta_i\|\theta\|_1)^2\leq C\|\theta\|^2\|\theta\|_1^2$. At the same time, $d_i^*\geq \theta_iP_{kk}({\bf 1}_k'\theta)\geq C\theta_i\|\theta\|_1$, where we have used the condition \eqref{condition1b}. As a result, $\sum_{i\in {\cal N}_k}(d_i^*)^2\geq C\|\theta\|_1^2\sum_{i\in {\cal N}_k}\theta_i^2\geq C\|\theta\|^2\|\theta\|_1^2$, where we have used \eqref{condition1b} again. Combining the above gives 
\beq \label{lem-converge-bound1}
{\bf 1}_k'(D^*)^2{\bf 1}_k\asymp \|\theta\|^2\|\theta\|^2_1. 
\eeq
Note that ${\bf{1}}_k' D^2{\bf 1}_k =\sum_{t\in {\cal N}_k}(\sum_{i:i\neq t}A_{it})^2=\sum_{i,j}\sum_{t\in {\cal N}_k\backslash\{i,j\}}A_{it}A_{jt}$. Similarly, ${\bf 1}_k'(D^*)^2{\bf 1}_k=\sum_{i,j}\sum_{t\in {\cal N}_k}\Omega_{it}\Omega_{jt}$. We now write
\begin{align*}
{\bf 1}_k'D^2{\bf 1}_k &= \sum_{i} \sum_{t\in {\cal N}_k\backslash\{i\}}A^2_{it} +2 \sum_{ i< j}\sum_{t\in {\cal N}_k\backslash\{i,j\}}A_{it}A_{jt},\cr
{\bf 1}_k'(D^*)^2{\bf 1}_k &=\sum_i \sum_{t\in {\cal N}_k}\Omega^2_{it} +2 \sum_{ i< j}\sum_{t\in {\cal N}_k}\Omega_{it}\Omega_{jt}. 
\end{align*}
Note that $\mathbb{E}[A^2_{it}]=\mathbb{E}[A_{it}]=\Omega_{it}$ and $\mathbb{E}[A_{it}A_{jt}]=\Omega_{it}\Omega_{jt}$. 
As a result, 
\begin{align*}
\bigl|\mathbb{E}[{\bf 1}_k'D^2{\bf 1}_k] - {\bf 1}_k'(D^*)^2{\bf 1}_k\bigr|& \leq \sum_i\sum_{t\in {\cal N}_k\backslash\{i\}}(\Omega_{it}-\Omega_{it}^2)+\sum_{i}\Omega_{ii}^2 + 2\sum_{i<j}(\Omega_{ii}\Omega_{ji}+\Omega_{ij}\Omega_{jj})\cr
&\leq C\sum_{i}\sum_{t\in {\cal N}_k}\theta_i\theta_t + \|\theta\|^2 + C\sum_{i,j}\theta_i^3\theta_j\cr
&\leq C\bigl(\|\theta\|_1^2+\|\theta\|^2+\|\theta\|_3^3 \|\theta\|_1)\cr
&\leq C\|\theta\|_1^2, 
\end{align*}
where the last line is because $\|\theta\|_3^3\leq \theta_{\max}^2\|\theta\|_1\leq C\|\theta\|_1$. 
Combining it with \eqref{lem-converge-bound1} gives 
\beq  \label{lem-converge-bound2}
\Bigl| \frac{\mathbb{E}[{\bf 1}_k'D^2{\bf 1}_k]}{{\bf 1}_k'(D^*)^2{\bf 1}_k}-1  \Bigr|\leq \frac{C\|\theta\|_1^2}{\|\theta\|^2\|\theta\|_1^2}=o(1). 
\eeq
We then compute the variance. Write for short $X=\sum_{ i< j}\sum_{t\in {\cal N}_k\backslash\{i,j\}}A_{it}A_{jt}$. Note that
\begin{align*}
\mathrm{Var}({\bf 1}_k'D^2{\bf 1}_k)& \leq  2\mathrm{Var}\Bigl( \sum_{i} \sum_{t\in {\cal N}_k\backslash\{i\}}A^2_{it}\Bigr) + 2\mathrm{Var}(2X)\cr
&\leq C \sum_{i} \sum_{t\in {\cal N}_k}\Omega_{it} + 8\mathrm{Var}(X)\cr
&\leq C\|\theta\|_1^2 + 8\mathrm{Var}(X). 
\end{align*}
Since 
$A_{it}A_{jt}=(\Omega_{it} + W_{it})(\Omega_{jt} + W_{jt})$, we write
\begin{align*}
  X&= \sum_{ i< j}\sum_{t\in {\cal N}_k\backslash\{i,j\}} \Omega_{it}\Omega_{jt}+
      2\sum_{j}\sum_{t\in {\cal N}_k\backslash\{j\}}\Bigl(\sum_{i: i\neq t,i<j} \Omega_{it}\Bigr)W_{jt}+\sum_{ i< j}\sum_{t\in {\cal N}_k\backslash\{i,j\}}W_{it}W_{jt}\cr
      &\equiv X_0+2X_1+X_2. 
\end{align*}
Here, $X_0$ is non-stochastic. Therefore, $\mathrm{Var}(X)=\mathrm{Var}(2X_1+X_2)\leq 8\mathrm{Var}(X_1)+2\mathrm{Var}(X_2)$. It is seen that $\mathrm{Var}(X_1)\leq \sum_{j}\sum_{t\in {\cal N}_k}(\sum_{i} \Omega_{it})^2\cdot \Omega_{jt}\leq C\sum_{j}\sum_{t\in {\cal N}_k}(\theta_t\|\theta\|_1)^2\cdot \theta_j\theta_t\leq C\|\theta\|_3^3\|\theta\|_1^3$. Additionally, the summands in $X_2$ are mutually uncorrelated, so $\mathrm{Var}(X_3)\leq \sum_{i<j}\sum_{t\in {\cal N}_k}\Omega_{it}\Omega_{jt}\leq C\sum_{i,j,t}\theta_i\theta_j\theta_t^2\leq C\|\theta\|_1^2\|\theta\|^2$. Combining the above gives
\[
\mathrm{Var}(X)\leq C\bigl( \|\theta\|_3^3\|\theta\|_1^3+\|\theta\|_1^2\|\theta\|^2 \bigr)\leq C\|\theta\|_3^3\|\theta\|_1^3, 
\]
where in the second inequality we have used $\|\theta\|^2\leq \|\theta\|_1\|\theta\|_3^3$, which is a direct consequence of the Cauchy-Schwarz inequality. We combine the above to get
\[
\mathrm{Var}({\bf 1}_k'D^2{\bf 1}_k)\leq C\bigl(\|\theta\|_1^2 +\|\theta\|_3^3\|\theta\|_1^3)\leq C(\|\theta\|_1^2 + \theta_{\max}\|\theta\|^2\|\theta\|_1^3), 
\]
where in the second inequality we have used $\|\theta\|_3^3\leq \theta_{\max}\|\theta\|^2$. 
Combining it with \eqref{lem-converge-bound1} gives 
\beq  \label{lem-converge-bound3}
\frac{\mathrm{Var}({\bf 1}_k'D^2{\bf 1}_k)}{[{\bf 1}_k'(D^*)^2{\bf 1}_k]^2}\leq \frac{C\|\theta\|_1^2}{\|\theta\|^4\|\theta\|_1^4}+ \frac{C\theta_{\max}\|\theta\|^2\|\theta\|_1^3}{\|\theta\|^4\|\theta\|_1^4}=o(1). 
\eeq
By \eqref{lem-converge-bound2} and \eqref{lem-converge-bound3}, we have $({\bf 1}_k'D^2{\bf 1}_k)/[{\bf 1}_k'(D^*)^2{\bf 1}_k]\stackrel{p}{\goto}1$. 

Last, we show the second claim in \eqref{lem-converge-claim2}. Since $\hat{\theta}_i=d_i \sqrt{{\bf 1}_k'A{\bf 1}_k}/({\bf 1}_k'A{\bf 1}_n)$, we have
\[
\|\hat{\theta}\|^2 = \sum_{k=1}^K  \frac{({\bf 1}_k'D^2{\bf 1}_k)({\bf 1}_k'A{\bf 1}_k)}{({\bf 1}_k'A{\bf 1}_n)^2}. 
\]
At the same time, ${\bf 1}_k'\Omega {\bf 1}_k=({\bf 1}_k'\theta)^2$ and ${\bf 1}_k'\Omega {\bf 1}_n=({\bf 1}_k'\theta)[\sum_{\ell=1}^KP_{k\ell}({\bf 1}_\ell'\theta)]$. Furthermore, for $i\in {\cal N}_k$, $d_i^*=(\Omega{\bf 1}_n)_i=\theta_i[\sum_{\ell=1}^KP_{k\ell}({\bf 1}_\ell'\theta)]$, and so ${\bf 1}_k'(D^*)^2{\bf 1}_n=({\bf 1}_k'\Theta^2 {\bf 1}_k)[\sum_{\ell=1}^KP_{k\ell}({\bf 1}_\ell'\theta)]^2$. Combining these equalities gives
\[
\|\theta\|^2 = \sum_{k=1}^K{\bf 1}_k'\Theta^2{\bf 1}_k = \sum_{k=1}^K \frac{[{\bf 1}_k'(D^*)^2{\bf 1}_k]({\bf 1}_k'\Omega{\bf 1}_k)}{({\bf 1}_k'\Omega{\bf 1}_n)^2}. 
\]
It follows that
\[
\frac{\|\hat{\theta}\|^2}{\|\theta\|^2}=\sum_{k=1}^K \tilde{\delta}_k\tilde{X}_k, \quad\mbox{where}\;\; \tilde{\delta}_k=\frac{ \frac{[{\bf 1}_k'(D^*)^2{\bf 1}_k]({\bf 1}_k'\Omega{\bf 1}_k)}{({\bf 1}_k'\Omega{\bf 1}_n)^2} }{\sum_{\ell=1}^K \frac{[{\bf 1}_\ell'(D^*)^2{\bf 1}_\ell]({\bf 1}_\ell'\Omega{\bf 1}_\ell)}{({\bf 1}_\ell'\Omega{\bf 1}_n)^2}}, 
\;\;\tilde{X}_k= \frac{{\bf 1}_k'D^2{\bf 1}_k}{{\bf 1}_k'(D^*)^2{\bf 1}_k} \frac{{\bf 1}_k'A{\bf 1}_k}{{\bf 1}_k'\Omega {\bf 1}_k} \frac{({\bf 1}_k'\Omega{\bf 1}_n)^2}{({\bf 1}_k'A{\bf 1}_n)^2}. 
\]
By the claims in \eqref{lem-converge-claim1} and the last claim in \eqref{lem-converge-claim2}, as well as the continuous mapping theorem,  we have $\tilde{X}_k\stackrel{p}{\goto}1$ for each $1\leq k\leq K$. Since $\sum_{k=1}^K \tilde{\delta}_k=1$, it follows that $\sum_{k=1}^K\tilde{\delta}_k\tilde{X}_k\stackrel{p}{\goto}1$. This proves that $\|\hat{\theta}\|^2/\|\theta\|^2\stackrel{p}{\goto}1$. By the continuous mapping theorem again, $\|\hat{\theta}\|/\|\theta\|\stackrel{p}{\goto}1$. 
\qed


\subsection{Proof of Lemma~\ref{prop:remainder0}}\label{subsec:proof-remainder0}
We introduce a notation $M_{ijk\ell}(X)=X_{ij}X_{jk}X_{k\ell}X_{\ell i}$, for any symmetric $n\times n$ matrix $X$ and distinct indices $(i,j,k,\ell)$. Using the definition in \eqref{def:proxies}, we can write 
\begin{align*}
& Q_n^{*(m,0)} - \widetilde{Q}_n^{(m,0)}\cr
=\qquad & \sum_{i_1, i_2, i_3, i_4(dist)}[M_{i_1i_2i_3i_4}(X^*)-M_{i_1i_2i_3i_4}(\widetilde{X})], \qquad \mbox{where}\quad \begin{cases}
X^*_{ij} = \widetilde{\Omega}^{(m,0)}_{ij}+W_{ij}+\delta_{ij}^{(m,0)},\\
\widetilde{X}_{ij} = \widetilde{\Omega}^{(m,0)}_{ij}+W_{ij}.
\end{cases}
\end{align*}
For the rest of the proof, we omit superscripts in $\widetilde{\Omega}_{ij}^{(m,0)}$ and $\delta_{ij}^{(m,0)}$ to simplify notations. From the expression of $X^*_{ij}$ and $\widetilde{X}_{ij}$, we notice that $[M_{i_1i_2i_3i_4}(X^*)-M_{i_1i_2i_3i_4}(\widetilde{X})]$ expands to $3^4-2^4=65$ terms. Consequently, there are 65 post-expansion sums in $Q_n^{*(m,0)} - \widetilde{Q}_n^{(m,0)}$, each with the form
\[
\sum_{i_1, i_2, i_3, i_4(dist)} a_{i_1i_2}b_{i_2i_3}c_{i_3i_4}d_{i_4i_1}, \qquad\mbox{where}\quad a,b,c,d\in \{\widetilde{\Omega}, W, \delta\}. 
\]
In the first 4 columns of Table \ref{tab:ProxySgnQsum}, we group these post-expansion sums into 15 distinct terms, where the second column shows the counts of each distinct term. For example, in the setting of Lemma~\ref{lemma:SgnQ1c}, $\widetilde{\Omega}$ reduces to a zero matrix. Therefore, any post-expansion sum that involves $\widetilde{\Omega}$ is zero. Then, it follows from Table~\ref{tab:ProxySgnQsum} that 
\beq \label{lem-remainder0-decomposition}
Q_n^{*(m,0)} - \widetilde{Q}_n^{(m,0)}=4Y_1+4Z_1+2Z_2+4T_1+F,
\eeq
where the expression of $(Y_1, Z_1, Z_2, T_1,F)$ are given in the fourth column of Table~\ref{tab:ProxySgnQsum}. Similarly, in the setting of Lemma~\ref{lemma:SgnQ2c}, we have $Q_n^{*(m,0)} - \widetilde{Q}_n^{(m,0)}=4Y_1+8Y_2+4Y_3+\cdots +4T_2+F$. These are elementary calculations.

\begin{table}[htb!]  
\caption{The $10$ types of the post-expansion sums for  $(Q_n^{*(m,0)} - \widetilde{Q}_n^{(m,0)})$. Notations: same as in Table \ref{tab:IdealSgnQsums}.}     \label{tab:ProxySgnQsum} 
\centering
\scalebox{.83}{
\begin{tabular}{lcclcr}
Type  & $\#$ &  Name &   Formula   &  Abs. Mean &  Variance \\
\hline 
Ia   & 4  & $Y_1$   & $\sum_{\substack{i_1,i_2,i_3,i_4\\ (dist)}} \delta_{i_1i_2} W_{i_2i_3} W_{i_3i_4} W_{i_4i_1}$ & 0  & $\leq C\|\theta\|^2\|\theta\|_3^6=o(\|\theta\|^8)$    \\  
Ib  &  8  & $Y_2$   & $\sum_{\substack{i_1,i_2,i_3,i_4\\ (dist)}}\delta_{i_1i_2} \widetilde{\Omega}_{i_2i_3}  W_{i_3i_4}  W_{i_4i_1}$& 0 &  $\leq C\tau^2\|\theta\|^4\|\theta\|_3^6=o(\|\theta\|^8)$  \\ 
   & 4  &  $Y_3$  & $\sum_{\substack{i_1,i_2,i_3,i_4\\ (dist)}}\delta_{i_1i_2} W_{i_2i_3}   \widetilde{\Omega}_{i_3i_4}  W_{i_4i_1}$& 0 & $\leq C\tau^2\|\theta\|^4\|\theta\|_3^6=o(\|\theta\|^8)$   \\ 
Ic & 8 & $Y_4$  & $\sum_{\substack{i_1,i_2,i_3,i_4\\ (dist)}}\delta_{i_1i_2}  \widetilde{\Omega}_{i_2i_3} \widetilde{\Omega}_{i_3i_4}W_{i_4i_1}$ & $\leq C\tau^2\|\theta\|^6$=$o(\tau^4\|\theta\|^8)$ & $\leq \frac{C\tau^4\|\theta\|^{10}\|\theta\|_3^3}{\|\theta\|_1}=o(\tau^6\|\theta\|^8\|\theta\|_3^6)$  \\   
  & 4 & $Y_5$ & $\sum_{\substack{i_1,i_2,i_3,i_4\\ (dist)}}\delta_{i_1i_2}  \widetilde{\Omega}_{i_2i_3} W_{i_3i_4}  \widetilde{\Omega}_{i_4i_1}$ & 0 & $\leq \frac{C\tau^4\|\theta\|^4\|\theta\|_3^9}{\|\theta\|_1}=o(\|\theta\|^8)$  \\   
Id & 4  & $Y_6$ & $\sum_{\substack{i_1,i_2,i_3,i_4\\ (dist)}}\delta_{i_1i_2}  \widetilde{\Omega}_{i_2i_3} \widetilde{\Omega}_{i_3i_4}  \widetilde{\Omega}_{i_4i_1}$  & 0  & $\leq \frac{C\tau^6\|\theta\|^{12}\|\theta\|_3^3}{\|\theta\|_1}=O(\tau^6\|\theta\|^8\|\theta\|_3^6)$  \\
IIa & 4 &  $Z_1$  & $\sum_{\substack{i_1,i_2,i_3,i_4\\ (dist)}}\delta_{i_1i_2} \delta_{i_2i_3} W_{i_3i_4}  W_{i_4i_1}$  &  $\leq C\|\theta\|^4$=$o(\tau^4\|\theta\|^8)$ &  $\leq C \|\theta\|^2\|\theta\|_3^6=o(\|\theta\|^8)$   \\  
     & 2 &     $Z_2$   & $\sum_{\substack{i_1,i_2,i_3,i_4\\ (dist)}}\delta_{i_1i_2} W_{i_2i_3} \delta_{i_3i_4}  W_{i_4i_1}$    &  $\leq C\|\theta\|^4$=$o(\tau^4\|\theta\|^8)$  &  $\leq \frac{C\|\theta\|^6\|\theta\|_3^3}{\|\theta\|_1}=o(\|\theta\|^8)$    \\ 
IIb & 8 &   $Z_3$  & $\sum_{\substack{i_1,i_2,i_3,i_4\\ (dist)}}\delta_{i_1i_2} \delta_{i_2i_3} \widetilde{\Omega}_{i_3i_4} W_{i_4i_1}$    & $0$ &  $\leq C\tau^2\|\theta\|^4\|\theta\|_3^6=o(\|\theta\|^8)$   \\ 
& 4 &    $Z_4$  & $\sum_{\substack{i_1,i_2,i_3,i_4\\ (dist)}}\delta_{i_1i_2}  \widetilde{\Omega}_{j k} \delta_{i_3i_4}  W_{i_4i_1}$    & $\leq C\tau\|\theta\|^4$=$o(\tau^4\|\theta\|^8)$  &  $\leq \frac{C\tau^2\|\theta\|^8\|\theta\|_3^3}{\|\theta\|_1} = o(\|\theta\|^8)$   \\
IIc &  4 &     $Z_5$  & $\sum_{\substack{i_1,i_2,i_3,i_4\\ (dist)}}\delta_{i_1i_2} \delta_{i_2i_3} \widetilde{\Omega}_{i_3i_4} \widetilde{\Omega}_{i_4i_1}$ &  $\leq C\tau^2\|\theta\|^6$=$o(\tau^4\|\theta\|^8)$ & $\leq\frac{C\tau^4\|\theta\|^{14}}{\|\theta\|_1^2}=o(\tau^6\|\theta\|^8\|\theta\|_3^6)$ \\ 
    &  2 &   $Z_6$    & $\sum_{\substack{i_1,i_2,i_3,i_4\\ (dist)}}\delta_{i_1i_2} \widetilde{\Omega}_{i_2i_3} \delta_{i_3i_4} \widetilde{\Omega}_{i_4i_1}$ &  $\frac{C\tau^2\|\theta\|^8}{\|\theta\|_1^2}$=$o(\tau^4\|\theta\|^8)$  & $\leq \frac{C\tau^4\|\theta\|^8\|\theta\|_3^6}{\|\theta\|_1^2}=o(\|\theta\|^8)$  \\ 
IIIa & 4 &  $T_1$  & $\sum_{\substack{i_1,i_2,i_3,i_4\\ (dist)}}\delta_{i_1i_2} \delta_{i_2i_3} \delta_{i_3i_4} W_{i_4i_1}$ &  $\leq C\|\theta\|^4$=$o(\tau^4\|\theta\|^8)$ & $\leq \frac{C\|\theta\|^6 \|\theta\|_3^3}{\|\theta\|_1}=o(\|\theta\|^8)$  \\ 
IIIb & 4 &  $T_2$ & $\sum_{\substack{i_1,i_2,i_3,i_4\\ (dist)}}\delta_{i_1i_2} \delta_{i_2i_3} \delta_{i_3i_4} \widetilde{\Omega}_{i_4i_1}$ &
$\leq \frac{C\tau\|\theta\|^6}{\|\theta\|_1^3}$=$o(\tau^4\|\theta\|^8)$   & $\leq \frac{C\tau^2 \|\theta\|^8 \|\theta\|_3^3}{\|\theta\|_1}=o(\|\theta\|^8)$  \\ 
IV & 1 & $F$  & $\sum_{\substack{i_1,i_2,i_3,i_4\\ (dist)}}\delta_{i_1i_2} \delta_{i_2i_3} \delta_{i_3i_4} \delta_{i_4i_1}$ &  $\leq C\|\theta\|^4$=$o(\tau^4\|\theta\|^8)$  & $\leq \frac{C\|\theta\|^{10}}{\|\theta\|_1^2}=o(\|\theta\|^8)$ \\ 
\hline 
\end{tabular} 
} 
\end{table}

To show the claim, we need to study the mean and variance of each post-expansion sum. We take $Y_1$ for example. Let ${\cal N}_1^{(m,0)},{\cal N}_2^{(m,0)},\ldots,{\cal N}_m^{(m,0)}$ be the pseudo-communities defined by $\Pi_0$. For each $1\leq i\leq n$, let $\tau(i)\in \{1,2,\ldots,m\}$ be the index of the pseudo-community that contains node $i$.  By \eqref{def:delta},  
\begin{align} \label{delta-expansion}
\delta_{i_1i_2}& =U_{\tau(i_1)\tau(i_2)}\bigl[ (\mathbb{E}d_{i_1})(\mathbb{E}d_{i_2}-d_{i_2}) + (\mathbb{E}d_{i_2})(\mathbb{E}d_{i_1}-d_{i_1}) \bigr]\cr
&= U_{\tau(i_1)\tau(i_2)} \cdot \mathbb{E}d_{i_1}\cdot \Bigl(-\sum_{j: j\neq i_2}W_{ji_2}\Bigr) +U_{\tau(i_1)\tau(i_2)} \cdot \mathbb{E}d_{i_2}\cdot \Bigl(-\sum_{\ell: \ell\neq i_1}W_{\ell i_1}\Bigr) \cr
&=-2 \sum_{j:j\neq i_2}U_{\tau(i_1)\tau(i_2)} \cdot \mathbb{E}d_{i_1}\cdot W_{ji_2}.  
\end{align}
It follows that 
\[
Y_1 = -2 \sum_{i_2,i_3,i_4, j} \Bigl(\sum_{i_1}U_{\tau(i_1)\tau(i_2)} \cdot \mathbb{E}d_{i_1}\Bigr) \cdot W_{ji_2}W_{i_2i_3}W_{i_3i_4}W_{i_4i_1},
\]
where we note that the indices $\{i_1,i_2,i_3,i_4,j\}$ have to satisfy the constraint that $i_1, i_2, i_3, i_4$ are distinct and that $j\neq i_2$. We can see that $Y_1$ is a weighted sum of $W_{ji_2}W_{i_2i_3}W_{i_3i_4}W_{i_4i_1}$, where the summands have zero mean and are mutually uncorrelated. The mean and variance of $Y_1$ can be calculated easily. We will use the same strategy to analyze each term in Table~\ref{tab:ProxySgnQsum}--- we use the expansion of $\delta_{ij}$ in \eqref{delta-expansion} to write each post-expansion sum as a weighted sum of monomials of $W$, and then we calculate the mean and variance. The calculations can become very tedious for some terms (e.g., $T_1$, $T_2$ and $F$), because of combinatorics. Fortunately, similar calculations were done in Section G of the supplement of \cite{SP2019}, where they analyzed a special case with $U_{k\ell}\equiv 1/v$ for all $1\leq k,\ell\leq m$. However, their proof does not rely on that $U_{k\ell}$'s are equal but only require that $U_{k\ell}$'s have a uniform upper bound. Essentially, they have proved the following lemma:
  
\begin{lemma} \label{lemma:proxySQ}
Consider a DCBM model where (\ref{condition1a})-(\ref{condition1b}) and (\ref{condition1d}) hold. Let $W=A-\Omega$ and $\Delta=\sum_{i_1,i_2,i_3,i_4(dist)}\bigl[ M_{i_1i_2i_3i_4}\bigl(\widetilde{\Omega}+W+\delta\bigr)-M_{i_1i_2i_3i_4}\bigl(\widetilde{\Omega}+W\bigr)\bigr]$, where $\widetilde{\Omega}$ is a non-stochastic symmetric matrix, $\delta_{ij}=v_{ij}\cdot[(\mathbb{E}d_i)(\mathbb{E}d_j-d_j)+(\mathbb{E}d_j)(\mathbb{E}d_i-d_i)]$, $\{v_{ij}\}_{1\leq i\neq j\leq n}$ are non-stochastic scalars, $d_i$ is the degree of node $i$, and $M_{i_1i_2i_3i_4}(\cdot)$ is as defined above. As $n\to\infty$, suppose there is a constant $C>0$ and a scalar $\alpha_n>0$ such that $\alpha_n\leq C$, $\alpha_n\|\theta\|\to\infty$, $|\widetilde{\Omega}_{ij}|\leq C\alpha_n\theta_i\theta_j$ and $|v_{ij}|\leq C\|\theta\|_1^{-1}$ for all $1\leq i,j\leq n$. Then, $|\mathbb{E}[\Delta]|= o(\alpha_n^4\|\theta\|^8)$ and $\mathrm{Var}(\Delta)\leq C\alpha_n^6\|\theta\|^8\|\theta\|_3^6+o(\|\theta\|^8)$. Furthermore, if $\widetilde{\Omega}$ is a zero matrix, then $|\mathbb{E}[\Delta]|\leq C\|\theta\|^4$ and $\mathrm{Var}(\Delta)=o(\|\theta\|^8)$. 
\end{lemma}

To apply Lemma~\ref{lemma:proxySQ}, we need to verify that $U_{k\ell}$ has a uniform upper bound for all $1\leq k,\ell\leq m$.  By Lemma~\ref{lemma:SgnQ2a}, $\tau\leq C$, $\tau\|\theta\|\to\infty$, and $|\widetilde{\Omega}_{ij}|\leq C\tau\theta_i\theta_j$. By \eqref{def:U},
\[
U_{k\ell}=({\bf 1}_k'\mathbb{E}[A] {\bf 1}_\ell)/[({\bf 1}_k'\mathbb{E}[d]) ({\bf 1}_\ell'\mathbb{E}[d]) ]. 
\]
where ${\bf 1}_k={\bf 1}_k^{(m,0)}$ is the same as in \eqref{def:1m0}. Since $\mathbb{E}[A_{ij}]=\Omega_{ij}\leq C\theta_i\theta_j$, we have $0\leq {\bf 1}_k'\mathbb{E}[A] {\bf 1}_\ell\leq C\|\theta\|_1^2$.  At the same time, by the NSP of SCORE, for each $1\leq k\leq m$, there is at least one true community ${\cal N}_{k^*}$ such that ${\cal N}_{k^*}\subset {\cal N}_k^{(m,0)}$. It follows that ${\bf 1}_k'\mathbb{E}[d]= \sum_{i\in {\cal N}_k^{(m,0)}}\sum_{j: j\neq i}\Omega_{ij}\geq \sum_{\{i,j\}\subset {\cal N}_{k^*}, i\neq j}\theta_i\theta_jP_{kk}=\|\theta^{(k)}\|_1^2[1+o(1)]\geq C\|\theta\|_1^2$, where the last inequality is from the condition \eqref{condition1b}. We plug these results into $U_{k\ell}$ to get
\beq \label{lem-remainder0-Ubound}
0\leq U_{k\ell}\leq C\|\theta\|_1^{-2}. 
\eeq
Then, the conditions of Lemma~\ref{lemma:proxySQ} are satisfied. We apply this lemma with $\alpha_n=\tau$ and $v_{ij}=U_{k\ell}$ for $i\in {\cal N}_k^{(m,0)}$ and $j\in {\cal N}_\ell^{(m,0)}$. It yields that, under the conditions of Lemma~\ref{lemma:SgnQ2c}, 
\[
\bigl|\mathbb{E}[Q_n^{*(m,0)} - \widetilde{Q}^{(m,0)}_n]\bigr|=o(\tau^4\|\theta\|^8), \qquad \mathrm{Var}\bigl( Q_n^{*(m,0)} - \widetilde{Q}^{(m,0)}_n \bigr)\leq C\tau^6\|\theta\|^8\|\theta\|_3^6 + o(\|\theta\|^8), 
\] 
and that under the conditions of Lemma~\ref{lemma:SgnQ1c} (where $\widetilde{\Omega}$ is a zero matrix)
\[
\bigl|\mathbb{E}[Q_n^{*(m,0)} - \widetilde{Q}^{(m,0)}_n]\bigr|\leq C\|\theta\|^4, \qquad \mathrm{Var}\bigl( Q_n^{*(m,0)} - \widetilde{Q}^{(m,0)}_n \bigr)\leq o(\|\theta\|^8). 
\]
This proves all the desirable claims except for the following one: Under conditions of Lemma~\ref{lemma:SgnQ1c}, 
\beq \label{lem-remainder0-goal}
\mathbb{E}[Q_n^{*(m,0)} - \widetilde{Q}^{(m,0)}_n]=b_n+o(\|\theta\|^4).
\eeq

We now show \eqref{lem-remainder0-goal}. By \eqref{lem-remainder0-decomposition}, we only need to calculate the expectations of $Y_1, Z_1, Z_2, T_1$ and $F$. From Table~\ref{tab:ProxySgnQsum}, $\mathbb{E}[Y_1]=0$. We now study $\mathbb{E}[Z_1]$. Recall that $\delta_{ij}=U_{\tau(i)\tau(j)}[(\mathbb{E}d_i)(\mathbb{E}d_j-d_j)+(\mathbb{E}d_j)(\mathbb{E}d_i-d_i)]$, where $\tau(i)$ is the index of pseudo-community defined by $\Pi_0$ that contains node $i$. We plug $\delta_{ij}$ into $Z_1$, by elementary calculations, 
\begin{align*}
Z_1 &= \sum_{i_1, i_2, i_3, i_4(dist)}U_{\tau(i_1)\tau(i_2)}U_{\tau(i_2)\tau(i_3)}(\mathbb{E}d_{i_1})(\mathbb{E}d_{i_2}-d_{i_2})^2(\mathbb{E}d_{i_3})W_{i_3i_4}W_{i_4i_1}\cr
&\qquad +2 \sum_{i_1, i_2, i_3, i_4(dist)}U_{\tau(i_1)\tau(i_2)}U_{\tau(i_2)\tau(i_3)}(\mathbb{E}d_{i_1})(\mathbb{E}d_{i_2}-d_{i_2})(\mathbb{E}d_{i_2})(\mathbb{E}d_{i_3}-d_{i_3})W_{i_3i_4}W_{i_4i_1}\cr
&\qquad + \sum_{i_1, i_2, i_3, i_4(dist)}U_{\tau(i_1)\tau(i_2)}U_{\tau(i_2)\tau(i_3)}(\mathbb{E}d_{i_1}-d_{i_1})(\mathbb{E}d_{i_2})^2(\mathbb{E}d_{i_3}-d_{i_3})W_{i_3i_4}W_{i_4i_1}. 
\end{align*}
We write it as $Z_1=Z_{11}+2Z_{12}+Z_{13}$. 
For $Z_{1k}$, we can further replace $\mathbb{E}d_i-d_i$ by $\sum_{j: j\neq i}W_{ji}$ and write $Z_{1k}$ as a weighted sum of monomials of $W$. Then, $\mathbb{E}[Z_{1k}]\neq 0$ if some of the monomials are $W_{i_3i_4}^2W^2_{i_4i_1}$. This will not happen in $Z_{11}$ and $Z_{12}$, and so only $Z_{13}$ has a nonzero mean. It is seen that 
\begin{align}\label{lem-remainder0-Z12}
\mathbb{E}[Z_{13}]&= \mathbb{E}\biggl[  \sum_{\substack{i_1, i_2, i_3, i_4\\(dist)}}U_{\tau(i_1)\tau(i_2)}U_{\tau(i_2)\tau(i_3)}\Bigl(\sum_{j: j\neq i_1}W_{ji_1}\Bigr)(\mathbb{E}d_{i_2})^2\Bigl(\sum_{k: k\neq i_3}W_{i_3k}\Bigr) W_{i_3i_4}W_{i_4i_1} \biggr]\cr
&=\mathbb{E}\biggl[  \sum_{i_1, i_2, i_3, i_4(dist)}U_{\tau(i_1)\tau(i_2)}U_{\tau(i_2)\tau(i_3)}\bigl(W_{i_4i_1}\bigr)(\mathbb{E}d_{i_2})^2(W_{i_3i_4})\cdot W_{i_3i_4}W_{i_4i_1} \biggr] \cr
&= \sum_{i_1, i_2, i_3, i_4(dist)}U_{\tau(i_1)\tau(i_2)}U_{\tau(i_2)\tau(i_3)}(\mathbb{E}d_{i_2})^2\cdot \mathbb{E}[W^2_{i_3i_4}W^2_{i_4i_1}]\cr
&=\sum_{k_1,k_2,k_3,k_4}\sum_{j=1}^4 \sum_{i_j\in {\cal N}_{k_j}}U_{k_1k_2}U_{k_2k_3}(\mathbb{E}d_{i_2})^2\cdot \mathbb{E}[W^2_{i_3i_4}W^2_{i_4i_1}].
\end{align}
Here, in the second line, we only keep $(j,k)=(i_4, i_4)$, because other $(j,k)$ only contribute zero means. Recall that we are considering the setting of Lemma~\ref{lemma:SgnQ1c}, where $m=K$ and $\Pi_0=\Pi$. 
In \eqref{def:Ustar}, we introduce a proxy of $U_{k\ell}$ as $U^*_{k\ell}=({\bf 1}_k'\Omega {\bf 1}_\ell)/[({\bf 1}_k'\Omega {\bf 1}_n)({\bf 1}_k'\Omega {\bf 1}_n)]$, for all $1\leq k,\ell\leq K$. Note that $\Omega_{ij}=\theta_i\theta_jP_{k\ell}$ for $i\in {\cal N}_k$ and $j\in {\cal N}_\ell$. At the same time, by  \eqref{def:bncn}, $g_k=({\bf 1}_k'\theta)/\|\theta\|_1$, and $V_{kk}=(\diag(Pg))_{kk}=[\sum_{\ell}P_{k\ell}({\bf 1}_\ell'\theta)]/\|\theta\|_1$.  It follows that 
\[
U^*_{k\ell} = \frac{P_{k\ell}({\bf 1}_k'\theta)({\bf 1}_\ell'\theta)}{({\bf 1}_k'\theta)[\sum_{k_1}P_{kk_1}({\bf 1}_{k_1}'\theta)]\cdot ({\bf 1}_\ell'\theta)[\sum_{\ell_1}P_{\ell \ell_1}({\bf 1}_{\ell_1}'\theta)]}=\frac{P_{k\ell}}{V_{kk}V_{\ell\ell} \|\theta\|_1^2}.
\]
Comparing $U_{k\ell}$ with $U^*_{k\ell}$ (see \eqref{def:Ustar}-\eqref{def:U}), the difference is negligible. (We can rigorously justify this by directly computing the difference caused by replacing $U_{k\ell}$ with $U^*_{k\ell}$, similarly as in the proof of $c_n=\tr(\widetilde{\Omega}^4)+o(\|\theta\|^8)$ in Section~\ref{subsec:SgnQ1a-proof}; see details therein. Such calculations are too elementary and so omitted.) We thus have
\beq \label{lem-remainder0-U}
U_{k\ell} = [1+o(1)]\cdot \frac{P_{k\ell}}{V_{kk}V_{\ell\ell} \|\theta\|_1^2}.
\eeq
Furthermore, for $i\in {\cal N}_k$, 
\beq \label{lem-remainder0-Ed}
\mathbb{E}[d_i]=[1+o(1)]\sum_{j=1}^n\Omega_{ij}=[1+o(1)] \cdot \theta_i\Bigl[\sum_{\ell=1}^K P_{k\ell}({\bf 1}_\ell'\theta)\Bigr]=[1+o(1)]\cdot\theta_i \|\theta\|_1 V_{kk}. 
\eeq
Also, $\mathbb{E}[W^2_{ij}]=\Omega_{ij}(1-\Omega_{ij})=\Omega_{ij}[1+o(1)]$. We plug these results into \eqref{lem-remainder0-Z12} to get 
\begin{align*}
\mathbb{E}[Z_{13}]&= [1+o(1)] \sum_{\substack{k_1,k_2,\\k_3,k_4}}\sum_{j=1}^4 \sum_{i_j\in {\cal N}_{k_j}}\frac{P_{k_1k_2} P_{k_2k_3}}{V_{k_1k_1}V_{k_2k_2}^2V_{k_3k_3} \|\theta\|_1^4}\cdot \bigl(\theta_{i_2}^2\|\theta\|_1^2 V^2_{k_2k_2}\bigr)\cdot\Omega_{i_3i_4}\Omega_{i_4i_1}\cr
&= [1+o(1)] \sum_{\substack{k_1,k_2,\\k_3,k_4}} \frac{P_{k_1k_2} P_{k_2k_3}P_{k_3k_4}P_{k_4k_1}}{V_{k_1k_1}V_{k_3k_3} \|\theta\|_1^2}  \Bigl(\sum_{i_j\in {\cal N}_{k_j}}\sum_{j=1}^4  \theta_{i_1}\theta^2_{i_2}\theta_{i_3} \theta_{i_4}^2\Bigr)\cr
&= [1+o(1)] \sum_{\substack{k_1,k_2,\\k_3,k_4}} \frac{P_{k_1k_2} P_{k_2k_3}P_{k_3k_4}P_{k_4k_1}}{V_{k_1k_1}V_{k_3k_3} \|\theta\|_1^2} \bigl(\|\theta\|^4\|\theta\|_1^2\cdot g_{k_1} g_{k_3} H^2_{k_2k_2} H^2_{k_4k_4}\bigr)\cr
&= [1+o(1)]\|\theta\|^4 \sum_{k_1,k_3}\frac{g_{k_1}}{V_{k_1k_1}}\Bigl( \sum_{k_2}P_{k_1k_2}H^2_{k_2k_2} P_{k_2k_3}\Bigr)\Bigl(\sum_{k_4}P_{k_3k_4}H^2_{k_4k_4}P_{k_4k_1}\Bigr)\frac{g_{k_3}}{V_{k_3k_3}}\cr
&= [1+o(1)]\|\theta\|^4 \sum_{k_1,k_3} (V^{-1}g)_{k_1} (PH^2P)_{k_1k_3} (PH^2P)_{k_3k_1}(V^{-1}g)_{k_3}\cr
&=[1+o(1)]\|\theta\|^4\cdot g'V^{-1}[(PH^2P)\circ(PH^2P)]V^{-1}g\cr
&= [1+o(1)]\cdot b_n/2, 
\end{align*}
where in the third line we have used the definition of $H$ which gives $H_{kk}=({\bf 1}_k'\Theta^2{\bf 1}_k)^{1/2}/\|\theta\|$. 
It follows that
\beq \label{lem-remainder0-Z1-mean}
\mathbb{E}[Z_1]=\mathbb{E}[Z_{13}]= [1+o(1)]\cdot b_n/2. 
\eeq
We then study $\mathbb{E}[Z_2]$. Similarly, we first plug in $\delta_{ij}=U_{\tau(i)\tau(j)}[(\mathbb{E}d_i)(\mathbb{E}d_j-d_j)+(\mathbb{E}d_j)(\mathbb{E}d_i-d_i)]$ and then plug in $d_i-\mathbb{E}d_i=\sum_{j\neq i}W_{ij}$. This allows us to write $Z_2$ as a weighted sum of monomials of $W$. When calculating $\mathbb{E}[Z_2]$, we only keep monomials of the form $W_{i_1i_4}^2W_{i_2i_3}^2$. It follows that 
\begin{align*} 
\mathbb{E}[Z_2]&=\mathbb{E}\biggl[2\sum_{i_1, i_2, i_3, i_4(dist)} U_{\tau(i_1)\tau(i_2)}(\mathbb{E}d_{i_1})(\mathbb{E}d_{i_2}-d_{i_2})W_{i_2i_3}U_{\tau(i_3)\tau(i_4)}(\mathbb{E}d_{i_3})(\mathbb{E}d_{i_4}-d_{i_4})W_{i_4i_1}\biggr]\cr
&=\mathbb{E}\biggl[2\sum_{i_1, i_2, i_3, i_4(dist)} U_{\tau(i_1)\tau(i_2)}(\mathbb{E}d_{i_1})W^2_{i_2i_3}U_{\tau(i_3)\tau(i_4)}(\mathbb{E}d_{i_3})W^2_{i_4i_1}\biggr]\cr
&=2\sum_{\substack{k_1,k_2,\\k_3,k_4}}\sum_{j=1}^4\sum_{i_j\in {\cal N}_j} U_{k_1k_2}U_{k_3k_4} 
(\mathbb{E}d_{i_1})(\mathbb{E}d_{i_3})W^2_{i_2i_3}W^2_{i_1i_4}\cr
&=2\, [1+o(1)]\sum_{\substack{k_1,k_2,\\k_3,k_4}}\sum_{j=1}^4\sum_{i_j\in {\cal N}_j} \frac{P_{k_1k_2}P_{k_3k_4}}{V_{k_1k_1}V_{k_2k_2}V_{k_3k_3}V_{k_4k_4}\|\theta\|_1^4}
\bigl(\theta_{i_1}\theta_{i_3}\|\theta\|_1^2V_{k_1k_1}V_{k_3k_3}\bigr)\cdot\Omega_{i_2i_3}\Omega_{i_1i_4}\cr
&=2\, [1+o(1)]\sum_{\substack{k_1,k_2,\\k_3,k_4}}\frac{P_{k_1k_2}P_{k_3k_4}P_{k_2k_3}P_{k_1k_4}}{V_{k_2k_2}V_{k_4k_4}\|\theta\|_1^2}
\Bigl(\sum_{j=1}^4\sum_{i_j\in {\cal N}_j}\theta^2_{i_1}\theta_{i_2}\theta^2_{i_3}\theta_{i_4}\Bigr)\cr
&=[1+o(1)]\cdot 2\|\theta\|^4 g'V^{-1}[(PH^2P)\circ(PH^2P)]V^{-1}g. 
\end{align*}
Here, the first two lines come from discarding terms with mean zero, the fourth line is because of \eqref{lem-remainder0-U}-\eqref{lem-remainder0-Ed}, and the last line is obtained similarly as in the equation above \eqref{lem-remainder0-Z1-mean}. Hence,
\beq \label{lem-remainder0-Z2-mean}
\mathbb{E}[Z_2]=b_n\cdot [1+o(1)]. 
\eeq
We then study $\mathbb{E}[T_1]$. We plug in $\delta_{ij}=U_{\tau(i)\tau(j)}[(\mathbb{E}d_i)(\mathbb{E}d_j-d_j)+(\mathbb{E}d_j)(\mathbb{E}d_i-d_i)]$ to get
\begin{align*}
T_1&=2\sum_{\substack{i_1,i_2,i_3,i_4\\(dist)}}U_{\tau(i_1)\tau(i_2)}U_{\tau(i_2)\tau(i_3)}U_{\tau(i_3)\tau(i_4)}\times \cr
&\qquad\qquad (\mathbb{E}d_{i_1})(\mathbb{E}d_{i_2}-d_{i_2})^2(\mathbb{E}d_{i_3})^2(\mathbb{E}d_{i_4}-d_{i_4}) W_{i_4i_1} + rem\cr
&\equiv 2T_{11}+rem. 
\end{align*}
We claim that
\[
|\mathbb{E}[rem]| = o(\|\theta\|^4). 
\]
The calculations here are similar to those in Equation (E.176) of \cite{SP2019}, where $T_1$ there (with a slightly different meaning) is decomposed into $2T_{1a} + 2T_{1b}+ 2T_{1c} + 2T_{1d}$. Here, $T_{11}$ is analogous to $T_{1d}$, and the remainder term is analogous to $2T_{1a}+2T_{1b}+2T_{1c}$. In \cite{SP2019}, it was shown that $|\mathbb{E}[T_{1a}]|+|\mathbb{E}[T_{1b}]|+|\mathbb{E}[T_{1c}]|=o(\|\theta\|^4)$; see Equations (E.179)-(E.181) in \cite{SP2019}. We can adapt their proof to show $|\mathbb{E}[rem]| = o(\|\theta\|^4)$. Since the calculations are elementary, we omit the details to save space. We then compute $\mathbb{E}[T_{11}]$. Since $\mathbb{E}d_i-d_i=-\sum_{j:j\neq i}W_{ji}$, it follows that 
\begin{align*}
\mathbb{E}[T_{11}]&=-\mathbb{E}\biggl[\sum_{\substack{k_1,k_2\\k_3,k_4}}\sum_{j=1}^4\sum_{i_j\in {\cal N}_{k_j}}U_{k_1k_2}U_{k_2k_3}U_{k_3k_4}(\mathbb{E}d_{i_1})\Bigl(\sum_{i_5:i_5\neq i_2}W_{i_2i_5}\Bigr)^2(\mathbb{E}d_{i_3})^2\Bigl(\sum_{i_6:i_6\neq i_4}W_{i_4i_6}\Bigr) W_{i_4i_1}\biggr]\cr
&= -\mathbb{E}\biggl[\sum_{\substack{k_1,k_2\\k_3,k_4}}\sum_{j=1}^4\sum_{i_j\in {\cal N}_{k_j}}U_{k_1k_2}U_{k_2k_3}U_{k_3k_4}(\mathbb{E}d_{i_1})\Bigl(\sum_{i_5:i_5\neq i_2}W^2_{i_2i_5}\Bigr)(\mathbb{E}d_{i_3})^2W_{i_4i_1}^2\biggr]\cr
&= -\sum_{\substack{k_1,k_2\\ k_3, k_4}}\sum_{j=1}^4\sum_{i_j\in {\cal N}_{k_j}} U_{k_1k_2}U_{k_2k_3}U_{k_3k_4}(\mathbb{E}d_{i_1})(\mathbb{E}d_{i_3})^2\mathbb{E}[W_{i_4i_1}^2]\Bigl(\sum_{i_5:i_5\neq i_2}\mathbb{E}[W_{i_2i_5}^2]\Bigr)\cr
&= - \sum_{\substack{k_1,k_2\\ k_3, k_4}}\sum_{j=1}^4\sum_{i_j\in {\cal N}_{k_j}} U_{k_1k_2}U_{k_2k_3}U_{k_3k_4}(\mathbb{E}d_{i_1})(\mathbb{E}d_{i_3})^2\mathbb{E}[W_{i_4i_1}^2]\cdot [1+o(1)] \Bigl(\theta_{i_2}\|\theta\|_1 \underbrace{\sum_{k_5}  P_{k_2k_5}g_{k_5}}_{V_{k_2k_2}} \Bigr)\cr
&=-[1+o(1)] \sum_{\substack{k_1,k_2\\k_3, k_4}} \frac{P_{k_1k_2}P_{k_2k_3}P_{k_3k_4}P_{k_1k_4}}{V_{k_2k_2}V_{k_4k_4}\|\theta\|_1^2}\Bigl( \sum_{j=1}^4\sum_{i_j\in {\cal N}_{k_j}}\theta^2_{i_1}\theta_{i_2}\theta_{i_3}^2\theta_{i_4} \Bigr) \cr
&= -[1+o(1)]\cdot \|\theta\|^4 g'V^{-1}[(PH^2P)\circ(PH^2P)]V^{-1}g, 
\end{align*}
where we have plugged in \eqref{lem-remainder0-U}-\eqref{lem-remainder0-Ed} in the second last line, and the last line can be derived similarly as in the equation above \eqref{lem-remainder0-Z1-mean}. We have proved $\mathbb{E}[T_{11}]=- [1+o(1)]\cdot b_n/2$. Then, 
\beq \label{lem-remainder0-T1-mean}
\mathbb{E}[T_1] = 2\mathbb{E}[T_{11}]+o(\|\theta\|^4) = - b_n\cdot[1+o(1)]. 
\eeq 
We then study $\mathbb{E}[F]$. Similar to the analysis of $T_1$, after plugging in $\delta_{ij}=U_{\tau(i)\tau(j)}[(\mathbb{E}d_i)(\mathbb{E}d_j-d_j)+(\mathbb{E}d_j)(\mathbb{E}d_i-d_i)]$, we can obtain that 
\begin{align*}
F &=rem+ 2 \sum_{i_1,i_2,i_3,i_4(dist)}U_{\tau(i_1)\tau(i_2)}U_{\tau(i_2)\tau(i_3)}U_{\tau(i_3)\tau(i_4)}U_{\tau(i_4)\tau(i_1)}\times \cr
&\qquad\qquad\qquad\qquad (\mathbb{E}d_{i_1})(\mathbb{E}d_{i_2}-d_{i_2})^2(\mathbb{E}d_{i_3})^2 (\mathbb{E}d_{i_4}-d_{i_4})^2(\mathbb{E}d_{i_1}), \cr
&\equiv rem + 2F_1, \qquad\qquad \mbox{where}\quad |\mathbb{E}[rem]|=o(\|\theta\|^4). 
\end{align*}
The proof of $ |\mathbb{E}[rem]|=o(\|\theta\|^4)$ is similar to the proof of (E.188)-(E.189) in \cite{SP2019}. There they analyzed a quantity $F$, which bears some similarity to the $F$ here, and decomposed $F=2F_a+12F_b+2F_c$, where $2F_a+12F_b$ is analogous to $rem$ here. They proved that $|\mathbb{E}[F_a]|+|\mathbb{E}[F_b]|=o(\|\theta\|^4)$. We can mimic their proof to show $ |\mathbb{E}[rem]|=o(\|\theta\|^4)$. By direct calculations, 
\begin{align*}
\mathbb{E}[F_1] &= \mathbb{E}\biggl[\sum_{\substack{k_1,k_2\\k_3,k_4}}\sum_{j=1}^4\sum_{i_j\in {\cal N}_{k_j}}U_{k_1k_2}U_{k_2k_3}U_{k_3k_4}U_{k_4k_1}(\mathbb{E}d_{i_1})^2(\mathbb{E}d_{i_3})^2 (\mathbb{E}d_{i_2}-d_{i_2})^2 (\mathbb{E}d_{i_4}-d_{i_4})^2\biggr]\cr
&=\mathbb{E}\biggl[\sum_{\substack{k_1,k_2\\k_3,k_4}}\sum_{j=1}^4\sum_{i_j\in {\cal N}_{k_j}}U_{k_1k_2}U_{k_2k_3}U_{k_3k_4}U_{k_4k_1}(\mathbb{E}d_{i_1})^2(\mathbb{E}d_{i_3})^2 \Bigl(\sum_{i_5: i_5\neq i_2}W^2_{i_2i_5}\Bigr) \Bigl(\sum_{i_6: i_6\neq i_4}W^2_{i_4i_6}\Bigr)\biggr]\cr
&=[1+o(1)]\sum_{\substack{k_1,k_2\\k_3,k_4}}\sum_{j=1}^4\sum_{i_j\in {\cal N}_{k_j}}\frac{P_{k_1k_2}P_{k_2k_3}P_{k_3k_4}P_{k_4k_1}\theta_{i_1}^2\theta_{i_3}^2}{V^2_{k_2k_2}V^2_{k_4k_4}\|\theta\|_1^4}\Bigl(\theta_{i_2}\|\theta\|_1\underbrace{\sum_{k_5}P_{k_2k_5}g_{k_5}}_{V_{k_2k_2}}\Bigr) \Bigl(\theta_{i_4}\|\theta\|_1\underbrace{\sum_{k_6}P_{k_4k_6}g_{k_6}}_{V_{k_4k_4}}\Bigr)\cr
&=[1+o(1)]\sum_{\substack{k_1,k_2\\k_3,k_4}}\frac{P_{k_1k_2}P_{k_2k_3}P_{k_3k_4}P_{k_4k_1}}{V_{k_2k_2}V_{k_4k_4}\|\theta\|_1^2}\Bigl( \sum_{j=1}^4\sum_{i_j\in {\cal N}_{k_j}} \theta_{i_1}^2\theta_{i_2}\theta_{i_3}^2\theta_{i_4} \Bigr)\cr
&= [1+o(1)]\cdot \|\theta\|^4 g'V^{-1}[(PH^2P)\circ(PH^2P)]V^{-1}g, 
\end{align*}
where in the second line we discard terms with mean zero, in the third line we plug in \eqref{lem-remainder0-U}-\eqref{lem-remainder0-Ed}, and in the last line we use elementary calculations similar to those in the equation above \eqref{lem-remainder0-Z1-mean}. It follows that $\mathbb{E}[F_1]=[1+o(1)]\cdot b_n/2$ and that
\beq\label{lem-remainder0-F-mean}
\mathbb{E}[F] = 2\mathbb{E}[F_1]+o(\|\theta\|^4) = [1+o(1)]\cdot b_n. 
\eeq
We now plug \eqref{lem-remainder0-Z1-mean}, \eqref{lem-remainder0-Z2-mean}, \eqref{lem-remainder0-T1-mean}, and \eqref{lem-remainder0-F-mean} into \eqref{lem-remainder0-decomposition} to get
\begin{align*}
\mathbb{E}[Q_n^{*(m,0)} -\widetilde{Q}_n^{(m,0)} ] & = 4\mathbb{E}[Z_1] + 2\mathbb{E}[ Z_2] + 4\mathbb{E}[T_1]+\mathbb{E}[F] \cr
&= [1+o(1)]\cdot [4(b_n/2) + 2b_n- 4b_n+b_n]\cr
&=[1+o(1)]\cdot b_n. 
\end{align*}
Since $b_n\asymp \|\theta\|^4$, \eqref{lem-remainder0-goal} follows immediately. \qed

\subsection{Proof of Lemma~\ref{prop:remainder1}} \label{subsec:proof-remainder1}
Similar to the proof of Lemma~\ref{prop:remainder0}, we use the notation $M_{ijk\ell}(X)=X_{ij}X_{jk}X_{k\ell}X_{\ell i}$. By \eqref{def:proxies}, 
\begin{align*}
 \widetilde{Q}_n^{*(m,0)} - Q_n^{*(m,0)}= & \sum_{i_1, i_2, i_3, i_4(dist)}[M_{i_1i_2i_3i_4}(\widetilde{X}^*)-M_{i_1i_2i_3i_4}(X^*)],\cr
& \qquad \mbox{where}\quad \begin{cases}
\widetilde{X}^*_{ij} = \widetilde{\Omega}^{(m,0)}_{ij}+W_{ij}+\delta_{ij}^{(m,0)}+\tilde{r}_{ij}^{(m,0)},\cr
X^*_{ij} = \widetilde{\Omega}^{(m,0)}_{ij}+W_{ij}+\delta_{ij}^{(m,0)}.
\end{cases}
\end{align*}
For the rest of the proof, we omit the superscripts $(m,0)$ in $(\widetilde{\Omega},\delta,\tilde{r})$. There are $4^4-3^4=175$ post-expansion sums in $\widetilde{Q}_n^{*(m,0)}-Q_n^{*(m,0)}$, each with the form
\beq \label{lem-remainder1-S}
S\equiv \sum_{i_1, i_2, i_3, i_4(dist)} a_{i_1i_2}b_{i_2i_3}c_{i_3i_4}d_{i_4i_1}, \qquad\mbox{where}\quad a,b,c,d\in \{\widetilde{\Omega}, W, \delta, \tilde{r}\}. 
\eeq
Here we use $S$ as a generic notation for any post-expansion sum. To show the claim, it suffices to bound $|\mathbb{E}[S]|$ and $\mathrm{Var}(S)$ for each post-expansion sum $S$. 

We now study $S$. Let ${\cal N}_1^{(m,0)}, {\cal N}_2^{(m,0)},\ldots,{\cal N}_m^{(m,0)}$ be the pseudo-communities defined by $\Pi_0$. By \eqref{def:delta} and \eqref{def:tilder}, for $i\in {\cal N}_k^{(m,0)}$ and $j\in {\cal N}_\ell^{(m,0)}$, 
\[
\delta_{ij}=U_{k\ell}\bigl[(\mathbb{E}d_i)(d_j-\mathbb{E}d_j)+(\mathbb{E}d_j)(d_i-\mathbb{E}d_i)\bigr],\qquad \tilde{r}_{ij} = -\widehat{U}_{k\ell}(d_i-\mathbb{E}d_i)(d_j-\mathbb{E}d_j). 
\]
The term $\widehat{U}_{k\ell}$ has a complicated correlation with each summand, so we want to ``replace'' it with ${U}_{k\ell}$. Introduce a proxy of $\tilde{r}_{ij}$ as
\beq \label{def:r}
r_{ij}=- U_{k\ell}(d_i-\mathbb{E}d_i)(d_j-\mathbb{E}d_j)
\eeq
We define a proxy of $S$ as
\beq \label{lem-remainder1-T}
T\equiv \sum_{i_1, i_2, i_3, i_4(dist)} a_{i_1i_2}b_{i_2i_3}c_{i_3i_4}d_{i_4i_1}, \qquad\mbox{where}\quad a,b,c,d\in \{\widetilde{\Omega}, W, \delta, r\}. 
\eeq
We note that $T$ is also a generic notation, and it has a one-to-one correspondence with $S$. For example, if $S=\sum_{i_1,i_2,i_3, i_4 (dist)}\delta_{i_1i_2}W_{i_2i_3}\widetilde{\Omega}_{i_3i_4}\tilde{r}_{i_4i_1}$, then $T=\sum_{i_1,i_2,i_3, i_4 (dist)}\delta_{i_1i_2}W_{i_2i_3}\widetilde{\Omega}_{i_3i_4}r_{i_4i_1}$; if $S=\sum_{i_1,i_2,i_3, i_4 (dist)}\delta_{i_1i_2}\tilde{r}_{i_2i_3}\tilde{r}_{i_3i_4}W_{i_4i_1}$, then $T=\sum_{i_1,i_2,i_3, i_4 (dist)}\delta_{i_1i_2}r_{i_2i_3}r_{i_3i_4}W_{i_4i_1}$. Therefore, to bound the mean and variance of $S$, we only need to study $T$ and $S-T$ separately.

First, we study the mean and variance of $T$. Since $d_i-\mathbb{E}d_i=\sum_{j:j\neq i}W_{ij}$, we can write $\delta_{ij}$ as a linear form of $W$ and $r_{ij}$ as a quadratic form of $W$. We then plug them into the expression of $T$ and write $T$ as a weighted sum of monomials of $W$. Take $T=\sum_{i_1,i_2,i_3,i_4(dist)}r_{i_1i_2}W_{i_2i_3}W_{i_3i_4}W_{i_4i_1}$ for example. It can be re-written as (note: $\tau(i)$ is the index of pseudo-community that contains node $i$)
\begin{align*}
T &= -  \sum_{i_1,i_2,i_3,i_4(dist)}U_{\tau(i_1)\tau(i_2)}\Bigl(\sum_{j_1:j_1\neq i_1}W_{i_1j_1}\Bigr)\Bigl(\sum_{j_2:j_2\neq i_2}W_{i_2j_2}\Bigr)W_{i_2i_3}W_{i_3i_4}W_{i_4i_1}\cr
&= -  \sum_{\substack{i_1,i_2,i_3,i_4(dist)\\j_1,j_2: j_1\neq i_1, j_2\neq i_2}}U_{\tau(i_1)\tau(i_2)}W_{i_1j_1}W_{i_2j_2}W_{i_2i_3}W_{i_3i_4}W_{i_4i_1}. 
\end{align*}
Then, we can compute the mean and variance of $T$ directly. We use the same strategy to analyze each of the 175 post-expansion sums of the form \eqref{lem-remainder1-T}. Similar calculations were conducted in the proof of Lemma E.11 of \cite{SP2019}. The setting of Lemma E.11 is a special case where $U_{k\ell}\equiv 1/v$ for a scalar $v$. However, their proof does not rely on that $U_{k\ell}$'s are equal to each other. Instead, their proof only requires a universal upper bound on $U_{k\ell}$. In fact, they have proved the following lemma:
\begin{lemma} \label{lemma:realSQ1}
Consider a DCBM model where (\ref{condition1a})-(\ref{condition1b}) and (\ref{condition1d}) hold. Let $W=A-\Omega$ and $\Delta=\sum_{i_1,i_2,i_3,i_4(dist)}\bigl[ M_{i_1i_2i_3i_4}\bigl(\widetilde{\Omega}+W+\delta+r\bigr)-M_{i_1i_2i_3i_4}\bigl(\widetilde{\Omega}+W+\delta\bigr)\bigr]$, where $\widetilde{\Omega}$ is a non-stochastic symmetric matrix, $\delta_{ij}=v_{ij}\cdot[(\mathbb{E}d_i)(\mathbb{E}d_j-d_j)+(\mathbb{E}d_j)(\mathbb{E}d_i-d_i)]$, $r_{ij}=-u_{ij}(d_i-\mathbb{E}d_i)(d_j-\mathbb{E}d_j)$, $\{v_{ij}, u_{ij}\}_{1\leq i\neq j\leq n}$ are non-stochastic scalars, $d_i$ is the degree of node $i$, and $M_{i_1i_2i_3i_4}(\cdot)$ is as defined above. As $n\to\infty$, suppose there is a constant $C>0$ and a scalar $\alpha_n>0$ such that $\alpha_n\leq C$, $\alpha_n\|\theta\|\to\infty$, $|\widetilde{\Omega}_{ij}|\leq C\alpha_n\theta_i\theta_j$, $|v_{ij}|\leq C\|\theta\|_1^{-1}$, and $|u_{ij}|\leq C\|\theta\|_1^{-1}$ for $1\leq i,j\leq n$. Let $T$ be an arbitrary post-expansion sum of $\Delta$. Then, $|\mathbb{E}[T]|\leq C\alpha_n^2\|\theta\|^6 +  o(\|\theta\|^4)$ and $\mathrm{Var}(T)= o\bigl(\alpha_n^6\|\theta\|^8\|\theta\|_3^6+\|\theta\|^8\bigr)$. 
\end{lemma}
\noindent
We apply Lemma~\ref{lemma:realSQ1} for $\alpha_n=\tau$ and $v_{ij}=u_{ij}=U_{\tau(i)\tau(j)}$. 
By Lemma~\ref{lemma:SgnQ2a}, $\tau\leq C$, $\tau\|\theta\|\to\infty$, and $|\widetilde{\Omega}_{ij}|\leq C\tau\theta_i\theta_j$. In \eqref{lem-remainder0-Ubound}, we have seen that $|U_{k\ell}|\leq C\|\theta\|_1^{-1}$. The conditions of Lemma~\ref{lemma:realSQ1} are satisfied. We immediately have: Under the conditions of Lemma~\ref{lemma:SgnQ2c} (note: $\tau\|\theta\|\to\infty$)
\beq \label{lem-remainder1-T-alt}
|\mathbb{E}[T]|\leq C\tau^2\|\theta\|^6+ o(\|\theta\|^4)=o(\tau^4\|\theta\|^8),\qquad \mathrm{Var}(T)= o\bigl(\tau^6\|\theta\|^8\|\theta\|_3^6 + \|\theta\|^8\bigr), 
\eeq
and under the conditions of Lemma~\ref{lemma:SgnQ1c} (i.e., $\widetilde{\Omega}$ is a zero matrix and $\tau=0$), 
\beq \label{lem-remainder1-T-null}
|\mathbb{E}[T]|=o(\|\theta\|^4),\qquad \mathrm{Var}(T)=o(\|\theta\|^8). 
\eeq

Next, we study the variable $(S-T)$. In \eqref{lem-remainder1-S} and \eqref{lem-remainder1-T}, if we group the summands based on pseudo-communities of $(i_1,i_2,i_3,i_4)$, then we have 
\[
S=\sum_{1\leq k_1,k_2,k_3,k_4\leq m}S_{k_1k_2k_3k_4} \qquad\mbox{and}\qquad T=\sum_{1\leq k_1,k_2,k_3,k_4\leq m}T_{k_1k_2k_3k_4},
\]
where $S_{k_1k_2k_3k_4}$ contains all the summands such that $i_s\in {\cal N}_{k_s}^{(m,0)}$ for $s=1,2,3,4$. By straightforward calculations and definitions of $(r_{ij}, \tilde{r}_{ij})$, we have
\begin{align*}
S_{k_1k_2k_3k_4} &= \widehat{U}^{\ell_a}_{k_1k_2}\widehat{U}^{\ell_b}_{k_2k_3}\widehat{U}^{\ell_c}_{k_3k_4}\widehat{U}^{\ell_d}_{k_4k_1} \sum_{s=1}^4\sum_{i_s\in {\cal N}_{k_s}^{(m,0)}}\tilde{a}_{i_1i_2}\tilde{b}_{i_2i_3}\tilde{c}_{i_3i_4}\tilde{d}_{i_4i_1},\cr
T_{k_1k_2k_3k_4}&= U^{\ell_a}_{k_1k_2} U^{\ell_b}_{k_2k_3} U^{\ell_c}_{k_3k_4}U^{\ell_d}_{k_4k_1} \sum_{s=1}^4\sum_{i_s\in {\cal N}_{k_s}^{(m,0)}}\tilde{a}_{i_1i_2}\tilde{b}_{i_2i_3}\tilde{c}_{i_3i_4}\tilde{d}_{i_4i_1},\cr
&\qquad \mbox{where}\quad \tilde{a}_{ij}, \tilde{b}_{ij}, \tilde{c}_{ij}, \tilde{d}_{ij}\in \bigl\{ \widetilde{\Omega}_{ij},\, W_{ij},\, \delta_{ij},\, -(d_i-\mathbb{E}d_i)(d_j-\mathbb{E}d_j)\bigr\}. 
\end{align*}
Here $\ell_a\in \{0,1\}$ is an indicator about whether $a_{ij}$ takes the value of $\tilde{r}_{ij}$ in $S$, and $(\ell_b,\ell_c,\ell_d)$ are similar. For example, if $S=\sum_{i_1,i_2,i_3, i_4 (dist)}\delta_{i_1i_2}W_{i_2i_3}\widetilde{\Omega}_{i_3i_4}\tilde{r}_{i_4i_1}$, then $(\ell_a,\ell_b,\ell_c,\ell_d)=(0,0,0,1)$; if $S=\sum_{i_1,i_2,i_3, i_4 (dist)}\delta_{i_1i_2}\tilde{r}_{i_2i_3}\tilde{r}_{i_3i_4}W_{i_4i_1}$, then $(\ell_a,\ell_b,\ell_c,\ell_d)=(0,1,1,0)$. For any post-expansion sum $S$ considered here, $1\leq \ell_a+\ell_b+\ell_c+\ell_d\leq 4$. To study the difference between $S_{k_1k_2k_3k_4}$ and $T_{k_1k_2k_3k_4}$, we introduce an intermediate term 
\[
R_{k_1k_2k_3k_4} = \Bigl(\frac{1}{\|\theta\|_1^2}\Bigr)^{\ell_a+\ell_b+\ell_c+\ell_d}\sum_{s=1}^4\sum_{i_s\in {\cal N}_{k_s}^{(m,0)}}\tilde{a}_{i_1i_2}\tilde{b}_{i_2i_3}\tilde{c}_{i_3i_4}\tilde{d}_{i_4i_1}. 
\]
In fact, $R_{k_1k_2k_3k_4}$ has a similar form as $T_{k_1k_2k_3k_4}$ except that the scalar $U_{k\ell}$ in the definition of $r_{ij}$ (see \eqref{def:r}) is replaced by $1/\|\theta\|_1^2$.  We apply Lemma~\ref{lemma:realSQ1} with $u_{ij}\equiv 1/\|\theta\|_1^2$. It yields that, under conditions of Lemma~\ref{lemma:SgnQ1c}, 
\[
|\mathbb{E}[R_{k_1k_2k_3k_4}]|=o(\|\theta\|^4), \qquad \mathrm{Var}(R_{k_1k_2k_3k_4})=o(\|\theta\|^8),
\]
and under conditions of Lemma~\ref{lemma:SgnQ2c},
\[
|\mathbb{E}[R_{k_1k_2k_3k_4}]|\leq C\tau^2\|\theta\|^6+ o(\|\theta\|^4), \qquad \mathrm{Var}(R_{k_1k_2k_3k_4})=o\bigl(\|\theta\|^8+\tau^6\|\theta\|^8\|\theta\|_3^6\bigr).
\]
Particularly, since $\mathbb{E}[X^2]= (\mathbb{E}[X])^2+\mathrm{Var}(X)$ for any variable $X$, we have
\begin{align} \label{lem-remainder1-Rsquare}
\|\theta\|^{-4}\, \mathbb{E}[R^2_{k_1k_2k_3k_4}]& \leq  
\begin{cases}
o(\|\theta\|^4), & \mbox{for setting of Lemma~\ref{lemma:SgnQ1c}},\cr
C\tau^4\|\theta\|^8+o\bigl(\|\theta\|^{4}+\tau^6\|\theta\|^4\|\theta\|_3^6\bigr), & \mbox{for setting of Lemma~\ref{lemma:SgnQ2c}},\cr
\end{cases}\cr
&= \begin{cases}
o(\|\theta\|^4), & \mbox{for setting of Lemma~\ref{lemma:SgnQ1c}},\cr
C\|\theta\|^8, & \mbox{for setting of Lemma~\ref{lemma:SgnQ2c}}.
\end{cases}
\end{align}
Note that in deriving \eqref{lem-remainder1-Rsquare} we have used $\tau\leq C$ and $\tau^6\|\theta\|^4\|\theta\|_3^6\leq \tau^6\|\theta\|^4\cdot\theta_{\max}^2\|\theta\|^4\leq C\|\theta\|^8$. 

We now investigate $(S_{k_1k_2k_3k_4}-T_{k_1k_2k_3k_4})$. By condition \eqref{condition1a}, $\sqrt{\log(n)} \ll \|\theta\|_1/\|\theta\|^2$. Hence, we can take a sequence of $x_n$, such that $\sqrt{\log(n)} \ll x_n \ll \|\theta\|_1/\|\theta\|^2$, and define the event $E_n$: 
\beq\label{def:Dn}
E_n = \bigg\{ |U_{k\ell}  - \widehat{U}_{k\ell}| \leq \frac{C_0 x_n}{\|\theta\|_1^3}, \quad\mbox{ for all $1\leq k,\ell\leq m$} \bigg\},
\eeq
where $C_0>0$ is a constant to be decided. 
To bound the probability of $E_n^c$, we recall that (by definitions in \eqref{def:hatU} and \eqref{def:U}) 
\[
\widehat{U}_{k\ell} = \frac{ {\bf 1}_k'A{\bf 1}_\ell}{({\bf 1}_k'd)({\bf 1}_\ell'd)}, \quad\mbox{ and } \quad U_{k\ell} = \frac{ {\bf 1}_k'\mathbb{E}[A]{\bf 1}_\ell}{({\bf 1}_k'\mathbb{E}[d])({\bf 1}_\ell'\mathbb{E}[d])}, 
\]
where ${\bf 1}_k$ is a shorthand notation for ${\bf 1}_k^{(m,0)}$ in \eqref{def:1m0}. Using Bernstein's inequality and mimicking the argument from (E.299)-(E.300) of \cite{SP2019}, we can easily show that, there is a constant $C_1>0$ such that, for any $1\leq k, \ell\leq m$, 
 \beq\label{lem-remainder1-deviation}
\mathbb{P}\Big( \bigl| {\bf 1}_k'A{\bf 1}_\ell- {\bf 1}_k'\mathbb{E}[A]{\bf 1}_\ell \bigr| > x_n \|\theta\|_1 \Big) \leq 2\exp(-C_1x_n^2).
\eeq
By probability union bound, with probability $1-2m^2\exp(-C_1x_n^2)$,  
\[
\max_{1\leq k,\ell\leq m}\bigl\{\bigl| {\bf 1}_k'A{\bf 1}_\ell- {\bf 1}_k'\mathbb{E}[A]{\bf 1}_\ell \bigr|\bigr\}\leq x_n\|\theta\|_1.
\] 
Furthermore, ${\bf 1}_k'd-{\bf 1}_k'\mathbb{E}[d]=\sum_{\ell=1}^m ({\bf 1}_k'A{\bf 1}_\ell - {\bf 1}_k'\mathbb{E}[A]{\bf 1}_\ell)$. So, with probability $1-2m^2\exp(-C_1x_n^2)$,
\[
\max_{1\leq k\leq m} \bigl\{\bigl|{\bf 1}_k'd-{\bf 1}_k'\mathbb{E}[d]\bigr|\bigr\} \leq m\cdot x_n\|\theta\|_1. 
\] 
At the same time, we know that ${\bf 1}_k'\mathbb{E}[A]{\bf 1}_\ell\asymp\|\theta\|_1^2$ and ${\bf 1}_k'\mathbb{E}[d]\asymp \|\theta\|_1^2$. We plug the above results into the expressions of $U_{k\ell}$ and $\widehat{U}_{k\ell}$ and can easily find that, with probability $1-2m^2\exp(-C_1 x_n^2)$,
\[
\max_{1\leq k,\ell\leq m}|\widehat{U}_{k\ell}-U_{k\ell}|\leq C_0 x_n/\|\theta\|_1^3,
\]
for some constant $C_0>0$ ($C_0$ still depends on $m$, but $m$ is bounded here). We use the same $C_0$ to define $E_n$. Then,
\beq \label{lem-remainder1-eventprob}
\mathbb{P}(E_n^c) \leq 2m^2\exp(- C_1 x_n^2)=o(n^{-L}), \qquad\mbox{for any fixed }L >0,  
\eeq
where the last equality is due to $x_n^2\gg\log(n)$. 
We aim to use \eqref{lem-remainder1-eventprob} to bound $\mathbb{E}[(S_{k_1k_2k_3k_4}-T_{k_1k_2k_3k_4})\cdot I_{E_n^c}]$. 
It is easy to see the trivial bound $|\widehat{U}_{k\ell}| \leq 1$ and $|{U}_{k\ell}|\leq 1$. Also, recall that $\tilde{a}_{ij}$ takes value in
$\{\widetilde{\Omega}_{ij}, W_{ij}, \delta_{ij},- (d_i-\mathbb{E}d_i)(d_{j}-\mathbb{E}d_j)\}$, and so $|a_{ij}| \leq n^2$; we have the same bound for $|\tilde{b}_{ij}|, |\tilde{c}_{ij}|, |\tilde{d}_{ij}|$. This gives a trivial bound
 \[
(S_{k_1k_2k_3k_4}-T_{k_1k_2k_3k_4})^2 \leq 2S_{k_1k_2k_3k_4}^2 + 2T_{k_1k_2k_3k_4}^2 \leq 2(n^4\cdot n^8)^2 + 2(n^4\cdot n^8)^2 = 4n^{24}. 
\]
Combining it with \eqref{lem-remainder1-eventprob}, we have 
\beq   \label{lem-remainder1-moment1}
\mathbb{E}[(T_{k_1k_2k_3k_4} - S_{k_1k_2k_3k_4})^2\cdot I_{E_n^c}] \leq 4n^{24}\cdot 2m^2 \exp(-C_1x_n^2) = o(1). 
\eeq 
At the same time, on the event $E_n$,
\begin{align*}
& |S_{k_1k_2k_3k_4}-T_{k_1k_2k_3k_4}|\cr
=\;\; & \bigl| \widehat{U}^{\ell_a}_{k_1k_2}\widehat{U}^{\ell_b}_{k_2k_3}\widehat{U}^{\ell_c}_{k_3k_4}\widehat{U}^{\ell_d}_{k_4k_1}-U^{\ell_a}_{k_1k_2} U^{\ell_b}_{k_2k_3} U^{\ell_c}_{k_3k_4}U^{\ell_d}_{k_4k_1} \bigr|\cdot \|\theta\|_1^{2(\ell_a+\ell_b+\ell_c+\ell_d)}|R_{k_1k_2k_3k_4}|\cr
\leq\;\; &C\Bigl(|U^{\ell_a}_{k_1k_2} U^{\ell_b}_{k_2k_3} U^{\ell_c}_{k_3k_4}U^{\ell_d}_{k_4k_1}|\max_{1\leq k,\ell\leq m} \bigl|\widehat{U}_{k\ell}/U_{k\ell} -1\bigr|\Bigr)\cdot \|\theta\|_1^{2(\ell_a+\ell_b+\ell_c+\ell_d)}|R_{k_1k_2k_3k_4}|\cr
\leq \;\; & C\|\theta\|_1^2 \cdot \max_{1\leq k,\ell\leq m}|\widehat{U}_{k\ell}-U_{k\ell}|\cdot |R_{k_1k_2k_3k_4}|\cr
\leq \;\; & Cx_n\|\theta\|_1^{-1} \cdot |R_{k_1k_2k_3k_4}|\cr
=\;\; & o(\|\theta\|^{-2})\cdot |R_{k_1k_2k_3k_4}|,
\end{align*} 
where the fourth line is because $\|\theta\|_1^{-2}\leq |U_{k\ell}|\leq C\|\theta\|_1^{-2}$ (e.g., see \eqref{lem-remainder0-Ubound}) and the last line is because $x_n\ll \|\theta\|_1/\|\theta\|^2$. It follows that
\beq   \label{lem-remainder1-moment2}
\mathbb{E}[(T_{k_1k_2k_3k_4} - S_{k_1k_2k_3k_4})^2\cdot I_{E_n}] =o(\|\theta\|^{-4})\cdot \mathbb{E}[R^2_{k_1k_2k_3k_4}].  
\eeq 
We combine \eqref{lem-remainder1-moment1} and \eqref{lem-remainder1-moment2} and plug in \eqref{lem-remainder1-Rsquare}. It follows that
\begin{align*}  
\mathbb{E}[(T_{k_1k_2k_3k_4} - S_{k_1k_2k_3k_4})^2]
&= o(\|\theta\|^{-4})\cdot \mathbb{E}[R^2_{k_1k_2k_3k_4}] + o(1)\cr
&=\begin{cases}
o(\|\theta\|^4), & \mbox{under conditions of Lemma~\ref{lemma:SgnQ1c}},\cr
o(\|\theta\|^8), & \mbox{under conditions of Lemma~\ref{lemma:SgnQ2c}}. 
\end{cases}
\end{align*}
Since $m$ is bound, we immediately know that 
\beq  \label{lem-remainder1-(S-T)}
\mathbb{E}[(S-T)^2] =\begin{cases}
o(\|\theta\|^4), & \mbox{under conditions of Lemma~\ref{lemma:SgnQ1c}},\cr
o(\|\theta\|^8), & \mbox{under conditions of Lemma~\ref{lemma:SgnQ2c}}. 
\end{cases}
\eeq

Last, we combine the results on $T$ and the results on $(S-T)$. By \eqref{lem-remainder1-T-alt}-\eqref{lem-remainder1-T-null} and \eqref{lem-remainder1-(S-T)}, 
\begin{align*}
|\mathbb{E}[S]| &\leq |\mathbb{E}[T]| + |\mathbb{E}[S-T]||\cr
&\leq |\mathbb{E}[T]| + \sqrt{\mathbb{E}[(S-T)^2]}\cr
&=\begin{cases}
o(\|\theta\|^4)+o(\|\theta\|^2)=o(\|\theta\|^4), & \mbox{for setting of Lemma~\ref{lemma:SgnQ1c}},\cr
o(\tau^4\|\theta\|^8)+o(\|\theta\|^4)=o(\tau^4\|\theta\|^8), & \mbox{for setting of Lemma~\ref{lemma:SgnQ2c}}. 
\end{cases} 
\end{align*}
Additionally,
\begin{align*}
\mathrm{Var}(S) &\leq 2\mathrm{Var}(T) + 2\mathrm{Var}(S-T)\cr
&\leq 2\mathrm{Var}(T) +  2\mathbb{E}[(S-T)^2]\cr
&\leq \begin{cases}
o(\|\theta\|^8)+o(\|\theta\|^4)=o(\|\theta\|^8), & \mbox{for setting of Lemma~\ref{lemma:SgnQ1c}},\cr
o\bigl(\|\theta\|^8 + \tau^6\|\theta\|^8\|\theta\|_3^6\bigr)+o(\|\theta\|^8)=o\bigl(\|\theta\|^8 + \tau^6\|\theta\|^8\|\theta\|_3^6\bigr), & \mbox{for setting of Lemma~\ref{lemma:SgnQ2c}}. 
\end{cases} 
\end{align*}
This gives the desirable claim. \qed


\subsection{Proof of Lemma~\ref{prop:remainder2}} \label{subsec:proof-remainder2}
Similar to the proof of Lemma~\ref{prop:remainder0}, we use the notation $M_{ijk\ell}(X)=X_{ij}X_{jk}X_{k\ell}X_{\ell i}$. By \eqref{def:proxies}, 
\begin{align*}
Q_n^{(m,0)}- \widetilde{Q}_n^{*(m,0)}= & \sum_{i_1, i_2, i_3, i_4(dist)}[M_{i_1i_2i_3i_4}(X)- M_{i_1i_2i_3i_4}(\widetilde{X}^*)],\cr
& \quad \mbox{where}\quad \begin{cases}
X_{ij} = \widetilde{\Omega}^{(m,0)}_{ij}+W_{ij}+\delta_{ij}^{(m,0)}+\tilde{r}_{ij}^{(m,0)}+\epsilon_{ij}^{(m,0)},\cr
\widetilde{X}^*_{ij} = \widetilde{\Omega}^{(m,0)}_{ij}+W_{ij}+\delta_{ij}^{(m,0)}+\tilde{r}_{ij}^{(m,0)}.
\end{cases}
\end{align*}
We shall omit the superscripts $(m,0)$ in $(\widetilde{\Omega},\delta,\tilde{r}, \epsilon)$. Let ${\cal N}_1^{(m,0)}, {\cal N}_2^{(m,0)},\ldots,{\cal N}_m^{(m,0)}$ be the pseudo-communities defined by $\Pi_0$. By \eqref{def:eps}, $\epsilon_{ij}=\tilde{\alpha}_{ij}+\tilde{\beta}_{ij}+\tilde{\gamma}_{ij}$, where for $i\in {\cal N}_k^{(m,0)}$ and $j\in {\cal N}_\ell^{(m,0)}$, 
\begin{align} \label{lem-remainder2-tabc}
\tilde{\alpha}_{ij}&= d_i^*d_j^*U^*_{k\ell}-(\mathbb{E}d_i)(\mathbb{E}d_j)U_{k\ell},\cr
\tilde{\beta}_{ij} &= (U_{k\ell}-\widehat{U}_{k\ell}) (\mathbb{E}d_i)(\mathbb{E}d_j),\cr
\tilde{\gamma}_{ij} &=  (U_{k\ell}-\widehat{U}_{k\ell})[(\mathbb{E}d_i) (d_j-\mathbb{E}d_j)+(\mathbb{E}d_j)(d_i-\mathbb{E}d_i)]. 
\end{align}
Therefore, we can write
\begin{align*}
& Q_n^{(m,0)}- \widetilde{Q}_n^{*(m,0)}\cr
= \;\; &\sum_{i_1, i_2, i_3, i_4(dist)}[M_{i_1i_2i_3i_4}(\widetilde{\Omega}+W+\delta+\tilde{r}+\tilde{\alpha}+\tilde{\beta}+\tilde{\gamma})- M_{i_1i_2i_3i_4}(\widetilde{\Omega}+W+\delta+\tilde{r})]. 
\end{align*}
There are $7^4-4^4=2145$ post-expansion sums. Let $S$ be the generic notation for any such post-expansion sum.
Similarly as in the proof of Lemma~\ref{prop:remainder1}, we group the summands according to which pseudo-communities $(i_1,i_2,i_3,i_4)$ belong to, i.e., we write $S=\sum_{1\leq k_1,k_2,k_3,k_4\leq m}S_{k_1k_2k_3k_4}$, where 
\beq \label{lem-remainder2-S}
S_{k_1k_2k_3k_4}=\sum_{j=1}^4\sum_{i_j\in {\cal N}_{k_j}^{(m,0)}}a_{i_1i_2}b_{i_2i_3}c_{i_3i_4}d_{i_4 i_1}, \quad \mbox{where}\;\; a,b,c,d\in \{\widetilde{\Omega}, W,\delta, \tilde{r}, \tilde{\alpha},\tilde{\beta},\tilde{\gamma}\}.  
\eeq
It suffices to study the mean and variance of each $S_{k_1k_2k_3k_4}$. 

Let $\tau$ and $r_{ij}$ be the same as in \eqref{def:tau} and \eqref{def:r}. Define
\begin{align} \label{lem-remainder2-abc}
\alpha_{ij}&=  \frac{\tau\|\theta\|_1}{\theta_{\max}} \bigl[d_i^*d_j^*U^*_{k\ell}-(\mathbb{E}d_i)(\mathbb{E}d_j)U_{k\ell}\bigr],\cr
\beta_{ij} &= \tau U_{k\ell} (\mathbb{E}d_i)(\mathbb{E}d_j),\cr
\gamma_{ij} &=  U_{k\ell}[(\mathbb{E}d_i) (d_j-\mathbb{E}d_j)+(\mathbb{E}d_j)(d_i-\mathbb{E}d_i)]. 
\end{align}
We introduce a proxy of $S_{k_1k_2k_3k_4}$ as 
\beq \label{lem-remainder2-Sstar}
S^*_{k_1k_2k_3k_4}=\sum_{j=1}^4\sum_{i_j\in {\cal N}_{k_j}^{(m,0)}}a_{i_1i_2}b_{i_2i_3}c_{i_3i_4}d_{i_4 i_1}, \quad \mbox{where}\;\; a,b,c,d\in \{\widetilde{\Omega}, W,\delta, r, \alpha,\beta,\gamma\}.  
\eeq
Reviewing the expressions of $(\widetilde{\Omega}_{ij}, W_{ij},\delta_{ij}, r_{ij}, \alpha_{ij},\beta_{ij},\gamma_{ij})$, we know that $S^*_{k_1k_2k_3k_4}$ can always be written as a weighted sum of monomials of $W$, and so we can calculate the mean and variance of $S^*_{k_1k_2k_3k_4}$ (the straightforward calculations are still tedious, but later we will introduce a simple trick to do that). 
Comparing \eqref{lem-remainder2-abc} with \eqref{lem-remainder2-tabc} and $r_{ij}$ with $\tilde{r}_{ij}$, we observe that, for $i\in {\cal N}_k^{(m,0)}$ and $j\in {\cal N}_\ell^{(m,0)}$,
\[
\tilde{r}_{ij}=\frac{\widehat{U}_{k\ell}}{U_{k\ell}}\, r_{ij},\qquad 
\tilde{\alpha}_{ij}= \frac{\theta_{\max}}{\tau\|\theta\|_1}\,\alpha_{ij}, \qquad \tilde{\beta}_{ij}=\frac{U_{k\ell}-\widehat{U}_{k\ell}}{\tau U_{k\ell}}\,\beta_{ij}, \qquad \tilde{\gamma}_{ij}=\frac{U_{k\ell}-\widehat{U}_{k\ell}}{U_{k\ell}}\,\gamma_{ij}. 
\]
We plug them into \eqref{lem-remainder2-S} to get
\beq \label{lem-remainder2-relateS}
S_{k_1k_2k_3k_4}= \Bigl( \frac{\widehat{U}_{k\ell}}{U_{k\ell}}\Bigr)^{N_{\tilde{r}}}\Bigl( \frac{\theta_{\max}}{\tau\|\theta\|_1^2} \Bigr)^{N_{\tilde{\alpha}}} \Bigl(\frac{U_{k\ell}-\widehat{U}_{k\ell}}{\tau U_{k\ell}} \Bigr)^{N_{\tilde{\beta}}}\Bigl(\frac{U_{k\ell}-\widehat{U}_{k\ell}}{U_{k\ell}} \Bigr)^{N_{\tilde{\gamma}}}S^*_{k_1k_2k_3k_4},
\eeq
where $N_{\tilde{r}}$ is the count of $\{a,b,c,d\}$ in \eqref{lem-remainder2-S} taking the value of $\tilde{r}$, and $(N_{\tilde{\alpha}}, N_{\tilde{\beta}}, N_{\tilde{\gamma}})$ are similar. 
For any post-expansion sum considered here, $1\leq N_{\tilde{\alpha}}+N_{\tilde{\beta}}+ N_{\tilde{\gamma}}\leq 4$. The notation $(\frac{\widehat{U}_{k\ell}}{U_{k\ell}})^{N_{\tilde{r}}}$ is interpreted in this way: For example, if in \eqref{lem-remainder2-S} only $a$ takes the value of $\tilde{r}$, then $N_{\tilde{r}}=1$ and $(\frac{\widehat{U}_{k\ell}}{U_{k\ell}})^{N_{\tilde{r}}}=\frac{\widehat{U}_{k_1k_2}}{U_{k_1k_2}}$; if $(a,b,c)$ take the value of $\tilde{r}$, then $N_{\tilde{r}}=3$ and $(\frac{\widehat{U}_{k\ell}}{U_{k\ell}})^{N_{\tilde{r}}}=\frac{\widehat{U}_{k_1k_2}}{U_{k_1k_2}}\frac{\widehat{U}_{k_2k_3}}{U_{k_2k_3}}\frac{\widehat{U}_{k_3k_4}}{U_{k_3k_4}}$. In \eqref{lem-remainder2-relateS}, $S^*_{k_1k_2k_3k_4}$ is a random variable whose mean and variance are relatively easy to calculate. The factor in front of $S^*_{k_1k_2k_3k_4}$ has a complicated correlation with the summands in $S^*_{k_1k_2k_3k_4}$, but fortunately we can apply a simple bound on this factor. Consider the event $E_n$ as in \eqref{def:Dn}. We have shown in \eqref{lem-remainder1-eventprob} that $\mathbb{P}(E_n^c)=o(n^{-L})$ for any fixed $L>0$. Therefore, the event $E_n^c$ has a negligible effect on the mean and variance of $S_{k_1k_2k_3k_4}$, i.e.,
\[
\mathbb{E}[S^2_{k_1k_2k_3k_4}\cdot I_{E_n^c}] = o(1). 
\]
On the event $E_n$, we have $\max_{k,\ell}\{|\widehat{U}_{k\ell}-U_{k\ell}|/U_{k\ell}\}\leq C_0 x_n/\|\theta\|_1$. It follows that  
\begin{align*} 
|S_{k_1k_2k_3k_4}| &\leq \Bigl( \max_{k,\ell}\frac{|\widehat{U}_{k\ell}|}{U_{k\ell}} \Bigr)^{N_{\tilde{\gamma}}} \Bigl( \frac{\theta_{\max}}{\tau\|\theta\|_1^2} \Bigr)^{N_{\tilde{\alpha}}} \Bigl( \max_{k,\ell}\frac{|U_{k\ell}-\widehat{U}_{k\ell}|}{\tau U_{k\ell}} \Bigr)^{N_{\tilde{\beta}}}\Bigl( \max_{k,\ell}\frac{|U_{k\ell}-\widehat{U}_{k\ell}|}{U_{k\ell}} \Bigr)^{N_{\tilde{\gamma}}}|S^*_{k_1k_2k_3k_4}|\cr
&\leq C\Bigl( \frac{\theta_{\max}}{\tau\|\theta\|_1} \Bigr)^{N_{\tilde{\alpha}}} \Bigl(\frac{x_n}{\tau \|\theta\|_1} \Bigr)^{N_{\tilde{\beta}}}\Bigl( \frac{x_n}{\|\theta\|_1} \Bigr)^{N_{\tilde{\gamma}}}|S^*_{k_1k_2k_3k_4}|.
\end{align*}
Since $x_n\ll\frac{\|\theta\|_1}{\|\theta\|^2}$ and $\tau\|\theta\|\to\infty$, we immediately have $\frac{x_n}{\|\theta\|_1}=o(\frac{1}{\|\theta\|^2})$, $\frac{x_n}{\tau\|\theta\|_1}=o( \frac{1}{\tau\|\theta\|^2})=o(\frac{1}{\|\theta\|})$ and $\frac{\theta_{\max}}{\tau\|\theta\|_1}\leq \frac{\theta_{\max}^2}{\tau\|\theta\|^2}=o(\frac{1}{\|\theta\|})$. It follows that
\[
|S_{k_1k_2k_3k_4}| = o(1)\cdot \|\theta\|^{-(N_{\tilde{\alpha}}+N_{\tilde{\beta}}+2N_{\tilde{\gamma}})}\cdot |S^*_{k_1k_2k_3k_4}|, \qquad\mbox{on the event }E_n. 
\]
Combining the above gives  
\begin{align} \label{lem-remainder2-relateSmoment}
\mathbb{E}[S^2_{k_1k_2k_3k_4}] &= \mathbb{E}[S^2_{k_1k_2k_3k_4}\cdot I_{E_n}] +\mathbb{E}[S^2_{k_1k_2k_3k_4}\cdot I_{E^c_n}] \cr
&= o(1)\cdot \|\theta\|^{-(2N_{\tilde{\alpha}}+2N_{\tilde{\beta}}+4N_{\tilde{\gamma}})}\cdot \mathbb{E}\bigl[(S^*_{k_1k_2k_3k_4})^2\bigr] + o(1). 
\end{align}

It remains to bound $\mathbb{E}\bigl[(S^*_{k_1k_2k_3k_4})^2\bigr]$. As we mentioned, we can write $S^*_{k_1k_2k_3k_4}$ as a weighted sum of monomials of $W$ and calculate its mean and variance directly. However, given that there are $2145$ types of $S^*_{k_1k_2k_3k_4}$, the calculation is still very tedious. We now use a simple trick to relate the $S^*_{k_1k_2k_3k_4}$ to the post-expansion sums we have analyzed in Lemmas~\ref{prop:remainder0}-\ref{prop:remainder1}. We first bound $|\alpha_{ij}|$ in \eqref{lem-remainder2-abc}. Since $d_i^* = \mathbb{E}[d_i] + \Omega_{ii}$, 
\[
|\alpha_{ij}|\leq \frac{\tau\|\theta\|_1}{\theta_{\max}}\Bigl( \mathbb{E}[d_i]\mathbb{E}[d_j]|U_{k\ell}^{*} - U_{k\ell}| +( \Omega_{ii}\mathbb{E}[d_j] +  \Omega_{jj}\mathbb{E}[d_i])U_{k\ell}^{*} + \Omega_{ii}\Omega_{jj} U_{k\ell}^{*}\Bigr). 
\]
By basic algebra, $| (x_1+x_2)/(y_1+y_2) - x_1/y_1| \leq |x_2|/(y_1+y_2) + |x_1y_2|/[(y_1+y_2)y_1]$. We apply it on \eqref{def:Ustar}-\eqref{def:U}  and note that ${\bf 1}_k'(\Omega-\mathbb{E}[A]){\bf1}_\ell={\bf 1}_k'\diag(\Omega){\bf 1}_\ell=O(\|\theta\|^2)$ and ${\bf 1}_k'(d^*-\mathbb{E}[d])={\bf 1}_k'\diag(\Omega){\bf 1}_n=O(\|\theta\|^2)$. It yields 
\begin{align*}
& |U_{k\ell}^{*} - U_{k\ell}| \\
\leq \;\; & \frac{|{\bf 1}_k'\Omega{\bf 1}_\ell-{\bf 1}_k'\mathbb{E}[A]{\bf 1}_\ell|}{({\bf 1}_k'd^*)({\bf 1}'_{\ell}d^*)} + \frac{  ({\bf{1}}_k' \mathbb{E}[A] {\bf{1}}_\ell)\, |({\bf 1}_k'd^*)({\bf 1}'_{\ell}d^*)- ({\bf 1}_k'\mathbb{E}[d])({\bf 1}'_{\ell}\mathbb{E}[d])|}{({\bf 1}_k'd^*)({\bf 1}'_{\ell}d^*)({\bf 1}_k'\mathbb{E}[d])({\bf 1}'_{\ell}\mathbb{E}[d])}\cr
\leq\;\; & C\|\theta\|_1^{-4}\cdot {\bf 1}_k'\diag(\Omega) {\bf 1}_n + C\|\theta\|_1^{-6}\cdot |({\bf 1}_k'd^*)({\bf 1}'_{\ell}d^*)- ({\bf 1}_k'\mathbb{E}[d])({\bf 1}'_{\ell}\mathbb{E}[d])|\cr
\leq \;\; & C\|\theta\|_1^{-4}\cdot \|\theta\|^2 + C\|\theta\|_1^{-6}\cdot \|\theta\|^2_1\|\theta\|^2\cr
 \leq \;\; & C\|\theta\|_1^{-3}\theta_{\max},
\end{align*}
where in the last line we have used $\|\theta\|^2\leq \theta_{\max}\|\theta\|_1$. 
Combining the above gives 
\begin{align*}
|\alpha_{ij}| &\leq  \frac{C\tau\|\theta\|_1}{\theta_{\max}} \Bigl[\theta_i\theta_j\|\theta\|_1^2\cdot \|\theta\|_1^{-3}\theta_{\max} + (\theta_i^2\theta_j\|\theta\|_1+\theta_j^2\theta_i\|\theta\|_1)\cdot \|\theta\|_1^{-2}+ \theta_i^2\theta_j^2\|\theta\|_1^{-2}\Bigr] \cr
&\leq \frac{C\tau\|\theta\|_1}{\theta_{\max}} \cdot \frac{\theta_i\theta_j\theta_{\max}}{\|\theta\|_1}\leq C\tau\theta_i\theta_j. 
\end{align*}
Additionally, in \eqref{lem-remainder2-abc}, we observe that $\gamma_{ij}=\delta_{ij}$. Since $|U_{k\ell}|\leq C\|\theta\|^{-1}$ and $\mathbb{E}[d_i]\leq C\theta_i\|\theta\|_1$, it is true that $|\beta_{ij}|\leq C\tau \theta_i\theta_j$. We summarize the results as
\beq \label{lem-remainder2-analogy}
|\alpha_{ij}|\leq C\tau\theta_i\theta_j, \qquad |\beta_{ij}|\leq C\tau\theta_i\theta_j, \qquad \gamma_{ij}=\delta_{ij}. 
\eeq
It says that $\gamma$ is the same as $\delta$, and $(\alpha, \beta)$ behave similarly as $\widetilde{\Omega}$. Consequently, the calculation of mean and variance of $S^*_{k_1k_2k_3k_4}$ in \eqref{lem-remainder2-Sstar} can be carried out by replacing $(\alpha,\beta,\gamma)$ with $(\widetilde{\Omega}, \widetilde{\Omega}, \delta)$. In other words, we only need to study a sum like
\[
S^{**}_{k_1k_2k_3k_4}=\sum_{j=1}^4\sum_{i_j\in {\cal N}_{k_j}^{(m,0)}}a_{i_1i_2}b_{i_2i_3}c_{i_3i_4}d_{i_4 i_1}, \quad \mbox{where}\;\; a,b,c,d\in \{\widetilde{\Omega}, W,\delta, r\}.  
\]
Let $(N_{\widetilde{\Omega}}, N_W, N_{\delta}, N_r, N_{\alpha}, N_{\beta}, N_{\gamma})$ be the count of different terms in $\{a,b,c,d\}$ determined by $S^*_{k_1k_2k_3k_4}$, where these counts sum to $4$. In $S^{**}_{k_1k_2k_3k_4}$, the counts become $N^*_{\widetilde{\Omega}}=N_{\widetilde{\Omega}}+N_\alpha+N_\beta$, $N^*_W=N_W$,  $N^*_\delta=N_{\delta}+N_\gamma$ and $N^*_r=N_r$. Luckily, anything like $S^{**}_{k_1k_2k_3k_4}$ has been analyzed in Lemmas~\ref{prop:remainder0}-\ref{prop:remainder1}. Especially, in light of \eqref{lem-remainder2-relateSmoment}, the mean and variance contributed by any post-expansion sum considered here must be dominated by the mean and variance of some post-expansion sum considered in Lemmas~\ref{prop:remainder0}-\ref{prop:remainder1}. We thus immediately obtain the claim, without any extra calculation.\qed

\section{Proof of the non-splitting property (NSP) of SCORE}  \label{sec:proof-B}
In this section, we prove the NSP of SCORE, i.e., Theorem~\ref{thm:SCORE2}. As explained in Section~\ref{sec:NSP}, the proof relies on a few technical results, including Lemmas~\ref{lem:entrywise-bound}-\ref{lem:d_m} and Theorem~\ref{thm:kmeans}, which we also prove in this section. 

\subsection{Proof of Theorem \ref{thm:SCORE2}}
\label{subsec:thmscore2pf} 
By Lemma~\ref{lem:entrywise-bound}, there is an event $E$, where $\mathbb{P}(E^c)=O(n^{-3})$, and on this event there exists a $(K-1)\times (K-1)$ orthogonal  matrix $\Gamma$ (which may depend on $n$ and $\widehat{R}^{(K)}$) such that 
\[
\max_{1\leq i\leq n}\|\hat{r}^{(m)}_i -r_i^{(m)}(\Gamma) \|\leq Cs_n^{-1}\sqrt{\log(n)}, \qquad\mbox{for all }1<m\leq K. 
\]
Fix $1<m\leq K$. By Lemma~\ref{lem:matR}, $r_i^{(m)}(\Gamma)=v_k^{(m)}(\Gamma)$ for each $i\in {\cal N}_k$ and $1\leq k\leq K$.  
Suppose $v_1^{(m)}(\Gamma),   \ldots,v_K^{(m)}(\Gamma)$ have $L$ distinct values, where $L$ may depend on $m$ and $\Gamma$ and $L \geq m$ by Lemma~\ref{lem:d_m}.  
Note that whenever two vectors (say) $v_1^{(m)}(\Gamma)$ and $v_2^{(m)}(\Gamma)$ are identical, 
we can always treat ${\cal N}_1$ and ${\cal N}_2$ as the same cluster before we apply  
Theorem \ref{thm:kmeans}. Therefore, without loss of generality, we assume $L = K$, so 
$v_1^{(m)}(\Gamma),  \ldots,v_K^{(m)}(\Gamma)$ are   distinct. 
It suffices to show that, on the event $E$, none of ${\cal N}_1, {\cal N}_2, \ldots, {\cal N}_K$ is split by the k-means.

We now apply Theorem~\ref{thm:kmeans} with $\hat{x}_i=\hat{r}^{(m)}_i$,  $x_i=r_i^{(m)}(\Gamma)$, $F_k = {\cal N}_k$, and $U = V^{(m)}(\Gamma)$. Note that by Lemma~\ref{lem:d_m}, $d_m(U)\geq C$.  
Also, in the proof of Lemma~\ref{lem:d_m}, we have shown that $\max_{1\leq k\leq K}\|v^{(m)}_k(\Gamma)\|\leq C$. It follows that the $\ell^2$-norm of each row of $U$ is bounded by $C\cdot d_m(U)$. Additionally, on the event $E$, $\max_{1\leq i\leq n}\|\hat{x}_i-x_i\|\leq Cs_n^{-1}\sqrt{\log(n)}$. As long as $s_n\geq C_0\sqrt{\log(n)}$ for a sufficiently large constant $C_0$, we have $\max_{1\leq i\leq n}\|\hat{x}_i-x_i\|\leq c\cdot d_m(U)$ for a sufficiently small constant $c$. The claim now follows by applying Theorem~\ref{thm:kmeans}. \qed

\subsection{Two useful lemmas}

We present two technical lemmas. The first lemma is about the eigenvalues and eigenvectors of $\Omega$. It is proved in Section~\ref{subsec:proof-Omega-eig}. 
\begin{lemma} \label{lemma:eigenvector-relation} 
Consider a DCBM where (\ref{condition1d}) holds and let $\lambda_k,\mu_k, \eta_k, \xi_k$ be defined as above. We have the following claims.  First, $\lambda_k = \|\theta\|^2 \mu_k$ for $1 \leq k \leq K$.  Second, the multiplicity of $\mu_1$ is $1$ and all entries of $\eta_1$ have the same sign, and the same holds for $\lambda_1$ and $\xi_1$.  Last,  if  $\eta_k$ is an eigenvector of $HPH$ corresponding to $\mu_k$,  then $\|\theta\|^{-1} \Theta \Pi H^{-1} \eta_k$ is an eigenvector of $\Omega$ corresponding to $\lambda_k$, and conversely,  if $\xi_k$ is an eigenvector of $\Omega$ corresponding to $\lambda_k$, then $\|\theta\|^{-1} H^{-1} \Pi' \Theta \xi_k$ is an eigenvector of $HPH$ corresponding to $\mu_k$. 
\end{lemma}

The second lemma characterizes the change of the k-means objective under perturbation of cluster assignment. Consider the problem of clustering points $y_1, y_2,\ldots,y_n$ into two disjoint clusters $A$ and $B$. The k-means objective is the residual sum of squares by setting the two cluster centers as the within-cluster means. Now, we move a subset $C$ from cluster $A$ to cluster $B$. The new clusters are $\tilde{A}=A\setminus C$ and $\tilde{B}=B\cup C$, and the cluster centers are updated accordingly. There is an explicit formula for the change of the k-means objective:

\begin{lemma} \label{lem:RSS}
For any $y_1,y_2,\ldots,y_n\in\mathbb{R}^d$ and subset $M\subset\{1,2,\ldots,n\}$, define $\bar{y}_M=\frac{1}{|M|}\sum_{i\in M}y_i$. Let $\{1,2,\ldots,n\}=A\cup B$ be a partition, and let $C$ be a strict subset of $A$. Write $\tilde{A}=A\backslash C$ and $\tilde{B}=B\cup C$.  Define 
\[
RSS = \sum_{i\in A}\|y_i-\bar{y}_A\|^2 + \sum_{i\in B}\|y_i-\bar{y}_B\|^2, \qquad \widetilde{RSS} = \sum_{i\in \tilde{A}}\|y_i-\bar{y}_{\tilde{A}}\|^2 + \sum_{i\in \tilde{B}}\|y_j-\bar{y}_{\tilde{B}}\|^2. 
\]
Then, 
\[
\widetilde{RSS} - RSS =  \frac{|B||C|}{|B|+|C|}\|\bar{y}_C - \bar{y}_B\|^2 - \frac{|A||C|}{|A|-|C|}\|\bar{y}_C-\bar{y}_A\|^2. 
\]
\end{lemma}

\noindent This lemma is proved in Section~\ref{subsec:proof-RSS}. It shows that the change of k-means objective depends on the distances from $\bar{y}_C$ to two previous cluster centers.

\subsection{Proof of Lemma~\ref{lem:entrywise-bound}} 
Since $\|\hat{r}^{(m)}-r_i^{(m)}(\Gamma)\|\leq \|\hat{r}^{(K)}-r_i^{(K)}(\Gamma)\|$, we only need to show the claim for $m=K$. 
Write $r_i^{(K)}(\Gamma)=r_i(\Gamma)$ for short. In the special case of $\Gamma=I_{K-1}$ (i.e., $\eta_k(\Gamma)=\eta_k$ for $2\leq k\leq K$), we further write $r_i=r_i(I_{K-1})$ for short. It is easy to see that
\[
r_i(\Gamma)=\Gamma'\cdot r_i, \qquad\mbox{for any orthogonal matrix $\Gamma\in\mathbb{R}^{(K-1)\times (K-1)}$}. 
\]
It suffices to show that with probability $1-O(n^{-3})$ there exists a $(K-1)\times (K-1)$ orthogonal matrix $\Gamma$, which may depend on $n$ and $\widehat{R}^{(K)}$, such that 
\[
\max_{1\leq i\leq n}\|\hat{r}_i-\Gamma'\cdot r_i\|\leq Cs_n^{-1}\sqrt{\log(n)}.
\]  
Such a bound was given by Theorem 4.1 of \cite{SCOREplus} (also, see Lemma 2.1 of \cite{Mixed-SCORE} for a special case where $\lambda_2,\ldots,\lambda_K$ are at the same order). \qed

\subsection{Proof of Lemma~\ref{lem:matR}} 
For convenience, we use $\xi_1,\xi_2^*,\ldots,\xi_K^*$ to denote a specific choice of eigenvectors and drop ``$\Gamma$" in $\xi_k(\Gamma)$, $1\leq k\leq K$. 
With these notations, we have
\[
[\eta_1, \eta_2, \ldots, \eta_K] = [\eta_1, \eta_2^*, \ldots, \eta^*_K]\begin{bmatrix}1\\&\Gamma\end{bmatrix}, \qquad  [\xi_1, \xi_2, \ldots, \xi_K] = [\xi_1, \xi_2^*, \ldots, \xi^*_K]\begin{bmatrix}1\\&\Gamma\end{bmatrix}.
\]
Here, $\eta_1, \eta_2^*,\ldots,\eta_K^*$ is a particular candidate of the eigenvectors of $HPH$ and $\xi_1, \xi_2^*,\ldots,\xi_K^*$ is linked to $\eta_1, \eta_2^*,\ldots,\eta_K^*$ through
\[
[\xi_1, \xi_2^*, \ldots,\xi_K^*] = \|\theta\|^{-1}\Theta\Pi H^{-1}[\eta_1, \eta_2^*, \ldots,\eta_K^*]. 
\]
It follows immediately that
\beq \label{lem-matR-1}
[\xi_1, \xi_2, \ldots,\xi_K] = \|\theta\|^{-1}\Theta\Pi H^{-1}[\eta_1, \eta_2, \ldots,\eta_K]. 
\eeq
As a result, for any true community ${\cal N}_k$, 
\[
\xi_\ell(i) = [\theta_i/(\|\theta\| H_{kk})]\cdot \eta_\ell(k), \qquad \mbox{for all }i\in {\cal N}_k. 
\]
We plug it into the definition of $R^{(m)}$ to get that for each $i\in {\cal N}_k$ and $1\leq \ell\leq m-1$, 
\[
R^{(m)}(i, \ell)=\frac{\xi_{\ell+1}(i)}{\xi_1(i)}=\frac{ [\theta_i/(\|\theta\| H_{kk})]\cdot \eta_{\ell+1}(k)}{ [\theta_i/(\|\theta\| H_{kk})]\cdot \eta_1(k)}=\frac{\eta_{\ell+1}(k)}{\eta_1(k)}=V^{(m)}(k,\ell). 
\] 
It follows that $r_i^{(m)}=v_k^{(m)}$ for each $i\in {\cal N}_k$. \qed

\subsection{Proof of Lemma~\ref{lem:d_m}}
The matrix $V^{(K)}(\Gamma)$ was studied in \cite{SCORE, Mixed-SCORE}. 
Since the pairwise distances for rows in $V^{(K)}(\Gamma)$ are invariant of $\Gamma$, the quantity $d_K(V^{(K)}(\Gamma))$ does not change with $\Gamma$ either. Using Lemma B.3 of \cite{SCORE}, we immediately know that $d_K(V^{(K)}(\Gamma))\geq \sqrt{2}$.

Below, we fix $1<m<K$ and a $(K-1)\times (K-1)$ orthogonal matrix $\Gamma$, and study $d_m(V^{(m)}(\Gamma))$. For notation simplicity, we drop ``$\Gamma$'' when there is no confusion. 

We apply a bottom up pruning procedure (same as in Definition~\ref{def:bottom-up-pruning}) to $V^{(m)}$. First, we find two rows $v_k^{(m)}$ and $v_\ell^{(m)}$ that attain the minimum pairwise distance (if there is a tie, pick the first pair in the lexicographical order) and change the $\ell$th row to $v_k^{(m)}$ (suppose $k<\ell$). Denote the resulting matrix by $V^{(m,K-1)}$. Next, we consider the rows of $V^{(m,K-1)}$ and similarly find two rows attaining the minimum pairwise distance and replace one row by the other. Denote the resulting matrix by $V^{(m,K-2)}$. We repeat these steps to get a sequence of matrices:
\[
V^{(m,K)}, \; V^{(m,K-1)}, \; V^{(m,K-2)},\;\dots, V^{(m,2)},\; V^{(m,1)}, 
\]
where $V^{(m,K)}=V^{(m)}$ and for each $1\leq k\leq K$, $V^{(m,k)}$ has at most $k$ distinct rows. Comparing it with the definition of $d_k(V^{(m)})$ (see Definition~\ref{def:bottom-up-pruning}), we find that $V^{(m,k-1)}$ differs from $V^{(m,k)}$ in only 1 row, and the difference on this row is a vector whose Euclidean norm is exactly $d_k(V^{(m)})$. As a result,
\beq \label{lem-mindist-1}
\| V^{(m,k)}-V^{(m,k-1)}\|= d_k(V^{(m)}), \qquad 2\leq k\leq K. 
\eeq
By triangle inequality and the fact that $d_k(V^{(m)})\leq d_{k-1}(V^{(m)})$, we have 
\[
\|V^{(m,K)}-V^{(m,m-1)}\|\leq \sum_{k=m}^K d_k(V^{(m)})\leq (K-m+1)\cdot d_m(V^{(m)}). 
\]
To show the claim, it suffices to show that
\beq \label{lem-mindist-2}
\|V^{(m,K)}-V^{(m,m-1)}\|\geq C. 
\eeq 

We now show \eqref{lem-mindist-2}. Introduce two matrices
\[
V_*^{(m,K)}=[{\bf1}_K, V^{(m,K)}], \qquad  V_*^{(m,m-1)}=[{\bf 1}_K, V^{(m,m-1)}], 
\]
where ${\bf 1}_K$ is the $K$-dimensional vector of $1$s. Adding the vector ${\bf 1}_K$ as the first column changes neither the number of distinct rows nor pairwise distances among rows. Additionally, 
\beq\label{lem-mindist-3}
\|V^{(m,K)}-V^{(m,m-1)}\| = \|V_*^{(m,K)}-V_*^{(m,m-1)}\|. 
\eeq
Let $\sigma_m(U)$ denote the $m$-th singular value of a matrix $U$. Since $V_*^{(m,m-1)}$ has at most $(m-1)$ distinct rows, its rank is at most $(m-1)$. As a result,
\beq \label{lem-mindist-4}
\sigma_m(V_*^{(m,m-1)})=0. 
\eeq
We then study $\sigma_m(V_*^{(m,K)})$. Note that
\beq \label{lem-mindist-temp}
V_*^{(m,K)} =[{\bf 1}_K, V^{(m)}] = \begin{bmatrix} 1 & v_1^{(m)}\\
\vdots & \vdots \\
1 & v_K^{(m)}\end{bmatrix} = [\diag(\eta_1)]^{-1}\cdot \bigl[\eta_1,\eta_2(\Gamma),\ldots,\eta_m(\Gamma)\bigr], 
\eeq
where $\eta_1,\eta_2(\Gamma),\ldots,\eta_K(\Gamma)$ is one choice of eigenvectors of $HPH$ indexed by $\Gamma$ (see the definitions in the paragraph above \eqref{DefineV}) and $\diag(\eta_1)$ is the diagonal matrix whose diagonal entries are from $\eta_1$. Write for short $Q=[\eta_1,\eta_2(\Gamma),\ldots,\eta_m(\Gamma)]$. We have
\[
(V_*^{(m,K)})'V_*^{(m,K)}  = Q' [\diag(\eta_1)]^{-2} Q. 
\]
By the last item of \eqref{condition1d} and that $\|\eta_1\|=1$, we conclude that $\eta_1(k)\asymp 1/\sqrt{K}$ for all $1\leq k\leq K$. In particular, there exists a constant $c>0$ such that $\bigl([\diag(\eta_1)]^{-2}-cI_K\bigr)$ is a positive semi-definite matrix. It follows that $\bigl(Q' [\diag(\eta_1)]^{-2} Q-c Q'Q\bigr)$ is a positive semi-definite matrix. Therefore, 
\beq \label{lem-dm-add}
\lambda_m\bigl((V_*^{(m,K)})'V_*^{(m,K)}\bigr)\geq  \lambda_m(cQ'Q)=c\cdot \lambda_m(Q'Q),
\eeq
where $\lambda_m(\cdot)$ denotes the $m$-th largest eigenvalue of a symmetric matrix. 
From the way $\eta_k$'s are defined (see the paragraph above \eqref{DefineV}), for some pre-specified choice of eigenvectors, $\eta_1,\eta_2^*,\ldots,\eta_K^*$, of $HPH$,  
\[
Q \mbox{ is the first $m$ columns of the matrix }[\eta_1,\eta_2^*,\ldots,\eta_K^*]\cdot\diag(1,\Gamma). 
\]
Note that $[\eta_1,\eta_2^*,\ldots,\eta_K^*]$ and $\diag(1,\Gamma)$ are both $K\times K$ orthogonal matrices. Then, their product is also an orthogonal matrix, and the columns in $Q$ are orthonormal. It follows that 
\[
Q'Q=I_m.
\] 
This shows that the right hand side of \eqref{lem-dm-add} is equal to $c$. The left hand side of \eqref{lem-dm-add} is equal to $\sigma_m^2(V_*^{(m,K)})$. It follows that 
\beq \label{lem-mindist-5}
\sigma_m(V_*^{(m,K)})\geq C. 
\eeq
We now combine \eqref{lem-mindist-4} and \eqref{lem-mindist-5}, and apply Weyl's inequality for singular values \citep[Corollary 7.3.5]{HornJohnson}. It gives
\[
C\leq \sigma_m(V_*^{(m,K)}) - \sigma_m(V_*^{(m,m-1)}) \leq \|V_*^{(m,K)} - V_*^{(m,m-1)}\|. 
\]
Combining it with \eqref{lem-mindist-3} gives \eqref{lem-mindist-2}. The claim follows immediately. \qed

{\bf Remark}. The proof of Theorem~\ref{thm:SCORE2} uses $\max_{1\leq k\leq K}\|v_k^{(m)}(\Gamma)\|\leq C$, and we prove this claim here. Note that $v_k^{(m)}(\Gamma)$ is a sub-vector of the $k$th row of $V_*^{(m,K)}$. In light of \eqref{lem-mindist-temp}, the row-wise $\ell_2$-norms of $V_*^{(m,K)}$ are uniformly bounded by $C \|\diag^{-1}(\eta_1)\|$. We have argued that $\eta_1(k)\asymp 1/\sqrt{K}\leq C$ for all $1\leq k\leq K$. As a result, $\max_{1\leq k\leq K}\|v_k^{(m)}(\Gamma)\|\leq C$.  

\subsection{Proof of Theorem~\ref{thm:kmeans} (NSP of $k$-means)} \label{subsec:thm-kmeans-proof}

Write for short $d_m = d_m(U)$ and $\delta =\max_{1\leq i\leq n}\|\hat{x}_i-x_i\|$. 
Given any partition $\{1,2,\ldots,n\}=\cup_{k=1}^m B_k$ and vectors $b_1,b_2,\ldots,b_m\in\mathbb{R}^d$, define
\beq \label{thm-km-defR}
R(B_1,\ldots,B_m; b_1,\ldots,b_m) = n^{-1}\sum_{k=1}^m \sum_{i\in B_k}\|x_i - b_k\|^2. 
\eeq
Fixing $B_1,\ldots,B_m$, the value of $R(B_1,\ldots,B_m; b_1,\ldots,b_m)$ is minimized when $b_k$ is the average of $x_i$'s within each $B_k$. 
When $b_1,\ldots,b_m$ take these special values, we skip them in the notation. Namely, define
\beq\label{thm-km-defR2}
R(B_1,\ldots,B_m) =
R(B_1,\ldots,B_m;\underline{x}_1,\ldots,\underline{x}_m),\quad\mbox{where}\;\; \underline{x}_k=|B_k|^{-1}\sum_{i\in B_k}x_i,\eeq
We define $\hat{R}(B_1,\ldots,B_m; b_1,\ldots,b_m)$ and $\hat{R}(B_1,\ldots,B_m)$ similarly but replace $x_i$ by $\hat{x}_i$.  
We shall prove the claim by contradiction. Suppose there is $1\leq k\leq K$ such that $F_k$ intersects with more than one $\hat{S}_j$. By pigeonhole principle, there exists $j_1$, such that $|F_k\cap \hat{S}_{j_1}| \geq m^{-1}|F_k|$. Let $\hat{S}_{j_2}$ be another cluster that intersects with $F_k$. We have  
\[
|F_k\cap \hat{S}_{j_1}|\geq m^{-1}\alpha_0n, \qquad F_k\cap \hat{S}_{j_2}\neq \emptyset, \qquad 
\]
Below, we aim to show: There exists $C_1=C_1(\alpha_0, C_0, m)$ such that 
\beq \label{thm-km-(a)}
\min_{\tilde{S}_1,\ldots,\tilde{S}_m}R(\tilde{S}_1,\ldots,\tilde{S}_m)\geq R(\hat{S}_1,\ldots,\hat{S}_m) - C_1\delta\cdot d_m,
\eeq
where the minimum on the left hand side is taken over possible partitions of $\{1,2,\ldots,n\}$ into $m$ clusters. 
We also aim to show that there exists $C_2=C_2(\alpha_0, C_0, m)$ such that we can construct a clustering structure $\tilde{S}_1, \tilde{S}_2, \ldots, \tilde{S}_m$ satisfying that  
\beq \label{thm-km-(b)}
R(\tilde{S}_1,\ldots,\tilde{S}_m)\leq R(\hat{S}_1, \ldots, \hat{S}_m)- C_2\cdot d_m^2.
\eeq 
Combining \eqref{thm-km-(a)}-\eqref{thm-km-(b)} gives
\[
R(\hat{S}_1,\ldots,\hat{S}_m) - C_1\delta\cdot d_m\leq R(\hat{S}_1, \ldots, \hat{S}_m)- C_2\cdot d_m^2
\]
This is impossible if $C_1\delta\cdot d_m < C_2\cdot d_m^2$. 
Hence, we can take 
\[
c(\alpha_0, C_0, m) < C_2/C_1. 
\]
There is a contradiction between \eqref{thm-km-(a)} and \eqref{thm-km-(b)} whenever $\delta \leq c\cdot d_m$. The claim follows.

It remains to prove \eqref{thm-km-(a)} and \eqref{thm-km-(b)}. Consider \eqref{thm-km-(a)}. For an arbitrary cluster structure $B_1,B_2,\ldots, B_m$, let $\hat{R}(B_1,\ldots,B_m)$, $R(B_1,\ldots,B_m)$, $\underline{\hat{x}}_k$ and $\underline{x}_k$ be defined as in \eqref{thm-km-defR2}. By direct calculations,  
\[
(\hat{x}_i - \underline{\hat{x}}_k)-(x_i-\underline{x}_k) =\frac{|B_k|-1}{|B_k|}(\hat{x}_i-x_i) - \frac{1}{|B_k|} \sum_{j\in B_k:j\neq i}(\hat{x}_j-x_j).
\]
Since 
$\|\hat{x}_j-x_j\|\leq \delta$ for all $1\leq j\leq n$, the above equality implies that $\|(\hat{x}_i - \underline{\hat{x}}_k)-(x_i-\underline{x}_k)\|\leq \delta$. As a result, $\|\hat{x}_i - \underline{\hat{x}}_k\|^2\leq \|x_i - \underline{x}_k\|^2 +2\delta\|x_i-\underline{x}_k\|+\delta^2$. It follows that 
\begin{align*}  
\hat{R}(B_1,\ldots,B_m) &\leq R(B_1,\ldots,B_m) +2\delta n^{-1}\sum_{k=1}^m \sum_{i\in B_k}\|x_i-\underline{x}_k\| +  \delta^2\cr
 &\leq R(B_1,\ldots,B_m) + 2\delta\sqrt{R(B_1,\ldots,B_m)} + \delta^2\cr 
 &\leq \bigl(\sqrt{R(B_1,\ldots,B_m)}+\delta\bigr)^2,
\end{align*} 
where the second line is from the Cauchy-Schwarz inequality. It follows that $\sqrt{\hat{R}(B_1,\ldots,B_m)}\leq \sqrt{R(B_1,\ldots,B_m)}+\delta$.
We can switch $\hat{R}(B_1,\ldots,B_m)$ and $R(B_1,\ldots,B_m)$ to get a similar inequality. Combining them gives
\beq  \label{thm-km-3}
\sqrt{R(B_1,\ldots,B_m)}-\delta \leq\sqrt{\hat{R}(B_1,\ldots,B_m)}\leq \sqrt{R(B_1,\ldots,B_m)}+\delta. 
\eeq
This inequality holds for an arbitrary partition $(B_1, B_2,\ldots,B_m)$. We now apply it to $(\hat{S}_1,\ldots,\hat{S}_m)$, which are the clusters obtained from applying k-means on $\hat{x}_1,\hat{x}_2, \ldots,\hat{x}_n$. We also consider applying k-means on $x_1,x_2, \ldots,x_n$ and let $S_1, S_2, \ldots,S_m$ denote the resultant clusters. 
By optimality of the k-means solutions,  
\[
\hat{R}(\hat{S}_1,\ldots, \hat{S}_m)\leq \hat{R}(S_1,\ldots,S_m). 
\]
Combining it with \eqref{thm-km-3} gives
\begin{align} \label{thm-km-4}
\sqrt{R(\hat{S}_1,\ldots,\hat{S}_m)} &\leq \sqrt{\hat{R}(\hat{S}_1,\ldots, \hat{S}_m)}+\delta\cr
&\leq \sqrt{\hat{R}(S_1,\ldots, S_m)}+\delta\cr
&\leq \sqrt{R(S_1,\ldots, S_m)}+2\delta. 
\end{align}
Since $\max_{1\leq i\leq n}\|x_i\|\leq C_0\cdot d_m$, we can easily see that $R(S_1,\ldots,S_m)\leq C^2_0\cdot d_m^2$. It follows that, as long as $\delta \leq d_m/4$, 
\begin{align*}  
R(\hat{S}_1,\ldots,\hat{S}_m) & \leq R(S_1,\ldots, S_m)+4\delta\sqrt{R(S_1,\ldots, S_m)} + 4\delta^2 \cr
&\leq R(S_1,\ldots, S_m) + 4C_0\delta\cdot d_m + \delta\cdot d_m\cr
&\leq R(S_1,\ldots, S_m) + (4C_0+1)\delta \cdot d_m. 
\end{align*}
As a result, 
\[
\min_{\tilde{S}_1,\ldots,\tilde{S}_m} R(\tilde{S}_1,\ldots,\tilde{S}_m)=R(S_1,\ldots, S_m)\geq R(\hat{S}_1,\ldots,\hat{S}_m)-(4C_0+1)\delta\cdot d_m. 
\]
This proves \eqref{thm-km-(a)} for $C_1=4(C_0+1)$.

Consider \eqref{thm-km-(b)}. Define 
\beq \label{thm-km-hat(w)}
w_j=|\hat{S}_j|^{-1}\sum_{i\in \hat{S}_j}x_i, \qquad \mbox{for each $1\leq j\leq m$}. 
\eeq
Using the notations in \eqref{thm-km-defR}-\eqref{thm-km-defR2}, we write $R(\hat{S}_1,\ldots,\hat{S}_m)=R(\hat{S}_1,\ldots,\hat{S}_m, w_1,\ldots,w_m)$.
We aim to construct $\{(\tilde{S}_j, \tilde{w}_j)\}_{1\leq j\leq m}$ such that
\beq \label{thm-km-defgoal}
R(\tilde{S}_1,\ldots,\tilde{S}_m,\tilde{w}_1,\ldots,\tilde{w}_m)\leq R(\hat{S}_1,\ldots,\hat{S}_m, w_1,\ldots,w_m) - C_2\cdot d_m^2. 
\eeq
Since $R(\tilde{S}_1,\ldots,\tilde{S}_m)=\min_{b_1,\ldots,b_m}R(\tilde{S}_1,\ldots,\tilde{S}_m, b_1,\ldots,b_m)$, we immediately have 
\[
R(\tilde{S}_1,\ldots,\tilde{S}_m) \leq  R(\tilde{S}_1,\ldots,\tilde{S}_m,\tilde{w}_1,\ldots,\tilde{w}_m)\leq R(\hat{S}_1,\ldots,\hat{S}_m, w_1,\ldots,w_m) - C_2\cdot d_m^2. 
\]
This proves \eqref{thm-km-(b)}. 

What remains is to construct $\{(\tilde{S}_j, \tilde{w}_j)\}_{j=1}^m$ so that \eqref{thm-km-defgoal} is satisfied. Let $\hat{w}_j=|\hat{S}_j|^{-1}\sum_{i\in \hat{S}_j}\hat{x}_i$, for $1\leq j\leq m$. Then, $\{(\hat{S}_j, \hat{w}_j)\}_{1\leq j\leq m}$ are the clusters and cluster centers obtained by applying the k-means algorithm on $\hat{x}_1,\hat{x}_2,\ldots,\hat{x}_n$. The k-means solution guarantees to assign each point to the closest center. Take $i\in F_k\cap \hat{S}_{j_1}$ and $i'\in F_k\cap\hat{S}_{j_2}$. It follows that 
\[
\|\hat{x}_i-\hat{w}_{j_1}\|\leq \|\hat{x}_i-\hat{w}_{j_2}\|, \qquad  \|\hat{x}_{i'}-\hat{w}_{j_2}\|\leq \|\hat{x}_{i'} -\hat{w}_{j_1}\|. 
\]
Since $x_i=x_{i'}=u_k$ and $\max\{\|\hat{x}_i-x_i\|, \|\hat{x}_{i'}-x_{i'}\|, \|\hat{w}_{j_1}-w_{j_1}\|, \|\hat{w}_{j_2}-w_{j_2}\|\}\leq \delta$, we have 
\[
\|u_k-w_{j_1}\|\leq \|\hat{x}_i-\hat{w}_{j_1}\| +2\delta \leq \|\hat{x}_i-\hat{w}_{j_2}\|+2\delta\leq \|u_k-w_{j_2}\|+4\delta.  
\]
Similarly, we can derive that $\|u_k-w_{j_2}\|\leq \|u_k-w_{j_1}\|+4\delta$. Combining them gives  
\beq \label{thm-km-key1}
\left|  \|u_k-w_{j_1}\|-\|u_k-w_{j_2}\| \right|\leq 4\delta. 
\eeq 
This inequality tells us that $\|u_k-w_{j_1}\|$ and $\|u_k-w_{j_2}\|$ are sufficiently close. Introduce
\[
C_3=\frac{m^{-1}\alpha_0}{36\times 12 C_0}.
\]
Below, we consider two cases: $\|u_k-w_{j_1}\|<C_3\cdot d_m$ and $\|u_k-w_{j_1}\|\geq C_3\cdot d_m$. 

In the first case, $\|u_k-w_{j_1}\|< C_3\cdot d_m$. The definition of $d_m$ guarantees that there are $m$ points from $\{u_1,u_2,\ldots,u_K\}$ such that their minimum pairwise distance is $d_m$. Without loss of generality, we assume these $m$ points are $u_1,u_2,\ldots, u_m$. If $k\in\{1,2,\ldots,m\}$, then the distance from $u_k$ to any of the other $(m-1)$ points is at least $d_m$. If $k\notin\{1,2,\ldots,m\}$, then $u_k$ cannot be simultaneously within a distance of $<d_m/2$ to two or more points of $u_1,u_2,\ldots,u_m$. In other words, there exists at least $(m-1)$ points from $u_1,u_2,\ldots,u_m$ whose distance to $u_k$ is at least $\geq d_m/2$. 
Combining the above situations, we conclude that there exist $(m-1)$ points from $\{u_1,u_2,\ldots,u_K\}$, which we assume to be $u_1,u_2,\ldots,u_{m-1}$ without loss of generality, such that
\beq \label{thm-km-(m-1)points}
\min_{1\leq \ell\neq s\leq m-1}\|u_\ell-u_s\|\geq d_m, \qquad \min_{1\leq \ell\leq m-1}\|u_\ell - u_k\|\geq d_m/2. 
\eeq 

We then consider two sub-cases. In the first sub-case, there exists $\ell\in \{1,2,\ldots,m-1\}$ such that $|F_\ell\cap (\hat{S}_{j_1}\cup \hat{S}_{j_2})|\geq m^{-1}\alpha_0n$. Then, at least one of $\hat{S}_{j_1}$ and $\hat{S}_{j_2}$ contains more than $(m^{-1}\alpha_0/2) n$ nodes from $F_\ell$. We only study the situation of $|F_\ell\cap \hat{S}_{j_2}|\geq (m^{-1}\alpha_0/2) n$. The proof for the situation of $|F_\ell\cap \hat{S}_{j_1}|\geq (m^{-1}\alpha_0/2)n$ is similar and omitted. We modify the clusters and cluster centers $\{(\hat{S}_j, w_j)\}_{1\leq j\leq m}$ as follows: 
\begin{itemize}
\item[(i)] Combine $\hat{S}_{j_2}\backslash F_\ell$ and $\hat{S}_{j_1}$ into one cluster and set the cluster center to be $w_{j_1}$. 
\item[(ii)] Create a new cluster as $\hat{S}_{j_2}\cap F_\ell$ and set the cluster center to be $u_\ell$.  
\end{itemize} 
The other clusters and cluster centers remain unchanged. Namely, we let
\[
\tilde{S}_j = \begin{cases}
\hat{S}_{j_1}\cup (\hat{S}_{j_2}\backslash F_\ell), & \mbox{if }j=j_1,\cr
\hat{S}_{j_2}\cap F_\ell,  &\mbox{if }j=j_2,\cr
\hat{S}_j, &\mbox{if }j\notin\{j_1,j_2\},
\end{cases}\qquad
\tilde{w}_j = \begin{cases}
u_\ell,  &\mbox{if }j=j_2,\cr
w_j, &\mbox{otherwise}. 
\end{cases}
\]
Recall that $n\cdot R(B_1,\ldots,B_m,b_1,\ldots,b_m)=\sum_{j=1}^m\sum_{i\in B_j}\|x_i-b_j\|^2$. 
By direct calculations,
\begin{align*}
\Delta &\equiv n\cdot R(\hat{S}_1,\ldots,\hat{S}_m, w_1,\ldots,w_m)- n\cdot R(\tilde{S}_1,\ldots,\tilde{S}_m,\tilde{w}_1,\ldots,\tilde{w}_m)\cr
&=\sum_{i\in (\hat{S}_{j_2}\cap F_\ell)}\bigl(\|x_i-w_{j_2}\|^2-\|x_i-u_\ell\|^2\bigr) - \sum_{i\in (\hat{S}_{j_2}\backslash F_\ell)}\bigl( \|x_i-w_{j_1}\|^2 - \|x_i-w_{j_2}\|^2 \bigr)\cr
&\equiv \Delta_2-\Delta_1. 
\end{align*}
Here $\Delta_1$ is the increase of the residual sum of squares (RSS) caused by the operation (i) and $\Delta_2$ is the decrease of RSS caused by the operation (ii).  
\begin{align*}
\Delta_1 
&= \sum_{i\in (\hat{S}_{j_2}\backslash F_\ell)}(\|x_i-w_{j_1}\|- \|x_i-w_{j_2}\|)(\|x_i-w_{j_1}\|+\|x_i-w_{j_2}\|)\cr
&\leq \sum_{i\in (\hat{S}_{j_2}\backslash F_\ell)}\|w_{j_1}-w_{j_2}\|\cdot (\|x_i-w_{j_1}\|+\|x_i-w_{j_2}\|)\cr
&\leq |\hat{S}_{j_2}\backslash F_\ell|\cdot \|w_{j_1}-w_{j_2}\|\cdot 4C_0\cdot d_m, 
\end{align*}
where the third line is from the triangle inequality and the last line is because $\max_{1\leq j\leq m}\|w_j\|\leq \max_{1\leq i\leq n}\|x_i\|\leq C_0\cdot d_m$. Note that $\|w_{j_1}-w_{j_2}\|\leq \|u_k-w_{j_1}\|+\|u_k-w_{j_2}\|$. We have assumed $\|u_k-w_{j_1}\|< C_3\cdot d_m$ in this case. Combing it with \eqref{thm-km-key1}, as long as $\delta<(C_3/4)\cdot d_m$, 
\[
\|w_{j_1}-w_{j_2}\|\leq 2\|u_k-w_{j_1}\|+4\delta\leq 3C_3\cdot d_m. 
\]
It follows that
\beq \label{thm-km-RSSplus1}
\Delta_1 \leq 12 C_0C_3\cdot nd^2_m. 
\eeq
Since $x_i=u_\ell$ for $i\in F_\ell$, we immediately have 
\[
\Delta_2 = |\hat{S}_{j_2}\cap F_\ell|\cdot \|u_\ell-w_{j_2}\|^2. 
\]
We have assumed $\|u_k-w_{j_1}\|\leq C_3\cdot d_m$ in this case. 
Combining it with \eqref{thm-km-key1} and \eqref{thm-km-(m-1)points} gives
\begin{align*}
\|u_\ell -w_{j_2}\|& \geq \|u_\ell-u_k\|-\|u_k-w_{j_2}\|\cr
&\geq \|u_\ell-u_k\| - \bigl( \|u_k-w_{j_1}\|+4\delta\bigr)\cr
&\geq d_m/2 - (C_3\cdot d_m+4\delta). 
\end{align*}
Recall that $C_3=\frac{m^{-1}\alpha_0}{36\times 12 C_0}<1/12$. Then, as long as $\delta<(1/48)d_m$, we have $\|u_\ell-w_{j_2}\|\geq d_m/3$. It follows that 
\beq \label{thm-km-RSSminus1}
\Delta_2\geq (m^{-1}\alpha_0/2)n\cdot (d_m/3)^2\geq \frac{m^{-1}\alpha_0 }{18}\cdot nd_m^2. 
\eeq
As a result, 
\[
\Delta=\Delta_2-\Delta_1 \geq \Bigl(\frac{m^{-1}\alpha_0 }{18}-12C_0C_3\Bigr) \cdot nd_m^2. 
\] 
We plug in the expression of $C_3$, the right hand side is $(m^{-1}\alpha_0/36)\cdot nd_m^2$. It follows that  
\beq \label{thm-km-case1-result}
R(\hat{S}_1,\ldots,\hat{S}_m, w_1,\ldots,w_m)- R(\tilde{S}_1,\ldots,\tilde{S}_m,\tilde{w}_1,\ldots,\tilde{w}_m)  \geq \frac{m^{-1}\alpha_0}{36}\cdot d_m^2. 
\eeq
This gives \eqref{thm-km-defgoal} in the first sub-case. 

In the second sub-case, $|F_\ell\cap (\hat{S}_{j_1}\cup \hat{S}_{j_2})|< m^{-1}\alpha_0n$ for all $1\leq \ell\leq m-1$.  
For each $F_\ell$, by pigeonhole principle, there exists at least one $j\in\{1,2,\ldots,m\}$ such that $|F_\ell\cap \hat{S}_j|\geq m^{-1}|F_\ell|\geq m^{-1}\alpha_0 n$. Denote such a $j$ by $j^*_\ell$; if there are multiple indices satisfying the requirement, we pick one of them. This gives
\[
j^*_1,\, j^*_2,\, \ldots,\, j^*_{m-1}\quad \in\quad \{1,2,\ldots,m\}\backslash\{j_1,j_2\}. 
\]
These $(m-1)$ indices take at most $(m-2)$ distinct values. By pigeonhole principle, there exist $1\leq \ell_1\neq \ell_2\leq m-1$ such that $j^*_{\ell_1}=j^*_{\ell_2}=j^*$, for some $j^*\notin\{j_1,j_2\}$. Recalling \eqref{thm-km-hat(w)}, we let $w_{j^*}$ denote the average of $x_i$'s in $\hat{S}_{j^*}$. 
 Since $\|u_{\ell_1}-u_{\ell_2}\|\geq d_m$, the point $w_{j^*}$  cannot be simultaneously within a distance of $d_m/2$ to both $u_{\ell_1}$ and $u_{\ell_2}$. Without loss of generality, suppose 
\[
\|u_{\ell_1} - w_{j^*}\|\geq d_m/2. 
\]
We modify the clusters and cluster centers $\{(\hat{S}_j, w_j)\}_{1\leq j\leq m}$ as follows: 
\begin{itemize}
\item[(i)] Combine $\hat{S}_{j_1}$ and $\hat{S}_{j_2}$ into one cluster and set the cluster center to be $w_{j_1}$. 
\item[(ii)] Split $\hat{S}_{j^*}$ into two clusters, where one is $(\hat{S}_{j^*}\cap F_{\ell_1})$, and the other is $( \hat{S}_{j^*}\backslash F_{\ell_1})$; the two cluster centers are set as $u_{\ell_1}$ and $w_{j^*}$, respectively. 
\end{itemize} 
The other clusters and cluster centers remain unchanged. Namely, we let
\[
\tilde{S}_j = \begin{cases}
\hat{S}_{j_1}\cup \hat{S}_{j_2},  &\mbox{if }j=j_1,\cr
\hat{S}_{j^*}\cap F_{\ell_1}, & \mbox{if }j=j_2,\cr
\hat{S}_{j^*}\backslash F_{\ell_1}, & \mbox{if }j=j^*,\cr
\hat{S}_j, &\mbox{if }j\notin\{j_1,j_2, j^*\},
\end{cases}\qquad
\tilde{w}_j = \begin{cases}
u_{\ell_1}, & \mbox{if }j=j_2,\cr
w_j, &\mbox{otherwise}. 
\end{cases}
\]
By direct calculations, 
\begin{align*}
\Delta &\equiv n\cdot R(\hat{S}_1,\ldots,\hat{S}_m, w_1,\ldots,w_m)- n\cdot R(\tilde{S}_1,\ldots,\tilde{S}_m,\tilde{w}_1,\ldots,\tilde{w}_m)\cr
&=\sum_{i\in (\hat{S}_{j^*}\cap F_{\ell_1})} \bigl(\|x_i-w_{j^*}\|^2-\|x_i-u_{\ell_1}\|^2\bigr)-  \sum_{i\in \hat{S}_{j_2}}\bigl( \|x_i-w_{j_1}\|^2-\|x_i-w_{j_2}\|^2 \bigr)\cr
&\equiv \Delta_2-\Delta_1, 
\end{align*}
where $\Delta_1$ is the increase of RSS caused by (i) and $\Delta_2$ is the decrease of RSS caused by (ii). We can bound $\Delta_1$ in a similar way as in the previous sub-case, and the details are omitted. It gives
\[
\Delta_1 \leq 12C_0C_3\cdot nd_m^2. 
\]
Since $x_i=u_{\ell_1}$ for all $i\in F_{\ell_1}$, we immediately have 
\[
\Delta_2 = |\hat{S}_{j^*}\cap F_{\ell_1}|\cdot \|u_{\ell_1}-w_{j^*}\|^2\geq (m^{-1}\alpha_0n)\cdot (d_m/2)^2\geq \frac{m^{-1}\alpha_0}{4}\cdot nd_m^2. 
\]
As a result, $\Delta\geq (\frac{m^{-1}\alpha_0}{4}-12C_0C_3)m^{-1}\alpha_0\cdot nd_m^2$. If we plug in the expression of $C_3$, it becomes $\geq (\frac{2}{9}m^{-1}\alpha_0)\cdot nd_m^2$. This gives
\beq \label{thm-km-case2-result}
R(\hat{S}_1,\ldots,\hat{S}_m, w_1,\ldots,w_m)- R(\tilde{S}_1,\ldots,\tilde{S}_m,\tilde{w}_1,\ldots,\tilde{w}_m) \geq  \frac{2m^{-1}\alpha_0}{9}\cdot d_m^2. 
\eeq
This gives \eqref{thm-km-defgoal} in the second sub-case. 

In the second case, $\|u_k-w_{j_1}\|\geq C_3\cdot d_m$. We recall that $|F_k\cap \hat{S}_{j_1}|\geq m^{-1}\alpha_0n$. Let $E$ be a subset of $F_k\cap \hat{S}_{j_1}$ such that $|E|=\lceil |F_k\cap \hat{S}_{j_1}|/2\rceil$. Note that $|\hat{S}_{j_1}\backslash E|\leq n$. We have 
\[
\hat{S}_{j_1}\backslash E\neq \emptyset, \qquad\mbox{and}\qquad \frac{|E|}{|\hat{S}_{j_1}\backslash E|}\geq m^{-1}\alpha_0/2. 
\] 
We now modify the clusters and cluster centers $\{(\hat{S}_j, w_j)\}_{1\leq j\leq m}$ as follows:
\begin{itemize}
\item Move the subset $E$ from $\hat{S}_{j_1}$ to $\hat{S}_{j_2}$, and update each cluster center to be the within cluster average of $x_i$'s. 
\end{itemize}
The other clusters and cluster centers are unchanged. 
Namely, we let
\[
\tilde{S}_j = \begin{cases}
\hat{S}_{j_1}\backslash E,  &\mbox{if }j=j_1,\cr
\hat{S}_{j_2}\cup E, & \mbox{if }j=j_2,\cr
\hat{S}_j, &\mbox{if }j\notin\{j_1,j_2\},
\end{cases}\qquad
\tilde{w}_j = \begin{cases}
\frac{1}{|\tilde{S}_j|} \sum_{i\in \tilde{S}_j}x_i, & \mbox{if }j\in \{j_1,j_2\},\cr
w_j, &\mbox{otherwise}. 
\end{cases}
\]
We apply Lemma~\ref{lem:RSS} to $A=\hat{S}_{j_1}$, $B=\hat{S}_{j_2}$, and $C=E$, and note that $x_i=u_k$ for all $i\in E$. It follows that 
\begin{align} \label{thm-km-RSSchange}
\Delta &\equiv n\cdot R(\hat{S}_1,\ldots,\hat{S}_m, w_1,\ldots,w_m)- n\cdot R(\tilde{S}_1,\ldots,\tilde{S}_m,\tilde{w}_1,\ldots,\tilde{w}_m)\cr
&= -\left( \frac{|\hat{S}_{j_2}|\cdot |E|}{|\hat{S}_{j_2}|+ |E|}\|u_k-w_{j_2}\|^2 - \frac{|\hat{S}_{j_1}|\cdot |E|}{|\hat{S}_{j_1}|-|E|} \|u_k-w_{j_1}\|^2\right)\cr
&= \frac{|E|^2\cdot (|\hat{S}_{j_1}|+|\hat{S}_{j_2}|)}{(|\hat{S}_{j_2}|+ |E|)(|\hat{S}_{j_1}|- |E|)}\, \|u_k-w_{j_1}\|^2 + \frac{|\hat{S}_{j_2}|\cdot |E|}{|\hat{S}_{j_2}|+ |E|} \bigl( \|u_k-w_{j_1}\|^2-\|u_k-w_{j_2}\|^2 \bigr)\cr
&\geq \frac{|E|^2}{|\hat{S}_{j_1}|- |E|}\, \|u_k-w_{j_1}\|^2 + \frac{|\hat{S}_{j_2}|\cdot |E|}{|\hat{S}_{j_2}|+ |E|} \bigl( \|u_k-w_{j_1}\|^2-\|u_k-w_{j_2}\|^2 \bigr). 
\end{align}
By \eqref{thm-km-key1}, $\|u_k-w_{j_2}\|\leq \|u_k-w_{j_1}\|+4\delta$. It follows that, as long as $\delta<(C_3/16)\cdot d_m$, 
\begin{align*}
\|u_k-w_{j_1}\|^2-\|u_k-w_{j_2}\|^2 & \geq -8\delta\cdot \|u_k-w_{j_1}\|-16\delta^2\cr
&\geq -9\delta\cdot \|u_k-w_{j_1}\|, 
\end{align*}
where the last line is because $16\delta^2\leq C_3\delta\cdot d_m \leq\delta\cdot \|u_k-w_{j_1}\|$. We plug it into \eqref{thm-km-RSSchange} to get
\begin{align*}
\Delta &\geq \frac{|E|^2}{|\hat{S}_{j_1}\backslash E|}\, \|u_k-w_{j_1}\|^2 - \frac{|\hat{S}_{j_2}|\cdot |E|}{|\hat{S}_{j_2}|+ |E|}\cdot 9\delta\cdot \|u_k-w_{j_1}\|\cr
&\geq |E|\cdot (m^{-1}\alpha_0/2)\cdot \|u_k-w_{j_1}\|^2 - |E|\cdot9\delta \cdot \|u_k-w_{j_1}\|\cr
&\geq |E|\cdot \|u_k-w_{j_1}\|\cdot\Bigl(\frac{C_3 m^{-1}\alpha_0}{2} d_m - 9\delta\Bigr),
\end{align*}
where the second line is because $|E|\geq (m^{-1}\alpha_0/2)\cdot |\hat{S}_{j_1}\backslash E|$ and the last line is because we have assumed $\|u_k-w_{j_1}\|\geq C_3\cdot d_m$ in the current case. As long as $\delta <\frac{C_3 m^{-1}\alpha_0}{27}\cdot d_m$, the number in brackets is $\geq \frac{C_3m^{-1}\alpha_0}{6}d_m$. We also plug in $|E|=\lceil m^{-1}\alpha_0/2\rceil n$  and $\|u_k-w_{j_1}\|\geq C_3\cdot d_m$ to get 
\[
\Delta \geq \frac{m^{-1}\alpha_0}{2} n \cdot C_3d_m \cdot \frac{C_3 m^{-1}\alpha_0}{6} d_m
\geq \frac{C_3^2m^{-2}\alpha_0^2}{12}\cdot nd_m^2. 
\]
It follows that
\beq \label{thm-km-case3-result}
R(\hat{S}_1,\ldots,\hat{S}_m, w_1,\ldots,w_m)- R(\tilde{S}_1,\ldots,\tilde{S}_m,\tilde{w}_1,\ldots,\tilde{w}_m)\geq \frac{C_3^2m^{-2}\alpha_0^2}{12}\cdot d_m^2. 
\eeq
This gives \eqref{thm-km-defgoal} in the second case. 
We combine \eqref{thm-km-case1-result}, \eqref{thm-km-case2-result} and \eqref{thm-km-case3-result}, and take the minimum of the right hand sides of three inequalities. Since $m^{-1}\alpha_0<1$ and $C_3^2<1/3$, we choose 
\[
C_2 =(1/12) C_3^2m^{-2}\alpha_0^2. 
\] 
Then, \eqref{thm-km-defgoal} is satisfied for all cases. This completes the proof of \eqref{thm-km-(b)}. 

We remark that the scalar $c=c(\alpha_0, C_0, m)$ is not exactly $C_2/C_1$. In the derivation of \eqref{thm-km-(a)} and \eqref{thm-km-(b)}, we have imposed other restrictions on $\delta$, which can be expressed as $\delta\leq C_4\cdot d_m$, where $C_4$ is determined by $(C_0, \alpha, m)$ and $(C_1, C_2, C_3)$. Since $(C_1, C_2, C_3)$ only depend on $(\alpha_0, C_0, m)$, $C_4$ is a function of $(\alpha_0, C_0, m)$ only. We take $c=\min\{C_2/C_1, \; C_4\}$. \qed

\subsection{Proof of Lemma~\ref{lemma:eigenvector-relation}} \label{subsec:proof-Omega-eig}
By definition of $H$, we have $\Pi^2\Theta\Pi=\|\theta\|^2\cdot H^2$. As a result, the matrix $U=\|\theta\|^{-1}\Theta\Pi H^{-1}$ satisfies that $U'U=I_K$. We now write
\[
\Omega = \Theta\Pi P \Pi'\Theta = \|\theta\|^2\cdot U\cdot (HPH)\cdot U', \qquad\mbox{where}\quad U'U=I_K. 
\] 
Since $U$ contains orthonormal columns, the nonzero eigenvalues of $\Omega$ are the nonzero eigenvalues of $\|\theta\|^2(HPH)$. This proves that $\lambda_k=\|\theta\|^2\mu_k$. Furthermore, there is a one-to-one correspondence between the eigenvectors of $\Omega$ and the eigenvectors of $HPH$ through
\[
[\xi_1,\xi_2,\ldots,\xi_k] = U[\eta_1,\eta_2,\ldots,\eta_K]. 
\]
It follows that $\xi_k=U\eta_k=\|\theta\|^{-1}\Theta\Pi H^{-1}\eta_k$. This proves the claim about $\xi_k$. We can multiply both sides of the equation $\xi_k=U\eta_k$ by $\|\theta\|^{-1}H^{-1}\Pi'\Theta$ from the left. It yields that 
\begin{align*}
\|\theta\|^{-1}H^{-1}\Pi'\Theta\xi_k&= (\|\theta\|^{-1}H^{-1}\Pi'\Theta)(\|\theta\|^{-1}\Theta\Pi H^{-1}\eta_k)\cr
&=\|\theta\|^{-2}H^{-1}(\Pi'\Theta^2\Pi) H^{-1}\eta_k\; =\; \eta_k.
\end{align*}
This proves the claim about $\eta_k$. Last, the condition \eqref{condition1d} ensures that the multiplicity of $\mu_1$ is 1 and that $\mu_1$ is a strictly positive vector. It follows that $\lambda_1$ has a multiplicity of 1. Note that $\xi_k=U\eta_k$ implies 
\[
\xi_1(i) = \|\theta\|^{-1}\theta_i\sum_{k=1}^K H_{kk}^{-1}\pi_i(k)\eta_1(k)\geq \|\theta\|^{-1}\theta_i\min_{1\leq k\leq K}\bigl\{H_{kk}^{-1}\eta_1(k)\bigr\}. 
\]
Since $\eta_1$ is a positive vector and $H$ is a positive diagonal matrix, we conclude that all entries of $\xi_1$ are positive. 
\qed

\subsection{Proof of Lemma~\ref{lem:RSS}} \label{subsec:proof-RSS}
Note that for any set $M\subset\{1,2,\ldots,n\}$ and $z\in\mathbb{R}^d$,
\begin{align*}
\sum_{i\in M}\|y_i-z\|^2 &= \sum_{i\in M}\|(y_i-\bar{y}_M)+(\bar{y}_M-z)\|^2\cr
&=\sum_{i\in M}\|y_i-\bar{y}_M\|^2 + 2 (\bar{y}_M-z)'\sum_{i\in M}(y_i-\bar{y}_M) + |M|\|\bar{y}_M-z\|^2\cr
&=\sum_{i\in M}\|y_i-\bar{y}_M\|^2 +|M|\|\bar{y}_M-z\|^2. 
\end{align*}
The clusters associated with $RSS$ are $A=\tilde{A}\cup C$ and $B$, and the clusters associated with $\widetilde{RSS}$ are $\tilde{A}$ and $\tilde{B}=C\cup B$. 
By direct calculations, 
\begin{align*}
RSS &= \sum_{i\in \tilde{A}}\|y_i-\bar{y}_A\|^2 +  \sum_{i\in C}\|y_i-\bar{y}_A\|^2+ \sum_{i\in B}\|y_i-\bar{y}_B\|^2\cr
&= \biggl(\sum_{i\in \tilde{A}}\|y_i-\bar{y}_{\tilde{A}}\|^2 + |\tilde{A}|\|\bar{y}_{\tilde{A}}-\bar{y}_A\|^2\biggr) +  \biggl(\sum_{i\in C}(y_i-\bar{y}_C)^2+|C|\|\bar{y}_C-\bar{y}_A\|^2\biggr)+ \sum_{i\in B}\|y_i-\bar{y}_B\|^2, \cr
\widetilde{RSS} &= \sum_{i\in \tilde{A}}\|y_i-\bar{y}_{\tilde{A}}\|^2 +  \sum_{i\in C}\|y_i-\bar{y}_{\tilde{B}}\|^2+ \sum_{i\in B}\|y_i-\bar{y}_{\tilde{B}}\|^2\cr
&= \sum_{i\in \tilde{A}}\|y_i-\bar{y}_{\tilde{A}}\|^2  + \biggl( \sum_{i\in C}\|y_i-\bar{y}_C\|^2+ |C|\|\bar{y}_C-\bar{y}_{\tilde{B}}\|^2\biggr) + \biggl(\sum_{i\in B}\|y_i-\bar{y}_B\|^2 + |B|\|\bar{y}_B-\bar{y}_{\tilde{B}}\|\biggr). 
\end{align*}
It follows that
\beq \label{prop-rss-1}
\widetilde{RSS}-RSS = \bigl(|B|\|\bar{y}_B-\bar{y}_{\tilde{B}}\|^2+|C|\|\bar{y}_C-\bar{y}_{\tilde{B}}\|^2\bigr)-\bigl(|\tilde{A}|\|\bar{y}_{\tilde{A}}-\bar{y}_A\|^2+|C|\|\bar{y}_C-\bar{y}_A\|^2\bigr). 
\eeq
By definition,
\[
\bar{y}_{A}=  \frac{|A|-|C|}{|A|}\bar{y}_{\tilde{A}}+\frac{|C|}{|A|}\bar{y}_{C}, \qquad \bar{y}_{\tilde{B}}=  \frac{|B|}{|B|+|C|}\bar{y}_B+\frac{|C|}{|B|+|C|}\bar{y}_{C}. 
\]
Re-arranging the terms, we have
\beq \label{prop-rss-2}
\bar{y}_{\tilde{A}}-\bar{y}_A= \frac{|C|}{|A|-|C|}(\bar{y}_A- \bar{y}_C), \quad \bar{y}_{\tilde{B}}-\bar{y}_B=\frac{|C|}{|B|+|C|}(\bar{y}_{C}-\bar{y}_B), \quad  \bar{y}_C-\bar{y}_{\tilde{B}}=\frac{|B|}{|B|+|C|}(\bar{y}_{C}-\bar{y}_B).
\eeq
We plug \eqref{prop-rss-2} into \eqref{prop-rss-1} to get
\begin{align*}
\widetilde{RSS}-RSS& =\left( |B|\cdot\frac{|C|^2}{(|B|+|C|)^2}+|C|\cdot\frac{|B|^2}{(|B|+|C|)^2}\right)  \|\bar{y}_{C}-\bar{y}_B\|^2\cr
&\qquad -\left( |\tilde{A}|\cdot\frac{|C|^2}{(|A|-|C|)^2}+ |C|\right)  \|\bar{y}_{C}-\bar{y}_A\|^2\cr
&=\frac{|B||C|}{|B|+|C|} \|\bar{y}_{C}-\bar{y}_B\|^2 - \frac{|A||C|}{|A|-|C|}\|\bar{y}_{C}-\bar{y}_A\|^2. 
\end{align*}
This proves the claim. \qed


\section{Proof of the lower bounds}\label{sec:proof-A}

\subsection{Proof of Theorem~\ref{thm:LB}}
First, we show the claims on $|\lambda_K|/\sqrt{\lambda_1}$. Define a diagonal matrix $H$ by $H_{kk}=\|\theta\|^{-1}\sqrt{\sum_{i: \ell_i=k}\theta_i^2}$, for $1\leq k\leq K$. Note that $H$ is also stochastic. 
By Lemma~\ref{lemma:eigenvector-relation}, 
the eigenvalues of $\Omega$ are equal to the eigenvalues of $\|\theta\|^2HPH$, i.e., 
\[
\lambda_k = \|\theta\|^2\cdot \lambda_k(HPH), \qquad 1\leq k\leq K. 
\]
It follows that
\beq \label{thm-LB0-1}
|\lambda_K|/\sqrt{\lambda_1} = \|\theta\|\cdot |\lambda_K(HPH)|/\sqrt{\lambda_1(HPH)}. 
\eeq
Below, we first study the matrix $H$ and then show the claims.

Consider the matrix $H$. Let $\widetilde{N}_1, \widetilde{N}_2,\ldots,\widetilde{N}_{K_0}$ be the (non-stochastic) communities of the DCBM with $K_0$ communities. For each $1\leq k\leq K_0$, let $\theta^{(k)}\in\mathbb{R}^n$ be such that $\theta_i^{(k)}=\theta_i\cdot 1\{i\in \widetilde{N}_k\}$. By definition,
\[
H^2_{kk} = 
\|\theta\|^{-2}\begin{cases}
\|\theta^{(k)}\|^2, & \mbox{for }1\leq k\leq K_0-1,\\
\sum_{i\in \widetilde{N}_{K_0}} \theta_i^2\cdot 1\{\ell_i=k\}, & \mbox{for }K_0\leq k\leq K_0+m. 
\end{cases}
\]
Since \eqref{condition1b} is satisfied, $\|\theta\|^2\geq \|\theta^{(k)}\|^2\geq C\|\theta\|^2$, for $1\leq k\leq K_0$. It implies that
\beq \label{thm-LB0-2}
C^{-1}\leq H_{kk}\leq C, \qquad \mbox{for }1\leq k\leq K_0-1.
\eeq
Fix $k\geq K_0$. The $n$ indicators $1\{\ell_i=k\}$ are $iid$ Bernoulli variables with a success probability of $\frac{1}{m+1}$. Therefore, $\mathbb{E}H^2_{kk}=\frac{1}{m+1}\|\theta\|^{-2}\|\theta^{(K_0)}\|^2$. 
Furthermore, by Hoeffding's inequality, 
\[
\mathbb{P}\Bigl( \bigl| \|\theta\|^2 (H^2_{kk}-\mathbb{E}H^2_{kk})\bigr|>t \Bigr)\leq 2\exp\Bigl(-\frac{t^2}{2\sum_{i\in \widetilde{N}_{K_0}}\theta_i^4}\Bigr). 
\]
By \eqref{condition1a}, $\theta_{\max}\sqrt{\log(n)}\to 0$. Hence, $\sum_{i\in \widetilde{N}_{K_0}}\theta_i^4\leq \theta_{\max}^2\|\theta^{(K_0)}\|^2\ll \|\theta\|^2/\log(n)$. Taking $t=\|\theta\|$ in the above equation yields $
\bigl| H^2_{kk}-\mathbb{E}H^2_{kk}\bigr|\leq \|\theta\|^{-1}$ with probability $1-o(n^{-1})$. We have seen that $\mathbb{E}H^2_{kk}=\frac{1}{m+1}\|\theta\|^{-2}\|\theta^{(K_0)}\|^2$, which is bounded above and below by constants. Additionally, $\|\theta\|^{-1}=o(1)$. Combining these results gives 
\beq \label{thm-LB0-3}
C^{-1}\leq H_{kk}\leq C, \qquad \mbox{with probability $1-o(n^{-1})$, for any $k\geq K_0$}. 
\eeq
It follows from \eqref{thm-LB0-2} and \eqref{thm-LB0-3} that  
\beq \label{thm-LB0-4}
\|H\|\leq C, \qquad \|H^{-1}\|\leq C, \qquad \mbox{with probability }1-o(n^{-1}). 
\eeq 

Consider the the upper bound for $|\lambda_K|/\sqrt{\lambda_1}$. It suffices to get an upper bound for $|\lambda_K(HPH)|$ and a lower bound for $\lambda_1(HPH)$. 
Note that $|\lambda_K(HPH)|$ is the smallest singular value of $HPH$, which can be different from the absolute value of the smallest eigenvalue. Therefore, we cannot use Cauchy's interlacing theorem \citep{HornJohnson} to relate $|\lambda_K(HPH)|$ to the smallest eigenvalue of $M$. We need a slightly longer proof.  
Write 
\[
P =
\begin{bmatrix}
S & \beta{\bf 1}_{m+1}' \\
{\bf 1}_{m+1} \beta' & {\bf 1}_{m+1}{\bf 1}_{m+1}'
\end{bmatrix}+ 
\begin{bmatrix}
{\bf 0}_{(K_0-1)\times (K_0-1)} & {\bf 0}_{(K_0-1)\times 1}\\
{\bf 0}_{1\times (K_0-1)} & \frac{m+1}{1+mb_n}M-{\bf 1}_{m+1}{\bf 1}_{m+1}' 
\end{bmatrix}\equiv P^*+\Delta.
\]
The matrix $P^*$ can be re-expressed as ($e_{K_0}$ is the $K_0$th standard basis of $\mathbb{R}^{K_0}$) 
\[
P^* =\begin{bmatrix}
I_{K_0} \\ {\bf 1}_{m} e'_{K_0}
\end{bmatrix}
  \begin{bmatrix} S & \beta\\\beta' & 1\end{bmatrix}
  \begin{bmatrix}
I_{K_0} &  e_{K_0}{\bf 1}'_{m}
\end{bmatrix}.
\]
Therefore, the rank of $P^*$ is only $K_0$. Then, $HP^*H$ is also a rank-$K_0$ matrix. Consequently, for $K=K_0+m$, 
\[
\lambda_K(HP^*H) = 0. 
\]
By Weyl's inequality \citep{HornJohnson}, $|\lambda_K(HPH)-\lambda_K(HP^*H)|\leq \|H\Delta H\|$. Combining these results gives
\beq \label{thm-LB0-5}
|\lambda_K(HPH)|\leq \|H\Delta H\|. 
\eeq
Note that $\|\Delta\|=\|\frac{m+1}{1+mb_n}M-{\bf 1}_{m+1}{\bf 1}_{m+1}'\|$. $M$ is a matrix whose diagonals are $1$ and off-diagonals are equal to $b_n$. As a result, $\Delta$ is a matrix whose diagonals are equal to $\frac{m(1-b_n)}{1+mb_n}$ and off-diagonals are equal to $\frac{-(1-b_n)}{1+mb_n}$. It follows immediately that
\[
\|\Delta\|\leq C(1-b_n). 
\]
We plug it into \eqref{thm-LB0-5} and apply \eqref{thm-LB0-4}. It yields that
\beq \label{thm-LB0-6}
|\lambda_K(HPH)|\leq C(1-b_n). 
\eeq
Furthermore, $\lambda_1(P)\geq P_{11}=1$ and $\lambda_1(P)\leq \|H^{-1}\|^2\lambda_1(HPH)$. Combining it with \eqref{thm-LB0-4} gives
\beq \label{thm-LB0-7}
\lambda_1(HPH)\geq C^{-1}. 
\eeq 
Note that \eqref{thm-LB0-6}-\eqref{thm-LB0-7} hold with probability $1-o(n^{-1})$, because their derivation utilizes \eqref{thm-LB0-4}. 
We plug \eqref{thm-LB0-6}-\eqref{thm-LB0-7} into \eqref{thm-LB0-1} to get $
|\lambda_K|/\sqrt{\lambda_1}\leq C \|\theta\|(1-b_n)$, with probability $1-o(n^{-1})$.  
This proves the upper bound of $|\lambda_K|/\sqrt{\lambda_1}$.

Consider the the lower bound for $|\lambda_K|/\sqrt{\lambda_1}$. Using \eqref{thm-LB0-4}, we have 
\beq \label{thm-LB0-add0}
|\lambda_K(HPH)|^{-1}=\|(HPH)^{-1}\|\leq \|H^{-1}\|^2\cdot \|P^{-1}\|\leq C\|P^{-1}\|. 
\eeq
We then bound $\|P^{-1}\|$. Write 
\[
P = A + B, \qquad\mbox{where}\quad A = \begin{bmatrix} S\\ & \frac{m+1}{1+mb_n} M\end{bmatrix} \quad \mbox{and}\quad B = \begin{bmatrix} {\bf 0} & \beta {\bf 1}_{m+1}'\\
{\bf 1}_{m+1}\beta' & {\bf 0} \end{bmatrix}. 
\]
The matrix $B$ is a rank-2 matrix, which can be re-expressed as 
\[   
B = XD^{-1}X', \qquad \mbox{where}\quad  X = \begin{bmatrix}
\beta & \beta\\
{\bf 1}_{m+1}  & -{\bf 1}_{m+1}
\end{bmatrix}\quad \mbox{and}\quad D = \begin{bmatrix}2\\ & -2 \end{bmatrix}. 
\]
We use the matrix inversion formula to get
\begin{align}  \label{thm-LB0-add1}
\|P^{-1}\| & =\| (A + XD^{-1}X')^{-1}\|\cr
& =\| A^{-1} - A^{-1}X(D + X'A^{-1}X)^{-1}X'A^{-1}\|\cr
&\leq \| A^{-1}\|\cdot \bigl( 1 + \|X (D+X'A^{-1}X)^{-1}X'A^{-1}\|\bigr)\cr
&= \| A^{-1}\|\cdot \bigl( 1 + \|(D+X'A^{-1}X)^{-1}(X'A^{-1}X)\|\bigr).
\end{align}
By direct calculations, writing $M_0=\frac{1+mb_n}{m+1}M$ and ${\bf 1}={\bf 1}_{m+1}$ for short, we have
\[
X'A^{-1}X = \begin{bmatrix}
\beta'S^{-1}\beta + {\bf 1}'M_0^{-1} {\bf 1} & \beta'S^{-1}\beta - {\bf 1}'M_0^{-1}{\bf 1}\\
\beta'S^{-1}\beta - {\bf 1}'M_0^{-1} {\bf 1} & \beta'S^{-1}\beta + {\bf 1}'M_0^{-1} {\bf 1} \end{bmatrix}. 
\]
Note that $M {\bf 1} = (1+mb_n) {\bf 1}$. It implies that $M^{-1}{\bf 1}=\frac{1}{1+mb_n}{\bf 1}$. As a result,  
\[
{\bf 1}'M_0^{-1}{\bf 1} = \frac{1+mb_n}{m+1}{\bf 1}'M_0^{-1}{\bf 1} = \frac{1+mb_n}{m+1}{\bf 1}'\Bigl( \frac{1}{1+mb_n}{\bf 1}\Bigr)= 1. 
\]
Plugging it into the expression of $X'A^{-1}X$ gives
\[
X'A^{-1}X = \begin{bmatrix}
\beta'S^{-1}\beta + 1 & \beta'S^{-1}\beta - 1\\
\beta'S^{-1}\beta - 1  & \beta'S^{-1}\beta + 1
\end{bmatrix}. 
\]
It follows from direct calculations that 
\beq \label{thm-LB0-add2}
(D+X'A^{-1}X)^{-1} (X'A^{-1}X) 
=
\frac{1}{2}\begin{bmatrix}
1 & -1\\
1 & \frac{3\beta'S^{-1}\beta+1}{\beta'S^{-1}\beta-1}
\end{bmatrix}. 
\eeq
Under the condition $|\beta'S^{-1}\beta-1|\geq C$, the absolute value of $\frac{3\beta'S^{-1}\beta+1}{\beta'S^{-1}\beta-1}$ is bounded by a constant. Therefore, the spectral norm of the matrix in \eqref{thm-LB0-add2} is bounded by a constant. We plug it into \eqref{thm-LB0-add1} to get 
\[
\|P^{-1}\|\leq C\|A^{-1}\|\leq C\max\bigl\{ |\lambda_{\min}(S)|^{-1},\; |\lambda_{\min}(M)|^{-1} \bigr\}.
\]
The minimum eigenvalue of $M$ is $(1-b_n)$. Hence, under the condition of $|\lambda_{\min}(S)|\gg 1-b_n$, we immediately have $\|P^{-1}\|\leq C(1-b_n)^{-1}$. We plug it into \eqref{thm-LB0-add0} to get
\beq \label{thm-LB0-add3}
|\lambda_K(HPH)|\geq C^{-1}(1-b_n). 
\eeq
Additionally, $\|\tilde{P}\|\leq C$ by \eqref{condition1a}. It follows from the connection between $P$ and $\tilde{P}$ in \eqref{LB-construct1} that $\|P\|\leq C$. Combining it with \eqref{thm-LB0-4} gives $\|HPH\|\leq C$, i.e., 
\beq \label{thm-LB0-add4}
\lambda_1(HPH)\leq C. 
\eeq
Here \eqref{thm-LB0-add3} and \eqref{thm-LB0-add4} are satisfied with probability $1-o(1)$, because their derivation uses \eqref{thm-LB0-4}. 
We plug \eqref{thm-LB0-add3}-\eqref{thm-LB0-add4} into \eqref{thm-LB0-1}. It yields that $
|\lambda_K|/\sqrt{\lambda_1}\geq C^{-1} \|\theta\|(1-b_n)$, with probability $1-o(1)$. 
This proves the lower bound of $|\lambda_K|/\sqrt{\lambda_1}$.

Next, we show that, if $\|\theta\|(1-b_n)\to 0$, the two random-label DCBM models associated with $m_1$ and $m_2$ are asymptotically indistinguishable. It is sufficient to show that each random-label DCBM is asymptotically indistinguishable from the (fixed-label) DCBM with $K_0$ communities.  

Fix $m\geq 1$. Let $f_0(A)$ and $f_1(A)$ be the respective likelihood of the (fixed-label) DCBM and the random-label DCBM. Write $\widetilde{\Omega}=\Theta\widetilde{\Pi}\widetilde{P}\widetilde{\Pi}'\Theta$ and $\Omega=\Theta\Pi P\Pi'\Theta$. It is seen that 
\[
f_0(A) = \prod_{1\leq i<j\leq n} \widetilde{\Omega}_{ij}^{A_{ij}}(1-\widetilde{\Omega}_{ij})^{1-A_{ij}}, \qquad 
f_1(A) = \int \prod_{1\leq i<j\leq n} \Omega_{ij}^{A_{ij}}(1-\Omega_{ij})^{1-A_{ij}}d\mathbb{P}(\Pi). 
\]
Recall that $\widetilde{\cal N}_1,\widetilde{\cal N}_2,\ldots,\widetilde{\cal N}_{K_0}$ are the (non-stochastic) communities in the first DCBM. 
We observe that $\widetilde{\Omega}_{ij}\neq  \Omega_{ij}$ only when both $i$ and $j$ are in $\widetilde{\cal N}_{K_0}$. Therefore, the likelihood ratio is
\beq \label{thm-LB-likelihood}
L(A)\equiv \frac{f_1(A)}{f_0(A)} = \int \prod_{\{i,j\}\subset\widetilde{N}_{K_0}, i<j} \Bigl(\frac{\Omega_{ij}}{\widetilde{\Omega}_{ij}}\Bigr)^{A_{ij}}\Bigl(\frac{1-\Omega_{ij}}{1-\widetilde{\Omega}_{ij}}\Bigr)^{1-A_{ij}}d\mathbb{P}(\Pi).
\eeq
When $i,j$ are both in $\widetilde{N}_{K_0}$, it is seen that 
\[
\widetilde{\Omega}_{ij}=\theta_i\theta_j, \qquad \Omega_{ij} =\theta_i\theta_j\cdot \pi_i'\Bigl(\frac{(m+1)}{1+mb_n} M\Bigr)\pi_j, 
\]
where $\pi_i=e_k$ if and only if $\ell_i=K_0-1+k$, $1\leq k\leq m+1$, and $e_1,e_2,\ldots,e_{m+1}$ are the standard bases of $\mathbb{R}^{m+1}$. Here we have mis-used the notation $\pi_i$; previously, we use $\pi_i'$ to denote the $i$-th row of $\Pi$, but currently, the $i$-th row of $\Pi$ is $({\bf 0}_{K_0-1}', \pi_i')'$. Define 
\[
z_i = \pi_i - \frac{1}{m+1}{\bf 1}_{m+1}, \qquad \mbox{for all }i\in \widetilde{\cal N}_{K_0}. 
\]
The random vectors $\{z_i\}_{i\in \widetilde{\cal N}_{K_0}}$ are independently and identically distributed,  satisfying $\mathbb{E}z_i={\bf 0}$ and $\|z_i\|\leq 1$. In the paragraph below \eqref{thm-LB0-5}, we have seen that 
\[
\frac{m+1}{1+mb_n}M= {\bf 1}_{m+1}{\bf 1}_{m+1}' + \frac{1-b_n}{1+mb_n}
 \begin{bmatrix}
 m & -1 & \cdots & -1\\
 -1 & m &\ddots & \vdots \\
 \vdots &\ddots & \ddots &-1\\
 -1 & \cdots &-1 & m
 \end{bmatrix}\equiv {\bf 1}_{m+1}{\bf 1}_{m+1}' + G. 
\]
The matrix $G$ satisfies that $G{\bf 1}_{m+1}={\bf 0}$ and $\|G\|\leq C(1-b_n)$. It follows that 
\begin{align}  \label{thm-LB-Omega(ij)}
\Omega_{ij} &= \theta_i\theta_j \cdot \pi_i'\left( {\bf 1}_{m+1}{\bf 1}_{m+1}'+G\right)\pi_j\cr
& =\theta_i\theta_j + \theta_i\theta_j (\pi_i'G\pi_j)\cr
&= \theta_i\theta_j + \theta_i\theta_j \Bigl(\frac{1}{m+1}{\bf 1}_{m+1}+z_i\Bigr)'G \Bigl(\frac{1}{m+1}{\bf 1}_{m+1}+z_j\Bigr)\cr
&=\theta_i\theta_j (1+z_i'Gz_j). 
\end{align}
We plug it into \eqref{thm-LB-likelihood} to get 
\beq \label{thm-LB-1}
L(A)\equiv \frac{f_2(A)}{f_1(A)} =\mathbb{E}_z\left\{ \prod_{\substack{i,j\in \widetilde{\cal N}_{K-1}\\i< j}} ( 1+z_i'Gz_j)^{A_{ij}}\left[ \frac{1-\theta_i\theta_j(1+z_i'Gz_j)}{1-\theta_i\theta_j}\right]^{1-A_{ij}}\right\}. 
\eeq
The $\chi^2$-distance between two models is $\mathbb{E}_{A\sim f_0}[(L(A)-1)^2]$.  
To show that the two models are asymptotically indistinguishable, it suffices to show that the $\chi^2$-distance is $o(1)$ \cite{tsybakov2008introduction}. Using the property that $\mathbb{E}_{A\sim f_0}[(L(A)-1)^2] = \mathbb{E}_{A\sim f_0}[L^2(A)]-1$, we only need to show 
\beq \label{thm-LB-2}
\mathbb{E}_{A\sim f_0}[L^2(A)] \leq 1+o(1). 
\eeq

We now show \eqref{thm-LB-2}. Write $L(A)=\mathbb{E}_z[g(A,z)]$, where $g(A, z)$ is the term inside the expectation in \eqref{thm-LB-1}. Let $\{\tilde{z}_i\}_{i\in \widetilde{\cal N}_{K_0}}$ be an independent copy of $\{z_i\}_{i\in \widetilde{\cal N}_{K_0}}$. Then, 
\beq \label{thm-LB-3}
\mathbb{E}_{A\sim f_0}[L^2(A)] = \mathbb{E}_{A\sim f_0}\Bigl\{   \mathbb{E}_z[g(A,z)]  \cdot \mathbb{E}_{\tilde{z}}[g(A,\tilde{z})]   \Bigr\} = \mathbb{E}_{z,\tilde{z}}\Bigl\{ \mathbb{E}_{A\sim f_0} [g(A,z)g(A,\tilde{z})] \Bigr\}.
\eeq
Using the expression of $g(A,z)$ in \eqref{thm-LB-1}, we have
\begin{align*}
g(A,z)g(A,\tilde{z}) &= \prod_{\substack{i,j\in \widetilde{\cal N}_{K_0}\\i< j}} \left[(1+z_i'Gz_j)(1+\tilde{z}_i'G\tilde{z}_j) \right]^{A_{ij}} \left\{ \frac{[1-\theta_i\theta_j(1+z_i'Gz_j)][1-\theta_i\theta_j(1+\tilde{z}_i'G\tilde{z}_j)]}{(1+\theta_i\theta_j)^2}\right\}^{1-A_{ij}}. 
\end{align*}
Here $A_{ij}$'s are independent Bernoulli variables, where $\mathbb{P}(A_{ij}=1)=\theta_i\theta_j$.   If we take expectation with respect to $A_{ij}$ in each term of the product, it gives
\begin{align*}
&(1+z_i'Gz_j)(1+\tilde{z}_i'G\tilde{z}_j) \cdot \mathbb{P}(A_{ij}=1) +  \frac{[1-\theta_i\theta_j(1+z_i'Gz_j)][1-\theta_i\theta_j(1+\tilde{z}_i'G\tilde{z}_j)]}{(1-\theta_i\theta_j)^2}\cdot \mathbb{P}(A_{ij}=0)\cr
=\;\; &  \theta_i\theta_j (1+z_i'Gz_j)(1+\tilde{z}_i'G\tilde{z}_j)  + \frac{[1-\theta_i\theta_j(1+z_i'Gz_j)][1-\theta_i\theta_j(1+\tilde{z}_i'G\tilde{z}_j)]}{1-\theta_i\theta_j} \cr
=\;\; &(1+z_i'Gz_j)(1+\tilde{z}_i'G\tilde{z}_j)\Bigl( \theta_i\theta_j+\frac{\theta^2_i\theta^2_j}{1-\theta_i\theta_j}\Bigr) + \frac{1-\theta_i\theta_j(1+z_i'Gz_j)-\theta_i\theta_j(1+\tilde{z}_i'G\tilde{z}_j)}{1-\theta_i\theta_j}\cr
=\;\; & (1+z_i'Gz_j)(1+\tilde{z}_i'G\tilde{z}_j)\frac{\theta_i\theta_j}{1-\theta_i\theta_j} + 1 - \frac{\theta_i\theta_j(1+z_i'Gz_j+\tilde{z}_i'G\tilde{z}_j) }{1-\theta_i\theta_j}\cr
=\;\; & 1+ \frac{\theta_i\theta_j}{1-\theta_i\theta_j} (z_i'Gz_j)(\tilde{z}_i'G \tilde{z}_j). 
\end{align*}
As a result,
\begin{align*}
\mathbb{E}_{A\sim f_0}[g(A,z)g(A,\tilde{z})] & = \prod_{\{i,j\}\subset \widetilde{\cal N}_{K_0}, i< j}\left[ 1+ \frac{\theta_i\theta_j}{1-\theta_i\theta_j} (z_i'Gz_j)(\tilde{z}_i'G \tilde{z}_j)  \right]\cr
&\leq \exp\left(  \sum_{\{i,j\}\subset \widetilde{\cal N}_{K_0}, i< j} \frac{\theta_i\theta_j}{1-\theta_i\theta_j} (z_i'Gz_j)(\tilde{z}_i'G \tilde{z}_j) \right), 
\end{align*}
where the second line is from the inequality that $1+x\leq e^{x}$ for all $x\in\mathbb{R}$. 
We plug it into \eqref{thm-LB-3}. Then, to show \eqref{thm-LB-2}, it suffices to show that 
\beq \label{thm-LB-4}
\mathbb{E}_{z,\tilde{z}}[\exp(Y)]\leq 1+o(1), \qquad \mbox{where}\quad Y\equiv  \sum_{\{i,j\}\subset \widetilde{\cal N}_{K_0}, i< j} \frac{\theta_i\theta_j}{1-\theta_i\theta_j} (z_i'Gz_j)(\tilde{z}_i'G \tilde{z}_j). 
\eeq

We now show \eqref{thm-LB-4}. We drop the subscript $\{i,j\}\subset \widetilde{\cal N}_{K_0}$ in most places to make notations simpler. The matrix $G$ can be re-written as 
\[
G = \frac{1-b_n}{1+mb_n}\Bigl[(m+1)I_{m+1}-{\bf 1}_{m+1}{\bf 1}_{m+1}'\Bigr]. 
\]
Additionally, $z_i'{\bf 1}_{m+1}\equiv 0$. It follows that $z_i'Gz_j=\frac{(m+1)(1-b_n)}{1+mb_n}(z_i'z_j)$. As a result, 
\begin{align*}
Y &= \frac{(m+1)^2(1-b_n)^2}{(1+mb_n)^2} \; \sum_{i<j}\frac{\theta_i\theta_j}{1-\theta_i\theta_j}(z_i'z_j)(\tilde{z}_i'\tilde{z}_j')\cr
&=\frac{1}{(m+1)^2}\sum_{1\leq k,\ell\leq m+1} \underbrace{\frac{(m+1)^4(1-b_n)^2}{(1+mb_n)^2} \sum_{i<j}\frac{\theta_i\theta_j}{1-\theta_i\theta_j}z_i(k)z_j(k)\tilde{z}_i(\ell)\tilde{z}_j(\ell)}_{\equiv Y_{k\ell}}. 
\end{align*}
By Jensen's inequality, $\exp(Y)=\exp\bigl(\frac{1}{(m+1)^2}\sum_{k,\ell}Y_{k\ell}\bigr)\leq \frac{1}{(m+1)^2}\sum_{k,\ell}\exp(Y_{k\ell})$. It follows that 
\[
\mathbb{E}_{z,\tilde{z}}[\exp(Y)]\leq \frac{1}{(m+1)^2}\sum_{1\leq k,\ell\leq m+1} \mathbb{E}_{z,\tilde{z}}\bigl[ \exp(Y_{k\ell})\bigr]\leq \max_{1\leq k,\ell\leq m+1}\mathbb{E}_{z,\tilde{z}}\bigl[\exp(Y_{k\ell})\bigr]. 
\]
Therefore, to show \eqref{thm-LB-4}, it suffices to show that, for each $1\leq k,\ell\leq m+1$, 
\beq \label{thm-LB-4(add)}
\mathbb{E}_{z,\tilde{z}}\bigl[\exp(Y_{k\ell})\bigr]\leq 1+o(1). 
\eeq

Fix $(k,\ell)$. We now show \eqref{thm-LB-4(add)}. Define $\sigma_i=z_i(k)\tilde{z}(\ell)$, for all $i\in \widetilde{N}_{K_0}$. Then,
\begin{align*}
Y_{k\ell} &= \frac{(m+1)^4(1-b_n)^2}{(1+mb_n)^2} \sum_{i<j}\frac{\theta_i\theta_j}{1-\theta_i\theta_j}\sigma_i\sigma_j \cr
&=\frac{(m+1)^4(1-b_n)^2}{(1+mb_n)^2} \sum_{i<j}\sum_{s=1}^\infty \theta_i^s\theta_j^s\sigma_i\sigma_j\cr
&=\sum_{s=1}^\infty \underbrace{(1-\theta_{\max}^2) \theta_{\max}^{2s-2}}_{\equiv w_s} \underbrace{ \frac{(m+1)^4(1-b_n)^2}{(1+mb_n)^2(1-\theta_{\max}^2)\theta_{\max}^{2s-2}} \sum_{i<j} \theta_i^s\theta_j^s\sigma_i\sigma_j}_{\equiv X_s}. 
\end{align*}
In the second line above, we used the Taylor expansion $\frac{\theta_i\theta_j}{1-\theta_i\theta_j}=\sum_{s=1}^\infty\theta_i^s\theta_j^s$. It is valid because $|\theta_i\theta_j|\leq \theta_{\max}^2=o(1)$. In the third line, we have switched the order of summation. It is valid because the double sum is finite if we take the absolute value of each summand. 
The numbers $\{w_s\}_{s=1}^\infty$ satisfy that $\sum_{s=1}^\infty w_s=1$. 
By Jenson's inequality,
\[
\exp(Y_{k\ell}) = \exp\Bigl( \sum_{s=1}^\infty w_s\cdot X_s \Bigr) \leq  \sum_{s=1}^\infty w_s \cdot \exp(X_s). 
\]
By Fatou's lemma, 
\beq \label{thm-LB-5}
\mathbb{E}_{\sigma}[\exp(Y_{k\ell})] \leq \sum_{s=1}^\infty w_s \cdot \mathbb{E}_{\sigma}[\exp(X_s)]\leq \max_{s\geq 1}\; \mathbb{E}_{\sigma}[\exp(X_s)]
\eeq
It remains to study $X_s$. Note that 
\begin{align*}
X_s &=\frac{(m+1)^4(1-b_n)^2}{(1+mb_n)^2(1-\theta_{\max}^2)\theta_{\max}^{2s-2}} \sum_{i<j} \theta_i^s\theta_j^s\sigma_i\sigma_j \cr
&= \frac{(m+1)^4(1-b_n)^2}{(1+mb_n)^2(1-\theta_{\max}^2)\theta_{\max}^{2s-2}} \left[\frac{1}{2}\sum_{i,j} \theta_i^s\theta_j^s\sigma_i\sigma_j -\sum_i\theta_i^{2s}\sigma_i^2\right] \cr
&\leq \frac{(m+1)^4(1-b_n)^2}{2(1+mb_n)^2(1-\theta_{\max}^2)\theta_{\max}^{2s-2}} \left(\sum_i \theta_i^s\sigma_i\right)^2. 
\end{align*}
Note that the summation is over $i\in \widetilde{N}_{K_0}$. Let $\theta^*\in\mathbb{R}^n$ be defined by $\theta_i^*=\theta_i\cdot 1\{i\in \widetilde{N}_{K_0}\}$.  
Since $1-\theta_{\max}^2\geq 1/2$ and $\|\theta^*\|_{2s}^{2s}\leq \theta^{2s-2}_{\max}\|\theta^*\|^2\leq  \theta^{2s-2}_{\max}\|\theta\|^2$, we have 
\beq \label{thm-LB-6}
X_s \leq \frac{a_0(1-b_n)^2\|\theta\|^2}{\|\theta^*\|_{2s}^{2s}} \left(\sum_i \theta_i^s\sigma_i\right)^2, 
\eeq
for a constant $a_0>0$. 
We apply Hoeffding's inequality to get that, for all $t>0$, 
\beq \label{thm-LB-7}
\mathbb{P}\Bigl( \Bigl| \sum_{i}\theta_i^s\sigma_i \Bigr|>t \Bigr)\leq 2\exp\Bigl(-\frac{t^2}{2\sum_{i}\theta^{2s}_i}\Bigr)=  2\exp\Bigl(-\frac{t^2}{2\|\theta^*\|^{2s}_{2s}}\Bigr). 
\eeq
For any nonnegative variable $X$, using the formula of integration by part, we can derive that $\mathbb{E}[\exp(aX)]=1+ a\int_{0}^\infty  \exp(at) \mathbb{P}(X>t)dt$. 
As a result,
\begin{align*}
\mathbb{E}_\sigma \bigl[\exp\bigl(X_s\bigr)\bigr] &\leq \mathbb{E}_\sigma \left\{ \exp\biggl[ \frac{a_0(1-b_n)^2\|\theta\|^2}{\|\theta^*\|_{2s}^{2s}} \Bigl(\sum_{i}\theta_i^s\sigma_i \Bigr)^2 \biggr]\right\}\cr
&= 1+ \frac{a_0\|\theta\|^2(1-b_n)^2}{\|\theta^*\|_{2s}^{2s}} \int_{0}^\infty \exp\left(\frac{a_0\|\theta\|^2(1-b_n)^2}{\|\theta^*\|_{2s}^{2s}} t\right)\cdot\mathbb{P}\biggl\{\Bigl(\sum_{i}\theta_i^m\sigma_i \Bigr)^2 >t \biggr\}dt\cr
&\leq 1+ \frac{a_0\|\theta\|^2(1-b_n)^2}{\|\theta^*\|_{2s}^{2s}} \int_{0}^\infty \exp\left(\frac{a_0\|\theta\|^2(1-b_n)^2}{\|\theta^*\|_{2s}^{2s}} t\right)\cdot\exp\left(-\frac{t}{2\|\theta^*\|_{2s}^{2s}}\right) dt\cr
&= 1+ \frac{a_0\|\theta\|^2(1-b_n)^2}{\|\theta^*\|_{2s}^{2s}} \int_0^\infty \exp \left(- \frac{1-2a_0\|\theta\|^2(1-b_n)^2}{2\|\theta^*\|_{2s}^{2s}}t \right)dt\cr
&= 1+ \frac{a_0\|\theta\|^2(1-b_n)^2}{\|\theta^*\|_{2s}^{2s}}\cdot  \frac{2\|\theta^*\|_{2s}^{2s}}{1-2a_0\|\theta\|^2(1-b_n)^2}\cr
&= 1+ \frac{2a_0\|\theta\|^2(1-b_n)^2}{1-2a_0\|\theta\|^2(1-b_n)^2}. 
\end{align*}
The right hand side does not depend on $s$, so the same bound holds for  $\max_{s\geq 1}\{\mathbb{E}_\sigma[\exp(X_s)]\}$. 
When $\|\theta\|^2(1-b)^2\to 0$, this upper bound is $1+o(1)$. Plugging it into \eqref{thm-LB-5} gives \eqref{thm-LB-4(add)}. Then, the second claim follows. \qed

\subsection{Proof of Theorem~\ref{thm:LB2}}
We show a slightly stronger argument. Given $1\leq K_1<K_2\leq m_0$, let ${\cal M}_n(K_1, K_2, a_n)$ be the sub-collection of ${\cal M}_n(m_0, a_n)$ corresponding to $K_1\leq K\leq K_2$. Note that
\[
\inf_{\hat{K}} \bigl\{ \sup_{{\cal M}_n(m_0, a_n)} \mathbb{P}(\hat{K}\neq K) \bigr\}\geq \inf_{\hat{K}} \bigl\{ \sup_{{\cal M}_n(K_1,K_2, a_n)} \mathbb{P}(\hat{K}\neq K) \bigr\}.
\]
It suffices to lower bound the right hand side. 

Fix an arbitrary DCBM model with $(K_1-1)$ communities. For each $1\leq m\leq K_2-K_1+1$, we use \eqref{LB-construct1}-\eqref{LB-construct2} to construct a random-label DCBM with $(K_1-1+m)$ communities, where $b_n=1-c\, \|\theta\|^{-1}a_n$, for a constant $c$ to be decided. Let $\mathbb{P}_k$ denote the probability measure associated with the $k$-community random-label DCBM, for $K_1\leq k\leq K_2$. By Theorem~\ref{thm:LB}, we can choose an appropriately small constant $c$ such that $|\lambda_K|/\sqrt{\lambda_1}\geq a_n$ with probability $1-o(n^{-1})$, under each $\mathbb{P}_k$.  
Additionally, using a proof similar to that of \eqref{thm-LB0-3}, we can show that \eqref{condition1a}-\eqref{condition1b} are satisfied with probability $1-o(n^{-1})$.   Therefore, under each $\mathbb{P}_k$, the realization of $(\Theta,\Pi, P)$  belongs to ${\cal M}_n(K_1,K_2, a_n)$ with probability $1-o(n^{-1})$. Then, for any $\hat{K}$, 
\beq \label{thm-LB-add0}
\sup_{{\cal M}_n(K_1,K_2, a_n)} \mathbb{P}(\hat{K}\neq K) \geq   \max_{K_1\leq k\leq K_2}\mathbb{P}_k(\hat{K}\neq K) + o(n^{-1}). 
\eeq
To bound the right hand side of \eqref{thm-LB-add0}, consider a multi-hypothesis testing problem:  Given an adjacency matrix $A$, choose one out of the models $\{\mathbb{P}_k\}_{K_1\leq k\leq K_2}$. For any test $\psi$, define
\[
\bar{p}(\psi) = \frac{1}{K_2-K_1+1}\sum_{k=K_1}^{K_2}\mathbb{P}_k(\psi\neq k).  
\]
We apply \cite[Proposition 2.4]{tsybakov2008introduction}. It yields that 
\[
\frac{1}{K_2-K_1}\sum_{k=K_1+1}^{K_2}\chi^2(\mathbb{P}_k,\mathbb{P}_{K_1})\leq \alpha^* \quad \Longrightarrow \quad \inf_\psi \bar{p}(\psi) \geq \sup_{0<\tau<1}\left\{\frac{\tau(K_2-K_1)}{1+\tau(K_2-K_1)}[1-\tau(\alpha^*+1)]\right\}. 
\] 
We have shown in Theorem~\ref{thm:LB} that $\alpha^*=o(1)$. By letting $\tau=1/2$ in the above, we immediately find that 
\beq \label{thm-LB-add1}
\inf_\psi \bar{p}(\psi) \gtrsim  \frac{K_2-K_1}{2+(K_2-K_1)}\Bigl(1-\frac{1+o(1)}{2}\Bigr)\geq 1/6+o(1). 
\eeq
Now, given any estimator $\hat{K}$, it defines a test $\psi_{\hat{K}}$, where $\psi_{\hat{K}}=\hat{K}$ if $K_1\leq \hat{K}\leq K_2$ and $\psi_{\hat{K}}=K_1$ otherwise. It is easy to see that 
\beq \label{thm-LB-add2}
\bar{p}(\psi_{\hat{K}})\leq \max_{K_1\leq k\leq K_2} \mathbb{P}_k(\hat{K}\neq k). 
\eeq
Combining \eqref{thm-LB-add1}-\eqref{thm-LB-add2} gives that $\max_{K_1\leq k\leq K_2} \mathbb{P}_k(\hat{K}\neq k)\geq 1/6+o(1)$. We plug it into \eqref{thm-LB-add0} to get the claim. \qed

\section{Proof of Lemma~\ref{lemma:complexity}} \label{sec:proof-compu}

For the goodness-of-fit test, it contains calculation of (a) $\widehat\Omega^{(m)}$ as the refitted $\Omega$, (b) $Q_n^{(m)}$ as the main term, (c) $B_n^{(m)}$ as the bias correction term and (d) $C_n$ as the variance estimator. 

For (a), it requires calculation of $d_i$ for $1\leq i \leq n$, and $\hat{\bf {1}}_k' A\hat{\bf {1}}_\ell$ and $\hat{\bf{1}}_k' A\hat{\bf{1}}_n$ for $1\leq k, \ell\leq m$ with $m\leq K.$ Since $d_i$ needs $O(d_i)$ operations, it takes $O(n\bar d)$ for calculating $d_i$, $1\leq i \leq n$.  Similarly, it takes $O(\hat{\bf {1}}_k' A\hat{\bf {1}}_\ell)$ to calculate $\hat{\bf {1}}_k' A\hat{\bf {1}}_\ell$ and $O(\hat{\bf {1}}_k' A\hat{\bf {1}}_n)$ to calculate $\hat{\bf {1}}_k' A\hat{\bf {1}}_n$, $1\leq k, \ell\leq m$. The total complexity is then $O(n\bar d).$ By \eqref{refitting2b}, \[
\widehat\Omega^{(m)}(i, j) = \widehat\theta^{(m)}(i)\widehat\theta^{(m)}(j)(\hat\pi^{(m)}_i)'\widehat{P}^{(m)}\hat\pi^{(m)}_j,
\]
 whose calculation takes $O(m^2)$ operations. Hence, calculation of $\widehat\Omega^{(m)}$ needs $O(m^2n^2)$ operations. Combining together, we conclude that  step (a) costs $O(m^2n^2)$. 

For (b), $Q_n^{(m)}$ can be calculated using the same form in Theorem 1.1 of \cite{SP2019}. As is shown there, this step requires $O(n^2\bar d)$ operations.

For (c), given $\widehat\Omega^{(m)}$ and $\widehat P^{(m)}$, the calculation of $\hat g^{(m)}$, $\widehat V^{(m)}$ and $\widehat H^{(m)}$ only takes  $O(n)$. By \eqref{refitting2c}, calculation of $B_n^{(m)}$ only involves calculate $\|\hat\theta\|$ and $\hat g'\widehat V^{-1}(\widehat P\widehat H^2\widehat P\circ \widehat P\widehat H^2\widehat P) \widehat V^{-1}\hat g$. 
The first part needs $O(n)$ operations. The second part only involves vectors in $\mathbb{R}^m$ and matrices in $\mathbb{R}^{m, m}$. Moreover since $m \leq K$ and $K$ is fixed, it takes at most $o(n)$ operations. Combining above, step (c) costs $O(n)$.

For (d), the calculation follows from Proposition A.1 of \cite{OGC}. It should be noted $C_n$ is denoted as $\widehat C_4$ there, and it requires calculation of (i) trace of a matrix,  (ii) $A^4$ for matrix $A$ and (iii) quadratic form of matrix $A$ and $A^2$. 
For (i), it only takes $O(n)$. For (iii), it takes at most $O(n^2)$.
For (ii), we can compute $A^k$ recursively from $A^k = A^{k-1}A$. it suffices to consider the complexity of computing
$BA$, for an arbitrary $n \times n$ matrix B. The $(i, j)$-th entry of $BA$ is $\sum_{\ell: A_{\ell j} \neq 0} B_{i\ell}A_{\ell j}$, where the total number of
nonzero $A_{\ell j}$ equals to $d_j$, the degree of node $j$. Hence, the complexity of computing the   $(i, j)$-th entry of $BA$ is $O(d_j).$ It follows that the complexity of computing $BA$ is $O(n^2
\bar d).$

Combining above, the goodness-of-fit test needs $O(n^2\bar d)$ operations. \qed


\spacingset{1}

\bibliographystyle{chicago}
\bibliography{network}

\end{document}